\documentclass[11pt,onecolumn,draftcls,romanappendices]{IEEEtran-kludge}
\usepackage{amssymb}
\usepackage{mathrsfs}   
\usepackage{amstext}
\usepackage{amsmath}
\usepackage{theorem}
\usepackage{stmaryrd}
\usepackage{graphicx}

\let\savedsection=\section

\title{
A Theory of Network Equivalence \\
{\LARGE Part I:  Point-to-Point Channels}
}

\author{
Ralf Koetter,~\IEEEmembership{Fellow,~IEEE,}
Michelle Effros,~\IEEEmembership{Fellow,~IEEE,} and 
Muriel M\'edard,~\IEEEmembership{Fellow,~IEEE}%
\thanks{R. Koetter was with the Technical University of Munich.  
        M. Effros is with the Department of Electrical Engineering, 
	California Institute of Technology, Pasadena, CA  91125 USA 
	e-mail:  effros@caltech.edu.  
	M. M\'edard is with the Department of Electrical Engineering
	and Computer Science, 
	Massachusetts Institute of Technology, Cambridge, MA  02139 USA 
	e-mail:  medard@mit.edu.}}%
\date{\bf \large Submitted to IEEE Transactions on Information Theory
      April 14, 2010}

\advance \topmargin by -\headheight
\advance \topmargin by -\headsep
     
\evensidemargin \oddsidemargin
\renewcommand{\Bbb}{\mathbb}
\newcommand{\R}{\Bbb{R}}
\newcommand{\deff}{\mbox{$\stackrel{\rm def}{=}$}}
\newcommand{\ceil}[1]{\left\lceil {#1} \right\rceil}
\newcommand{\Proof}{\hspace*{10pt}{{\it Proof. }}}
\newcommand{\script}[1]{{\mathscr #1}}
\newcommand{\cA}{{\cal A}}    
\newcommand{\cC}{{\cal C}}  
\newcommand{\cE}{{\cal E}}  \newcommand{\setE}{{\script{E}}}
\newcommand{\cM}{{\cal M}}  
\newcommand{\cN}{{\cal N}}  
\newcommand{\cR}{{\cal R}}  \newcommand{\setR}{{\script{R}}}
\newcommand{\cS}{{\cal S}}  
\newcommand{\cU}{{\cal U}}  
\newcommand{\cV}{{\cal V}}  
\newcommand{\cW}{{\cal W}}  
\newcommand{\cX}{{\cal X}}  
\newcommand{\cY}{{\cal Y}}  
\newcommand{\cZ}{{\cal Z}}  

\newcommand{\ar}[2]{(#1 \rightarrow #2)}
\theoremstyle{plain}
\theorembodyfont{\normalfont\slshape}

\newtheorem{thm}{Theorem}

\newtheorem{lem}[thm]{Lemma}

\newtheorem{cor}[thm]{Corollary}

\newtheorem{defi}{Definition}

\theorembodyfont{\normalfont}

\newtheorem{exam}{Example}

\newtheorem{algo}{Algorithm}

\newtheorem{rem}{Remark}

\makeatletter

\gdef\@punct{.\ \ }  
\def\@sect#1#2#3#4#5#6[#7]#8{%
  \ifnum #2>\c@secnumdepth
     \def\@svsec{}
  \else
     \refstepcounter{#1}\edef\@svsec{%
     \ifnum #2>0{{\csname the#1\endcsname}}.\fi%
    \hskip .5em}
  \fi
  \@tempskipa #5\relax
  \ifdim \@tempskipa>\z@
     \begingroup #6\relax
       \@hangfrom{\hskip #3\relax\@svsec}{\interlinepenalty \@M #8\par}
     \endgroup
     \csname #1mark\endcsname{#7}
     \addcontentsline{toc}{#1}{\ifnum #2>\c@secnumdepth\else
          \protect\numberline{\csname the#1\endcsname}\fi#7}
  \else
     \def\@svsechd{#6\hskip #3\@svsec #8\@punct\csname
#1mark\endcsname{#7}
     \addcontentsline{toc}{#1}{\ifnum #2>\c@secnumdepth \else
          \protect\numberline{\csname the#1\endcsname}\fi#7}}
  \fi
  \@xsect{#5}}

\def\@ssect#1#2#3#4#5{\@tempskipa #3\relax
  \ifdim \@tempskipa>\z@
     \begingroup #4\@hangfrom{\hskip #1}{\interlinepenalty \@M
#5\par}\endgroup
  \else \def\@svsechd{#4\hskip #1\relax #5\@punct}\fi
  \@xsect{#3}}

\makeatother

\newcommand{\ub}[1]{{\underline{#1}}}

\newcommand{\hp}{\hat{p}}
\newcommand{\Ruv}{R^{\ar{u}{v}}}
\newcommand{\tR}{\tilde{R}}
\newcommand{\tcR}{\tilde{\cR}}
\newcommand{\tRuv}{\tilde{R}^{\ar{u}{v}}}
\newcommand{\ucS}{\ub{\cS}}
\newcommand{\bx}{{\bf x}}
\newcommand{\ubx}{\ub{\bx}}
\newcommand{\bX}{{\bf X}}
\newcommand{\ubX}{\ub{\bX}}
\newcommand{\hPr}{\widehat{\Pr}}
\newcommand{\hx}{\hat{x}}
\newcommand{\tx}{\tilde{x}}
\newcommand{\tX}{\tilde{X}}
\newcommand{\hX}{\hat{X}}
\newcommand{\vX}{X^{(v)}}
\newcommand{\vcX}{\cX^{(v)}}
\newcommand{\ux}{\ub{x}}
\newcommand{\uX}{\ub{X}}
\newcommand{\ucX}{\ub{\cX}}
\newcommand{\uvX}{\uX^{(v)}}
\newcommand{\uvx}{\ux^{(v)}}
\newcommand{\uvcX}{\ucX^{(v)}}
\newcommand{\uxV}{\ux^{(i,1)}}

\newcommand{\uhX}{\ub{\hX}}
\newcommand{\XV}{X^{(i,1)}}
\newcommand{\xV}{x^{(i,1)}}
\newcommand{\tXV}{\tX^{(i,1)}}
\newcommand{\txV}{\tx^{(i,1)}}
\newcommand{\utXV}{\ub{\tX}^{(i,1)}}
\newcommand{\utxV}{\ub{\tx}^{(i,1)}}
\newcommand{\XVm}{X^{-(i,1)}}
\newcommand{\xVm}{x^{-(i,1)}}

\newcommand{\uxVm}{\ux^{-(i,1)}}
\newcommand{\cXV}{\cX^{(i,1)}}
\newcommand{\cXVm}{\cX^{-(i,1)}}
\newcommand{\uXV}{\uX^{(i,1)}}
\newcommand{\ucXV}{\ucX^{(i,1)}}
\newcommand{\ucXVm}{\ucX^{-(i,1)}}
\newcommand{\uhcX}{\ub{\hcX}}

\newcommand{\ucY}{\ub{\cY}}
\newcommand{\by}{{\bf y}}
\newcommand{\uby}{\ub{\by}}
\newcommand{\bY}{{\bf Y}}
\newcommand{\ubY}{\ub{\bY}}
\newcommand{\vY}{Y^{(v)}}
\newcommand{\jYr}{Y^{(j,r)}}
\newcommand{\uy}{\ub{y}}
\newcommand{\uY}{\ub{Y}}
\newcommand{\uvY}{\ub{Y}^{(v)}}
\newcommand{\ty}{\tilde{y}}
\newcommand{\tY}{\tilde{Y}}
\newcommand{\hvcY}{\hcY^{(v)}}

\newcommand{\tcY}{\tilde{\cY}}
\newcommand{\YV}{Y^{(j,1)}}
\newcommand{\tYV}{\tY^{(j,1)}}
\newcommand{\tyV}{\ty^{(j,1)}}
\newcommand{\utyV}{\ub{\ty}^{(j,1)}}
\newcommand{\utYV}{\ub{\tY}^{(j,1)}}
\newcommand{\yV}{y^{(j,1)}}
\newcommand{\YVm}{Y^{-(j,1)}}
\newcommand{\yVm}{y^{-(j,1)}}

\newcommand{\uyVm}{\uy^{-(j,1)}}
\newcommand{\cYV}{\cY^{(j,1)}}
\newcommand{\cYVm}{\cY^{-(j,1)}}
\newcommand{\vcY}{\cY^{(v)}}
\newcommand{\uhy}{\ub{\hy}}
\newcommand{\uhY}{\ub{\hY}}
\newcommand{\uYV}{\uY^{(j,1)}}
\newcommand{\uyV}{\uy^{(j,1)}}
\newcommand{\ucYV}{\ucY^{(j,1)}}
\newcommand{\ucYVm}{\ucY^{-(j,1)}}
\newcommand{\uhcY}{\ub{\hcY}}
\newcommand{\uvcY}{\ub{\cY}^{(v)}}
\newcommand{\Wuv}{W^{\ar{u}{v}}}
\newcommand{\Wvu}{W^{\ar{v}{u}}}
\newcommand{\Wvo}{W^{\ar{v}{1}}}
\newcommand{\Wvm}{W^{\ar{v}{m}}}
\newcommand{\cWuv}{\cW^{\ar{u}{v}}}
\newcommand{\cWvp}{\cW^{\ar{v}{v'}}}

\newcommand{\bw}{\hat{w}}
\newcommand{\uw}{\ub{w}}
\newcommand{\ubw}{\ub{\bw}}
\newcommand{\uW}{\ub{W}}
\newcommand{\uWuv}{\uW^{\ar{u}{v}}}
\newcommand{\uWvo}{\uW^{\ar{v}{1}}}
\newcommand{\uWvm}{\uW^{\ar{v}{m}}}
\newcommand{\ucW}{\ub{\cW}}
\newcommand{\ucWuv}{\ucW^{\ar{u}{v}}}
\newcommand{\ucWvp}{\ucW^{\ar{v}{v'}}}
\newcommand{\ucWvu}{\ucW^{\ar{v}{u}}}
\newcommand{\bW}{\hat{W}}
\newcommand{\bWuv}{\bW^{\ar{u}{v}}}
\newcommand{\ubW}{\ub{\bW}}
\newcommand{\ubWuv}{\ubW^{\ar{u}{v}}}
\newcommand{\ubWvu}{\ubW^{\ar{v}{u}}}
\newcommand{\tw}{\tilde{w}}
\newcommand{\tW}{\tilde{W}}

\newcommand{\tcW}{\tilde{\cW}}
\newcommand{\tcWuv}{\tcW^{\ar{u}{v}}}

\newcommand{\tcWvp}{\tcW^{\ar{v}{v'}}}
\newcommand{\utW}{\ub{\tW}}

\newcommand{\utWuv}{\utW^{\ar{u}{v}}}
\newcommand{\buw}{\underline{\bw}}
\newcommand{\buW}{\underline{\bW}}

\newcommand{\buWuv}{\hat{\uW}^{\ar{u}{v}}}

\newcommand{\utWvo}{\utW^{\ar{v}{1}}}
\newcommand{\utWvm}{\utW^{\ar{v}{m}}}

\newcommand{\utcW}{\ub{\tcW}}
\newcommand{\utcWuv}{\utcW^{\ar{u}{v}}}
\newcommand{\btw}{\hat{\tw}}
\newcommand{\btW}{\hat{\tW}}

\newcommand{\btWuv}{\btW^{\ar{u}{v}}}

\newcommand{\ubtW}{\hat{\ub{\tW}}}
\newcommand{\ubtWuv}{\ubtW^{\ar{u}{v}}}

\newcommand{\ucN}{\ub{\cN}}
\newcommand{\hcN}{\hat{\cN}}
\newcommand{\uhcN}{\ub{\hcN}}

\newcommand{\eps}{\epsilon}
\newcommand{\typ}{A_{\eps,t}^{(N)}}
\newcommand{\rtyp}{\hat{A}_{\eps,t}^{(N)}}
\newcommand{\rtypp}{\hat{A}_{\eps,t'}^{(N)}}
\newcommand{\rtypo}{\hat{A}_{\eps,1}^{(N)}}
\newcommand{\bad}{B_t^{(N)}}
\newcommand{\badp}{B_{t'}^{(N)}}
\newcommand{\bado}{B_1^{(N)}}

\begin{document}
\maketitle

\begin{abstract}
A family of equivalence tools for bounding network capacities is introduced.  
Part~I treats networks built from point-to-point channels.  
Part~II generalizes the technique to networks containing 
wireless channels such as broadcast, multiple access, 
and interference channels.  
The main result of part~I is roughly as follows.  
Given a network of noisy, independent, memoryless point-to-point channels,
a collection of demands can be met on the given network
if and only if it can be met on another network
where each noisy channel is replaced by a noiseless bit pipe
with throughput equal to the noisy channel capacity.
This result was known previously 
for the case of a single-source multicast demand.  
The result given here treats general demands --  
including, for example, multiple unicast demands --
and applies even when the achievable rate region 
for the corresponding demands is unknown 
in both the noisy network and its noiseless counterpart.  
\end{abstract}

{\bf Keywords:  Capacity, network coding, equivalence, component models}

\section{Introduction}

The study of network communications has two natural facets 
reflecting different approaches to thinking about networks. 
On the one hand, networks are considered in 
the graph theoretic setup consisting of nodes connected by links.  
The links are typically not noisy channels but noise-free bit pipes 
that can be used error free up to a certain capacity.  
Typical concepts include information flows and routing issues. 
On the other hand, multiterminal information theory addresses 
information transmission through networks by studying noisy channels, 
or rather the stochastic relationship between input and
output signals at devices in a network. Here the questions typically 
concern fundamental limits of communication. The capacity regions
of broadcast, multiple access, and interference channels 
are all examples of questions that are addressed in the 
context of multiterminal information theory.  
These questions appear to have no obvious equivalent 
in networks consisting of error free bit pipes.
Nevertheless, these two views of networking are two natural facets of 
the same problem, namely communication through networks. 
This paper explores the relationship between these two worlds.

Establishing viable bridges between these two areas shows to be
surprisingly fertile. For example, questions about feedback in
multiterminal systems are quite nicely expressed in networks of error
free bit-pipes. Separation issues --- in particular separation
between network coding and channel coding -- have natural answers,
revealing many network capacity problems as combinatorial rather than
statistical, even when communication occurs 
across networks of noisy channels. Most importantly, 
bounding general network capacities reduces to solving a central network
coding problem described as follows: Given a network of {\it error
free} rate-constrained bit pipes, is a given set of demands 
(e.g., a collection of unicast and multicast connections) 
simultaneously satisfiable or not.  
In certain situations, most notably a single multicast
demand, this question is solved, 
and the answer is easily characterized~\cite{AhlswedeC:00}.  
Unfortunately, the general case is wide open and suspected 
to be hard.  (Currently, NP hardness is
only established for linear network coding~\cite{LangbergS:08}.)
While it appears that fully characterizing the combinatorial network
coding problem is out of reach~\cite{ChanG:08}, 
moderate size networks can be solved quite efficiently, 
and there are algorithms available that, with running time
that is exponential in the number of nodes, 
treat precisely this problem for 
general demands~\cite{SongY:03,HarveyK:06,SubramanianT:08}. 
The possibility of characterizing, in principle, the
rate region of a combinatorial network coding problem will be a corner
stone for our investigations.

The combinatorial nature of the network coding problem 
creates a situation not unlike issues in complexity theory.  
In that case, since precise expressions as to how difficult a problem is 
in absolute terms are difficult to derive, 
research is instead devoted to showing that one problem 
is essentially as difficult as another one 
(even though precise characterizations are not available for either).  
Inspired by this analogy, we here take a similar approach, 
characterizing the relationship between arbitrary network capacity problems 
and the central combinatorial network coding capacity problem.  
This characterization is, in fact, all we need if we want to
address separation issues in networks.  
It also opens the door to other questions, such
as degree-of-freedom or high signal to noise ratio analyses, 
which reveal interesting insights.

It is interesting to note the variety of new tools 
generated in recent years 
for studying network capacities 
(e.g.,~\cite{AhlswedeC:00,Borade:02,KoetterM:03,SongY:03,HarveyK:05,
 KramerS:06a,KramerS:06b,HarveyK:06,AvestimehrD:08,ChanG:08}).  
The reduction of a network information
theoretic question to its combinatorial essence is also at the heart
of some of these publications (see, e.g.~\cite{AvestimehrD:08}). 
Our approach is very different in terms of technique and also results, 
focusing not on the solution of network capacities 
when good outer bounds are available 
but on proving relationships between capacity regions 
even (or especially) when these capacity regions remain 
inaccessible using available analytical techniques.  
Nonetheless, we believe it to be
no coincidence that the reduction of a problem to its
combinatorial essence plays a central role 
in a variety of techniques for studying network capacities.  

\section{Intuition and Summary of Results}

The goal of finding capacities for general networks under general demands 
is currently out of reach.  
Establishing connections 
between the networking and information theoretic views 
of network communications simplifies the task 
by allowing us to identify both the stochastic and the combinatorial 
facets of the communication problem 
and to apply the appropriate tools to each.  
For example, consider a network of independent, memoryless, noisy 
point-to-point channels.  
To derive the multicast capacity of this network, 
Borade~\cite{Borade:02} and Song, Yeung, and Cai~\cite{SongY:06} 
first find the noisy network's cut-set outer bound 
and then demonstrate the achievability of that bound 
by applying a multicast network code over 
point-to-point channels made reliable 
using independent channel coding on each point-to-point channel.  
The resulting separation theorem establishes one tight connection 
between the two natural views of communication networks.  
This paper considers whether similar connections 
can be established for general demands.  
Relating the capacity of stochastic networks 
to the network coding capacity 
allows us to apply analytical and computational tools 
from the network coding literature 
(e.g.,~\cite{SongY:03,HarveyK:06,SubramanianT:08})
to bound the capacity of networks of stochastic channels.  

While it is tempting to believe that the separation result 
derived for a single-source multicast demand in~\cite{Borade:02,SongY:06} 
should also apply under general demands, 
it is clear that the proof technique does not.  
That is, first establishing a tight outer bound 
and then showing that that outer bound can be achieved 
by separate network and channel coding 
is not a feasible strategy for treating all possible demand types 
over all possible network topologies.  
The proof is further complicated by the observation 
that joint channel and network codes 
have a variety of clear advantages over separated codes 
even when separated strategies suffice to achieve the network capacity.  
Example~\ref{ex:par} illustrates one such advantage, 
showing that operating channels above their respective capacities 
can improve communication reliability across the network as a whole.  
It remains to be determined whether 
operating channels above their capacities 
can also increase the achievable rate region 
for cases beyond the single-source multicast demand 
studied in~\cite{Borade:02,SongY:06}.  

\begin{exam}\label{ex:par}
Consider the problem of establishing a unicast connection over 
the two-node network shown in Figure~\ref{fig:par}(a).  
\begin{figure}
  \begin{center}
  \begin{picture}(370,95)(-20,-7.5)
      \thicklines
      \put(-20,40){\circle*{5}}
      \put(-20,36){\makebox(0,0)[ct]{\tiny 1}}
      \put(-20,40){\vector(2,1){40}}
      \put(-20,40){\vector(2,-1){40}}
      \put(0,56){\makebox(0,0)[cb]{\small $X^{(1,1)}$}}
      \put(0,24){\makebox(0,0)[ct]{\small $X^{(1,2)}$}}
      \put(20,52.5){\framebox(70,15)[cc]{\small $p(y^{(2,1)}|x^{(1,1)})$}}
      \put(20,12.5){\framebox(70,15)[cc]{\small $p(y^{(2,2)}|x^{(1,2)})$}}
      \put(90,60){\vector(2,-1){40}}
      \put(90,20){\vector(2,1){40}}
      \put(110,56){\makebox(0,0)[cb]{\small $Y^{(2,1)}$}}
      \put(110,24){\makebox(0,0)[ct]{\small $Y^{(2,2)}$}}
      \put(130,40){\circle*{5}}
      \put(130,36){\makebox(0,0)[ct]{\tiny 2}}
      \put(50,-10){\makebox(0,0)[ct]{(a)}}
      \thinlines
      \multiput(0,30)(35,0){2}{
		\put(170,24){\vector(1,0){30}}
		\put(170,24){\vector(-1,0){0}}
		\put(170,27){\makebox(30,0)[cb]{\tiny $n$}}
      	\multiput(170,9)(0,3){4}{
			\put(0,0){\framebox(30,1){}}}
		\multiput(170,3)(0,1.5){3}{\put(0,0){\makebox(30,1){.}}}
      	\put(170,0){\framebox(30,1){}}}
		\put(165,30){\vector(0,1){21}}
		\put(165,30){\vector(0,-1){0}}
		\put(163,30){\makebox(0,21)[cr]{\tiny $2^{nR}$}}
		\put(240,30){\vector(0,1){21}}
		\put(240,30){\vector(0,-1){0}}
		\put(242,30){\makebox(0,21)[cl]{\tiny $2^{nR}$}}
	      \put(207.5,-10){\makebox(0,0)[ct]{(b)}}
		\put(290,84){\vector(1,0){60}}
		\put(290,84){\vector(-1,0){0}}
		\put(290,87){\makebox(60,0)[cb]{\tiny $2n$}}
      	\multiput(290,9)(0,3){24}{
			\put(0,0){\framebox(60,1){}}}
		\multiput(290,3)(0,1.5){3}{\put(0,0){\makebox(60,1){.}}}
      	\put(290,0){\framebox(60,1){}}
		\put(285,0){\vector(0,1){81}}
		\put(285,0){\vector(0,-1){0}}
		\put(283,0){\makebox(0,81)[cr]{\tiny $2^{(2n)R}$}}
	      \put(320,-10){\makebox(0,0)[ct]{(c)}}
      \end{picture} 
  \end{center}
\caption{(a) The network discussed in the comparison 
	of separate network and channel coding 
	to joint network and channel coding in Example~\ref{ex:par}.  
	(b) A pair of $(2^{nR},n)$ channel codes;  
	each is used to reliably transmit 
	$nR$ bits over $n$ uses of a single channel 
	in the separated strategy.  
	(c) A single $(2^{(2n)R},2n)$ channel code;  
	this is used to reliably transmit information 
	across $n$ uses of the pair of channels 
	in the joint coding strategy.  
	The joint coding strategy achieves twice the error exponent 
	by operating each channel at roughly twice its capacity.}
\label{fig:par}
\end{figure}
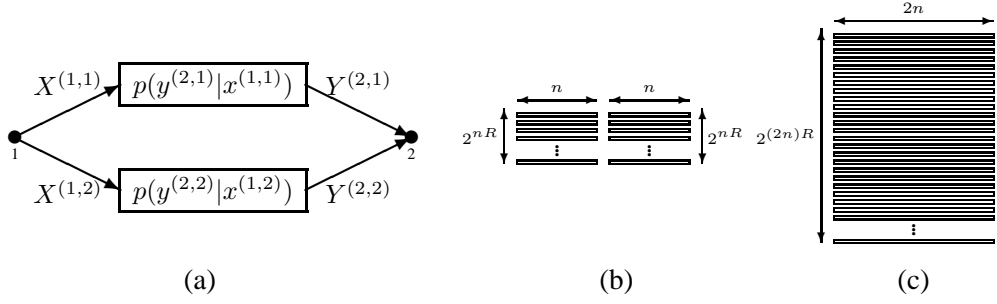
Node~1 transmits a pair of channel inputs $x^{(1)}=(x^{(1,1)},x^{(1,2)})$.  
Node~2 receives  a pair of channel outputs $y^{(2)}=(y^{(2,1)},y^{(2,2)})$.  
The inputs and outputs are stochastically related 
through a pair of independent but identical channels, 
thus 
\[
p(y^{(2,1)},y^{(2,2)}|x^{(1,1)},x^{(1,2)})
=p(y^{(2,1)}|x^{(1,1)})p(y^{(2,2)}|x^{(1,2)})
\]
for all $(x^{(1,1)},x^{(1,2)},y^{(2,1)},y^{(2,2)})
\in\cX^{(1,1)}\times\cX^{(1,2)}\times\cY^{(2,1)}\times\cY^{(2,2)}$ 
while $p(y^{(2,1)}|x^{(1,1)})=p(y^{(2,2)}|x^{(1,2)})$ 
for all $(x^{(1,1)},y^{(2,1)})=(x^{(1,2)},y^{(2,2)})$.  
For each rate $R<C=\max_{p(x)}I(X;Y)$ and each blocklength $n$, 
we compare two strategies for reliably communicating from node~1 to node~2.  
The first (see Figure~\ref{fig:par}(b)) 
is an optimal separate network and channel code 
that reliably communicates across each channel 
using an optimal $(2^{nR},n)$ channel code.  
The second strategy (see Figure~\ref{fig:par}(c)) 
applies a single optimal $(2^{2nR},2n)$ channel code 
across the pair of channels, 
sending the first $n$ symbols of each codeword across the first channel 
and the remaining $n$ symbols across the second channel.  
The decoder observes the outputs of both channels 
and reliably decodes using its blocklength-$2n$ channel decoder.  
Using this approach, each channel has $2^{2nR}$ possible inputs.  
Thus when $R$ is close to $C$, 
this joint channel and network code operates each channel 
at roughly twice its capacity -- 
making reliable transmission across each channel alone impossible.  
Since the joint code operates a $2n$ dimensional code 
over $n$ time steps, 
it achieves a better error exponent 
than the separated code.  \IEEEQED
\end{exam}

Our main result is roughly as follows.  
An arbitrary collection of demands can be met on 
a network of noisy, independent, memoryless point-to-point channels, 
if and only if the same demands can be met on another network
where each noisy channel is replaced by a noiseless bit pipe 
of the corresponding capacity.  
This result agrees with~\cite{Borade:02,SongY:06} 
in the case of multicast demands.  

Our proof introduces a new technique 
for bounding the capacity region of one network 
in terms of the capacity region of another network.  
Critically, this approach can be employed 
even when the capacity regions of both networks are unknown.  
We prove equivalence by first showing 
that the rate region for the noiseless bit-pipe network 
is a subset of that for the network of noisy channels 
and then showing that the rate region for the network of noiseless bit pipes 
is a superset of that for the network of noisy channels.  
In each case, we show the desired relationship 
by demonstrating that codes that can be operated reliably 
on one network can be operated with similar error probability 
on the other network.  
Codes for the bit-pipe network 
can be operated across the network of noisy channels 
using an independent channel code across each channel.  
Operating codes for the network of noisy channels 
across the bit-pipe network is more difficult 
since networks of noisy channels allow a far richer algorithmic behavior 
than networks of noiseless bit pipes.  
While it is known that a noiseless bit-pipe of a given
throughput can emulate any discrete memoryless channel
of lesser capacity~\cite{BennettS:02}, 
applying this result seems to be difficult.  
Difficulties arise with continuous random variables, 
timing questions, and proving continuity of rate regions 
in the channel statistics.  
Worst of all, since we do not know 
which strategy achieves the network capacity, 
we must be able to emulate all of them.  
We therefore prove our main claim directly, 
without exploiting~\cite{BennettS:02}.  
We use a source coding argument 
to show that we can emulate each noisy channel  
across the corresponding noiseless bit pipe 
to sufficient accuracy that any code designed for the network of noisy channels 
can be operated across the noiseless bit-pipe network 
with a similar error probability.  
It is important to note that the given approach 
does not require knowing the rate region of either network 
nor what the optimal codes look like, 
and it never answers the question of whether a particular
rate point is in the rate region or not.  
The proofs only demonstrate that any rate point in the interior 
of the rate region for one network must also be in the interior 
of the rate region for the other network.  

The given relationship between networks of noisy point-to-point channels 
and networks of noiseless bit-pipes 
has a number of surprisingly powerful consequences.  
For example, it demonstrates that at its core 
characterizing network capacity 
is a combinatoric problem 
rather than a probabilistic one:  
Shannon's channel coding theorem 
tells us everything that we need to know 
about the noise in independent, point-to-point channels.  
Understanding the relationship between 
the two facets of network communications 
likewise lends insight 
into a variety of network information theoretic questions.  
For example, the classical result that feedback does not
increase the capacity of a point-to-point channel 
now can be proven in two ways.
The first is the classical information theoretic argument 
that shows that the channel has no information
that is useful to the transmitter that the transmitter
does not already know.
The second observes that the min-cut 
between the transmitter and the receiver in the equivalent network 
is the same with or without feedback;  
therefore feedback does not increase capacity.  
While both proofs lead to the same well-known result, 
the latter is easier to generalize.  
For example, the following result is an immediate consequence 
of the given network equivalence 
and the well-known characterization 
of the multicast capacity in network coding~\cite{AhlswedeC:00}.  
Given any network of noisy, memoryless, point-to-point channels  
and any multicast demand, 
feedback increases the multicast capacity 
if and only if it increases the min-cut 
on the equivalent deterministic network.  
Likewise, since capacities 
are known for a variety of network coding problems~\cite{KoetterM:03}, 
we can immediately determine whether feedback increases 
the achievable rate regions for a variety of other demand types 
(e.g., multiple-source multicast demands, 
       single-source non-overlapping demands, 
       and single-source non-overlapping plus multicast demands).  

\section{The Setup}\label{sec:setup}
Our notation is similar to that of 
Cover and Thomas~\cite[Section 15.10]{CoverT:06}. 
A multiterminal network is defined by a vertex set 
$\cV=\{1,\ldots,m\}$ 
with associated random variables $\vX\in\cX^{(v)}$ 
which are transmitted from node $v$ and $\vY\in\cY^{(v)}$ which are
received at node $v$. 
The alphabets $\cX^{(v)}$ and $\cY^{(v)}$ 
may be discrete or continuous.  
They may also be vectors or scalars.  
For example, if node $v$ transmits information 
over $k$ binary symmetric channels, 
then $\cX^{(v)}=\{0,1\}^k$.  
The network is assumed
to be memoryless and characterized by a conditional
probability distribution 
\[
p(\by|\bx)=p(y^{(1)},\ldots,y^{(m)}|x^{(1)},\ldots, x^{(m)}).  
\]
Note that for continuous random variables 
this assumption implies that we restrict our attention to cases where 
this conditional distribution 
(in this case a conditional prbability density function) exists.  
A code of blocklength $n$ operates the network over $n$
time steps with the goal of communicating, 
for each distinct pair of nodes $u$ and $v$, 
message 
\[
\Wuv\in\cWuv\deff \{1,\ldots,2^{n\Ruv}\} 
\]
from source node $u$ to sink node $v$.  
The messages $\Wuv$ are independent and uniformly distributed 
by assumption (the proof also goes through unchanged 
if the same message is available 
at more than one node in the network).  
The vector of messages $\Wuv$ is denoted by $W$.  
The constant $\Ruv$ is called the rate of the transmission, 
and the vector of rates $\Ruv$ is denoted by $\cR$.  
Since no message is required from a node $u$ to itself, 
$R^{\ar{u}{u}}=0$, 
and $\cR$ is treated as a $m(m-1)$-dimensional vector.  
By~\cite{DoughertyZ:06}, 
for any network coding problem with generic demands, 
we can construct a multiple unicast problem such that 
the given demands can be met in the original network 
if and only if the unicast demands can be met 
in the constructed network.  
This argument generalizes immediately 
to the network model presented here.  
Therefore, there is no loss of generality 
(and considerable simplification of notation) 
in describing messages for all node pairs 
($(u,v)\in\{1,\ldots,m\}^2$ such that $u\neq v$)
rather than messages for all possible multicasts 
($(u,B)$ with $u\in\{1,\ldots,m\}$ 
and $B\subseteq\{1,\ldots,m\}\setminus u$).  

We denote the random variable transmitted by node $v$ at time $t$ 
as $\vX_t$ and the full vector of time-$t$ transmissions 
by all nodes as ${\bf X}_t$.   
We likewise denote the random variable received by node $v$ at time $t$ 
by $\vY_t$ and the full vector of time-$t$ channel outputs by ${\bf Y}_t$.  
A network is written as a triple
\begin{equation}\label{eq:network_triple}
\left(\prod_{v=1}^m\cX^{(v)},p(\by|\bx),\prod_{v=1}^m\cY^{(v)}\right)
\end{equation}
with the additional constraint that random variable $\vX_t$ is a function of 
random variables 
\[
\{\vY_1,\ldots,\vY_{t-1},\Wvo,\ldots,\Wvm\}
\]
alone. 

\begin{figure}
\begin{center}
\begin{picture}(200,100)(0,0)
\thicklines
\qbezier(35,25)(38,-15)(65,18)
\qbezier(65,18)(78,-5)(100,15)
\qbezier(100,15)(128,-15)(145,15)
\qbezier(145,15)(210,-10)(185,40)
\qbezier(185,40)(210,50)(190,60)
\qbezier(190,60)(210,110)(160,85)
\qbezier(160,85)(145,110)(122,90)
\qbezier(122,90)(110,100)(100,90)
\qbezier(100,90)(80,110)(60,85)
\qbezier(60,85)(50,100)(40,80)
\qbezier(40,80)(-10,110)(15,60)
\qbezier(15,60)(-10,50)(20,40)
\qbezier(20,40)(0,25)(35,25)
\qbezier(105,68)(98,55)(75,60)
\qbezier(75,60)(85,45)(75,35)
\qbezier(75,35)(105,45)(115,25)
\qbezier(115,25)(125,42)(145,35)
\qbezier(145,35)(135,48)(150,58)
\qbezier(150,58)(130,60)(130,68)
\qbezier(130,68)(120,55)(105,68)
\put(73,48){\circle*{4}}\put(73,45){\makebox(0,0)[ct]{\tiny $i$}}
\put(73,48){\vector(1,0){15}}
\put(88,41){\framebox(50,14){\tiny $p(y^{(j,1)}|x^{(i,1)})$}}
\put(138,48){\vector(1,0){15}}
\put(153,48){\circle*{4}}\put(153,45){\makebox(0,0)[ct]{\tiny $j$}}
\put(18,50){\circle*{4}}\put(18,47){\makebox(0,0)[ct]{\tiny 1}}
\put(23,82){\circle*{4}}\put(23,79){\makebox(0,0)[ct]{\tiny 2}}
\put(42,40){\circle*{4}}\put(42,37){\makebox(0,0)[ct]{\tiny 3}}
\put(48,10){\circle*{4}}
\put(61,81){\circle*{4}}
\put(85,72){\circle*{4}}
\put(85,22){\circle*{4}}
\put(118,85){\circle*{4}}
\put(125,8){\circle*{4}}
\put(142,30){\circle*{4}}
\put(143,64){\circle*{4}}
\put(148,85){\circle*{4}}
\put(168,50){\circle*{4}}
\put(173,18){\circle*{4}}\put(173,15){\makebox(0,0)[ct]{\tiny $m$$-$$2$}}
\put(182,80){\circle*{4}}\put(182,77){\makebox(0,0)[ct]{\tiny $m$$-$$1$}}
\put(188,55){\circle*{4}}\put(188,52){\makebox(0,0)[ct]{\tiny $m$}}
\end{picture}
\caption{An $m$-node network 
         containing a channel $p(y^{(j,1)}|x^{(i,1)})$ 
	 from node $i$ to node $j$.  
	 Here $x^{(i)}=(x^{(i,1)},x^{(i,2)})$, 
	      $y^{(j)}=(y^{(j,1)},y^{(j,2)})$, 
	 and the distribution 
	 $p(y^{(1)},\ldots,y^{(j-1)},y^{(j,2)},y^{(j+1)},\ldots,y^{(m)}|
	    x^{(1)},\ldots,x^{(i-1)},x^{(i,2)},x^{(i+1)},\ldots,x^{(m)})$
	 on the remaining channel outputs given 
	 the remaining channel inputs is arbitrary.}\label{fig:net}
\end{center}
\end{figure}
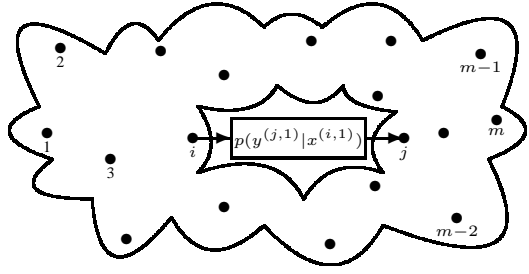

While this characterization is very general, it does not
exploit any information about the network's structure.  
Later discussion treats networks 
made entirely from point-to-point channels, 
but we begin by considering a network 
that is completely arbitrary 
except for its inclusion of a point-to-point channel 
from node $i$ to node $j$, as shown in Figure~\ref{fig:net}, 
that is independent of the remainder of the network distribution.  
Precisely, the conditional distribution 
on all channel outputs given all channel inputs factors as 
\begin{eqnarray*}
\lefteqn{p(\by|\bx)}  \\
&=& p(\yV|\xV)
   p(y^{(1)},\ldots,y^{(j-1)},y^{(j,2)},y^{(j+1)},\ldots,y^{(m)}
    |x^{(1)},\ldots,x^{(i-1)},x^{(i,2)},x^{(i+1)},\ldots,x^{(m)}), 
\end{eqnarray*}
where $\cX^{(i)}=\cX^{(i,1)}\times\cX^{(i,2)}$, 
      $\cY^{(j)}=\cY^{(j,1)}\times\cY^{(j,2)}$, 
$(\cXV,p(\yV|\xV),\cYV)$ is the point-to-point channel, and 
\begin{eqnarray*}
(\cX^{(i,2)}\times\prod_{v\neq i}\vcX, &
p(y^{(1)},..,y^{(j-1)},y^{(j,2)},y^{(j+1)},..,y^{(m)}
    |x^{(1)},..,x^{(i-1)},x^{(i,2)},x^{(i+1)},..,x^{(m)}), \\
&    \cY^{(j,2)}\times\prod_{v\neq j}\vcY)
\end{eqnarray*}
is the remainder of the network.  
As mentioned previously, 
for continuous-alphabet channels we restrict our attention 
to networks for which 
the above conditional probability density functions exist.  
We also restrict our attention to channels 
for which the input distribution 
that achieves the capacity of channel $(\cXV,p(\yV|\xV),\cYV)$ in isolation 
has a probability density function.  
This includes most of the continuous channels 
studied in the literature.  

The notation 
${\bf X}=(\XV,\XVm)$ and ${\bf Y}=(\YV,\YVm)$ 
is sometimes useful to succinctly distinguish 
the input and output of the point-to-point channel 
from the remainder of the network channel inputs and outputs.  
Using this notation, an $m$-node network 
containing an independent point-to-point channel from node $i$ to node $j$ 
is written as 
\begin{equation}\label{eqn:indep}
\cN = \left(\cXVm\times\cXV,p(\yV|\xV)p(\yVm|\xVm),\cYVm\times\cYV\right).
\end{equation}
Figure~\ref{fig:par}(a) shows one example 
where the remainder of the network 
is itself a point-to-point channel.  
In this paper we want to investigate 
some information theoretic aspects 
of replacing factor $p(y^{(j,1)}|x^{(i,1)})$.  

\begin{rem}
The given definitions are sufficiently general 
to model a wide variety of memoryless channel types.  
For example, the distribution $p(\yVm|\xVm)$ 
may model wireless components like broadcast, 
multiple access, and interference channels.  
If $\cX^{(i,1)}$ and $\cY^{(j,1)}$ are vector alphabets, 
then the channel from node $i$ to node $j$ 
is a point-to-point MIMO channel.  
In some situations it is important to be able to embed the transmissions of
various nodes in a schedule which may or may not depend on the messages to be
sent and the symbols that were received in the network. It is straightforward
to model such a situation in the above setup by including in the input and
output alphabets symbols for the case when nothing was sent on a particular 
node input. In this way we can assume that at each time $t$ random
variables $\vX_t$ and $\vY_t$ are given.  
\end{rem}

\begin{defi}\label{def:network_solution}
Let a network 
\[
\cN\ \deff\ \left(\prod_{v=1}^m\vcX,p(\by|\bx),
\prod_{v=1}^m\vcY\right)
\]
be given.  
A blocklength-$n$ solution $\cS(\cN)$ to
this network is defined as a set of encoding and decoding functions:
\begin{eqnarray*} 
\vX_t: &&   (\vcY)^{t-1}\times\prod_{v'=1}^m\cWvp\rightarrow\vcX \\
\bWuv: && (\vcY)^{n}\times\prod_{v'=1}^m\cWvp\rightarrow\cWuv 
\end{eqnarray*}
mapping $(\vY_1,\ldots,\vY_{t-1},\Wvo,\ldots,\Wvm)$ to $\vX_t$ 
for each $v\in V$ and $t\in\{1,\ldots,n\}$ and mapping 
$(\vY_1,\ldots,\vY_{n}, \Wvo,\ldots, \Wvm)$ to $\bWuv$ 
for each $u,v\in V$.  
The solution $\cS(\cN)$ is called a $(\lambda,\cR)$-solution, denoted 
$(\lambda,\cR)$-$\cS(\cN)$, if $\Pr(\Wuv\not = \bWuv)<\lambda$
for all source and sink pairs $u,v$ using the specified encoding and decoding functions.  
\end{defi}

\begin{defi} The rate region $\setR(\cN) \subset \R_+^{m(m-1)}$ of a network
  $\cN$ is the closure of all rate vectors $\cR$ 
  such that for any $\lambda>0$ and all $n$ sufficiently large, 
  there exists a $(\lambda,\cR)$-$\cS(\cN)$ solution of blocklength $n$.  
  We use $\mbox{int}(\setR(\cN))$ to denote the interior 
  of rate region $\setR(\cN)$.  
\end{defi}

The goal of this paper is not to give the capacity regions of networks with
respect to various demands, which is an intractable problem.  
Rather, we wish to develop equivalence relationships 
between capacity regions of
networks. Given the existence of a solution 
$(\lambda,\cR)$-$\cS(\cN)$ of some blocklength $n$ 
for a network $\cN$ we will try to imply statements
for the existence of a solution $(\lambda',\cR')$-$\cS(\cN')$ 
of some blocklength $n'$ for a network $\cN'$.  

To make this precise, consider a memoryless network $\cN$ 
containing an independent channel from node $i$ to node $j$.  
Then 
\[
p(\by|\bx)=p(\yV|\xV)p(\yVm|\xVm).  
\]
Let another network $\cN'$ be given with random variables
$(\tXV,\tYV)$ replacing $(\XV,\YV)$ in $\cN$. 
We have replaced the point-to-point channel characterized by
$p(\yV|\xV)$ with another point-to-point channel characterized by
$\tilde{p}(\tyV|\txV)$.  
When $I(\XV;\YV)<I(\tXV;\tYV)$, 
we want to prove that the existence of a 
$(\lambda,\cR)$-$\cS(\cN)$ solution implies the existence of a
$(\lambda',\cR')$-$\cS(\cN')$ solution, where $\lambda'$ can be made
arbitrarily small if $\lambda$ can.  
Since node $j$ need not decode $\YV$, 
channel capacity is not necessarily a relevant characterization 
of the channel's behavior.  
For example a Gaussian channel from $i$ to $j$ 
might contribute a real-valued estimation of the input random variable;  
a binary erasure channel that replaces it 
cannot immediately deliver the same functionality.  

Our proof does not invent a coding scheme.  
Instead, we demonstrate a technique for operating any coding scheme 
for $\cN$ on the network $\cN'$.  
Since there exists a coding scheme for $\cN$ 
that achieves any point in the interior of $\setR(\cN)$, 
showing that we can operate all codes for $\cN$ on $\cN'$ 
proves that $\setR(\cN)\subseteq\setR(\cN')$.  
We do not know the form of an optimal code for $\cN$.  
Therefore, our method must work for all possible codes on $\cN$.  
For example, it must succeed even when the code for $\cN$ is time-varying.  
As a result, we cannot apply typicality arguments across time.  
We introduce instead a notion of {\em stacking} 
in order to exploit averaging arguments across multiple uses 
of the network 
rather than trying to apply such arguments across time.  

\section{Stacked Networks and Stacked Solutions}\label{sec:stack}

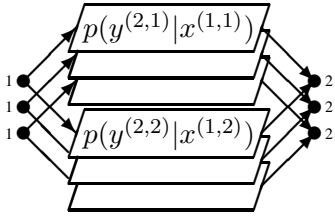
\begin{figure}
  \begin{center}
  \begin{picture}(110,55)(20,-7.5)
      \thicklines
      \put(20,40){\circle*{5}}
      \put(16,40){\makebox(0,0)[cr]{\tiny 1}}
      \put(20,40){\vector(1,1){20}}
      \put(20,40){\vector(1,-1){20}}
      \put(40,52.5){\makebox(70,15)[cc]{$p(y^{(2,1)}|x^{(1,1)})$}}
      \put(40,12.5){\makebox(70,15)[cc]{$p(y^{(2,2)}|x^{(1,2)})$}}
      \put(37,11){\line(1,0){70}}
      \put(37,11){\line(1,3){6}}
      \put(43,29){\line(1,0){70}}
      \put(107,11){\line(1,3){6}}
      \put(37,51){\line(1,0){70}}
      \put(37,51){\line(1,3){6}}
      \put(43,69){\line(1,0){70}}
      \put(107,51){\line(1,3){6}}
      \put(110,60){\vector(1,-1){20}}
      \put(110,20){\vector(1,1){20}}
      \put(130,40){\circle*{5}}
      \put(134,40){\makebox(0,0)[cl]{\tiny 2}}
      \put(20,30){\circle*{5}}
      \put(16,30){\makebox(0,0)[cr]{\tiny 1}}
      \put(20,30){\vector(1,1){20}}
      \put(20,30){\line(1,-1){17}}
      \put(37,1){\line(1,0){70}}
      \put(37,1){\line(1,3){3.4}}
      \put(113,19){\line(-1,0){3}}
      \put(107,1){\line(1,3){6}}
      \put(37,41){\line(1,0){70}}
      \put(37,41){\line(1,3){3.4}}
      \put(113,59){\line(-1,0){3.4}}
      \put(107,41){\line(1,3){6}}
      \put(110,50){\vector(1,-1){20}}
      \put(110,10){\vector(1,1){20}}
      \put(130,30){\circle*{5}}
      \put(134,30){\makebox(0,0)[cl]{\tiny 2}}
      \put(20,20){\circle*{5}}
      \put(16,20){\makebox(0,0)[cr]{\tiny 1}}
      \put(20,20){\vector(1,1){20}}
      \put(20,20){\line(1,-1){17}}
      \put(37,-9){\line(1,0){70}}
      \put(37,-9){\line(1,3){3.4}}
      \put(113,9){\line(-1,0){3}}
      \put(107,-9){\line(1,3){6}}
      \put(37,31){\line(1,0){70}}
      \put(37,31){\line(1,3){3.4}}
      \put(113,49){\line(-1,0){3.4}}
      \put(107,31){\line(1,3){6}}
      \put(110,40){\vector(1,-1){20}}
      \put(110,0){\vector(1,1){20}}
      \put(130,20){\circle*{5}}
      \put(134,20){\makebox(0,0)[cl]{\tiny 2}}
      \end{picture} 
  \end{center}
\caption{The 3-fold stacked network $\ucN$ 
         for the network $\cN$ in Figure~\ref{fig:par}(a).}\label{fig:stacked}
\end{figure}

An $N$-fold stacked network $\ucN_N$ 
is the network $\cN$ repeated $N$ times. 
That is, $\ucN_N$ has $N$ copies of each vertex $v\in\{1,\ldots,m\}$ 
and $N$ copies of the channel $p(\by|\bx)$.  
Figure~\ref{fig:stacked} shows the 3-fold stacked network 
for the network in Figure~\ref{fig:par}.
We abuse notation by simplifying $\ucN_N$ to $\ucN$ throughout, 
specifying the number layers in the stack ($N$) by context.  
Eventually, $N$ will be allowed to grow without bound 
in order to exploit asymptotic typicality arguments.  
The $N$-fold stacked network 
is used to deliver $N$ independent messages $\Wuv$ 
from each transmitter node $u$ to each receiver node $v$.  
All copies of a node can, at each time $t$, 
collaborate in determining their channel inputs $\vX_t$.  
Likewise, all copies of a node $v$ can collaborate 
in reconstructing messages $\Wuv$.  
This potential for collaboration across the layers of the stack 
seems to make the $N$-fold stacked network $\ucN$ 
considerably more powerful than the network $\cN$ 
from which it was derived.  
However, the increase in the number of degrees of freedom 
in a stacked network solution 
is accompanied by an increased burden in the reconstruction constraint.  
A code for the stacked network 
is successful only if it decodes without error in every layer.  
This becomes difficult as $N$ grows without bound.  

Since the $N$-fold stacked network contains $N$ copies of $\cN$, 
it does not meet the definition of a network 
(for example, its vertex set is a multiset and not a set\footnote{The 
vertex set is a multiset since 
it contains $N$ copies of each element $\{1,\ldots,m\}$.}).  
Thus new definitions are required.  
We carry over notation and variable definitions 
from the network $\cN$ to the stacked network $\ucN$ 
by underlining the variable names.  
So for any distinct $u,v\in \{1,\ldots,m\}$, 
$\uWuv\in\ucWuv\deff(\cWuv)^N$ 
is the $N$-dimensional vector of messages 
that the $N$ copies of node $u$ 
send to the corresponding copies of node $v$, and 
$\uvX_t\in\uvcX\deff(\vcX)^N$ and $\uvY_t\in\uvcY\deff(\vcY)^N$ 
are the $N$-dimensional vectors of channel inputs and channel outputs, 
respectively,  
for node $v$ at time $t$.  
The variables in the $\ell$-th layer of the stack 
are denoted by an argument $\ell$, for
example $\uWuv(\ell)$ is the message 
from node $u$ to node $v$ in the $\ell$-th layer of the stack 
and $\uvX_t(\ell)$ is the layer-$\ell$ channel input 
from node $v$ at time $t$.  
Since $\uWuv$ is an $N$-dimensional vector of messages, 
when $\Wuv\in\cWuv\ \deff\ \{1,\ldots,2^{n\Ruv}\}$ in $\cN$, 
$\uWuv\in\ucWuv\ \deff\ \{1,\ldots,2^{n\Ruv}\}^N$ in $\ucN$.  
We therefore define the rate $\Ruv$ for a stacked network 
to be $(\log|\ucWuv|)/(nN)$;  
this normalization makes rate regions in a network 
and its corresponding stacked network comparable.  

\begin{defi}\label{def:snetwork_solution}
Let a network 
\[
\cN \deff \left(\prod_{v=1}^m\vcX,p(\by|\bx),\prod_{v=1}^m\vcY\right)
\]
be given.  
Let $\ucN$ be the $N$-fold stacked network for $\cN$. 
A blocklength-$n$ solution $\cS(\ucN)$ to 
this network is defined as a set of encoding and decoding functions
\begin{eqnarray*} 
\uvX_t: & & {(\uvcY)}^{t-1}\times\prod_{v'=1}^m\ucWvp \rightarrow \uvcX \\
\ubWuv: & & (\uvcY)^n\times\prod_{v'=1}^m\ucWvp\rightarrow\ucWuv 
\end{eqnarray*}
mapping $(\uvY_1,\ldots,\uvY_{t-1}, \uWvo,\ldots,\uWvm)$ 
to $\uvX_t$ for each $t\in\{1,\ldots,n\}$ and 
$v\in\{1,\ldots,m\}$ and mapping 
$(\uvY_1,\ldots,\uvY_n,\uWvo,\ldots,\uWvm)$
to $\ubWuv$ for each $u,v\in\{1,\ldots,m\}$.  
The solution $\cS(\ucN)$ is called a $(\lambda,\cR)$-solution for $\ucN$, 
denoted $(\lambda,\cR)$-$\cS(\ucN)$, 
if the encoding and decoding functions imply
\[
\Pr(\uWuv\neq\ubWuv)<\lambda
\]
for all source and sink pairs $u,v$.
\end{defi}

\begin{defi} The rate region $\setR(\ucN) \subset \R_+^{m(m-1)}$ 
of a stacked network $\ucN$ is the closure of all rate vectors $\cR$ 
such that a $(\lambda,\cR)$-$\cS(\ucN)$ solution exists 
for any $\lambda>0$ and all $N$ sufficiently large.
\end{defi}

Theorem~\ref{thm:stacked}, below, shows that the rate regions 
for a network $\cN$ and its corresponding stacked network $\ucN$ 
are identical.  
That result further demonstrates that the error probability 
for the stacked network 
can be made to decay exponentially in the number of layers $N$.  
The proof builds a blocklength-$n$ solution for network $\ucN$ 
by first using a channel code to map each message 
$\uWuv\in\ucWuv\ \deff\ \{1,\ldots,2^{n\Ruv}\}^N$ 
to a message in alphabet $\utcWuv\ \deff\ \{1,\ldots,2^{n\tRuv}\}^N$ 
for some $\tRuv>\Ruv$ 
and then applying the same blocklength-$n$ solution for network $\cN$ 
independently in each layer of the stack.  
We call such a solution a stacked solution.  

\begin{defi}\label{def:stacked_sol}
Let a network 
$\cN\ \deff\ (\prod_{v=1}^m\vcX,p(\by|\bx),\prod_{v=1}^m\vcY)$ 
be given.  
Let $\ucN$ be the $N$-fold stacked network for $\cN$.  
A blocklength-$n$ stacked solution $\ucS(\ucN)$ to network $\ucN$ 
is defined as a set of mappings 
\begin{eqnarray*} 
\utWuv:&&\ucWuv\rightarrow\utcWuv  \\
\vX_t: &&(\vcY)^{t-1}\times\prod_{v'=1}^m\tcWvp\rightarrow\vcX \\
\btWuv:&&(\vcY)^{n}\times\prod_{v'=1}^m\tcWvp\rightarrow\tcWuv \\
\ubWuv:&& \utcWuv\rightarrow\ucWuv 
\end{eqnarray*} 
such that 
\begin{eqnarray*}
\utWuv & = & \utWuv(\uWuv) \\
\uvX_t(\ell) & = & \vX_t\left(\uvY_1(\ell),\ldots,\uvY_{t-1}(\ell),
                              \utWvo(\ell),\ldots,\utWvm(\ell)\right) \\
\ubtWuv(\ell) & = & \btWuv\left(\uvY_1(\ell),\ldots,\uvY_n(\ell),
                                \utWvo(\ell),\ldots,\utWvm(\ell)\right)\\
\ubWuv & = & \ubWuv(\ubtWuv)
\end{eqnarray*}
for each $u,v\in\{1,\ldots,m\}$, $t\in\{1,\ldots,n\}$, 
and $\ell\in\{1,\ldots,N\}$.  
The solution $\ucS(\ucN)$ is called a stacked $(\lambda,\cR)$-solution, 
denoted $(\lambda,\cR)$-$\ucS(\ucN)$, 
if the specified mappings imply 
\[
\Pr(\uWuv\neq\ubWuv)<\lambda
\]
for all source and sink pairs $u,v$.
\end{defi}

\begin{figure}
  \begin{center}
    \begin{picture}(200,175)(-50,-125)
      \thinlines
      \put(0,40){\framebox(30,10){\tiny $\uY_t^{(v)}(1)$}}
      \put(0,30){\framebox(30,10){\tiny $\uY_t^{(v)}(2)$}}
      \put(0,20){\framebox(30,10){\tiny $\uY_t^{(v)}(3)$}}
      \put(0,17){\makebox(30,0)[cc]{$\cdot$}}
      \put(0,15){\makebox(30,0)[cc]{$\cdot$}}
      \put(0,13){\makebox(30,0)[cc]{$\cdot$}}
      \put(0,0){\framebox(30,10){\tiny $\uY_t^{(v)}(N)$}}
      \thicklines
      \put(30,45){\vector(1,0){20}}
      \put(30,35){\vector(1,0){20}}
      \put(30,25){\vector(1,0){20}}
      \put(30,5){\vector(1,0){20}}
      \put(50,45){\circle*{5}}
      \put(50,35){\circle*{5}}
      \put(50,25){\circle*{5}}
      \put(50,17){\makebox(0,0)[cc]{$\cdot$}}
      \put(50,15){\makebox(0,0)[cc]{$\cdot$}}
      \put(50,13){\makebox(0,0)[cc]{$\cdot$}}
      \put(50,5){\circle*{5}}
      \put(50,42){\makebox(0,0)[ct]{\tiny $v$}}
      \put(50,32){\makebox(0,0)[ct]{\tiny $v$}}
      \put(50,22){\makebox(0,0)[ct]{\tiny $v$}}
      \put(50, 2){\makebox(0,0)[ct]{\tiny $v$}}
      \put(50,45){\vector(1,0){20}}
      \put(50,35){\vector(1,0){20}}
      \put(50,25){\vector(1,0){20}}
      \put(50,5){\vector(1,0){20}}
      \thinlines
      \put(70,40){\framebox(30,10){\tiny $\uX_t^{(v)}(1)$}}
      \put(70,30){\framebox(30,10){\tiny $\uX_t^{(v)}(2)$}}
      \put(70,20){\framebox(30,10){\tiny $\uX_t^{(v)}(3)$}}
      \put(70,17){\makebox(30,0)[cc]{$\cdot$}}
      \put(70,15){\makebox(30,0)[cc]{$\cdot$}}
      \put(70,13){\makebox(30,0)[cc]{$\cdot$}}
      \put(70,0){\framebox(30,10){\tiny $\uX_t^{(v)}(N)$}}
      \put(50,-20){\makebox(0,0)[cb]{\small $\uX_t^{(v)}=\uX_t^{(v)}(\uvY_1,\ldots,\uvY_{t-1},\uW^{\ar{v}{1}}\ldots,\uW^{\ar{v}{m}})$}}
      \put(50,-35){\makebox(0,0)[cb]{(a)}}
      \put(-170,-70){\framebox(40,10){\tiny $Y^{(v)}_{(N-1)t+1}$}}
      \put(-130,-70){\framebox(40,10){\tiny $Y^{(v)}_{(N-1)t+2}$}}
      \put(-90,-70){\framebox(40,10){\tiny $Y^{(v)}_{(N-1)t+3}$}}
      \put(-50,-70){\makebox(40,10){\tiny $\dots$}}
      \put(-10,-70){\framebox(40,10){\tiny $Y^{(v)}_{Nt}$}}
      \thicklines
      \put(30,-65){\vector(1,0){20}}
      \put(50,-65){\circle*{5}}
      \put(50,-65){\vector(1,0){20}}
      \thinlines
      \put(70,-70){\framebox(40,10){\tiny $X^{(v)}_{(N-1)t+1}$}}
      \put(110,-70){\framebox(40,10){\tiny $X^{(v)}_{(N-1)t+2}$}}
      \put(150,-70){\framebox(40,10){\tiny $X^{(v)}_{(N-1)t+3}$}}
      \put(190,-70){\makebox(40,10){\tiny $\dots$}}
      \put(240,-70){\framebox(40,10){\tiny $X^{(v)}_{Nt}$}}
      \put(-170,-90){\makebox(0,0)[lb]
	{\small $(X^{(v)}_{(t-1)N+1},\ldots,X^{(v)}_{tN})^T$}}
      \put(-150,-110){\makebox(0,0)[lb]
	{\small $=\uX_t^{(v)}((Y^{(v)}_1,\ldots,Y^{(v)}_N)^T,\ldots,
	                      (Y^{(v)}_{(t-2)N+1},\ldots,Y^{(v)}_{(t-1)N})^T,
	  f^{\ar{v}{1}}(W^{\ar{v}{1}}),\ldots,f^{\ar{v}{m}}(W^{\ar{v}{m}}))$}}
      \put(50,-125){\makebox(0,0)[cb]{(b)}}
    \end{picture}
    \caption{
      A blocklength-$n$ solution $\cS(\ucN)$ for network $\ucN$ 
      can be operated with the same error probability 
      over $nN$ time steps in $\cN$.
      (a)~Inputs and outputs at time $t$ of the $N$ copies of node $v$ 
      in $\ucN$.  
      (b)~Inputs and outputs of node $v$ 
      at times $(N-1)t+1,\ldots,Nt$ in $\cN$.  
      Vectors $(X^{(v)}_{(t-1)N+1},\ldots,X^{(v)}_{tN})^T$ 
      and     $(Y^{(v)}_{(t-1)N+1},\ldots,Y^{(v)}_{tN})^T$ in $\cN$ 
      play the same role as vectors $\uX^{(v)}_t$ and $\uY^{(v)}_t$ in $\ucN$.}
    \label{fig:unravel}
  \end{center}
\end{figure}

\begin{thm}\label{thm:stacked} 
The rate regions $\setR(\cN)$ and $\setR(\ucN)$ are identical, 
and for each $\cR\in\mbox{int}(\setR(\ucN))$, 
there exists a sequence of $(2^{-N\delta},\cR)$-$\ucS(\ucN)$ 
stacked solutions for $\ucN$ for some $\delta>0$.  

\Proof 
We first show that $\setR(\ucN)\subseteq\setR(\cN)$.  
Perhaps surprisingly, this turns out to be the easier part of the proof.  
Let $\cR\in\mbox{int}(\setR(\ucN))$.  
Then for any $\lambda\in(0,1]$, 
there exists a $(\lambda,\cR)$-$\cS(\ucN)$ solution 
to the stacked network $\ucN$.  
Let $n$ be the blocklength of $\cS(\ucN)$.  
The argument that follows uses $\cS(\ucN)$ to build a blocklength $nN$ 
$(\lambda,\cR)$-$\cS(\cN)$ solution for network $\cN$.  
Roughly, the operations performed at time $t$ 
by the $N$ copies of node $v$ in $\cS(\ucN)$ 
are performed by the single copy of node $v$ 
at times $(t-1)N+1,\ldots,tN$ in $\cS(\cN)$, 
as shown in Figure~\ref{fig:unravel}.  
This gives the desired result since 
the error probability and rate of $\cS(\cN)$ on $\cN$ 
equal the error probability and rate of $\cS(\ucN)$ on $\ucN$.  

To make the argument formal, for each $(u,v)$, let 
\[
f^{\ar{u}{v}}:\{1,\ldots,2^{Nn\Ruv}\}
       \rightarrow\{1,\ldots,2^{n\Ruv}\}^N 
\]
be the natural one-to-one mapping 
from a single sequence of $Nn\Ruv$ bits 
to $N$ consecutive subsequences each of $n\Ruv$ bits.  
Let $g^{\ar{u}{v}}$ be the inverse of $f^{\ar{u}{v}}$.  
We use $f^{\ar{u}{v}}$ to map messages from the message alphabet 
of the rate-$\Ruv$ blocklength-$Nn$ code $\cS(\cN)$ 
to the message alphabet 
for the $N$-layer, rate-$\Ruv$, 
blocklength-$n$ code $\cS(\ucN)$.  
The mapping is one-to-one since in each scenario 
the total number of bits transmitted 
from node $u$ to node $v$ is $Nn\Ruv$.  
For each $t\in\{1\ldots,n\}$, 
let 
\begin{eqnarray*}
\vX(t) & = & (\vX_{(t-1)N+1},\ldots,\vX_{tN})^T \\
\vY(t) & = & (\vY_{(t-1)N+1},\ldots,\vY_{tN})^T 
\end{eqnarray*}
denote vectors containing the channel inputs and outputs at node $v$ 
for $N$ consecutive time steps beginning at time $(t-1)N+1$.  
This is a simple blocking of symbols into vectors, 
with superscript $T$ denoting vector transpose.  
We define the solution $\cS(\cN)$ as 
\begin{eqnarray*}
\vX(t) & = & \uvX_t(\vY(1),\ldots,\vY(t-1),
             f^{\ar{v}{1}}(\Wvo),\ldots,f^{\ar{v}{m}}(W^{\ar{v}{m}}))\\
\bWuv & = & g^{\ar{u}{v}}(\ubWuv(\vY(1),\ldots,\vY(n),
            f^{\ar{v}{1}}(\Wvo),\ldots,f^{\ar{v}{m}}(\Wvm))).
\end{eqnarray*}
Since $\cS(\cN)$ satisfies the causality constraints 
and operates precisely the mappings from $\cS(\ucN)$ on $\cN$, 
the solution $\cS(\cN)$ achieves the same rate and error probability on $\cN$ 
as the solution $\cS(\ucN)$ achieves on $\ucN$.  

For the converse, the job is more difficult.  
A solution $(\lambda,\cR)$-$\cS(\cN)$ needs to achieve 
an error probability of at most $\lambda$ 
for every $(u,v)$ pair in a network.  
A solution $(\lambda,\cR)$-$\cS(\ucN)$ also needs to achieve 
an error probability of at most $\lambda$ for each $(u,v)$, 
but here the error event is a union 
over errors in each of the $N$ layers 
with $N$ growing arbitrarily large.  

Let $\cR\in\mbox{int}(\setR(\cN))$, 
and fix some $\tilde{\cR}\in\mbox{int}(\setR(\cN))$ for which 
$\tRuv>\Ruv$ for all $u,v$.  
We use a solution of rate $\tilde{\cR}$ on $\cN$ 
to build a stacked solution of rate $\cR$ on $\ucN$.  
Set $\rho=\min_{u,v}(\tRuv-\Ruv)$.  
For any $p\in[0,1]$, let $h(p)\ \deff\ -p\log p-(1-p)\log(1-p)$ 
be the binary entropy function.  
For reasons that will become clear later, 
we wish to find constants $\lambda$ and $n$ satisfying 
\[
\max_{u,v}\tRuv\lambda+h(\lambda)/n<\rho.  
\]
such that there exists a $(\lambda,\tilde{\cR})$-$\cS(\cN)$ solution 
of blocklength $n$.  
This is possible because $\tilde{\cR}\in\mbox{int}(\setR(\cN))$ 
implies that for any $\lambda\in(0,1]$ and all $n$ sufficiently large 
there exists a blocklength-$n$ $(\lambda,\tilde{\cR})$-$\cS(\cN)$ solution.  
We therefore meet the desired constraint 
by choosing $\lambda$ to be small 
(say $\lambda=\rho/(2\max_{i,j}\tRuv)$)
and then choosing $n$ sufficiently large.  
The chosen $n$ will be the blocklength of our code 
for all values of $N$.  

Fix a $(\lambda,\tilde{\cR})$-$\cS(\cN)$ solution of blocklength $n$.  
For the $(\lambda,\tilde{\cR})$-$\cS(\cN)$ solution, 
denote the message set by $\tcWuv\ \deff\ \{1,\ldots,2^{n\tRuv}\}$, 
and let $\check{W}^{\ar{u}{v}}$ 
and $\hat{\check{W}}^{\ar{u}{v}}$ 
be the message and its reconstruction, respectively, 
using the fixed $(\lambda,\tilde{\cR})$-$\cS(\cN)$ solution.  
We use $\cS(\cN)$ as the solution applied independently 
in each layer of our stacked solution.  

While solution $\cS(\cN)$ 
yields error probability no greater than $\lambda$ 
in each layer of the stack, 
the error probability over all $N$ layers may still be high.  
The stacked solution's channel codes are included to remedy this problem.  
For each $(u,v)$, the layers of the stack behave 
like $N$ independent instances of channel 
$(\tcWuv,p(\hat{\check{w}}^{\ar{u}{v}}|\check{w}^{\ar{u}{v}}),\tcWuv)$, 
where $p(\hat{\check{w}}^{\ar{u}{v}}|\check{w}^{\ar{u}{v}})=
\Pr(\hat{\check{W}}=\hat{\check{w}}
|\check{W}^{\ar{u}{v}}=\check{w}^{\ar{u}{v}})$ 
under solution $\cS(\cN)$.  
By assumption, 
$\check{W}^{\ar{u}{v}}$ is uniformly distributed on $\tcWuv$, 
so this channel has mutual information 
\begin{eqnarray*}
I(\check{W}^{\ar{u}{v}};\hat{\check{W}}^{\ar{u}{v}})
& = & n\tRuv-H(\check{W}^{\ar{u}{v}};\hat{\check{W}}^{\ar{u}{v}}) \\
& > & n\tRuv-(\lambda n\tRuv+h(\lambda))
\end{eqnarray*} 
by Fano's inequality.  
Note that the desired rate per channel use $n\Ruv$ 
is strictly less than the channel's mutual information, 
precisely 
\[
I(\check{\uW}^{\ar{u}{v}};\hat{\check{\uW}}^{\ar{u}{v}})-n\Ruv 
> n\rho-(\lambda n\tRuv+h(\lambda))  >0, 
\]
owing to our earlier choice of $\lambda$ and $n$.  
We therefore design a $(2^{N(n\Ruv)},N)$ channel code 
for each $(u,v)$ by choosing $2^{N(n\Ruv)}$ blocklength-$N$ codewords 
uniformly from $\utcWuv$, 
where $\utcWuv\ \deff\ (\tcWuv)^N$.  
The channel encoder and channel decoder 
specify the mappings $\utWuv$ and $\ubtWuv$, respectively, 
for our stacked solution.  
Applying the strong coding theorem for discrete memoryless 
channels~\cite[Theorem~5.6.2]{Gallager:68}, 
the expected error probability 
of this randomly drawn code is $2^{-N\delta}$.  
The value $\delta$ is an increasing function of the gap 
$\min_{u,v} [I(\check{\uW}^{\ar{u}{v}};\hat{\check{\uW}}^{\ar{u}{v}})-n\Ruv]$. 
Since the expected error probability 
(with respect to the random channel code designs for all messages $\uWuv$)
decays as $2^{-N\delta}$, 
there exists a single instance of all channel codes 
that does at least as well.  
Thus the stacked solution $\ucS(\ucN)$ 
that first channel codes each message $\uWuv$ to $\utWuv$ 
and then applies the blocklength-$n$ solution $\cS(\cN)$ 
independently in each layer of the stack 
achieves error probability no greater than $2^{-N\delta}$ for $N$ sufficiently large.  
\IEEEQED
\end{thm}

Since the proof of Theorem~\ref{thm:stacked} shows that 
stacked solutions can obtain all rates in the interior of $\setR(\ucN)$, 
we restrict our attention to stacked codes going forward;  
there is no loss of generality in this restriction.  

The arguments that build on Theorem~\ref{thm:stacked} 
later in the paper employ 
not the single instance of the code 
chosen at the end of the proof 
but the random code design that precedes it.  
This random code design is combined 
with a collection of other random code designs.  
Choosing the instances of all random codes 
jointly guarantees good end-to-end performance.  
To understand the implications 
of the given random code design, 
let 
\begin{eqnarray*}
\ubX_t(\ell) & \deff & (\uX^{(1)}_t(\ell), \ldots, \uX^{(m)}_t(\ell)) \\
\ubY_t(\ell) & \deff & (\uY^{(1)}_t(\ell), \ldots, \uY^{(m)}_t(\ell)) 
\end{eqnarray*}
be the vectors of all channel inputs 
and all channel outputs 
in layer $\ell$ of the stacked network at time $t$.  
For each $(u,v)$, the messages $\uWuv(1), \ldots,\uWuv(N)$ 
input to the stacked solution 
are independent and identically distributed (i.i.d.).  
Since each channel code's codewords are drawn 
from the uniform distribution on $\ucWuv$, 
the coded messages $\utWuv(1),\ldots,\utWuv(N)$ 
for a random code are also i.i.d.\ and uniform.  
Finally, since the solutions in the layers of $\ucN$ 
are independent and identical, 
\[
(\ubX_t(1),\ubY_t(1)),\ldots,(\ubX_t(N),\ubY_t(N))
\]
are also i.i.d.\ for each $t$.  
This structure allows us to apply typicality arguments 
across the layers of the network for a fixed time $t$.

\section{Network Equivalence}\label{sec:equiv}

\begin{figure}
\begin{center}
\begin{picture}(200,100)(0,0)
\thicklines
\qbezier(35,25)(38,-15)(65,18)
\qbezier(65,18)(78,-5)(100,15)
\qbezier(100,15)(128,-15)(145,15)
\qbezier(145,15)(210,-10)(185,40)
\qbezier(185,40)(210,50)(190,60)
\qbezier(190,60)(210,110)(160,85)
\qbezier(160,85)(145,110)(122,90)
\qbezier(122,90)(110,100)(100,90)
\qbezier(100,90)(80,110)(60,85)
\qbezier(60,85)(50,100)(40,80)
\qbezier(40,80)(-10,110)(15,60)
\qbezier(15,60)(-10,50)(20,40)
\qbezier(20,40)(0,25)(35,25)
\qbezier(105,68)(98,55)(75,60)
\qbezier(75,60)(85,45)(75,35)
\qbezier(75,35)(105,45)(115,25)
\qbezier(115,25)(125,42)(145,35)
\qbezier(145,35)(135,48)(150,58)
\qbezier(150,58)(130,60)(130,68)
\qbezier(130,68)(120,55)(105,68)
\put(73,48){\circle*{4}}\put(73,45){\makebox(0,0)[ct]{\tiny $i$}}
\put(73,48){\vector(1,0){15}}
\put(88,41){\framebox(50,14){\tiny $p(y^{(j,1)}|x^{(i,1)})$}}
\put(138,48){\vector(1,0){15}}
\put(153,48){\circle*{4}}\put(153,45){\makebox(0,0)[ct]{\tiny $j$}}
\put(18,50){\circle*{4}}\put(18,47){\makebox(0,0)[ct]{\tiny 1}}
\put(23,82){\circle*{4}}\put(23,79){\makebox(0,0)[ct]{\tiny 2}}
\put(42,40){\circle*{4}}\put(42,37){\makebox(0,0)[ct]{\tiny 3}}
\put(48,10){\circle*{4}}
\put(61,81){\circle*{4}}
\put(85,72){\circle*{4}}
\put(85,22){\circle*{4}}
\put(118,85){\circle*{4}}
\put(125,8){\circle*{4}}
\put(142,30){\circle*{4}}
\put(143,64){\circle*{4}}
\put(148,85){\circle*{4}}
\put(168,50){\circle*{4}}
\put(173,18){\circle*{4}}\put(173,15){\makebox(0,0)[ct]{\tiny $m$$-$$2$}}
\put(182,80){\circle*{4}}\put(182,77){\makebox(0,0)[ct]{\tiny $m$$-$$1$}}
\put(188,55){\circle*{4}}\put(188,52){\makebox(0,0)[ct]{\tiny $m$}}
\put(113,-12){\makebox(0,0)[cb]{(a)}}
\end{picture}
\hspace{.2in}
\begin{picture}(200,100)(0,0)
\thicklines
\qbezier(35,25)(38,-15)(65,18)
\qbezier(65,18)(78,-5)(100,15)
\qbezier(100,15)(128,-15)(145,15)
\qbezier(145,15)(210,-10)(185,40)
\qbezier(185,40)(210,50)(190,60)
\qbezier(190,60)(210,110)(160,85)
\qbezier(160,85)(145,110)(122,90)
\qbezier(122,90)(110,100)(100,90)
\qbezier(100,90)(80,110)(60,85)
\qbezier(60,85)(50,100)(40,80)
\qbezier(40,80)(-10,110)(15,60)
\qbezier(15,60)(-10,50)(20,40)
\qbezier(20,40)(0,25)(35,25)
\qbezier(105,68)(98,55)(75,60)
\qbezier(75,60)(85,45)(75,35)
\qbezier(75,35)(105,45)(115,25)
\qbezier(115,25)(125,42)(145,35)
\qbezier(145,35)(135,48)(150,58)
\qbezier(150,58)(130,60)(130,68)
\qbezier(130,68)(120,55)(105,68)
\put(73,48){\circle*{4}}\put(73,45){\makebox(0,0)[ct]{\tiny $i$}}
\put(73,48){\vector(1,0){80}}
\put(113,50){\makebox(0,0)[cb]{$R$}}
\put(153,48){\circle*{4}}\put(153,45){\makebox(0,0)[ct]{\tiny $j$}}
\put(18,50){\circle*{4}}\put(18,47){\makebox(0,0)[ct]{\tiny 1}}
\put(23,82){\circle*{4}}\put(23,79){\makebox(0,0)[ct]{\tiny 2}}
\put(42,40){\circle*{4}}\put(42,37){\makebox(0,0)[ct]{\tiny 3}}
\put(48,10){\circle*{4}}
\put(61,81){\circle*{4}}
\put(85,72){\circle*{4}}
\put(85,22){\circle*{4}}
\put(118,85){\circle*{4}}
\put(125,8){\circle*{4}}
\put(142,30){\circle*{4}}
\put(143,64){\circle*{4}}
\put(148,85){\circle*{4}}
\put(168,50){\circle*{4}}
\put(173,18){\circle*{4}}\put(173,15){\makebox(0,0)[ct]{\tiny $m$$-$$2$}}
\put(182,80){\circle*{4}}\put(182,77){\makebox(0,0)[ct]{\tiny $m$$-$$1$}}
\put(188,55){\circle*{4}}\put(188,52){\makebox(0,0)[ct]{\tiny $m$}}
\put(113,-12){\makebox(0,0)[cb]{(b)}}
\end{picture}
\caption{(a) A network $\cN$ and (b) the corresponding network $\cN_R$ 
         that replaces channel 
	 $(\cX^{(i,1)},p(y^{(j,1)}|x^{(i,1)}),\cY^{(j,1)})$ 
	 by a capacity-$R$ noiseless bit pipe 
	 $(\{0,1\}^R,\delta(\ty^{(j,1)}-\tx^{(i,1)}),\{0,1\}^R)$.
}\label{fig:netR}
\end{center}
\end{figure}
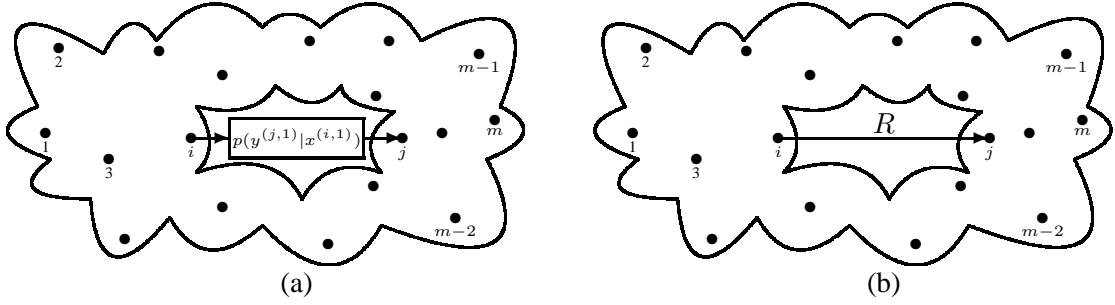

The equivalence result derived in this section 
relates the rate region of a network 
\[
\cN=\left(\cXVm\times\cXV,p(\yV|\xV)p(\yVm|\xVm),\cYVm\times\cYV\right)
\]
to that of a network $\cN_R$ 
that replaces channel $(\cXV,p(\yV|\xV),\cYV)$ 
by a capacity-$R$ noiseless bit pipe, 
here denoted by 
$(\{0,1\}^R,\delta(\tyV-\txV),\{0,1\}^R)$.
Thus 
\[
\cN_R
\ \deff\ \left(\cXVm\times\{0,1\}^R,\delta(\tyV-\txV)p(\yVm|\xVm),\cYVm\times\{0,1\}^R
\right)
\]
Employing a common abuse of notation, 
we allow non-integer values of $R$ 
to designate capacitated bit pipes 
that require more than a single channel use 
to deliver some integer number of bits.  
Applying the stacking approach of Theorem~\ref{thm:stacked}, 
the arguments that follow 
transmit information over the $N$ copies of each channel 
in an $N$-fold stacked network;  
thus we have channel $(\{0,1\}^{NR},\delta(\uxV-\uyV),\{0,1\}^{NR})$ 
in the $N$-fold stacked network $\ucN_R$.  
As usual, $N$ is allowed to grow without bound, 
so the transmission of $\lfloor NR\rfloor$ bits 
over $N$ channel uses achieves rate arbitrarily close to $R$.  

Before turning to the equivalence result, 
we prove the continuity of capacity region $\setR(\cN_R)$ in $R$ 
for all $R>0$.  
Precisely, for any $R>0$ and $\delta<R$, we define 
\[
\eps(\delta)\ \deff\ 
\max_{\cR\in\setR(\ucN_{R+\delta})}
\min_{\tcR\in\setR(\ucN_{R-\delta})}||\cR-\tcR||_\infty
\]
to be the worst-case $\ell_\infty$-norm 
between a point in $\setR(\ucN_{R+\delta})$ 
and its closest point in $\setR(\ucN_{R-\delta})$.  
We then show that for any $\eps>0$, 
there exists a $\delta>0$ for which $\eps(\delta)\leq\eps$.  
Continuity of the rate region at $R=0$ remains an open problem 
for most networks~\cite{GuE:09,Gu:09}.  
The subtle underlying question here is whether 
a number of bits that grows sublinearly in the coding dimension 
can change the network capacity.  

\begin{lem}\label{lem:cont}
Rate region $\setR(\cN_R)$ is continuous 
in $R$ for all $R>0$. 

\Proof
By Theorem~\ref{thm:stacked} it suffices to prove that 
$\setR(\ucN_R)$ is continuous in $R$.  
Note that $\setR(\ucN_R)$ is non-decreasing in $R$; 
that is $\tR<R$ implies $\setR(\ucN_{\tR})\subseteq\setR(\ucN_R)$ 
since any $(\lambda,\cR)$-$\cS(\ucN_{\tR})$ solution 
for $N$-fold stacked network $\ucN_{\tR}$ 
can be run with the same error probability 
on $N$-fold stacked network $\ucN_R$.  
Thus for any $R>0$ and any $\delta\in(0,R)$, 
$\setR(\ucN_{R-\delta})
\subseteq\setR(\ucN_R)
\subseteq\setR(\ucN_{R+\delta})$.  
Fix any $\delta>0$ and $\cR\in\mbox{int}(\setR(\ucN_{R+\delta}))$.  
For any $\lambda>0$ and all $N$ sufficiently large 
there exists a $(\lambda,\cR)$-$\cS(\ucN_{R+\delta})$ 
solution for the $N$-fold stacked network $\ucN_{R+\delta}$.  
Recall that $\cN_{R+\delta}$ and $\cN_{R-\delta}$ 
differ only in the capacity of the bit pipe from node $i$ to node $j$.  
Thus any solution $\cS(\ucN_{R+\delta})$ 
that achieves error probability $\lambda$ 
on $N$-fold stacked network $\ucN_{R+\delta}$ 
can be run with the same error probability 
on $\tilde{N}$-fold stacked network $\ucN_{R-\delta}$ 
provided 
\[
\tilde{N}(R-\delta)\geq N(R+\delta).  
\]
This is accomplished by operating solution $\cS(\ucN_{R+\delta})$ unchanged 
across the first $N$ copies of the channel $(\cXVm,p(\yVm|\xVm),\cYVm)$ 
in $\ucN_{R-\delta}$ and sending the $N(R+\delta)$ bits 
intended for transmission across $N$ bit pipes of rate $R+\delta$ 
in $\ucN_{R+\delta}$ 
across the $\tilde{N}$ bit pipes of rate $R-\delta$ in $\ucN_{R-\delta}$.  
Set $\tilde{N}=\lceil N(R+\delta)/(R-\delta)\rceil$.  
Then the rate of the resulting code is 
\[
\tcR=\frac{\cR N}{\tilde{N}}\geq\cR\frac{N(R-\delta)}{N(R+\delta)+R-\delta}.  
\]
Since $\cR$ and $R$ are fixed, the difference 
\[
\cR-\tcR\leq \cR\frac{2N\delta+R-\delta}{N(R+\delta)+R-\delta}.  
\]
can be made arbitrarily small 
by letting $N$ grow and $\delta$ approach 0.  
Since $\cR$ is arbitrary, we have the desired result.  
\IEEEQED
\end{lem}

The following lemma derives a lower bound on $\setR(\cN)$.  

\begin{lem}\label{lem:lb}
Consider a pair of networks 
\begin{eqnarray*}
\cN & = & \left(\cXVm\times\cXV,p(\yV|\xV)p(\yVm|\xVm),\cYVm\times\cYV\right)\\
\cN_C & = & \left(\cXVm\times\{0,1\}^C,\delta(\tyV-\txV)p(\yVm|\xVm),\cYVm
	\times\{0,1\}^C\right),
\end{eqnarray*}
where $C=\max_{p(\xV)}I(\XV;\YV)$ is the capacity of channel 
$(\cXV,p(\yV|\xV),\cYV)$.  
Then 
\[
\setR(\cN_C)\subseteq\setR(\cN).
\]
\Proof
The following proof shows that $\setR(\cN_R)\subseteq\setR(\cN)$ 
for all $R<C$.  
This shows that $\cup_{R<C}\setR(\cN_R)\subseteq\setR(\cN)$, 
which gives the desired result by Lemma~\ref{lem:cont} 
and the closure in the definition of $\setR(\cN)$.  
Applying Theorem~\ref{thm:stacked}, 
for each $R<C$ we show that $\setR(\cN_R)\subseteq\setR(\cN)$ 
by showing that $\setR(\ucN_R)\subseteq\setR(\ucN)$.  

Fix any $R<C$, $\cR\in\mbox{int}(\setR(\ucN_R))$, and $\lambda>0$.  
We first use the argument from the proof of Theorem~\ref{thm:stacked} 
to build a sequence of rate-$\cR$ solutions $\ucS(\ucN_R)$ 
with error probability no greater than $2^{-N\delta}$ for all $N$ sufficiently large.  
Recall that only the channel code on the messages $\uWuv$ changes with $N$.  
Thus for all $N\geq 1$, solution $\ucS(\ucN_R)$ 
applies the same solution $\cS(\cN_R)$ in each layer of the stack. 
Let $n$ be the blocklength of code $\cS(\cN_R)$ 
(and therefore the blocklength of $\ucS(\ucN_R)$ for all $N$).  

Since $R<C$, $\lambda>0$, and $n$ are fixed, 
there exists a sequence of channel codes 
$\{(\alpha_N,\beta_N)\}_{N=1}^\infty$ for channel $(\cXV,p(\yV|\xV),\cYV)$ 
with encoders $\alpha_N$, decoders $\beta_N$, 
and maximal error probability 
$\max_{\uw}\Pr(\beta_N(\uYV)\neq \uw|\uXV=\alpha_N(P_e^{(N)}))<\lambda/(2n)$ 
for all $N$ sufficiently large.\footnote{We here divide by $n$ 
since the channel code will be applied $n$ times, 
once for each instant in time for this blocklength-$n$ code.  
Application of the union bound then gives 
an error probability over these $n$ time steps.}

\begin{figure}
  \begin{center}
    \begin{picture}(460,130)(10,-10)
      \thicklines
      \put(10,60){\vector(1,0){30}}
      \put(40,40){\framebox(40,40)[cc]{$\uX_t^{(i)}$}}
      \put(60,-5){\vector(0,1){45}}
      \put(80,75){\vector(1,0){120}}
      \put(80,45){\vector(1,0){120}}
      \put(180,95){\vector(1,0){20}}
      \put(200,70){\framebox(80,30)[cc]{$p(\uyVm|\uxVm)$}}
      \put(200,30){\framebox(80,30)[cc]{$\delta(\utyV-\utxV)$}}
      	\thinlines
		\put(195,10){\framebox(90,95)[cb]{}}
		\put(195,12){\makebox(90,0)[cb]{$\ucN_R$}}
		\thicklines
      \put(280,95){\vector(1,0){20}}
      \put(280,75){\vector(1,0){120}}
      \put(280,45){\vector(1,0){120}}
      \put(400,40){\framebox(40,40)[cc]{$\uX_{t+1}^{(j)}$}}
      \put(420,-5){\vector(0,1){45}}
      \put(440,60){\vector(1,0){30}}
      \put(20,57){\makebox(0,0)[ct]{$\uY_{t-1}^{(i)}$}}
      \put(60,0){\makebox(0,0)[ct]{$\uW^{\ar{i}{1}},\ldots,\uW^{\ar{i}{m}}$}}
      \put(100,43){\makebox(0,0)[ct]{$\utXV_t$}}
      \put(100,73){\makebox(0,0)[ct]{$\uX^{(i,2)}_t$}}
      \put(178,95){\makebox(0,0)[cr]{$(\uX_t^{(v)}:v\neq i)$}}
      \put(303,95){\makebox(0,0)[cl]{$(\uY_t^{(v)}:v\neq j)$}}
      \put(380,73){\makebox(0,0)[ct]{$\uY^{(j,2)}_t$}}
      \put(380,43){\makebox(0,0)[ct]{$\utYV_t$}}
      \put(420,0){\makebox(0,0)[ct]{$\uW^{\ar{j}{1}},\ldots,\uW^{\ar{j}{m}}$}}
      \put(460,57){\makebox(0,0)[ct]{$\uX_{t+1}^{(j)}$}}
      \thinlines
      \put(35,5){\framebox(90,80)[cb]{}}
      \put(120,6){\makebox(0,0)[rb]{\tiny NODE $i$, TIME $t$}}
      \put(355,5){\framebox(90,80)[cb]{}}
      \put(360,6){\makebox(0,0)[lb]{\tiny NODE $j$, TIME $t+1$}}
      \put(115,0){\makebox(250,0)[ct]{(a)}}
    \end{picture}
    \begin{picture}(460,140)(10,-10)
      \thicklines
      \put(10,60){\vector(1,0){30}}
      \put(40,40){\framebox(40,40)[cc]{$\uX_t^{(i)}$}}
      \put(60,-5){\vector(0,1){45}}
      \put(80,75){\vector(1,0){120}}
      \put(80,45){\vector(1,0){40}}
      \put(120,25){\framebox(40,40)[cc]{$\alpha_{N,t}$}}
      \put(180,95){\vector(1,0){20}}
      \put(160,45){\vector(1,0){40}}
      \put(200,70){\framebox(80,30)[cc]{$p(\uyVm|\uxVm)$}}
      \put(200,30){\framebox(80,30)[cc]{$p(\uyV|\uxV)$}}
      	\thinlines
		\put(195,25){\framebox(90,95)[cb]{}}
		\put(195,117){\makebox(90,0)[ct]{$\ucN$}}
		\thicklines
      \put(280,95){\vector(1,0){20}}
      \put(280,75){\vector(1,0){120}}
      \put(280,45){\vector(1,0){40}}
      \put(320,25){\framebox(40,40)[cc]{$\beta_{N,t}$}}
      \put(360,45){\vector(1,0){40}}
      \put(400,40){\framebox(40,40)[cc]{$\uX_{t+1}^{(j)}$}}
      \put(420,-5){\vector(0,1){45}}
      \put(440,60){\vector(1,0){30}}
      \put(20,57){\makebox(0,0)[ct]{$\uY_{t-1}^{(i)}$}}
      \put(60,0){\makebox(0,0)[ct]{$\uW^{\ar{i}{1}},\ldots,\uW^{\ar{i}{m}}$}}
      \put(100,73){\makebox(0,0)[ct]{$\uX^{(i,2)}_t$}}
      \put(100,43){\makebox(0,0)[ct]{$\utXV_t$}}
      \put(178,95){\makebox(0,0)[cr]{$(\uX_t^{(v)}:v\neq i)$}}
      \put(180,43){\makebox(0,0)[ct]{$\uXV_t$}}
      \put(380,73){\makebox(0,0)[ct]{$\uY^{(j,2)}_t$}}
      \put(300,43){\makebox(0,0)[ct]{$\uYV_t$}}
      \put(303,95){\makebox(0,0)[cl]{$(\uY_t^{(v)}:v\neq j)$}}
      \put(380,43){\makebox(0,0)[ct]{$\utYV_t$}}
      \put(420,0){\makebox(0,0)[ct]{$\uW^{\ar{j}{1}},\ldots,\uW^{\ar{j}{m}}$}}
      \put(460,57){\makebox(0,0)[ct]{$\uX_{t+1}^{(j)}$}}
      \thinlines
      \put(35,5){\framebox(130,80)[cb]{}}
      \put(90,6){\makebox(0,0)[cb]{\tiny NODE $i$, TIME $t$}}
      \put(315,5){\framebox(130,80)[cb]{}}
      \put(395,6){\makebox(0,0)[cb]{\tiny NODE $j$, TIME $t+1$}}
      \put(115,10){\framebox(250,57)[cb]{}}
      \put(115,12){\makebox(250,0)[cb]{\tiny BIT PIPE EMULATOR}}
      \put(115,-5){\makebox(250,0)[ct]{(b)}}
    \end{picture}
  \end{center}
\caption{Operation of node $i$ at time $t$ 
         and node $j$ at time $t+1$ in solutions 
	 (a)~$\ucS(\ucN_R)$ and (b)~$\cS(\ucN)$.
         We show the nodes at different times 
         since the output $\utXV_t$ from node $i$ at time $t$ 
         cannot influence the encoder at node $j$ 
	 until time $t+1$ (due to the causality constraint).}\label{fig:lb}
\end{figure}

The next step is to build a solution $\cS(\ucN)$ for $N$-fold stacked network $\ucN$.  
Solution $\cS(\ucN)$ operates $\ucS(\ucN_R)$ across $\ucN$  
by using channel code $(\alpha_N,\beta_N)$ at each time $t$ 
to transmit across the $N$ copies of channel $(\cXV,p(\yV|\xV),\cYV)$ in $\cN$,  
as shown in Figure~\ref{fig:lb}.
Precisely, at time $t$, node $v$ performs any necessary channel decoding 
on the channel output to give 
\[
\tilde{\uY}^{(v)}_t = 
\left\{\begin{array}{ll}
(\beta_N(\uY^{(j,1)}_t),\uY^{(j,2)}_t) & v=j \\
\uvY_t & v\neq j, 
\end{array}\right.
\]
then applies the node encoders from $\ucS(\ucN_R)$ as 
\[
\tilde{\uX}^{(v)}_t = 
\uvX_t(\tilde{\uY}^{(v)}_1,\ldots,\tilde{\uY}^{(v)}_{t-1},\uWvo,\ldots,\uWvm),
\]
and finally applies any necessary channel encoding as 
\[
\uvX_t = \left\{\begin{array}{ll} 
	(\alpha_N(\tilde{\uX}_t^{(i,1)}),
                  \tilde{\uX}_t^{(i,2)}) & \mbox{if $v=i$} \\
	          \tilde{\uX}_t^{(v)} & \mbox{if $v\neq i$.}
   \end{array}\right.
\]
before transmission across the channel.  
At time $n$, node $v$ applies the decoder from $\ucS(\ucN_R)$ to give 
\[
\ubWuv=\ubWuv(\tilde{\uY}_1^{(v)},\ldots,\tilde{\uY}_n^{(v)},
        \uWvo,\ldots,\uWvm).
\]

To bound the error probability, 
note that two things can go wrong.  
Either the channel code fails at one or more time steps 
or the channel code succeeds at all $n$ time steps 
but the code fails anyway.  
If the channel code $(\alpha_N,\beta_N)$ 
succeeds at all times $t\in\{1,\ldots,n\}$, 
then the conditional probability of an error given $\uW=\uw$ 
is precisely what it would have been for the original code.  
Let $E_t$ denote the event that 
the channel code fails at time $t$.  
Then we bound the error probability as 
\begin{eqnarray*}
\Pr(\ubW\neq\uW) 
& \stackrel{(a)}{\leq} & \sum_{t=1}^n\Pr(E_t)+\sum_{\uw}
\Pr(\ubW\neq \uW|\{\uW=\uw\}\cap\cap_{t=1}^nE_t^c)\Pr(\{\uW=\uw\}\cap\cap_{t=1}^nE_t^c) \\
& \stackrel{(b)}{\leq} & 
	\left(\sum_{t=1}^n\frac{\lambda}{2n}\right)+2^{-N\delta},
\end{eqnarray*}
which is less than $\lambda$ for all $N$ sufficiently large.  
Inequality $(a)$ follows from the union bound.  
Inequality $(b)$ follows from the error probability bound 
for the channel code and from the observation 
that $\Pr(\{\uW=\uw\}\cap\cap_{t=1}^nE_t^c)\leq \Pr(\uW=\uw)$ 
for all $\uw$.  
\IEEEQED
\end{lem}

Lemma~\ref{lem:lb} applies channel coding 
to emulate a noiseless bit pipe $(\{0,1\}^R,\delta(\tyV|\txV),\{0,1\}^R)$ 
across a noisy channel $(\cXV,p(\yV|\xV),\cYV)$ 
so that a code for $\cN_R$ can be run across $\cN$ 
with the aid of the channel code.  
Theorem~\ref{thm:main} employs a code 
that emulates noisy channel $(\cXV,p(\yV|\xV),\cYV)$ 
across noiseless bit pipe $(\{0,1\}^R,\delta(\tyV|\txV),\{0,1\}^R)$ 
so that a code for $\cN$ can be run across $\cN_R$ 
with similar error probability.  

\begin{thm}\label{thm:main}
Consider a pair of networks 
\begin{eqnarray*}
\cN & = & \left(\cXVm\times\cXV,p(\yV|\xV)p(\yVm|\xVm),\cYVm\times\cYV\right)\\
\cN_R & = & \left(\cXVm\times\{0,1\}^R,\delta(\tyV-\txV)p(\yVm|\xVm),\cYVm
	\times\{0,1\}^R\right),
\end{eqnarray*}
where $(\cXV,p(\yV|\xV),\cYV)$ is a channel of capacity 
$C\ \deff\ \max_{p(\xV)}I(\XV;\YV)$.  

If $R>C$, then 
\[
\setR(\cN)\subseteq\setR(\cN_R).
\]

\Proof  
By Theorem~\ref{thm:stacked} it suffices to show that 
$\setR(\ucN)\subseteq\setR(\ucN_R)$.  
Fix an arbitrary point $\cR\in\mbox{int}(\setR(\cN))$ 
and any $\lambda>0$.  
The argument that follows shows that 
for all $N$ sufficiently large 
there exists a $(\lambda,\cR)$ solution 
$\cS(\ucN_R)$ for $N$-fold stacked network $\ucN_R$.  
We first define a random code design algorithm 
and bound the expected error probability 
with respect to the random design.  
This random design includes random selection of $m(m-1)$ channel codes 
and random design of channel emulators for each time step.  
In order to ensure good end-to-end performance, 
we do not choose a single instance of any of the randomly designed codes 
until all of the codes are in place.  
At that point, we choose all codes simultaneously.  

\underline{Step 1 -  Choose code $\cS(\cN)$ 
and define distributions $p_t(\xV,\yV)$:}\mbox{}\\
Recall from the proof of Theorem~\ref{thm:stacked} 
that there exists a rate-$\cR$ 
solution $\cS(\cN)$ of some finite blocklength $n$ 
from which good stacked solutions $\ucS(\ucN)$ 
for $N$-fold stacked network $\ucN$ 
can be built for all $N$ sufficiently large.  
The stacked solution applies an independent random channel code 
to each message $\uWuv$ 
and then applies $\cS(\cN)$ independently in each layer of $\ucN$.  
Taking an expectation over the random channel code designs 
yields expected error probability no larger than $2^{-N\delta}$ 
for all $N$ sufficiently large.  
We therefore begin by fixing a solution $\cS(\cN)$ 
as in Theorem~\ref{thm:stacked} 
and building the corresponding stacked solution $\ucS(\ucN)$.  
For each $t\in\{1,\ldots,n\}$, 
let $p_t(\xV)$ be the distribution established 
on the input to channel $(\cXV,p(\yV|\xV),\cYV)$ at time $t$ 
by solution $\cS(\cN)$.  
Distribution $p_t(\xV)$ may vary with $t$ 
due, for example, to feedback in the network.  
Then $p_t(\uxV,\uyV)\ \deff\ 
\prod_{\ell=1}^Np_t(\uxV(\ell))p(\uyV(\ell)|\uxV(\ell))$ 
is the time-$t$ distribution across 
$(\ucXV,p(\uyV|\uxV),\ucYV)$ under solution $\ucS(\ucN)$.  

\underline{Step 2 - Typical set definitions and properties:}\mbox{}\\
Let $\eps=(\eps(1),\ldots,\eps(n))$ be a vector 
of positive constants,\footnote{Our parameter choice 
in the typical set definition varies with $t$ 
both to accommodate variation in $p_t(\xV,\yV)$ 
and to handle the cumulative impact of channel emulation 
at multiple time steps.} 
and for each $t$ define 
\begin{eqnarray*}
\typ
& \deff & \left\{(\uxV,\uyV)\in\ucXV\times\ucYV: \right. \\
&&\left|-\frac1N\log p_t(\uxV)-H(\XV_t)\right|\leq \eps(t) \\
&&\left|-\frac1N\log p_t(\uyV)-H(\YV_t)\right|\leq a(\eps,t) \\
&&\left.\left|-\frac1N\log p_t(\uxV,\uyV)-H(\XV_t,\YV_t))\right|
\leq a(\eps,t)\right\},
\end{eqnarray*}
where $H(\XV_t)$, $H(\YV_t)$, and $H(\XV_t,\YV_t)$ 
are the entropies on $\XV_t$, $\YV_t$, and $(\XV_t,\YV_t)$ 
under $p_t(\xV,\yV)$,\footnote{We 
use notation $H(\cdot)$ for both discrete and differential entropy.  
We assume that $H(\XV_t,\YV_t)<\infty$.}
and 
\begin{eqnarray}
a(\eps,t) & \deff & (1+\eps(t))\cdot\inf\left\{\eps'>0: \right.
\Pr\left(\left|-\frac1N\log p_t(\uYV_t)-H(\YV_t)\right|>\eps'
\vee\right.\label{eqn:atdef} \\
& & \left.\left|-\frac1N\log p_t(\uXV_t,\uY_t)
-H(\XV_t,\YV_t)\right|>\eps'\right) 
\leq 2^{-N6\eps(t)}
\left.\mbox{ for all $N$ sufficiently large}\right\}.  \nonumber
\nonumber
\end{eqnarray}
(This infimum is shown to be well defined 
in the proof of Lemma~\ref{lem:p2ptyp} in Appendix~\ref{app:p2ptyp}.)  
Define set 
\[
\rtyp\ \deff\ \left\{(\uxV,\uyV)\in\typ:
p\left(\left.(\typ)^c\right|\uxV\right)\leq 2^{-3N\eps(t)}\right\},
\]
where $p_t((\typ)^c|\uxV)\ \deff\ 
\sum_{\uyV:(\uxV,\uyV)\not\in\rtyp}p(\uyV|\uxV)$.  
We henceforth call $\rtyp$ the typical set.  
This typical set definition 
restricts attention to those typical channel inputs $\uxV$ 
that are most likely to yield jointly typical channel outputs.  
This restriction is later useful 
for showing that the number of jointly typical channel outputs 
for each typical channel input is roughly the same.  
Such a result could be obtained more directly 
for finite-alphabet channels if we used strong typicality, 
but we here treat the general case.  
Lemma~\ref{lem:p2ptyp} in Appendix~\ref{lem:p2ptyp} shows that 
\begin{equation}\label{eqn:p2ptyp}
p_t((\rtyp)^c) < 2^{-Nc(\eps,t)} 
\end{equation}
for some constant $c(\eps,t)$ that goes to zero as $\eps(t)$ goes to zero 
and grows large as $\eps(t)$ grows large.  

\underline{Step 3 - Design of channel emulators:}\mbox{} \\
We next design codes $(\alpha_{N,t},\beta_{N,t})$, $t\in\{1,\ldots,n\}$.  
The goal of the code design is to build a collection of devices 
for emulating $N$ independent uses of channel $(\cXV,p(\yV|\xV),\cYV)$ 
over $N$ independent uses of bit pipe 
$(\{0,1\}^R,\delta(\tyV-\txV),\{0,1\}^R)$.  
The code for time $t$ emulates the channel 
under input distribution $p_t(\uxV)$.  
Code $(\alpha_{N,t},\beta_{N,t})$ 
has encoder $\alpha_{N,t}: \ucXV\rightarrow\{0,1\}^{NR}$ 
and decoder $\beta_{N,t}: \{0,1\}^{NR}\rightarrow\ucYV$.  
Thus $(\alpha_{N,t},\beta_{N,t})$ 
is effectively a lossy source code with rate $R$ and blocklength $N$.  
This source code differs from traditional source codes 
in that a good reproduction is not a value $\hat{\uX}_t^{(i,1)}$
that reproduces $\uXV_t$ to low distortion 
but a value $\uYV_t$ that is similar statistically 
to the vector of outputs observed 
when $\uXV_t$ is transmitted across $(\ucXV,p(\uyV|\uxV),\ucYV)$.  
Since the channel usually maps typical inputs 
to jointly typical outputs, we design our source code to do the same.  

First, randomly design decoder 
$\beta_{N,t}:\{1,\ldots,2^{NR}\}\rightarrow\ucYV$ 
by drawing codewords 
\begin{equation}
\beta_{N,t}(1),\ldots,\beta_{N,t}(2^{NR})\sim\mbox{ i.i.d. }p_t(\uyV).  
\label{eqn:p2pdec}
\end{equation}
Then, design encoder $\alpha_{N,t}:\ucXV\rightarrow\{1,\ldots,2^{NR}\}$ as 
\begin{equation}
\alpha_{N,t}(\uxV)=\left\{\begin{array}{ll} 
      k & \mbox{if $(\uxV,\beta_{N,t}(k))\in\rtyp$} \\
      1 & \mbox{if $\not\exists k$ s.t.\ $(\uxV,\beta_{N,t}(k))\in\rtyp$}.
\end{array}\right.
\label{eqn:p2penc}
\end{equation}
When there is more than one index $k$ 
for which $(\uxV,\beta_{N,t}(k))\in\rtyp$, 
the encoder design chooses uniformly at random among them.  

\begin{figure}
  \begin{center}
    \begin{picture}(460,130)(10,-10)
      \thicklines
      \put(10,60){\vector(1,0){30}}
      \put(40,40){\framebox(40,40)[cc]{$\uX_t^{(i)}$}}
      \put(60,-5){\vector(0,1){45}}
      \put(80,75){\vector(1,0){120}}
      \put(80,45){\vector(1,0){120}}
      \put(180,95){\vector(1,0){20}}
      \put(200,70){\framebox(80,30)[cc]{$p(\uyVm|\uxVm)$}}
      \put(200,30){\framebox(80,30)[cc]{$p(\uyV|\uxV)$}}
      	\thinlines
		\put(195,10){\framebox(90,95)[cb]{}}
		\put(195,12){\makebox(90,0)[cb]{$\ucN$}}
		\thicklines
      \put(280,95){\vector(1,0){20}}
      \put(280,75){\vector(1,0){120}}
      \put(280,45){\vector(1,0){120}}
      \put(400,40){\framebox(40,40)[cc]{$\uX_{t+1}^{(j)}$}}
      \put(420,-5){\vector(0,1){45}}
      \put(440,60){\vector(1,0){30}}
      \put(20,57){\makebox(0,0)[ct]{$\uY_{t-1}^{(i)}$}}
      \put(60,0){\makebox(0,0)[ct]{$\uW^{\ar{i}{1}},\ldots,\uW^{\ar{i}{m}}$}}
      \put(100,43){\makebox(0,0)[ct]{$\uXV_t$}}
      \put(100,73){\makebox(0,0)[ct]{$\uX^{(i,2)}_t$}}
      \put(178,95){\makebox(0,0)[cr]{$(\uX_t^{(v)}:v\neq i)$}}
      \put(303,95){\makebox(0,0)[cl]{$(\uY_t^{(v)}:v\neq j)$}}
      \put(380,73){\makebox(0,0)[ct]{$\uY^{(j,2)}_t$}}
      \put(380,43){\makebox(0,0)[ct]{$\uYV_t$}}
      \put(420,0){\makebox(0,0)[ct]{$\uW^{\ar{j}{1}},\ldots,\uW^{\ar{j}{m}}$}}
      \put(460,57){\makebox(0,0)[ct]{$\uX_{t+1}^{(j)}$}}
      \thinlines
      \put(35,5){\framebox(90,80)[cb]{}}
      \put(120,6){\makebox(0,0)[rb]{\tiny NODE $i$, TIME $t$}}
      \put(355,5){\framebox(90,80)[cb]{}}
      \put(360,6){\makebox(0,0)[lb]{\tiny NODE $j$, TIME $t+1$}}
      \put(115,0){\makebox(250,0)[ct]{(a)}}
    \end{picture}
    \begin{picture}(460,140)(10,-10)
      \thicklines
      \put(10,60){\vector(1,0){30}}
      \put(40,40){\framebox(40,40)[cc]{$\uX_t^{(i)}$}}
      \put(60,-5){\vector(0,1){45}}
      \put(80,75){\vector(1,0){120}}
      \put(80,45){\vector(1,0){40}}
      \put(120,25){\framebox(40,40)[cc]{$\alpha_{N,t}$}}
      \put(180,95){\vector(1,0){20}}
      \put(160,45){\vector(1,0){40}}
      \put(200,70){\framebox(80,30)[cc]{$p(\uyVm|\uxVm)$}}
      \put(200,30){\framebox(80,30)[cc]{$\delta(\utxV-\utyV)$}}
      	\thinlines
		\put(195,25){\framebox(90,95)[cb]{}}
		\put(195,118){\makebox(90,0)[ct]{$\ucN_R$}}
		\thicklines
      \put(280,95){\vector(1,0){20}}
      \put(280,75){\vector(1,0){120}}
      \put(280,45){\vector(1,0){40}}
      \put(320,25){\framebox(40,40)[cc]{$\beta_{N,t}$}}
      \put(360,45){\vector(1,0){40}}
      \put(400,40){\framebox(40,40)[cc]{$\uX_{t+1}^{(j)}$}}
      \put(420,-5){\vector(0,1){45}}
      \put(440,60){\vector(1,0){30}}
      \put(20,57){\makebox(0,0)[ct]{$\uY_{t-1}^{(i)}$}}
      \put(60,0){\makebox(0,0)[ct]{$\uW^{\ar{i}{1}},\ldots,\uW^{\ar{i}{m}}$}}
      \put(100,42){\makebox(0,0)[ct]{$\uXV_t$}}
      \put(100,73){\makebox(0,0)[ct]{$\uX^{(i,2)}_t$}}
      \put(178,95){\makebox(0,0)[cr]{$(\uX_t^{(v)}:v\neq i)$}}
      \put(180,42){\makebox(0,0)[ct]{$\utXV_t$}}
      \put(380,73){\makebox(0,0)[ct]{$\uY^{(j,2)}_t$}}
      \put(300,42){\makebox(0,0)[ct]{$\utYV_t$}}
      \put(303,95){\makebox(0,0)[cl]{$(\uY_t^{(v)}:v\neq j)$}}
      \put(380,42){\makebox(0,0)[ct]{$\uYV_t$}}
      \put(420,0){\makebox(0,0)[ct]{$\uW^{\ar{j}{1}},\ldots,\uW^{\ar{j}{m}}$}}
      \put(460,57){\makebox(0,0)[ct]{$\uX_{t+1}^{(j)}$}}
      \thinlines
      \put(35,5){\framebox(130,80)[cb]{}}
      \put(90,6){\makebox(0,0)[cb]{\tiny NODE $i$, TIME $t$}}
      \put(315,5){\framebox(130,80)[cb]{}}
      \put(395,6){\makebox(0,0)[cb]{\tiny NODE $j$, TIME $t+1$}}
      \put(115,10){\framebox(250,57)[cb]{}}
      \put(115,12){\makebox(250,0)[cb]{\tiny CHANNEL EMULATOR}}
      \put(115,-5){\makebox(250,0)[ct]{(b)}}
    \end{picture}
  \end{center}
\caption{Operation of node $i$ at time $t$ 
         and node $j$ at time $t+1$ in solutions 
	 (a)~$\ucS(\ucN)$ and (b)~$\cS(\ucN_R)$.
         We show the nodes at different times 
         since the output $\uXV_t$ from node $i$ at time $t$ 
         cannot influence the encoder at node $j$ 
	 until time $t+1$ (due to the causality constraint).}\label{fig:sc}
\end{figure}

\underline{Step 4 - Definition of solution $\cS(\ucN_R)$:}\mbox{} \\
The next step is to employ codes $\{(\alpha_{N,t},\beta_{N,t})\}_{t=1}^n$ 
to operate $\ucS(\ucN)$ across network $\ucN_R$.  
We begin with an informal description of the resulting code, 
here denoted by $\cS(\ucN_R)$.  
For each node $v\not\in\{i,j\}$, 
the operation of node $v$ in $\cS(\ucN_R)$ 
is identical to the operation of node $v$ in $\ucS(\ucN)$.  
Node $i$ applies its node encoder from $\ucS(\ucN)$ as usual 
and then source codes the resulting channel input transmission;  
the node decoder at node $i$ is unchanged.  
Node $j$ source decodes the bit-pipe output 
before applying its usual encoder and decoder from $\ucS(\ucN)$.  
Figure~\ref{fig:sc} illustrates these operations, 
defined formally below.  

For each $v\in V$ and $t\in\{1,\ldots,n\}$, let $\ub{\tY}^{(v)}_t$ 
be the time-$t$ channel output at node $v$ in $\cS(\ucN_R)$.  
Each node $v$ applies its node encoder as 
\[
\uvX_t=\uvX_t(\uvY_1,\ldots,\uvY_{t-1},\uWvo,\ldots,\uWvm),
\]
which it encodes (if necessary) as 
\[
\underline{\tilde{X}}^{(v)}_t = \left\{\begin{array}{ll} 
   (\alpha_{N,t}(\uX^{(i,1)}_t),\uX^{(i,2)}) & \mbox{if $v=i$} \\
   \uvX_t & \mbox{if $v\neq i$}.  
   \end{array}\right.
\]
before transmission.  
Here $\uvY_t$ designates the channel output 
after any necessary decoding, giving 
\[
\uvY_t= \left\{\begin{array}{ll} 
  (\beta_{N,t}(\underline{\tilde{Y}}^{(j,1)}_t),\underline{\tilde{Y}}^{(j,2)}_t) 
   	& \mbox{if $v=j$} \\
  \underline{\tilde{Y}}^{(v)}_t & \mbox{if $v\neq j$.} 
    \end{array}\right.
\]
Finally, node $v$ applies the decoders from $\ucS(\ucN)$ as 
\[
\ubWuv=\ubWuv(\uvY_1,\ldots,\uvY_n,\uWvo,\ldots,\uWvm).
\]
Solution $\cS(\ucN_R)$ is not a stacked solution 
since each $(\alpha_{N,t},\beta_{N,t})$ 
operates across the layers of the stack.  

\underline{Step 5 - Characterizing the behavior of $\cS(\ucN_R)$:}\mbox{} \\
In order to analyze the error probability of code $\cS(\ucN_R)$ 
we first characterize its statistical behavior.  
Table~\ref{tab:solN} summarizes the random variables 
used in the definition of the solution $\ucS(\ucN)$ 
from which $\cS(\ucN_R)$ is built.  
\begin{table}
\caption{Summary of notation for solution $\ucS(\ucN_R)$}\label{tab:solN}
\begin{center}
\begin{tabular}{|l|l|}
\hline
Variable & Meaning \\ \hline\hline 
$\cS(\cN)$ & solution used in each layer of $\ucS(\ucN)$ \\\hline
$n$        & blocklength of solutions $\cS(\cN)$ and $\ucS(\ucN)$ \\ \hline
$\uW=(\uWuv:(u,v)\in\{1,\ldots,m\})$   & messages \\ \hline
$\ubX_t=(\uX^{(v)}:v\in\{1,\ldots,m\})$   & network inputs at time $t$ \\\hline
$\ubY_t=(\uY^{(v)}:v\in\{1,\ldots,m\})$   & network outputs at time $t$\\\hline
$\buW =(\buWuv:(u,v)\in\{1,\ldots,m\}$ & reconstruction of messages \\ \hline
\end{tabular}
\end{center}
\end{table}
Applying solution $\ucS(\ucN)$ on $N$-fold stacked network $\ucN$ 
yields joint distribution 
\[
p(\uw,\ubx^n,\uby^n,\buw)= p(\uw)
    \left[\prod_{t=1}^{n}p(\ubx_t|\uby^{t-1},\uw)\right]
    \left[\prod_{t=1}^{n}p(\uby_t|\ubx_t)\right]
    p(\buw|\uw,\uby^n).  
\]
Here $p(\uw)$ is the distribution on messages; 
$p(\ubx_t|\uby^{t-1},\uw)$ results from 
the operation of all node encoders at time $t$, 
each of which maps its received channel outputs and outgoing messages 
to channel inputs; 
$p(\uby_t|\ubx_t)$ is the memoryless channel distribution; and 
$p(\buw|\uw,\uby^n)$ results from 
the operation of all node decoders, 
each of which maps its received channel outputs and outgoing messages 
to reproductions of its incoming messages.  
Here $p(\ubx_t|\uby^{t-1},\uw)$ and $p(\buw|\uw,\uby^n)$ 
capture both the distribution over channel codes 
and the deterministic operation 
of the node encoders from $\cS(\cN)$. 

The corresponding distribution 
for solution $\cS(\ucN_R)$ on $\ucN_R$ is similar.  
In particular, since the distribution on messages is given 
and we employ all of the same codes, 
distributions $p(\uw)$, $p(\ubx_t|\uby^{t-1},\uw)$, and $p(\buw|\uw,\uy^n)$ 
remain unchanged.  
The only difference between $\ucS(\ucN)$ and $\cS(\ucN_R)$ 
is the replacement of channel $(\ucXV,p(\uyV|\uxV),\ucYV)$ 
by the random channel emulator.  
Thus at time $t$, solution $\cS(\ucN_R)$ replaces the channel distribution 
\[
p(\uyV|\uxV)=\prod_{\ell=1}^Np(\uyV(\ell)|\uxV(\ell))
\]
by the emulation distribution 
\[
\hp_t(\uyV|\uxV)\ \deff\ \Pr(\beta_{N,t}(\alpha_{N,t}(\uxV))=\uyV).  
\]
(Note that the channel emulator eventually applied is a deterministic source code.  
The given distribution reflects only the random code design.)  
Thus $\cS(\ucN_R)$ achieves distribution 
\begin{equation}\label{eqn:fullchar}
\hp(\uw,\ubx^n,\uby^n,\buw)
= p(\uw)
    \left[\prod_{t=1}^{n}p(\ubx_t|\uby^{t-1},\uw)\right]
    \left[\prod_{t=1}^{n}\hp_t(\uyV_t|\uxV_t)p(\uyVm|\uxVm)\right]
    p(\buw|\uw,\uby^n).  
\end{equation}
In general, $\hp_t(\uyV|\uxV)$ will not be precisely equal to 
the channel distribution $p(\uyV|\uxV)$ that it was designed to emulate. 
Lemma~\ref{lem:p2pconprob} in Appendix~\ref{app:p2pconprob} shows 
\begin{equation}\label{eqn:p2pconprob}
\hp_t(\uyV|\uxV) \leq p(\uyV|\uxV)2^{N(4a(\eps,t)+2\eps(t)+1/N)}
\end{equation}
for all $(\uxV_t,\uyV_t)\in\rtyp$.  
Let 
\[
\hp_t((\rtyp)^c|\uxV_t)\ \deff\ 
\Pr((\uXV_t,\uYV_t)\not\in\rtyp|\uXV_t=\uxV_t) 
\]
denote the conditional probability that $(\uXV_t,\uYV_t)\not\in\rtyp$ 
given $\uXV_t=\uxV_t$ under operation of code $\cS(\ucN_R)$.  
Using a proof similar to that for the rate-distortion theorem, 
Lemma~\ref{lem:p2pa1} in Appendix~\ref{app:p2pa1} shows 
\begin{equation}\label{eqn:p2pa1}
\hp_t((\rtyp)^c|\uxV_t)
\leq p_t((\rtyp)^c|\uxV_t)+e^{-2^{N(R-I(\XV_t;\YV_t)-2a(\eps,t)-\eps(t))}}.
\end{equation}

\underline{Step 6 - Bounding the expected error probability:  }\mbox{} \\
The following error analysis relies on both probabilities 
resulting from the operation of $\ucS(\ucN)$ on $\ucN$ 
and probabilities resulting from 
the operation of random code $\cS(\ucN_R)$ on $\ucN_R$.  
To avoid confusion between the two, 
we use $\Pr$ in the former case and $\hPr$ in the latter case.  

Define
\begin{equation} \label{eqn:baddef}
\bad\ \deff\ \left\{(\uxV,\uyV):\Pr\left(\left.\ubW\neq\uW\right|
                          (\uXV_t,\uYV_t)=(\uxV,\uyV)\right)\geq 2^{-N\delta/2} \right\}
\end{equation}
to be the set of input-output pairs 
on channel $(\ucXV,p(\uyV|\uxV),\ucYV)$ at time $t$ 
that are most likely to lead to errors 
in the operation of $\ucS(\ucN)$ on $\ucN$;  
we think of $\bad$ as the ``bad'' set.
For each $t\in\{1,\ldots,n\}$ 
we treat $(\uXV_t,\uYV_t)\not\in\rtyp\setminus\bad$ as an error event.  
We therefore define $G_t\subseteq(\ucXV\times\ucYV)^n$ as 
\[
G_t\ \deff\ \cup_{t'=1}^t\{(\uXV_{t'},\uYV_{t'})\in\rtypp\setminus\badp\}\ \ \mbox{for each } 
t\in\{1,\ldots,n\}
\]
and $G_0\ \deff\ (\ucXV\times\ucYV)^n$ 
to be the event that none of these error events 
has occurred in the first $t$ time steps;  
we think of each $G_t$ as a ``good'' set since it describes the event 
that channel input-output pairs up to time $t$ were typical and not ``bad.''  
Since $(G_n)^c=\cup_{t=1}^n((G_{t-1}\cap(\rtyp)^c)\cup(G_{t-1}\cap\rtyp\cap\bad))$, 
the union bound gives 
\[
\hPr\left(\ub{W}\neq\ub{\hat{W}}\right)
 \leq \sum_{t=1}^n\left[\hPr(G_{t-1}\cap(\rtyp)^c)
              +\hPr(G_{t-1}\cap\rtyp\cap\bad)\right]
       + \hPr\left(G_n\cap \{\uW\neq\ubW\}\right).
\]
This is an expected error probability since $\hPr(\cdot)$ 
captures the random code design in addition to 
the random message choice and random action 
of the channel $(\ucXVm,p(\uyVm|\uxVm),\ucYVm)$.  
To bound the first two terms in this sum,  
note that by~(\ref{eqn:fullchar}) and~(\ref{eqn:p2pconprob}), 
\begin{eqnarray}
\lefteqn{\hPr(G_{t-1}\cap\{\uXV_t=\uxV\})} \nonumber \\
& \leq & \sum_{(\uw,\ubx^{t-1},\uby^{t-1},\uxVm_t):
    (\uxV_{t'},\uyV_{t'})\in\rtyp\ \forall t'<t}
    p(\uw)\left[\prod_{t'=1}^{t}p(\ubx_{t'}|\uby^{t'-1},\uw)\right] \nonumber\\
&&\cdot\left[\prod_{t'=1}^{t-1}\hp_t(\uyV_{t'}|\uxV_{t'})
      p(\uyVm_{t'}|\uxVm_{t'})\right]\nonumber \\
&\leq& \sum_{(\uw,\ubx^{t-1},\uby^{t-1},\uxVm_t):
      (\uxV_{t'},\uyV_{t'})\in\rtyp\ \forall t'<t}
     2^{N\sum_{t'=1}^{t-1}(4a(\eps,t')+2\eps(t')+1/N)}
     p(\uw,\ubx^t,\uby^{t-1}) \nonumber\\
&\leq& 2^{N\sum_{t'=1}^{t-1}(4a(\eps,t')+2\eps(t')+1/N)}p_t(\uxV) 
\label{eqn:good}
\end{eqnarray}
for each $\uxV\in\ucXV$.  
This bound captures how the input distribution to node $i$ at time $t$ 
is affected by the replacement of the channel by its emulator 
in all previous time steps.  
Applying~(\ref{eqn:p2pa1}),~(\ref{eqn:good}), and~(\ref{eqn:p2ptyp}) gives 
\begin{eqnarray}
\lefteqn{\hPr(G_{t-1}\cap(\rtyp)^c)} \nonumber \\
& = & \sum_{\uxV\in\ucXV}\hPr(G_{t-1}\cap\{\uXV_t=\uxV\})\hp_t((\rtyp)^c|\uxV)
	\nonumber \\
& \leq & \left[\sum_{\uxV\in\ucXV}\hPr(G_{t-1}\cap\{\uXV_t=\uxV\})p_t((\rtyp)^c|\uxV)
	\right] 
	+e^{-2^{N(R-I(\XV_t;\YV_t)-2a(\eps,t)-\eps(t))}} \nonumber \\
& \leq & \sum_{\uxV\in\ucXV}2^{N\sum_{t'=1}^{t-1}(4a(\eps,t')+2\eps(t')+1/N)}
	p_t(\uxV) 
	p_t((\rtyp)^c|\uxV) 
	+e^{-2^{N(R-I(\XV_t;\YV_t)-2a(\eps,t)-\eps(t))}} \nonumber \\
& \leq & 2^{-N(c(\eps,t)-\sum_{t'=1}^{t-1}(4a(\eps,t')+2\eps(t')+1/N)}
	+e^{-2^{N(R-I(\XV_t;\YV_t)-2a(\eps,t)-\eps(t))}}. \label{eqn:nottyp} 
\end{eqnarray}

To bound $\hPr(G_{t-1}\cap\rtyp\cap\bad)$, 
recall that for all $N$ is sufficiently large 
$\ucS(\ucN)$ is a $(2^{-N\delta},\cR)$ solution for $\ucN$ 
and that there are fewer than $m^2$ messages to transmit.  
Thus for solution $\ucS(\ucN)$ on $\ucN$, 
$\Pr(\ubW\neq\uW) < m^2 2^{-N\delta}$ 
by the union bound, giving 
\begin{eqnarray*}
m^22^{-N\delta} 
&> & \Pr(\ubW\neq\uW) \\
&\geq & \sum_{(\uxV,\uyV)\in\bad}p_t(\uxV,\uyV)
             \Pr(\ubW\neq\uW|(\uxV,\uyV))\\
&\geq & 2^{-N\delta/2}p_t(\bad), 
\end{eqnarray*}
which implies $p_t(\bad)<m^22^{-N\delta/2}$ on $\ucN$.  
Thus for solution $\cS(\ucN_R)$ on $\ucN_R$,
\begin{eqnarray}
\lefteqn{\hPr(G_{t-1}\cap\rtyp\cap\bad)} \nonumber \\
& = & \sum_{\uxV\in\ucXV}\hPr(G_{t-1}\cap\{\uXV_t=\uxV\})
\hp_t(\rtyp\cap\bad|\uXV_t=\uxV) 	\nonumber\\
& \stackrel{(a)}{\leq} & 2^{N\sum_{t'=1}^{t-1}(4a(\eps,t')+2\eps(t')+1/N)}\sum_{\uxV\in\ucXV}
	p_t(\uxV)\hp_t(\rtyp\cap\bad|\uXV_t=\uxV) \nonumber \\
& \stackrel{(b)}{\leq} & 2^{N\sum_{t'=1}^{t}(4a(\eps,t')+2\eps(t')+1/N)}\sum_{\uxV\in\ucXV}
	p_t(\uxV)p_t(\rtyp\cap\bad|\uXV_t=\uxV) \nonumber \\
& < & m^22^{-N(\delta/2-\sum_{t'=1}^t(4a(\eps,t')+2\eps(t')+1/N))},
\label{eqn:bad}
\end{eqnarray}
where $(a)$ follows from~(\ref{eqn:good}), 
and $(b)$ follows from~(\ref{eqn:p2pconprob}).  Finally, 
\begin{eqnarray}
\lefteqn{\hPr\left(G_n\cap\{\uW\neq\ubW\}\right)} \nonumber \\
& \stackrel{(a)}{<} 
& \sum_{(\uw,\buw,\ubx^n,\uby^n):
	\uw\neq\ubw,(\uxV_t,\uyV_t)\in\rtyp\setminus\bad}p(\uw)
    \left[\prod_{t=1}^{n}p(\ubx_t|\uby^{t-1},\uw)\right] \nonumber \\
&&\cdot 2^{N\sum_{t=1}^{n}(4a(\eps,t)+2\eps(t)+1/N)}
\left[\prod_{t=1}^{n}p(\uby_t|\ubx_t)\right]p(\buw|\uw,\uby^n)\nonumber \\
&\stackrel{(b)}{\leq} & 2^{N\sum_{t=1}^n(4a(\eps,t)+2\eps(t)+1/N)}
      \sum_{(\uw,\ubw,\uxV_1,\uyV_1):\uw\neq\ubw,(\uxV_1,\uyV_1)\in\rtypo\setminus\bado}
	p(\uw,\uxV_1,\uyV_1,\ubw) \nonumber \\
& = & 2^{N\sum_{t=1}^n(4a(\eps,t)+2\eps(t)+1/N)}
		\sum_{(\uxV_1,\uyV_1)\in\rtypo\setminus\bado}p_1(\uxV_1,\uyV_1)
		\Pr(\ubW\neq\uW|(\uxV_1,\uyV_1)) \nonumber \\
&\stackrel{(c)}{<} & 2^{N\sum_{t=1}^n(4a(\eps,t)+2\eps(t)+1/N)}2^{-N\delta/2}
\label{eqn:error}
\end{eqnarray}
Equation ($a$) follows from~(\ref{eqn:fullchar}) and~(\ref{eqn:p2pconprob}).  
In ($b$), we sum $\ucXV\times\ucYV$ rather than $\rtyp\setminus\bad$ 
for all $t>1$.  
Equation ($c$) follows 
from the definition of $\bad$ in (\ref{eqn:baddef}) 
and the bound $p_1(\rtypo\setminus\bado)\leq 1$.  

\underline{Step 7 - Parameter choice:}\mbox{} \\
We finally show that we can choose typical set parameters 
$\eps=(\eps(1),\ldots,\eps(n))$ 
such that $\hPr(\uW\neq\ubW)<\lambda$ for all $N$ sufficiently large.  
Since $n$ is fixed and finite,~(\ref{eqn:nottyp}),~(\ref{eqn:bad}), 
and~(\ref{eqn:error}) 
imply that the expected error probability of $\cS(\ucN_R)$ 
goes to zero provided 
\begin{eqnarray*}
\sum_{t'=1}^{t-1}(4a(\eps,t')+2\eps(t')+1/N) & < & c(\eps,t) \\
2a(\eps,t)+\eps(t) & < & R-I(\XV_t;\YV_t) \\
\sum_{t=1}^n (4a(\eps,t')+2\eps(t')+1/N) & < & \delta/2.
\end{eqnarray*}
Recall that constants $a(\eps,t)$ (defined in (\ref{eqn:atdef})) and 
$c(\eps,t)$ (defined in Lemma~\ref{lem:p2ptyp} in Appendix~\ref{app:p2ptyp}) 
depend only on distribution $p_t(\uxV,\uyV)$ and the value $\eps(t)$.  
Each goes to 0 as $\eps(t)$ approaches 0.  
The following sequential choice of $\eps(n),\ldots,\eps(1)$ 
yields the desired result.  
Set $\eps(n)$ such that $4a(\eps,n)+2\eps(n)\leq \delta/(4n)$.  
Then for each subsequent $t$, 
set $\eps(t)$ such that 
\[
2a(\eps,t)+\eps(t)
\leq \min\left\{\frac{\delta}{4n},\frac{R-I_t(\XV_t;\YV_t)}{2},
	\frac{c(\eps,t+1)}{t+1},\ldots,\frac{c(\eps,n)}{n}\right\}.  
\]
This gives the desired result since 
$R>I(\uXV_t;\uYV_t)$ (by the theorem assumption and definition of capacity) 
and $\delta>0$.  

Since the expected error probability 
with respect to the given distribution over codes 
approaches zero as $N$ grows without bound, 
there must exist a single instance of the code $\cS(\ucN_R)$ 
that does at least as well.  
\IEEEQED  
\end{thm}

\begin{rem}
It is interesting to specify the choice of parameters 
in Theorems~\ref{thm:stacked} and~\ref{thm:main} 
required to guarantee the existence of a 
$(\tilde{\lambda},\cR)$-$\cS(\cN_R)$ solution 
for an arbitrary $\tilde{\lambda}>0$
and $\cR\in \mbox{int}(\setR(\cN))$. 
Since we have $\cR\in \mbox{int}(\setR(\cN))$ there exists
a $\tilde{\cR}\in \mbox{int}(\setR(\cN))$ with $\tilde{\cR}>\cR$.  
We choose $\rho$ in Theorem~\ref{thm:stacked} accordingly 
as $\min_{u,v}\{\tRuv-\Ruv\}$. 
Once $\rho$ is chosen, 
we choose $\lambda$ and $n$ so that the condition 
$\rho>\max_{u,v}\{\tRuv\}\lambda+h(\lambda)/n$ is satisfied 
for a $({\lambda},\cR)$-$\cS(\cN)$ solution of blocklength $n$. 
Note that $\Ruv$ is less than the capacity of the channel 
$p(\bWuv|\Wuv)$ imposed by this solution, so $\delta>0$. 
Fixing $\cS(\cN)$ fixes distributions $p_t(\xV)$.  
We next choose $\eps$ as specified above 
and design source code $(\alpha_{N,t},\beta_{N,t})$ 
for $N$ sufficiently large.  
The resulting code can be run on $\cN_R$ 
(rather than $\ucN_R$) as described 
in the proof of Theorem~\ref{thm:stacked}.  
\end{rem}

Corollary~\ref{cor:p2pequiv} finally proves network equivalence 
for point-to-point channels.  

\begin{cor}\label{cor:p2pequiv} 
Consider a pair of networks 
\begin{eqnarray*}
\cN & = & \left(\cXVm\times\cXV,p(\yV|\xV)p(\yVm|\xVm),\cYVm\times\cYV\right)\\
\cN_C & = & \left(\cXVm\times\{0,1\}^C,\delta(\tyV-\txV)p(\yVm|\xVm),\cYVm
	\times\{0,1\}^C\right),
\end{eqnarray*}
where $(\cXV,p(\yV|\xV),\cYV)$ is a channel of capacity 
$C=\max_{p(\xV)}I(\XV;\YV)>0$.  Then  
\[
\setR(\cN)=\setR(\cN_C).
\]
\Proof The result is immediate from 
Lemmas~\ref{lem:cont} and~\ref{lem:lb} and Theorem~\ref{thm:main}.  
\IEEEQED
\end{cor}

\section{Conclusions}\label{concl}

The preceding results show that the capacity of a memoryless network 
containing an independent point-to-point channel 
equals the capacity of another network 
where that noiseless channel 
is replaced by a noiseless bit pipe of the same capacity;  
thus any collection of demands 
(e.g., a collection of unicast demands) 
can be met on the first network if and only if it can be met on the second.  
Sequentially applying this result 
to each channel in a network of point-to-point channels 
proves that the capacity of a network 
of independent, memoryless, point-to-point channels 
equals the capacity of a network of noiseless bit-pipes 
of the corresponding capacities.  
This also implies that the capacity of a network of 
independent, memoryless, point-to-point channels 
equals the capacity of any other network 
of independent, memoryless, point-to-point channels 
of the same capacities.  
Thus, from the perspective of capacity, 
a Gaussian channel is no different from 
a binary erasure channel of the same capacity, 
despite the Gaussian channel's far broader range of possible behaviors.  
The given equivalence result proves the optimality 
of coding strategies that separate joint source and network coding 
from channel coding;  
there is no loss in capacity associated with performing 
independent channel coding on every point-to-point channel.  
The result also opens the way to the analysis of noisy networks 
using analytical and computational tools 
built for characterizing network coding capacities.  

In addition to proving the equivalence 
between networks of noisy channels 
and networks of point-to-point bit pipes, 
the other main contribution of this work 
is the introduction of a new strategy 
for tackling networks of noisy components.  
Lemma~\ref{lem:lb} and Theorem~\ref{thm:main} 
show that the capacity of one network 
is a subset of the capacity of another network 
by showing that any code that can be run 
with asymptotically negligible error probability 
on the first network can be run on the second network 
with similar error probability.  
In part~II of this paper, 
we apply the same approach 
in bounding the capacities of more general networks.  
This approach represents one step towards the goal 
of building computational tools for bounding capacities of networks 
using deterministic models of the network's component channels.  

\appendices

\section{Lemma~\ref{lem:p2ptyp}}\label{app:p2ptyp}
Lemma~\ref{lem:p2ptyp} proves that $p_t((\rtyp)^c)$ 
decays exponentially to zero.  
Using the notation of Section~\ref{sec:stack}, 
$\uXV=(\uXV(1),\ldots,\uXV(N))$ and $\uYV=(\uYV(1),\ldots,\uYV(N))$ 
denote $N$-dimensional vectors 
corresponding to the $N$-fold stacked network.  

\begin{lem}\label{lem:p2ptyp}
Let $(\uXV,\uYV)$ be drawn i.i.d.\ $p_t(\xV,\yV)$.
Then there exists a constant $c(\eps,t)>0$ for which 
\[
p_t((\rtyp)^c) < 2^{-Nc(\eps,t)} 
\]
for all $N$ sufficiently large.  
Constant $c(\eps,t)$ approaches 0 as $\eps(t)>0$ approaches 0.  

\Proof The result follows from Chernoff's bound 
which we apply to averages of i.i.d.\ random variables.  
Chernoff's bound states that for any i.i.d.\ random variables 
$A(1),A(2),\ldots,A(N)$, 
\[
\Pr\left(\frac1N\sum_{\ell=1}^N A(\ell)>a \right) 
\leq e^{N\min_{s>0}[M(s)-sa]},
\]
where $M(s)\deff\ln E[e^{sA}]$ and  
$\min_{s>0}[M(s)-sa]\leq 0$ for all $a\geq E[A]$ 
with equality if and only if $a=E[A]$ 
(see, for example, \cite[pp.482-484]{Madhow:08}).
Note that $|\min_{s>0}[M(s)-sa]|$ grows without bound 
as $a$ increases 
while $|\min_{s>0}[M(s)-sa]|$ approaches 0 
as $a$ approaches $E[A]$.  

We begin by applying the Chernoff bound to the following 
sequence of random variables 
\[
-\log p_t(\uXV(1)),\ldots,-\log p_t(\uXV(N)).
\]
We then negate the sequence and apply the Chernoff bound again.  
Combining these results with the union bound gives 
\[
p_t\left(\left|-\frac1N\log p_t(\uXV)-H(\XV_t)\right|>\eps(t)\right)
\leq 2^{-Nb_0+1}
\]
for some $b_0>0$ as discussed above.  
Likewise, for any $\eps'>0$, 
\begin{eqnarray*}
p_t\left(\left|-\frac1N\log p_t(\uYV)-H(\YV_t)\right|>\eps'\right) 
& \leq & 2^{-Nb_1+1}\\
p_t\left(\left|-\frac1N\log p_t(\uXV,\uYV)-H(\XV_t,\YV_t)\right|>\eps'\right)
& \leq & 2^{-Nb_2+1}
\end{eqnarray*}
for some $b_1,b_2>0$.  
Since $b_1$ and $b_2$ can be made arbitrarily large 
by choosing $\eps'$ large enough, 
the infimum in the definition of $a(\eps,t)$ is well-defined.  

Now note that 
\begin{eqnarray*}
p_t((\typ)^c)
&\leq& p_t\left(\left|-\frac1N\log p_t(\uXV)-H(\XV_t)\right|>\eps(t)\right) \\
& & +  p_t\left(\left|-\frac1N\log p_t(\uYV)-H(\YV_t)\right|>a(\eps,t)\right.\\
&&\left.\vee\left|-\frac1N\log p_t(\uXV,\uYV)-H(\XV_t,\YV_t)\right|>a(\eps,t)
         \right) \\
& \leq & 2^{-Nb_0+1}+2^{-N6\eps(t)}
\end{eqnarray*}
where the first inequality applies the union bound 
and the second inequality follows from our first Chernoff bound 
and the definition of $a(\eps,t)$.
Let 
\begin{eqnarray*}
C_t^{(N)} & \deff & \left\{\uxV\in\ucXV: 
\left|-\frac1N\log p_t(\uxV)-H(\XV_t)\right|\leq \eps(t), \right. \\
&& \left.p_t\left(\left.(\typ)^c\right|\uXV_t=\uxV\right)>2^{-3N\eps(t)}
\right\}.  
\end{eqnarray*}
Then 
\begin{eqnarray*}
p_t((\rtyp)^c) & = & p_t\left((\typ)^c\right) 
+p_t\left(\{(\uxV,\uyV)\in\typ:\uxV\in C_t^{(N)}\}\right) \\
& = & p_t\left((\typ)^c\right) +\sum_{\uxV\in C_t^{(N)}}p_t(\uxV)
      p\left(\left.\typ\right|\uXV_t=\uxV\right) \\
& \leq & p_t\left((\typ)^c\right) +p_t\left(C_t^{(N)}\right) \\
& \leq & 2^{-Nb_0+1}+2^{-N6\eps(t)}+p_t\left(C_t^{(N)}\right).
\end{eqnarray*}
To bound $p_t(C_t^{(N)})$, note that from the definitions of $a(\eps,t)$ 
and $C_t^{(N)}$, 
\begin{eqnarray*}
2^{-N6\eps(t)}
& \geq & \sum_{\uxV\in C_t^{(N)}}p_t(\uxV)p\left(\left.
	\left|-\frac1N\log p_t(\uYV)-H(\YV_t)\right|>a(\eps,t)
	\right.\right. \\
&&	\left.\left.\vee\left|-\frac1N\log p_t(\uXV,\uYV)-H(\XV_t,\YV_t)\right|
	>a(\eps,t)\right|\uXV_t=\uxV\right) \\
& > & p_t(C_t^{(N)})2^{-3N\eps(t)}.  
\end{eqnarray*}
Thus $p_t(C_t^{(N)}) < 2^{-N3\eps(t)}$, which gives the desired result.  
\IEEEQED
\end{lem}

\section{Lemma~\ref{lem:p2pconprob}}\label{app:p2pconprob}

Lemma~\ref{lem:p2pconprob} bounds 
the distribution $\hat{p}_t(\uyV|\uxV)$ 
obtained by random source code $(\alpha_{N,t},\beta_{N,t})$.  
Our restriction on the typical set 
is useful for that proof.  
The randomness in $\hat{p}_t(\uyV|\uxV)$ 
results from the random source code choice.  
Lemmas~\ref{lem:phat} and~\ref{lem:qt}
are intermediate steps used in the proof of Lemma~\ref{lem:p2pconprob}.  

Let function $K_t(\uxV,\uyV)$  be defined as 
\begin{equation}
K_t(\uxV,\uyV)\ \deff\ \left\{\begin{array}{ll} 
     1 & \mbox{if $(\uxV,\uyV)\in\rtyp$} \\
     0 & \mbox{otherwise}
\end{array}\right.\label{eqn:Kt}
\end{equation}
(cf. \cite[steps 10.93-10.102]{CoverT:06}).  
Lemma~\ref{lem:phat}, below, characterizes $\hat{p}_t(\uyV|\uxV)$ 
as a function of the probability 
\[
q_t(\uxV)\deff\sum_{\uyV\in\ucYV}K_t(\uxV,\uyV)p_t(\uyV)  
\]
that a single codeword drawn at random 
is jointly typical with $\uxV$.  
Precisely, the lemma shows that 
$p_t(\uyV)/q_t(\uxV)$ is the probability 
that $\uxV$ is mapped to $\uyV$ 
given that there is at least one codeword in the codebook 
that is typical with $\uxV$.  
Lemma~\ref{lem:qt} then bounds $q_t(\uxV)$ 
for all $\uxV$ satisfying the conditions of $\rtyp$.  

\begin{lem}\label{lem:phat}
Let $(\alpha_{N,t},\beta_{N,t})$ be the random source code 
defined in (\ref{eqn:p2pdec}) and (\ref{eqn:p2penc}).  
Then for any $(\uxV,\uyV)\in\rtyp$, 
\[
\hat{p}_t(\uyV|\uxV)= p_t(\uyV)\frac{1-(1-q_t(\uxV))^{2^{NR}}}{q_t(\uxV)}.
\]
\Proof
Recall that $q_t(\uxV)$ 
is the probability that a single randomly drawn codeword $\uYV$ 
satisfies $(\uxV,\uYV)\in\rtyp$.  
Using the given random code design, 
for any $(\uxV,\uyV)\in\rtyp$, 
\begin{eqnarray*}
\lefteqn{\hat{p}_t(\uyV|\uxV)} \\
& = &  \sum_{j=1}^{2^{NR}}\sum_{k=1}^j
		\left(\begin{array}{c} 2^{NR} \\ j \end{array}\right)
		\left(\begin{array}{c} j \\ k \end{array}\right) 
		(1-q_t(\uxV))^{2^{NR}-j} 
		(q_t(\uxV)-p_t(\uyV))^{j-k} (p_t(\uyV))^k\frac{k}{j} \\
& = & p_t(\uyV)\sum_{j=1}^{2^{NR}}
	\left(\begin{array}{c} 2^{NR} \\ j \end{array}\right)\frac1{j}
		(1-q_t(\uxV))^{2^{NR}-j} 
\sum_{k=1}^j	\left(\begin{array}{c} j \\ k \end{array}\right)
		[a^{j-k}kb^{k-1}].  
\end{eqnarray*}
Here 
$j$ is the number of codewords 
that are jointly typical with $\uxV$, 
$k$ is the number of those codewords that equal $\uyV$, 
and term $k/j$ follows from the uniform distribution 
over jointly typical codewords in the encoder design.  
In the second equality, $a=q_t(\uxV)-p_t(\uyV)$ and $b=p_t(\uyV)$.  
Thus 
\begin{eqnarray*}
\hat{p}_t(\uyV|\uxV) 
& = &  p_t(\uyV) \sum_{j=1}^{2^{NR}}
		\left(\begin{array}{c} 2^{NR} \\ j \end{array}\right)
		\frac1{j} (1-q_t(\uxV))^{2^{NR}-j} 
\frac{\partial}{\partial b}[(a+b)^j-a^j] \\
& = &  p_t(\uyV) \sum_{j=1}^{2^{NR}}
		\left(\begin{array}{c} 2^{NR} \\ j \end{array}\right)
		\frac1{j} (1-q_t(\uxV))^{2^{NR}-j} 
j(q_t(\uxV))^{j-1} \\
& = &  p_t(\uyV)\frac{1-(1-q_t(\uxV))^{2^{NR}}}{q_t(\uxV)}.
\end{eqnarray*}	
\IEEEQED
\end{lem}

\begin{lem}\label{lem:qt}
Given $\uxV\in\ucXV$, if 
$\left|-\frac1N\log p_t(\uxV)-H(\XV_t)\right|\leq\eps(t)$ and 
$p_t((\typ)^c|\uxV) < 2^{-3N\eps(t)}$,
then 
\[
q_t(\uxV)\geq 2^{-N(I(\XV_t;\YV_t)+\eps(t)+2a(\eps,t)+\frac1N)}
\]
for all $N$ sufficiently large.  

\Proof
For any $\uxV$ satisfying the given constraints, 
we first derive a bound on the number of $\uyV$ values 
for which $(\uxV,\uyV)\in\rtyp$.  
This is obtained by drawing a random variable $\uYV$ 
according to conditional distribution 
$\prod_{\ell=1}^N p_t(\uyV(\ell)|\uxV(\ell))$ 
and showing that $(\uxV,\uYV)\in\rtyp$ with probability approaching 1.  
Since all $\uyV$ that are jointly typical with $\uxV$ 
are approximately equally probable, this probability bound 
leads to a bound on the number of $\uyV$ vectors that are jointly typical with $\uxV$ 
and then to a bound on the desired probability.  

By the lemma assumptions, 
\[
p_t((\uXV,\uYV)\not\in\typ|\uXV=\uxV)<2^{-3N\eps(t)}, 
\]
which approaches 0 as $N$ grows without bound. 
Let $F_t(\uxV)\deff\{\uyV:(\uxV,\uyV)\in\typ\}$.  
Then for $N$ sufficiently large, 
\begin{eqnarray*}
\frac12 
\leq 1-2^{-3N\eps(t)} 
&\leq & \sum_{\uyV\in F_t(\uxV)} p(\uyV|\uxV) \\
& = & \sum_{\uyV\in F_t(\uxV)} \frac{p_t(\uxV,\uyV)}{p_t(\uxV)} \\
&\leq& |F_t(\uxV)| 2^{-N(H(\YV_t|\XV_t)-a(\eps,t)-\eps(t))},
\end{eqnarray*}
where the last inequality follows 
from the usual probability bounds for typical strings.  
Thus 
\[
|F_t(\uxV)|\geq 2^{N(H(\YV_t|\XV_t)-a(\eps,t)-\eps(t)-1/N)}, 
\] 
which we apply to bound $q_t(\uxV)$ as 
\begin{eqnarray*}
q_t(\uxV) & = & \sum_{\uyV\in F_t(\uxV)} p_t(\uyV) \\
         & \geq & |F_t(\uxV)| 2^{-N(H(\YV_t)+a(\eps,t))} \\
         & \geq & 2^{-N(I(\XV_t;\YV_t)+2a(\eps,t)+\eps(t)+1/N)}.  
\end{eqnarray*}
\IEEEQED
\end{lem}

\begin{lem}\label{lem:p2pconprob}
For all $(\uxV,\uyV)\in\rtyp$, 
\[
\hp_t(\uyV|\uxV)\leq p(\uyV|\uxV)2^{N(4a(\eps,t)+2\eps(t)+1/N)}.
\]

\Proof  
By Lemmas~\ref{lem:phat} and~\ref{lem:qt} 
and the usual bounds on the probabilities of typical elements, 
\begin{eqnarray*}
\hp_t(\uyV|\uxV) 
& = &  p_t(\uyV)\frac{1-(1-q_t(\uxV))^{2^{NR}}}{q_t(\uxV)} 
\leq \frac{p_t(\uyV)}{q_t(\uxV)} \\
&\leq&  p(\uyV|\uxV)\frac{p_t(\uxV)p_t(\uyV)}{p_t(\uxV,\uyV)} 
 	\frac{1}{2^{-N(I(\XV_t;\YV_t)+2a(\eps,t)+\eps(t)+1/N)}} \\
&\leq& p(\uyV|\uxV)\frac{2^{-N(I(\XV_t;\YV_t)-2a(\eps,t)-\eps(t))}} 
			{2^{-N(I(\XV_t;\YV_t)+2a(\eps,t)+\eps(t)+1/N)}} \\
&=& p(\uyV|\uxV)2^{N(4a(\eps,t)+2\eps(t)+1/N)}.  
\end{eqnarray*}	
\IEEEQED
\end{lem}

\section{Lemma~\ref{lem:p2pa1}}\label{app:p2pa1}

Lemma~\ref{lem:p2pa1} bounds the conditional probability that 
$(\uXV_t,\uYV_t)$ is not jointly typical under the operation of $\cS(\uhcN)$.  

\begin{lem}\label{lem:p2pa1}
For all $\uxV\in\ucXV$, 
\begin{eqnarray*}
\hp_t((\rtyp)^c|\uxV)
& \leq & p_t((\rtyp)^c|\uxV)+e^{-2^{N(R-I(\XV_t;\YV_t)-2a(\eps,t)-\eps(t))}}.
\end{eqnarray*}
\Proof  
If $|-(1/N)\log p_t(\uxV)-H(\XV_t)|>\eps(t)$ 
or $p_t((\typ)^c|\uxV)>2^{-3N\eps(t)}$, 
then 
\[
\hp_t((\rtyp)^c|\uxV)=p_t((\rtyp)^c|\uxV)=1 
\]
by definition of $\rtyp$.  
Otherwise, $(\uxV_t,\uyV_t)\not\in\rtyp$ when 
none of the $2^{NR}$ codewords of $\beta_{N,t}$ is jointly typical with $\uxV_t$.  
In this case, using definition (\ref{eqn:Kt}) 
and following the proof of the rate-distortion theorem, 
\begin{eqnarray*}
\hp_t((\rtyp)^c|\uxV)
& = & \left(1-\sum_{\uyV}p_t(\uyV)K_t(\uxV,\uyV)\right)^{2^{NR}} \\
& \stackrel{(a)}{\leq} & 1-\sum_{\uyV} p(\uyV|\uxV)K_t(\uxV,\uyV)
	+e^{-2^{N(R-I(\XV_t;\YV_t)-2a(\eps,t)-\eps(t))}} \\
& = & p_t((\rtyp)^c|\uXV=\uxV)+e^{-2^{N(R-I(\XV_t;\YV_t)-2a(\eps,t)-\eps(t))}}.  
\end{eqnarray*} 
where $(a)$ follows from $(1-ab)^k\leq 1-a+e^{-bk}$~\cite[Lemma 10.5.3]{CoverT:06} 
and the usual bounds on probabilities of typical strings 
\begin{eqnarray*}
p_t(\uyV) & = & p(\uyV|\uxV)\frac{p_t(\uyV)p_t(\uxV)}{p_t(\uxV,\uyV)} \\
             &\geq& p(\uyV|\uxV)2^{-N(I(\XV_t;\YV_t)+2a(\eps,t)+\eps(t))}. 
\end{eqnarray*}
for all $\uxV\in\ucXV$.  
\IEEEQED
\end{lem}

\bibliographystyle{ieeetr}



\clearpage
\setcounter{page}{1}
\setcounter{equation}{0}
\setcounter{section}{0}
\setcounter{rem}{0}
\setcounter{thm}{0}
\setcounter{exam}{0}
\setcounter{algo}{0}
\setcounter{defi}{0}
\setcounter{figure}{0}
\setcounter{table}{0}

\let\section=\savedsection





\title{
A Theory of Network Equivalence \\
{\LARGE Part II:  Multiterminal Channels}
}

\def\Ruv{R^{\ar{u}{v}}}
\def\hPr{\widehat{\Pr}}
\def\tX{\tilde{X}}
\def\vX{X^{(v)}}
\def\vcX{\cX^{(v)}}
\def\ubX{\ub{\bf X}}
\def\uvX{\uX^{(v)}}
\def\uvx{\ux^{(v)}}
\def\uvcX{\ucX^{(v)}}
\def\xC{x^{V_1}}
\def\XC{X^{V_1}}
\def\uxC{\ux^{V_1}}
\def\uXC{\uX^{V_1}}
\def\tXC{\tX^{V_1}}
\def\txC{\tx^{V_1}}
\def\tXI{\tX^{V_o}}
\def\txI{\tx^{V_o}}
\def\utX{\ub{\tX}}
\def\uvtX{\utX^{(v)}}
\def\utXC{\utX^{V_1}}
\def\utx{\ub{\tx}}
\def\utxC{\utx^{V_1}}
\def\uhtxC{\hat{\utx}^{V_1}}
\def\XCm{X^{-V_1}}
\def\xCm{x^{-V_1}}
\def\uXCm{\uX^{-V_1}}
\def\uxCm{\ux^{-V_1}}
\def\cXC{\cX^{V_1}}
\def\tcXC{\tcX^{V_1}}
\def\tcXI{\tcX^{V_o}}
\def\tcXiB{\tcX^{\ar{\{i\}}{B}}}
\def\tcXAB{\tcX^{\ar{A}{B}}}
\def\utcX{\ub{\tcX}}
\def\utcXC{\utcX^{V_1}}
\def\utcXiB{\utcX^{\ar{\{i\}}{B}}}
\def\utcXAB{\utcX^{\ar{A}{B}}}
\def\cXCm{\cX^{-V_1}}
\def\ucXC{\ucX^{V_1}}
\def\ucXCm{\ucX^{-V_1}}
\def\vY{Y^{(v)}}
\def\ubY{\ub{\bf Y}}
\def\uvY{\ub{Y}^{(v)}}
\def\tY{\tilde{Y}}
\def\hvcY{\hcY^{(v)}}
\def\YC{Y^{V_2}}
\def\tYC{\tY^{V_2}}
\def\tyC{\ty^{V_2}}
\def\utyC{\ub{\ty}^{V_2}}
\def\utYC{\ub{\tY}^{V_2}}
\def\tYI{\tY^{V_o}}
\def\tyI{\ty^{V_o}}
\def\utY{\ub{\tY}}
\def\uvtY{\utY^{(v)}}
\def\yC{y^{V_2}}
\def\YCm{Y^{-V_2}}
\def\yCm{y^{-V_2}}
\def\uYCm{\uY^{-V_2}}
\def\uyCm{\uy^{-V_2}}
\def\cYC{\cY^{V_2}}
\def\tcYC{\tilde{\cY}^{V_2}}
\def\tcYI{\tilde{\cY}^{V_o}}
\def\cYCm{\cY^{-V_2}}
\def\vcY{\cY^{(v)}}
\def\uYC{\uY^{V_2}}
\def\uyC{\uy^{V_2}}
\def\ucYC{\ucY^{V_2}}
\def\ucYCm{\ucY^{-V_2}}
\def\uvcY{\ub{\cY}^{(v)}}
\def\Wuv{W^{\ar{u}{v}}}
\def\Wvu{W^{\ar{v}{u}}}
\def\Wvo{W^{\ar{v}{1}}}
\def\Wvm{W^{\ar{v}{m}}}
\def\cWuv{\cW^{\ar{u}{v}}}
\def\cWvp{\cW^{\ar{v}{v'}}}
\def\uWuv{\uW^{\ar{u}{v}}}
\def\uWvo{\uW^{\ar{v}{1}}}
\def\uWvm{\uW^{\ar{v}{m}}}
\def\ucWuv{\ucW^{\ar{u}{v}}}
\def\ucWvp{\ucW^{\ar{v}{v'}}}
\def\ucWvu{\ucW^{\ar{v}{u}}}
\def\bWuv{\bW^{\ar{u}{v}}}
\def\ubWuv{\ubW^{\ar{u}{v}}}
\def\ubWvu{\ubW^{\ar{v}{u}}}
\def\tcW{\tilde{\cW}}
\def\tcWuv{\tcW^{\ar{u}{v}}}
\def\tcWvp{\tcW^{\ar{v}{v'}}}
\def\utWuv{\utW^{\ar{u}{v}}}
\def\utWvo{\utW^{\ar{v}{1}}}
\def\utWvm{\utW^{\ar{v}{m}}}
\def\utcW{\ub{\tcW}}
\def\utcWuv{\utcW^{\ar{u}{v}}}
\def\btWuv{\btW^{\ar{u}{v}}}
\def\ubtWuv{\ubtW^{\ar{u}{v}}}
\def\rtyp{\hat{A}_{\eps}^{(N)}}
\def\bad{B_t^{(N)}}
\def\badp{B_{t'}^{(N)}}
\def\bado{B_1^{(N)}}

\def\hp{\hat{p}}
\def\hcC{\hat{\cC}}
\def\cRC{\cR_{\cC}}
\def\cIC{\cI_{\cC}}
\def\cRL{\cR_L}
\def\cRU{\cR_U}
\def\cRCL{\cR_{\cC,L}}
\def\cRCU{\cR_{\cC,U}}
\def\cRCLk{\cR_{\cC,L,k}}
\def\cRCUk{\cR_{\cC,U,k}}
\def\tC{\tilde{\cC}}
\def\ucS{\ub{\cS}}
\def\bx{{\bf x}}
\def\bX{{\bf X}}
\def\ubx{\ub{\bf x}}
\def\hx{\hat{x}}
\def\cx{\check{x}}
\def\tx{\tilde{x}}
\def\tixo{\tx^{(i_1,1)}}
\def\tixt{\tx^{(i_2,1)}}
\def\hX{\hat{X}}
\def\uhX{\ub{\hX}}
\def\uhXC{\ub{\hX}^{V_1}}
\def\tixs{\tx^{(i,s)}}
\def\tcX{\tilde{\cal X}}
\def\tciXo{\tcX^{(i_1,1)}}
\def\tciXt{\tcX^{(i_2,1)}}
\def\hcX{\hat{\cal X}}
\def\iX{X^{(i)}}
\def\ixs{x^{(i,1)}}
\def\cixs{\cx^{(i,1)}}
\def\iXs{X^{(i,1)}}
\def\ciXs{\cX^{(i,1)}}
\def\tciXs{\tcX^{(i,1)}}
\def\ixo{x^{(i_1,1)}}
\def\iXo{X^{(i_1,1)}}
\def\ciXo{\cX^{(i_1,1)}}
\def\ixt{x^{(i_2,1)}}
\def\iXt{X^{(i_2,1)}}
\def\ciXt{\cX^{(i_2,1)}}
\def\ux{\ub{x}}
\def\uX{\ub{X}}
\def\ucX{\ub{\cX}}
\def\uiX{\uX^{(i)}}
\def\uhiX{\uhX^{(i)}}
\def\uciX{\ucX^{(i)}}
\def\uhcX{\ub{\hcX}}

\def\ucY{\ub{\cY}}
\def\uicY{\ucY^{(i)}}
\def\uicYo{\ucY^{(i_1)}}
\def\uicYt{\ucY^{(i_2)}}
\def\by{{\bf y}}
\def\bY{{\bf Y}}
\def\uby{\ub{\bf y}}
\def\iY{Y^{(i)}}
\def\jYr{Y^{(j,1)}}
\def\uy{\ub{y}}
\def\uY{\ub{Y}}
\def\uiY{\ub{Y}^{(i)}}
\def\jyr{y^{(j,1)}}
\def\cjyr{\cy^{(j,1)}}
\def\jyo{y^{(j_1,1)}}
\def\jyt{y^{(j_2,1)}}
\def\tjyo{\ty^{(j_1,1)}}
\def\tjyt{\ty^{(j_2,1)}}
\def\jYo{Y^{(j_1,1)}}
\def\jYt{Y^{(j_2,1)}}
\def\cy{\check{y}}
\def\hy{\hat{y}}
\def\ty{\tilde{y}}
\def\hY{\hat{Y}}
\def\hjyr{\hy^{(j,1)}}
\def\tjyr{\ty^{(j,1)}}
\def\hjyo{\hy^{(j_1,1)}}
\def\hjyt{\hy^{(j_2,1)}}
\def\hcY{\hat{\cal Y}}
\def\hcjYo{\hcY^{(j_1,1)}}
\def\hcjYt{\hcY^{(j_2,1)}}
\def\tcY{\tilde{\cY}}
\def\tcjYr{\tcY^{(j,1)}}
\def\tcjYo{\tcY^{(j_1,1)}}
\def\tcjYt{\tcY^{(j_2,1)}}
\def\ujyo{\uy^{(j_1,1)}}
\def\ujyt{\uy^{(j_2,1)}}
\def\ujYr{\ub{Y}^{(j,1)}}
\def\ujYo{\ub{Y}^{(j_1,1)}}
\def\ujYt{\ub{Y}^{(j_2,1)}}
\def\cjYo{\cY^{(j_1,1)}}
\def\cjYt{\cY^{(j_2,2)}}
\def\uhy{\ub{\hy}}
\def\uhY{\ub{\hY}}
\def\uhcY{\ub{\hcY}}
\def\bw{\hat{w}}
\def\uw{\ub{w}}
\def\ubw{\ub{\bw}}
\def\uW{\ub{W}}
\def\ucW{\ub{\cW}}
\def\bW{\hat{W}}
\def\ubW{\ub{\bW}}
\def\tw{\tilde{w}}
\def\tW{\tilde{W}}
\def\ctW{\tilde{\cW}}
\def\utW{\ub{\tW}}
\def\uctW{\ub{\ctW}}
\def\btw{\hat{\tw}}
\def\btW{\hat{\tW}}
\def\ubtW{\hat{\ub{\tW}}}
\def\ucN{\ub{\cN}}
\def\ucjYr{\ucY^{(j,1)}}
\def\cjYr{\cY^{(j,1)}}

\def\uu{\ub{u}}
\def\uU{\ub{U}}
\def\uv{\ub{v}}
\def\uV{\ub{V}}
\def\hu{\hat{u}}
\def\hU{\hat{U}}
\def\ucU{\ub{\cU}}
\def\uhU{\ub{\hU}}
\def\uo{u_1}
\def\ut{u_2}
\def\Uo{U_1}
\def\Ut{U_2}
\def\uuo{\uu_1}
\def\uut{\uu_2}
\def\uUo{\uU_1}
\def\uUt{\uU_2}
\def\ucUo{\ucU_1}
\def\ucUt{\ucU_2}

\def\eps{\epsilon}
\def\typ{A_{\eps}^{(N)}}

%
%
%

\maketitle

\begin{abstract}
The equivalence tools used in Part~I 
to study networks of independent, noisy, memoryless, point-to-point channels 
are here extended to networks containing more general channel types.  
Definitions of upper and lower bounding channel models are introduced.  
By these definitions, 
a collection of communication demands can be met 
on a network of independent channels 
if it can be met on a network
where each channel is replaced by 
its lower bounding model 
and only if it can be met on a network 
where each channel is replaced by 
its upper bounding model.  
This work derives general conditions 
under which a network of noiseless bit pipes 
is an upper or lower bounding model for a multiterminal channel.  
Example upper and lower bounding models 
for broadcast, multiple access, and interference channels 
are given.  
It is then shown that bounding the difference 
between the upper and lower bounding models for a given channel 
yields bounds on the accuracy 
of network capacity bounds derived using those models.  
By bounding the capacity of a network of independent noisy channels 
by the network coding capacity of a network of noiseless bit pipes, 
this approach represents one step towards the goal 
of building computational tools for bounding network capacities.  
\end{abstract}

{\bf Keywords:  Capacity, network coding, equivalence, component models}

\section{Introduction}

This work is motivated by the desire to build computational tools 
for characterizing the capacities of networks.  
Traditionally, the information theoretic 
investigation of network capacities 
has proceeded largely by studying example networks.  
Shannon's original proof of the capacity 
of a network described by 
a single point-to-point channel~\cite{ZZZ-Shannon:48} 
was followed by 
Ahlswede's~\cite{ZZZ-Ahlswede:71} and Liao's~\cite{ZZZ-Liao:72} 
capacity derivations for a single multiple access channel, 
Cover's early work on a single 
broadcast channel~\cite{ZZZ-Cover:72}, 
and so on.  
While the solution to one network capacity problem 
may lend some insight into future problems, 
deriving the capacity of each new network 
is often difficult.  
As a result, even the capacities for three-node networks 
remain incompletely solved.  

The problem is further complicated 
by the fact that the capacities of individual channels 
can vastly underestimate the rates that those channels can carry 
in larger networks.  
For example, consider the network in Figure~\ref{ZZZ-fig:bcmac}(a), 
where a broadcast channel $p(y^{(2)},y^{(3)}|x^{(1)})$ 
is followed by a multiple access channel $p(y^{(4)}|x^{(2)},x^{(3)})$.  
The two channels are independent, giving 
\[
p(y^{(2)},y^{(3)},y^{(4)}|x^{(1)},x^{(2)},x^{(3)})
=p(y^{(2)},y^{(3)}|x^{(1)})p(y^{(4)}|x^{(2)},x^{(3)}).  
\]
Example~\ref{ZZZ-ex:bcmac} shows that the maximal rate 
for a single unicast demand from source node~1 to sink node~4 
can far exceed the maximal sum-rate 
in the broadcast channel's capacity region.  
Example~\ref{ZZZ-ex:bcmacind} provides another related example.  
Both examples show that reliable transmission across a network 
does not require reliable transmission 
across each channel in the network 
and that restricting each component to transmit reliably -- 
that is employing a separated network and channel coding strategy 
that makes each channel individually reliable -- 
sometimes decreases the network capacity.  
\begin{exam}\label{ZZZ-ex:bcmac}
Figure~\ref{ZZZ-fig:bcmac}(a) shows a four-node network 
comprising a Gaussian broadcast channel 
followed by a real additive multiple access channel.  
\begin{figure}
  \begin{center}
  \begin{picture}(172,99)(5,-32)
      \thicklines
      \put(5,20){\circle*{5}}
      \put(5,16){\makebox(0,0)[ct]{\tiny 1}}
      \put(5,20){\line(1,0){35}}
      \put(20,22){\makebox(0,0)[cb]{\small $X^{(1)}$}}
      \put(40,20){\vector(1,1){10}}
      \put(40,20){\vector(1,-1){10}}
      \put(53.5,33.5){\circle{10}}
      \put(53.5, 6.5){\circle{10}}
      \put(53.5,33.5){\makebox(0,0)[cc]{$+$}}
      \put(53.5,6.5){\makebox(0,0)[cc]{$+$}}
      \put(53.5,48.5){\vector(0,-1){10}}
      \put(53.5,-8.5){\vector(0,1){10}}
      \put(53.5,50){\makebox(0,0)[cb]{\small $Z^{(2)}$}}
      \put(53.5,-10){\makebox(0,0)[ct]{\small $Z^{(3)}$}}
      \put(57,37){\vector(1,1){30}}
      \put(57, 3){\vector(1,-1){30}}
      \put(72,53){\makebox(0,0)[lt]{\small $Y^{(2)}$}}
      \put(72,-10){\makebox(0,0)[lb]{\small $Y^{(3)}$}}
      \put(33,-20){\framebox(35,80){}}
      \put(33,-24){\makebox(35,0)[ct]{BC}}
      \put(87,67){\circle*{5}}
      \put(87,-27){\circle*{5}}
      \put(87,63){\makebox(0,0)[ct]{\tiny 2}}
      \put(87,-23){\makebox(0,0)[cb]{\tiny 3}}
      \put(95,60){\makebox(0,0)[lb]{\small $X^{(2)}$}}
      \put(93,-20){\makebox(0,0)[lt]{\small $X^{(3)}$}}
      \put(87,67){\vector(1,-1){42}}
      \put(87,-27){\vector(1,1){42}}
      \put(132,20){\circle{10}}
      \put(132,20){\makebox(0,0)[cc]{$+$}}
      \put(110,-20){\framebox(35,80){}}
      \put(110,-24){\makebox(35,0)[ct]{MAC}}
      \put(137,20){\vector(1,0){40}}
      \put(177,20){\circle*{5}}
      \put(177,16){\makebox(0,0)[ct]{\tiny 4}}
      \put(160,22){\makebox(0,0)[cb]{\small $Y^{(4)}$}}
      \put(87,-43){\makebox(0,0){(a)}}
      \end{picture} 
      \hspace{.3in}
  \begin{picture}(172,99)(5,-32)
      \thicklines
      \put(5,20){\circle*{5}}
      \put(5,16){\makebox(0,0)[ct]{\tiny 1}}
      \put(5,20){\line(1,0){35}}
      \put(20,22){\makebox(0,0)[cb]{\small $X^{(1)}$}}
      \put(40,20){\vector(1,1){10}}
      \put(40,20){\vector(3,1){10}}
      \put(40,20){\vector(1,-1){10}}
      \put(53.5,33.5){\circle{10}}
      \put(53.5,23.5){\circle{10}}
      \put(53.5, 6.5){\circle{10}}
      \put(53.5,33.5){\makebox(0,0)[cc]{$+$}}
      \put(53.5,23.5){\makebox(0,0)[cc]{$+$}}
      \put(53.5,17){\makebox(0,0)[cc]{$.$}}
      \put(53.5,15){\makebox(0,0)[cc]{$.$}}
      \put(53.5,13){\makebox(0,0)[cc]{$.$}}
      \put(53.5,6.5){\makebox(0,0)[cc]{$+$}}
      \put(53.5,48.5){\vector(0,-1){10}}
      \put(67,37){\vector(-1,-1){10}}
      \put(53.5,-8.5){\vector(0,1){10}}
      \put(53.5,50){\makebox(0,0)[cb]{\tiny $Z^{(2)}$}}
      \put(69,37){\makebox(0,0)[cb]{\tiny $Z^{(3)}$}}
      \put(53.5,-10){\makebox(0,0)[ct]{\tiny $Z^{(m+1)}$}}
      \put(57,37){\vector(1,1){30}}
      \put(58,26){\vector(3,1){28}}
      \put(57, 3){\vector(1,-1){30}}
      \put(85,63){\makebox(0,0)[rb]{\small $Y^{(2)}$}}
      \put(85,-20){\makebox(0,0)[rt]{\small $Y^{(m+1)}$}}
      \put(33,-20){\framebox(43,80){}}
      \put(33,-24){\makebox(45,0)[lt]{BC}}
      \put(87,67){\circle*{5}}
      \put(87,35.5){\circle*{5}}
      \put(87,-27){\makebox(0,62.5)[cc]{$\vdots$}}
      \put(87,-27){\circle*{5}}
      \put(87,63){\makebox(0,0)[ct]{\tiny 2}}
      \put(87,31.5){\makebox(0,0)[ct]{\tiny 3}}
      \put(87,-30){\makebox(0,0)[ct]{\tiny $m+1$}}
      \put(91,63){\makebox(0,0)[lb]{\small $X^{(2)}$}}
      \put(91,36){\makebox(0,0)[lb]{\small $X^{(3)}$}}
      \put(91,-20){\makebox(0,0)[lt]{\small $X^{(m+1)}$}}
      \put(87,67){\vector(1,-1){42}}
      \put(87,35.5){\vector(3,-1){40}}
      \put(87,-27){\vector(1,1){42}}
      \put(132,20){\circle{10}}
      \put(132,20){\makebox(0,0)[cc]{$+$}}
      \put(110,-20){\framebox(43,80){}}
      \put(110,-24){\makebox(43,0)[rt]{MAC}}
      \put(137,20){\vector(1,0){40}}
      \put(177,20){\circle*{5}}
      \put(177,16){\makebox(0,0)[ct]{\tiny $m+2$}}
      \put(157,22){\makebox(0,0)[lb]{\small $Y^{(m+2)}$}}
      \put(87,-43){\makebox(0,0){(b)}}
      \end{picture} 
  \end{center}
\caption{Separate network and channel coding 
         fails to achieve the unicast capacity 
	 of (a) a four-node network
	 with dependent noise at the receivers of the broadcast channel 
	 and (b) an $(m+2)$-node network 
	 with independent noise at the receivers of the broadcast channel.}
\label{ZZZ-fig:bcmac}
\end{figure}
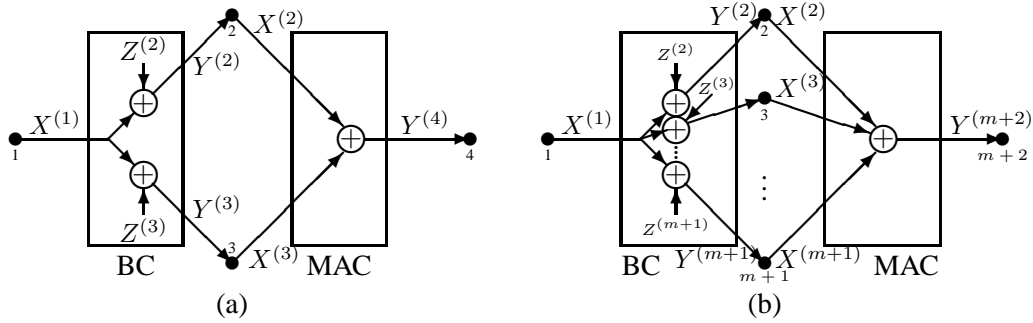
The broadcast channel has power constraint $E[(X^{(1)})^2]\leq P$ 
and channel outputs $Y^{(2)}=X^{(1)}+Z^{(2)}$ and $Y^{(3)}=X^{(1)}+Z^{(3)}$, 
where $Z^{(2)}$ and $Z^{(3)}$ are statistically dependent 
mean-0, variance-$N$ random variables with $Z^{(2)}=-Z^{(3)}$, 
and $P$ and $N$ are real-valued positive constants.  
The multiple access channel has power constraints 
$E[(X^{(2)})^2],E[(X^{(3)})^2]\leq P+N$ at each transmitter 
and output $Y^{(4)}=X^{(2)}+X^{(3)}$.  
We consider a single unicast demand, 
where node~1 wishes to reliably transmit information to node~4. 
If we channel code to make each channel reliable 
and then apply network coding, 
the achievable rate cannot exceed the broadcast channel's maximal sum rate 
\[
\max_{\alpha}\left[\frac12\log\left(1+\frac{\alpha P}{N}\right)+
\frac12\log\left(1+\frac{(1-\alpha)P}{\alpha P+ N}\right)\right] =
\frac12\log\left(1+\frac{P}{N}\right).  
\]
Yet the network's unicast capacity is infinite 
since nodes~2 and~3 can simply retransmit their channel outputs uncoded 
to give output $Y^{(4)}=(X^{(1)}+Z^{(2)})+(X^{(1)}+Z^{(3)})=2X^{(1)}$ 
at node~4.  \IEEEQED
\end{exam}
It is tempting to believe 
that the gap between the optimal performance 
and the performance achieved 
by separate network and channel coding in  Example~\ref{ZZZ-ex:bcmac} 
arises due to the unusual statistical dependence in the noise.  
Unfortunately, similar phenomena can also arise 
when the noise at the receivers of a broadcast channel is independent, 
as shown in Example~\ref{ZZZ-ex:bcmacind}
\begin{exam}\label{ZZZ-ex:bcmacind}
Figure~\ref{ZZZ-fig:bcmac}(b) shows a $(m+2)$-node network 
made from a Gaussian broadcast channel 
and a real additive multiple access channel.  
The broadcast channel has power constraint $E[(X^{(1)})^2]\leq P$ 
and channel outputs $Y^{(i)}=X^{(1)}+Z^{(i)}$, $i\in\{2,\ldots,m+1\}$, 
where $Z^{(i)}$ are independent mean-0, variance-$N$ 
Gaussian random variables,  
and $P$ and $N$ are real-valued positive constants.  
The multiple access channel has power constraint 
$E[(X^{(i)})^2]\leq P+N$ at each transmitter $i\in\{2,\ldots,m+1\}$ 
and output $Y^{(m+2)}=\sum_{i=2}^{m+1}X^{(i)}$.  
We consider a single unicast demand, 
where node~1 wishes to reliably transmit information to node~$(m+2)$. 
The maximal achievable unicast rate 
using separate network and channel codes 
is bounded by the broadcast channel's maximal sum rate 
\[
\max_{\alpha_2,\ldots,\alpha_{m+1}}
\sum_{i=2}^{m+1}
\frac12\log\left(1+\frac{\alpha_i P}{\sum_{j=2}^{i-1}\alpha_jP+N}\right)=
\frac12\log\left(1+\frac{P}{N}\right).  
\]
The unicast capacity of the network is greater than or equal to 
\[
\frac12\log\left(1+\frac{mP}{N}\right)
\]
since nodes~2 through $m+1$ 
can simply retransmit their channel outputs uncoded 
to give output 
\[
Y^{(m+2)}=\sum_{i=2}^{m+1}(X^{(1)}+Z^{(i)})=mX^{(1)}+\sum_{i=2}^{m+1}Z^{(i)}, 
\]
which is a Gaussian channel with power $E[(mX^{(1)})^2]=m^2 P$ 
and noise variance $E[(\sum_{i=2}^{m+1}Z^{(i)})^2]=m N$.  
Thus the gap between the optimal performance 
and the lower bound achieved through the use of a separated strategy 
is sometimes large even in networks with independent noise.  
\IEEEQED
\end{exam}

Given the difficulty of solving network capacities even for small networks 
and the failure of individual channel capacities to predict the capacity 
of networks made from those channels, 
the gap between the size of the networks whose capacities we can analyze 
and the size of the networks over which we communicate in practice 
seems to be growing  ever larger.  
To address this challenge, 
we here propose a strategy for bounding 
the behaviors of individual channels 
in a manner that captures their full range of behaviors 
in larger network systems.  
That is, we derive upper and lower bounding models 
on individual channels 
such that the capacity region of any network that contains the given channel 
is bounded below by the capacity region of a network 
that replaces that component by its lower bounding model 
and bounded above by the capacity region of a network 
that replaces that component by its upper bounding model.  
Thus, an arbitrary collection of demands 
(e.g., a collection of unicasts) 
can be met on a given network 
if it can bet met on the network that replaces channels 
by their lower bounding models and 
only if it can be met on the network that replaces channels 
by their upper bounding models. 

We focus on upper and lower bounding models 
comprised of noiseless bit pipes.  
Using such models, 
we can bound the capacity of a network of noisy channels 
by the network coding capacity 
of the network that replaces each channel by its noiseless model.  
While network coding capacities are not solved in the general case, 
a variety of computational tools can be used to bound them.  
(See, for example,~\cite{ZZZ-SongY:03,ZZZ-HarveyK:06,ZZZ-SubramanianT:08}.)  

Part~I~\cite{ZZZ-KoetterE:09a} in this two-part series 
derived upper and lower bounding models for point-to-point channels.  
In that case, the upper and lower bounds were identical.  
We here derive upper and lower bounds for more general channel types 
using the same basic strategy:  
We demonstrate that the capacity region 
of one network is a subset of that of another network 
by showing that solutions for the first network 
can be run on the second network.  
Sections~\ref{ZZZ-sec:setup} and~\ref{ZZZ-sec:models} 
include the problem setup and channel model definitions.  
Section~\ref{ZZZ-sec:equiv} derives sufficient conditions 
for upper and lower boundng models.  
We derive upper and lower bounding models 
for broadcast, multiple access, and interference channels 
as examples.  
When a channel's upper and lower bounding models differ, 
we bound the accuracy of the resulting capacity bounds 
by comparing the upper and lower bounding models.  
Such accuracy bounds may be useful both directly 
and for determining which larger network components 
should be modeled in the future.

\section{The Setup}\label{ZZZ-sec:setup}
We use the notation established in~\cite{ZZZ-KoetterE:10a}.  
Network $\cN$ has $m$ nodes, $\cV=\{1,\ldots,m\}$.  
Each node transmits an input random variable $\vX\in\cX^{(v)}$ 
and receives an output random variable $\vY\in\vcY$.  
We use $\bX=(\vX:v\in\cV)$ and $\bY=(\vY:v\in\cV)$ 
to denote the vectors of network inputs and outputs.  
The alphabets may be discrete or continuous.  
The network is assumed to be memoryless 
and to be characterized by a conditional 
probability distribution 
\[
p(\by|\bx)=p(y^{(1)},\ldots,y^{(m)}|x^{(1)},\ldots, x^{(m)}). 
\]
Applying a result from~\cite{ZZZ-DoughertyZ:06}, 
we characterize rate regions for arbitrary demands 
by characterizing the multiple unicast rate region.
This choice simplifies the notation 
and yields no loss of generality (see~\cite{ZZZ-KoetterE:10a}).  
Thus a blocklength-$n$ code communicates message 
\[
\Wuv\in\cWuv\deff \{1,\ldots,2^{n\Ruv}\} 
\]
from node $u$ to node $v$ for each $u,v\in\{1,\ldots,m\}$.  
Messages $W=(\Wuv:(u,v)\in\{1,\ldots,m\}^2)$ 
are independent and uniformly distributed 
(though the proof goes through 
if the same message is available at multiple nodes).  
By assumption, $\cR=(\Ruv:(u,v)\in\{1,\ldots,m\}^2)$ 
satisfies $R^{\ar{v}{v}}=0$ for all $v$.  

At time $t$, node $v$ transmits $\vX_t$ and receives $\vY_t$.  
We therefore describe the network by a triple
\begin{equation}\label{ZZZ-eq:network_triple2}
\left(\prod_{v=1}^m\cX^{(v)},p(\by|\bx),\prod_{v=1}^m\cY^{(v)}\right)
\end{equation}
with the causality constraint that $\vX_t$ is a function only of 
\[
\{\vY_1,\ldots,\vY_{t-1},\Wvo,\ldots,\Wvm\}.  
\]

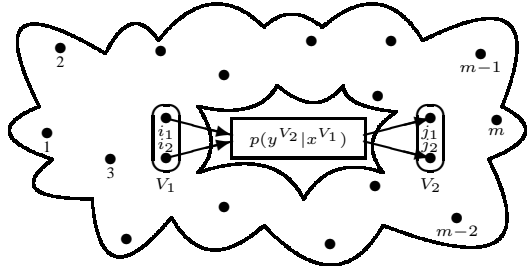
\begin{figure}
\begin{center}
\begin{picture}(200,100)(0,0)
\thicklines
\qbezier(35,25)(38,-15)(65,18)
\qbezier(65,18)(78,-5)(100,15)
\qbezier(100,15)(128,-15)(145,15)
\qbezier(145,15)(210,-10)(185,40)
\qbezier(185,40)(210,50)(190,60)
\qbezier(190,60)(210,110)(160,85)
\qbezier(160,85)(145,110)(122,90)
\qbezier(122,90)(110,100)(100,90)
\qbezier(100,90)(80,110)(60,85)
\qbezier(60,85)(50,100)(40,80)
\qbezier(40,80)(-10,110)(15,60)
\qbezier(15,60)(-10,50)(20,40)
\qbezier(20,40)(0,25)(35,25)
\qbezier(105,68)(98,55)(75,60)
\qbezier(75,60)(85,45)(75,35)
\qbezier(75,35)(105,45)(115,25)
\qbezier(115,25)(125,42)(145,35)
\qbezier(145,35)(135,48)(150,58)
\qbezier(150,58)(130,60)(130,68)
\qbezier(130,68)(120,55)(105,68)
\put(63,55.5){\circle*{4}}
\put(63,53){\makebox(0,0)[ct]{\tiny $i_1$}}
\put(63,55.5){\vector(4,-1){25}}
\put(63,40.5){\circle*{4}}
\put(63,43){\makebox(0,0)[cb]{\tiny $i_2$}}
\put(63,40.5){\vector(4,1){25}}
\put(63,48){\oval(10,25)}
\put(63,33){\makebox(0,0)[ct]{\tiny $V_1$}}
\put(88,41){\framebox(50,14){\tiny $p(y^{V_2}|x^{V_1})$}}
\put(138,49.25){\vector(4,1){25}}
\put(138,46.75){\vector(4,-1){25}}
\put(163,55.5){\circle*{4}}
\put(163,53){\makebox(0,0)[ct]{\tiny $j_1$}}
\put(163,43){\makebox(0,0)[cb]{\tiny $j_2$}}
\put(163,40.5){\circle*{4}}
\put(163,48){\oval(10,25)}
\put(163,33){\makebox(0,0)[ct]{\tiny $V_2$}}
\put(18,50){\circle*{4}}\put(18,47){\makebox(0,0)[ct]{\tiny 1}}
\put(23,82){\circle*{4}}\put(23,79){\makebox(0,0)[ct]{\tiny 2}}
\put(42,40){\circle*{4}}\put(42,37){\makebox(0,0)[ct]{\tiny 3}}
\put(48,10){\circle*{4}}
\put(61,81){\circle*{4}}
\put(85,72){\circle*{4}}
\put(85,22){\circle*{4}}
\put(118,85){\circle*{4}}
\put(125,8){\circle*{4}}
\put(142,30){\circle*{4}}
\put(143,64){\circle*{4}}
\put(148,85){\circle*{4}}
\put(173,18){\circle*{4}}\put(173,15){\makebox(0,0)[ct]{\tiny $m$$-$$2$}}
\put(182,80){\circle*{4}}\put(182,77){\makebox(0,0)[ct]{\tiny $m$$-$$1$}}
\put(188,55){\circle*{4}}\put(188,52){\makebox(0,0)[ct]{\tiny $m$}}
\end{picture}
\caption{An $m$-node network 
         containing a channel 
	 $p(y^{V_2}|x^{V_1})
	 =p(y^{(j_1,1)},y^{(j_2,1)}|x^{(i_1,1)},x^{(i_2,2)})$ 
	 from nodes $V_1=\{i_1,i_2\}$ to node $V_2=\{j_1,j_2\}$.  
	 The distribution 
	 $p(y^{-V_2}|x^{-V_1})$ 
	 on the remaining channel outputs given 
	 the remaining channel inputs is arbitrary.}\label{ZZZ-fig:net}
\end{center}
\end{figure}

For the purposes of this paper, 
network $\cN$ is arbitrary 
except for its inclusion of an independent channel $\cC$, 
as shown in Figure~\ref{ZZZ-fig:net}.  
To make this precise, 
let $V_1,V_2\subset\{1,\ldots,m\}$, $V_1\cap V_2=\emptyset$, 
denote the nodes transmitting to 
and receiving from channel $\cC$, respectively.  
For example, a broadcast channel $\cC$ 
has a single transmitter $V_1=\{i\}$ 
and multiple receivers $V_2=\{j_1,\ldots,j_k\}$, 
a multiple access channel has 
multiple transmitters $V_1=\{i_1,\ldots,i_k\}$ 
and a single receiver $V_2=\{j\}$, and so on.  
Since each node $v\in V_1$ may transmit 
over both $\cC$ and the remainder of the network 
and each node $v\in V_2$ may receive information 
both from $\cC$ and from the remainder of the network, 
we define $\vcX\ \deff\ \cX^{(v,1)}\times\cX^{(v,2)}$ for $v\in V_1$ and 
$\vcY\ \deff\ \cY^{(v,1)}\times\cY^{(v,2)}$ for $v\in V_2$.  
We then use $\XC\in\cXC$ and $\YC\in\cYC$ 
to denote the input and output to channel $\cC$ 
and $\XCm\in\cXCm$ and $\YCm\in\cYCm$ 
to denote the input and output to remainder of the network.  
The respective alphabets are given by 
\begin{eqnarray*}
\cXC = \prod_{v\in V_1}\cX^{(v,1)} & \hspace{.5in}&
\cXCm= \left(\prod_{v\not\in V_1}\vcX\right)
                 \times\left(\prod_{v\in V_1}\cX^{(v,2)}\right) \\
\cYC = \prod_{v\in V_2}\cY^{(v,1)} & &
\cYCm= \left(\prod_{v\not\in V_2}\vcY\right)
                 \times\left(\prod_{v\in V_2}\cY^{(v,2)}\right).
\end{eqnarray*}
The independence of channel $\cC$ from the rest of the network 
implies a factorization of the conditional distribution $p(\by|\bx)$,
giving network characterization 
\[
\cN = \left(\cXCm\times\cXC,p(\yCm|\xCm)p(\yC|\xC),
\cYCm\times\cYC\right),
\]
again with the constraint 
that random variable $\vX_t$ is a function of 
random variables $\{\vY_1,\ldots,\vY_{t-1},$
$\Wvo,\ldots,\Wvm\}$ alone. 

The following definitions are identical to those in~\cite{ZZZ-KoetterE:10a}, 
which describes them in greater detail.  
\begin{defi}\label{ZZZ-def:network_solution}
Let a network 
\[
\cN\ \deff\ \left(\prod_{v=1}^m\vcX,p(\by|\bx),
\prod_{v=1}^m\vcY\right)
\]
be given.  
A blocklength-$n$ solution $\cS(\cN)$ for this network 
is a set of encoding and decoding functions:
\begin{eqnarray*} 
\vX_t: &&   (\vcY)^{t-1}\times\prod_{v'=1}^m\cWvp\rightarrow\vcX \\
\bWuv: && (\vcY)^{n}\times\prod_{v'=1}^m\cWvp\rightarrow\cWuv 
\end{eqnarray*}
mapping $(\vY_1,\ldots,\vY_{t-1},\Wvo,\ldots,\Wvm)$ to $\vX_t$ 
for each $v\in V$ and $t\in\{1,\ldots,n\}$ and mapping 
$(\vY_1,\ldots,\vY_{n}, \Wvo,\ldots, \Wvm)$ to $\bWuv$ 
for each $u,v\in V$.  
The solution $\cS(\cN)$ is called a $(\lambda,\cR)$-solution, denoted 
$(\lambda,\cR)$-$\cS(\cN)$, if $\Pr(\Wuv\not = \bWuv)<\lambda$
for all source and sink pairs $u,v$ using the specified encoding and decoding functions.  
\end{defi}

\begin{defi} The rate region $\setR(\cN) \subset \R_+^{m(m-1)}$ of a network
  $\cN$ is the closure of all rate vectors $\cR$ 
  such that for any $\lambda>0$ and all $n$ sufficiently large, 
  there exists a $(\lambda,\cR)$-$\cS(\cN)$ solution of blocklength $n$.  
  We use $\mbox{int}(\setR(\cN))$ to denote the interior 
  of rate region $\setR(\cN)$.  
\end{defi}

Given a network $\cN$, the $N$-fold stacked network $\ucN$ 
contains $N$ copies of $\cN$ 
and delivers $N$ independent messages $\Wuv$ for each $(u,v)$.  
We carry over notation and variable definitions 
from the network $\cN$ to the stacked network $\ucN$ 
by underlining the variable names.  
So $\uWuv\in\ucWuv\deff(\cWuv)^N$ 
is the $N$-dimensional vector of messages 
that the $N$ copies of node $u$ 
send to the corresponding copies of node $v$, and 
$\uvX_t\in\uvcX\deff(\vcX)^N$ and $\uvY_t\in\uvcY\deff(\vcY)^N$ 
are the $N$-dimensional vectors of network inputs and network outputs, 
respectively,  
for node $v$ at time $t$.  
The variables in the $\ell$-th layer of the stack 
are denoted by an argument $\ell$, for
example $\uWuv(\ell)$ is the message 
from node $u$ to node $v$ in the $\ell$-th layer of the stack 
and $\uvX_t(\ell)$ is the layer-$\ell$ channel input 
from node $v$ at time $t$.  
The rate $\Ruv$ for a stacked network equals $(\log|\ucWuv|)/(nN)$;  
this normalization makes rate regions in a network 
and its corresponding stacked network comparable.

\begin{defi}\label{ZZZ-def:snetwork_solution}
Let a network 
\[
\cN\ \deff\ \left(\prod_{v=1}^m\vcX,p(\by|\bx),\prod_{v=1}^m\vcY\right)
\]
be given.  
Let $\ucN$ be the $N$-fold stacked network for $\cN$. 
A blocklength-$n$ solution $\cS(\ucN)$ to 
this network is defined as a set of encoding and decoding functions
\begin{eqnarray*} 
\uvX_t: & & {(\uvcY)}^{t-1}\times\prod_{v'=1}^m\ucWvp \rightarrow \uvcX \\
\ubWuv: & & (\uvcY)^n\times\prod_{v'=1}^m\ucWvp\rightarrow\ucWuv 
\end{eqnarray*}
mapping $(\uvY_1,\ldots,\uvY_{t-1}, \uWvo,\ldots,\uWvm)$ 
to $\uvX_t$ for each $t\in\{1,\ldots,n\}$ and 
$v\in\{1,\ldots,m\}$ and mapping 
$(\uvY_1,\ldots,\uvY_n,\uWvo,\ldots,\uWvm)$
to $\ubWuv$ for each $u,v\in\{1,\ldots,m\}$.  
The solution $\cS(\ucN)$ is called a $(\lambda,\cR)$-solution for $\ucN$, 
denoted $(\lambda,\cR)$-$\cS(\ucN)$, 
if the encoding and decoding functions imply
$\Pr(\uWuv\neq\ubWuv)<\lambda$ 
for all source and sink pairs $u,v$.
\end{defi}

\begin{defi} The rate region $\setR(\ucN) \subset \R_+^{m(m-1)}$ 
of a stacked network $\ucN$ is the closure of all rate vectors $\cR$ 
such that a $(\lambda,\cR)$-$\cS(\ucN)$ solution exists 
for any $\lambda>0$ and all $N$ sufficiently large.
\end{defi}

Theorem~\ref{ZZZ-thm:stacked} from~\cite{ZZZ-KoetterE:10a}, reproduced below, 
shows that if the messages $\uWuv$ are channel coded before transmission, 
then any rate $\cR$ that can be achieved across a stacked network 
can be achieved by a code that applies the same solution 
independently in each layer.  
Such solutions are called stacked solutions.  
A formal definition of stacked solutions follows.  
Since stacked solutions are optimal by Theorem~\ref{ZZZ-thm:stacked}, 
there is no loss of generality in restricting our attention to stacked solutions 
going forward.  

\begin{defi}\label{ZZZ-def:stacked_sol}
Let a network 
$\cN\ \deff\ (\prod_{v=1}^m\vcX,p(\by|\bx),\prod_{v=1}^m\vcY)$ 
be given.  
Let $\ucN$ be the $N$-fold stacked network for $\cN$.  
A blocklength-$n$ stacked solution $\ucS(\ucN)$ to network $\ucN$ 
is defined as a set of mappings 
\begin{eqnarray*} 
\utWuv:&&\ucWuv\rightarrow\utcWuv  \\
\vX_t: &&(\vcY)^{t-1}\times\prod_{v'=1}^m\tcWvp\rightarrow\vcX \\
\btWuv:&&(\vcY)^{n}\times\prod_{v'=1}^m\tcWvp\rightarrow\tcWuv \\
\ubWuv:&& \utcWuv\rightarrow\ucWuv 
\end{eqnarray*} 
such that 
\begin{eqnarray*}
\utWuv & = & \utWuv(\uWuv) \\
\uvX_t(\ell) & = & \vX_t\left(\uvY_1(\ell),\ldots,\uvY_{t-1}(\ell),
                              \utWvo(\ell),\ldots,\utWvm(\ell)\right) \\
\ubtWuv(\ell) & = & \btWuv\left(\uvY_1(\ell),\ldots,\uvY_n(\ell),
                                \utWvo(\ell),\ldots,\utWvm(\ell)\right)\\
\ubWuv & = & \ubWuv(\ubtWuv)
\end{eqnarray*}
for each $u,v\in\{1,\ldots,m\}$, $t\in\{1,\ldots,n\}$, 
and $\ell\in\{1,\ldots,N\}$.  
Here $(\utWuv,\ubWuv)$ is the blocklength-$N$ channel code 
for the message from $u$ to $v$, 
$\vX_t$ is the node-$v$ single-layer encoder at time $t$, 
and $\btWuv$ is the node-$v$ single-layer decoder at time $t$.  
The solution $\ucS(\ucN)$ is called a stacked $(\lambda,\cR)$-solution, 
denoted $(\lambda,\cR)$-$\ucS(\ucN)$, 
if the specified mappings imply $\Pr(\uWuv\neq\ubWuv)<\lambda$ 
for all pairs $(u,v)\in\cV^2$.
\end{defi}

\begin{defi} The rate region $\setR(\ucN) \subset \R_+^{m(m-1)}$ 
of a stacked network $\ucN$ is the closure of all rate vectors $\cR$ 
such that a $(\lambda,\cR)$-$\cS(\ucN)$ solution exists 
for any $\lambda>0$ and all $N$ sufficiently large.
\end{defi}

\begin{thm}\cite[Theorem~1]{ZZZ-KoetterE:10a}\label{ZZZ-thm:stacked} 
The rate regions $\setR(\cN)$ and $\setR(\ucN)$ are identical, 
and for each $\cR\in\mbox{int}(\setR(\ucN))$, 
there exists a sequence of blocklength-$n$ 
$(2^{-N\delta},\cR)$-$\ucS(\ucN)$ stacked solutions 
for $\ucN$ for some $n\geq 1$ and $\delta>0$.  
\IEEEQED
\end{thm}

\section{Bit-Pipe Models}\label{ZZZ-sec:models}

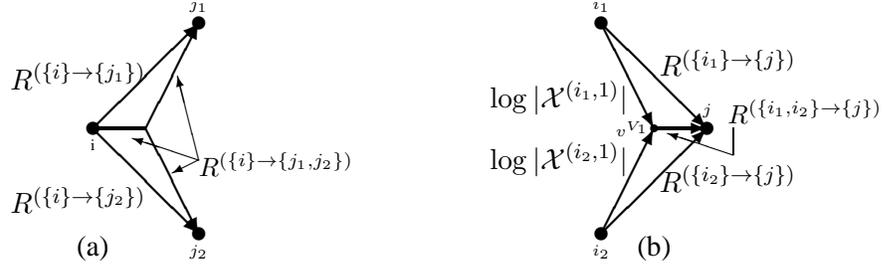
\begin{figure}
  \begin{center}
  \begin{picture}(40,90)(20,-30)
      \thicklines
      \put(20,20){\circle*{5}}
      \put(20,16){\makebox(0,0)[ct]{\tiny i}}
      \put(20,20){\vector(1,1){40}}
      \put(20,20){\line(1,0){20}}
      \put(20,20){\vector(1,-1){40}}
      \put(40,20){\vector(1,2){20}}
      \put(40,20){\vector(1,-2){20}}
      \put(40,35){\makebox(0,0)[rb]{$R^{\ar{\{i\}}{\{j_1\}}}$}}
      \put(40,-12){\makebox(0,0)[rb]{$R^{\ar{\{i\}}{\{j_2\}}}$}}
      \put(60,60){\circle*{5}}
      \put(60,-20){\circle*{5}}
      \put(60,64){\makebox(0,0)[cb]{\tiny $j_1$}}
      \put(60,-24){\makebox(0,0)[ct]{\tiny $j_2$}}
      \put(60,2){\makebox(0,0)[lb]{\small $R^{\ar{\{i\}}{\{j_1,j_2\}}}$}}
      \thinlines
      \put(60,8){\vector(-2,-1){10}}
      \put(60,8){\vector(-1,4){8}}
      \put(60,8){\vector(-3,1){25}}
      \put(20,-30){\makebox(0,0)[cb]{(a)}}
  \end{picture} 
  \hspace{2in}
  \begin{picture}(40,90)(0,-30)
    \thicklines
      \put(0,60){\circle*{5}}
      \put(0,-20){\circle*{5}}
      \put(0,64){\makebox(0,0)[cb]{\tiny $i_1$}}
      \put(0,-24){\makebox(0,0)[ct]{\tiny $i_2$}}
      \put(0,60){\vector(1,-1){40}}
      \put(22,40){\makebox(0,0)[lb]{$R^{\ar{\{i_1\}}{\{j\}}}$}}
      \put(0,60){\vector(1,-2){20}}
      \put(10,38){\makebox(0,0)[rt]{$\log|\cX^{(i_1,1)}|$}}
      \put(10,2){\makebox(0,0)[rb]{$\log|\cX^{(i_2,1)}|$}}
      \put(0,-20){\vector(1,2){20}}
      \put(0,-20){\vector(1,1){40}}
      \put(22,6){\makebox(0,0)[lt]{$R^{\ar{\{i_2\}}{\{j\}}}$}}
      \put(20,20){\circle*{3}}
      \put(18,20){\makebox(0,0)[rc]{\tiny $v^{V_1}$}}
      \put(20,20){\vector(1,0){20}}
      \put(40,20){\circle*{5}}
      \put(40,24){\makebox(0,0)[cb]{\tiny $j$}}
      \put(48,22){\makebox(0,0)[lb]{\small $R^{\ar{\{i_1,i_2\}}{\{j\}}}$}}
      \thinlines
      \put(50,20){\line(0,-1){10}}
      \put(50,10){\vector(-3,1){25}}
      \put(20,-30){\makebox(0,0)[cb]{(b)}}
      \end{picture} %
  \end{center}
\caption{Bit-pipe models $\cC(\cRC)$ 
  for (a) the broadcast channel with $V_1=\{i\}$ and $V_2=\{j_1,j_2\}$, 
  and (b) the multiple access channel with $V_1=\{i_1,i_2\}$ and $V_2=\{j\}$.  
  For the broadcast channel, 
  $\cRC=(R^{\ar{\{i\}}{\{j_1,j_2\}}},
  R^{\ar{\{i\}}{\{j_1\}}},R^{\ar{\{i\}}{\{j_2\}}})$ 
  describes a common information rate to be delivered to both receivers 
  and a private information rate for each receiver.  
  For the multiple access channel, 
  $\cRC=(R^{\ar{\{i_1\}}{\{j\}}},
  R^{\ar{\{i_2\}}{\{j\}}},R^{\ar{\{i_1,i_2\}}{\{j\}}})$ 
  describes an individual information rate from each transmitter 
  and a shared information rate from the pair of transmitters.}
  \label{ZZZ-fig:bitpipes}
\end{figure}

The equivalence tools derived below 
relate the rate region of a network $\cN$ 
to those of a network $\cN(\cRC)$ 
in which channel $\cC$ is replaced by a bit-pipe model $\cC(\cRC)$ 
corresponding to some rate vector $\cRC$.  
We here define $\cRC$ and $\cC(\cRC)$ 
for a generic channel $\cC$ with input nodes $V_1$ 
and output nodes $V_2$.  
Figure~\ref{ZZZ-fig:bitpipes} illustrates these definitions 
for two example channels.  
Let 
\begin{eqnarray*}
\cM & \deff & \{(A,B):A\subseteq V_1,B\subseteq V_2,A,B\neq\emptyset\} \\
\cRC & \deff & (R^{\ar{A}{B}}:(A,B)\in\cM).  
\end{eqnarray*}
For each $(A,B)\in\cM$, 
bit-pipe model $\cC(\cRC)$, defined formally below, 
delivers rate $R^{\ar{A}{B}}$ from transmitter set $A$ to receiver set $B$.  
When $|A|=1$, $A$ transmits directly to each node in $B$.  
When $|A|>1$, each node $i\in A$ delivers $\log|\cX^{(i,1)}|$ bits 
(i.e., a symbol from alphabet $\cX^{(i,1)}$) to an internal node $v^A$, 
which delivers $R^{\ar{A}{B}}$ bits to each node in $B$.  

\begin{defi}
The bit-pipe model $\cC(\cRC)$ is defined as 
\begin{equation}\label{ZZZ-eqn:crc}
\cC(\cRC)\ \deff\ 
\left(\tcXI\times\tcXC, p(\tyI,\tyC|\txI,\txC),\tcYI\times\tcYC\right),
\end{equation}
where $\txC$ and $\tyC$ are the network inputs 
and outputs for the nodes in $V_1$ and $V_2$, 
$\txI$ and $\tyI$ are the network inputs 
and outputs for the internal nodes 
$V_o=\{v^A:A\subseteq V_1,|A|>1\}$.  
For each $A\subseteq V_1$ with $|A|>1$ and $i\in A$, 
node $v^A$ receives copy $\ty^{(v^A,i)}$ of $x^{(i,1)}$.  
For each $(A,B)\in\cM$ and $j\in B$, 
node $j$ receives copy $\ty^{\ar{A}{B},j}$ of $\tx^{\ar{A}{B}}$.  
Therefore 
\[
\begin{array}{rclcrcl}
\tcXC & \deff & \prod_{i\in V_1}\tcX^{(i,1)} &\hspace{.5in}& 
\tcYC & \deff & \prod_{j\in V_2}\tcY^{(j,1)} \\
\tcX^{(i,1)}
&\deff&\cX^{(i,1)}\times\prod_{(\{i\},B)\in\cM}\tcX^{\ar{\{i\}}{B}} &&
\tcY^{(j,1)}&\deff&\prod_{(A,B)\in\cM:j\in B}\tcX^{\ar{A}{B}} \\
\tcXAB&\deff &\{0,1\}^{R^{\ar{A}{B}}} &&
\tcYI& \deff & \prod_{A\subseteq V_1:|A|>1}\tcY^{(v^A)} \\
\tcXI& \deff & \prod_{A\subseteq V_1:|A|>1}\tcX^{(v^A)} &&
\tcY^{(v^A)}&\deff&\prod_{i\in A}\cX^{(i,1)} \\
\tcX^{(v^A)}&\deff&\prod_{B\subseteq V_2}\tcX^{\ar{A}{B}} \\
\end{array}	
\]
\[
p\left(\left.\tyI,\tyC\right|\txI,\txC\right)
\ \deff\ \left(\prod_{(A,B)\in\cM:|A|>1}
   \prod_{i\in A}\delta(\ty^{(v^A,i)}-x^{(i,1)})\right)
   \left(\prod_{(A,B)\in\cM}
   \prod_{j\in B}\delta(\ty^{\ar{A}{B},j}-\tx^{\ar{A}{B}})\right).
\]
\end{defi}

Since any network $\cN(\cRC)$ interacts with $\cC(\cRC)$ 
only through nodes $V_1$ and $V_2$ 
and does not have direct access to the nodes in $V_o$, 
the remainder of this paper abuses notation by 
replacing~(\ref{ZZZ-eqn:crc}) by 
\begin{equation}\label{ZZZ-eqn:crc2}
\cC(\cRC)=\left(\tcXC, p(\tyC|\txC),\tcYC\right).  
\end{equation}
In another common abuse of notation, 
we allow non-integer values of $R^{\ar{A}{B}}$ 
to designate capacitated bit-pipes 
that require more than a single channel use 
to deliver some integer number of bits.  
Applying the stacking approach from the previous section, 
the arguments that follow transmit information over $N$ copies 
of each bit pipe in the stacked network, 
giving alphabet $\utcXAB\ \deff\ \{0,1\}^{NR^{\ar{A}{B}}}$.  

\begin{defi}
Bit-pipe model $\cC(\cRC)=(\tcXC,p(\tyC|\txC),\tcYC)$ 
is a lower-bounding model for channel $\cC=(\cXC,p(\yC|\xC),\cYC)$, 
written $\cC(\cRC)\subseteq\cC$, 
if and only if $\setR(\cN(\cRC))\subseteq\setR(\cN)$ 
for all 
\begin{eqnarray*}
\cN       & = & (\cXC\times\cXCm,p(\yC|\xC)p(\yCm|\xCm),\cYC\times\cYCm) \\
\cN(\cRC) & = & (\tcXC\times\cXCm,p(\tyC|\txC)p(\yCm|\xCm),\tcYC\times\cYCm).  
\end{eqnarray*}
\end{defi}

\begin{defi}
Bit-pipe model $\cC(\cRC)=(\tcXC,p(\tyC|\txC),\tcYC)$ 
is an upper-bounding model for channel $\cC=(\cXC,p(\yC|\xC),\cYC)$, 
written $\cC\subseteq\cC(\cRC)$, 
if and only if $\setR(\cN)\subseteq\setR(\cN(\cRC))$ 
for all 
\begin{eqnarray*}
\cN       & = & (\cXC\times\cXCm,p(\yC|\xC)p(\yCm|\xCm),\cYC\times\cYCm) \\
\cN(\cRC) & = & (\tcXC\times\cXCm,p(\tyC|\txC)p(\yCm|\xCm),\tcYC\times\cYCm).  
\end{eqnarray*}
\end{defi}

The following lemma shows the continuity of network capacity 
in the rate of any bit pipes it contains.  

\begin{lem}\cite[Lemma~2]{ZZZ-KoetterE:10a}\label{ZZZ-lem:cont}
Consider any network 
\[
\cN_R=\left(\cXCm\times\cXC,p(\yCm|\xCm)p(\yC|\xC),\cYCm\times\cYC\right)
\]
with $V_1=\{i\}$ and $V_2=\{j\}$ 
connected by a rate-$R$ bit pipe 
\[
(\cXC,p(\yC|\xC),\cYC)=(\{0,1\}^R,\delta(y^{(j,1)}-x^{(i,1)}),\{0,1\}^R).  
\]
Rate region $\setR(\cN_R)$ is continuous 
in $R$ for all $R>0$. 
\IEEEQED
\end{lem}

\section{The Equivalence Tools}\label{ZZZ-sec:equiv}

Given any network $\cN$ containing channel $\cC$, 
let $\cN(\cRC)$ be the network achieved 
by replacing $\cC$ by $\cC(\cRC)$ in $\cN$.  
We here derive conditions under which 
$\setR(\cN(\cRC))\subseteq\setR(\cN)$ 
(i.e., $\cC(\cRC)$ is a lower bounding model for $\cC$) 
or $\setR(\cN)\subseteq\setR(\cN(\cRC))$ 
(i.e., $\cC(\cRC)$ is an upper bounding model for $\cC$).

Lemma~\ref{ZZZ-lem:lb}, below, uses channel coding arguments 
to derive lower bounding models.  
The proof runs a code $\ucS(\ucN(\cRC))$ across network $\ucN$ 
with the aid of a rate-$\cRC$ channel code for $\cC$.  
The resulting error probability approximates 
the error probability of $\ucS(\ucN(\cRC))$ on $\ucN(\cRC)$ 
provided that the probability of channel coding error is small.  
We therefore begin by defining channel codes 
for a generic channel $\cC$.  

Given a channel $\cC$ with input nodes $V_1$ and output nodes $V_2$, 
a channel code for $\cC$ is a mechanism for reliably delivering 
some collection of rates $(R^{\ar{\{i\}}{B}}:i\in V_1,B\subseteq V_2)$ 
from each transmitter $i\in V_1$ 
to each subset of receivers $B\subseteq V_2$.  
For example, a channel code for broadcast channel $\cC$ 
with transmitter $V_1=\{i\}$ and receivers $V_2=\{j_1,j_2\}$ 
delivers common information at rate $R^{\ar{\{i\}}{\{j_1,j_2\}}}$ 
and private information at rates $R^{\ar{\{i\}}{\{j_2\}}}$ 
and $R^{\ar{\{i\}}{\{j_2\}}}$ for some 
$R^{\ar{\{i\}}{\{j_1,j_2\}}},R^{\ar{\{i\}}{\{j_2\}}},R^{\ar{\{i\}}{\{j_2\}}}\geq 0$.  
Since there is no mechanism for delivering messages 
from a set of transmitters, 
we define channel codes only for rates $\cRC$ 
that satisfy $R^{\ar{A}{B}}=0$ for all $(A,B)\in\cM$ 
with $|A|>1$.\footnote{Nonzero values of $R^{\ar{A}{B}}$ 
are useful for upper bounding models derived later in the paper.}

\begin{defi}
Given a channel $\cC=(\cXC,p(\yC|\xC),\cYC)$, 
let $\cRC$ be a rate vector with $R^{\ar{A}{B}}=0$ 
for all $(A,B)\in\cM$ with $|A|>1$.  
For any $N\geq 1$, a $(2^{N\cRC},N)$ channel code 
$(\alpha_N,\beta_N)$ for channel $\cC$ 
defines a collection of encoding functions 
$\alpha_N=(\alpha_N^{(i)}:i\in V_1)$ 
and decoding functions 
$\beta_N=(\beta_N^{\ar{\{i\}}{B},j}:(\{i\},B)\in\cM,j\in B)$ 
with
\begin{eqnarray*}
\alpha_N^{(i)}: & & \prod_{B\subseteq V_2}\utcXiB
	    \rightarrow\ucX^{(i,1)} \\
\beta_N^{\ar{\{i\}}{B},j}: & & \ucY^{(j,1)}\rightarrow\utcXiB.
\end{eqnarray*}
Let $\ucW\ \deff\ \prod_{(\{i\},B)\in\cM}\utcXiB$.  
The code's average error probability is 
\begin{eqnarray*}
P_e^{(N)}&\deff& \frac1{|\ucW|}\sum_{\uw\in\ucW}
	\Pr\left(\bigcup_{(\{i\},B)\in\cM}\bigcup_{j\in B}
      \beta_N^{\ar{\{i\}}{B},j}(\uY^{(j,1)})\neq \uw^{\ar{\{i\}}{B}}\right| \\
&&\left.\uX^{(i,1)}=\alpha_N^{(i)}(\uw^{\ar{\{i\}}{B}}:B\subseteq V_2)
	\forall i\in V_1\right).
\end{eqnarray*}
\end{defi}

\begin{defi}
The capacity region $\setR(\cC)$ of channel $\cC$ 
is the closure of all rate vectors $\cRC$ 
such that for any $\lambda>0$ and all $N$ sufficiently large, 
there exists a $(2^{N\cRC},N)$ channel code for channel $\cC$ 
with average error probability $P_e^{(N)}<\lambda$.  
\end{defi}

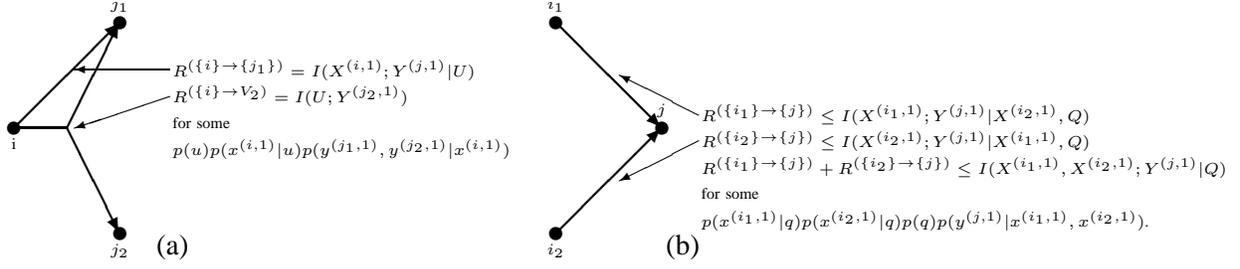
\begin{figure}
  \begin{center}
  \begin{picture}(460,100)(-20,-30)
    \thicklines
      \put(-20,20){\circle*{5}}
      \put(-20,16){\makebox(0,0)[ct]{\tiny i}}
      \put(-20,20){\vector(1,1){40}}
      \put(-20,20){\line(1,0){20}}
      \put(0,20){\vector(1,2){20}}
      \put(0,20){\vector(1,-2){20}}
      \put(20,60){\circle*{5}}
      \put(20,-20){\circle*{5}}
      \put(20,64){\makebox(0,0)[cb]{\tiny $j_1$}}
      \put(20,-24){\makebox(0,0)[ct]{\tiny $j_2$}}
      \put(40,-30){\makebox(0,0)[cb]{(a)}}
      \thinlines
      \put(40,42){\makebox(0,0)[lc]
	{\tiny $R^{\ar{\{i\}}{\{j_1\}}}=I(X^{(i,1)};Y^{(j,1)}|U)$}}
      \put(39,42){\vector(-1,0){36}}
      \put(40,32){\makebox(0,0)[lc]
	{\tiny $R^{\ar{\{i\}}{V_2}}=I(U;Y^{(j_2,1)})$}}
      \put(39,32){\vector(-3,-1){36}}
      \put(40,22){\makebox(0,0)[lc]{\tiny for some }}
      \put(40,12){\makebox(0,0)[lc]
	{\tiny $p(u)p(x^{(i,1)}|u)p(y^{(j_1,1)},y^{(j_2,1)}|x^{(i,1)})$}}  
      \thicklines
      \put(185,60){\circle*{5}}
      \put(185,-20){\circle*{5}}
      \put(185,64){\makebox(0,0)[cb]{\tiny $i_1$}}
      \put(185,-24){\makebox(0,0)[ct]{\tiny $i_2$}}
      \put(185,60){\vector(1,-1){40}}
      \put(185,-20){\vector(1,1){40}}
      \put(225,20){\circle*{5}}
      \put(225,24){\makebox(0,0)[cb]{\tiny $j$}}
      \put(233,-30){\makebox(0,0)[cb]{(b)}}
      \thinlines
      \put(240,25){\makebox(0,0)[lc]{\tiny $R^{\ar{\{i_1\}}{\{j\}}}
	  \leq I(X^{(i_1,1)};Y^{(j,1)}|X^{(i_2,1)},Q)$}}
      \put(239,25){\vector(-2,1){30}}
      \put(240,15){\makebox(0,0)[lc]{\tiny $R^{\ar{\{i_2\}}{\{j\}}}
	  \leq I(X^{(i_2,1)};Y^{(j,1)}|X^{(i_1,1)},Q)$}}
      \put(239,15){\vector(-2,-1){30}}
      \put(240,5){\makebox(0,0)[lc]
	{\tiny $R^{\ar{\{i_1\}}{\{j\}}}+R^{\ar{\{i_2\}}{\{j\}}}$
	$\leq I(X^{(i_1,1)},X^{(i_2,1)};Y^{(j,1)}|Q)$}}
      \put(240,-5){\makebox(0,0)[lc]{\tiny for some}}
      \put(240,-15){\makebox(0,0)[lc]{\tiny $p(x^{(i_1,1)}|q)
	  p(x^{(i_2,1)}|q)p(q)p(y^{(j,1)}|x^{(i_1,1)},x^{(i_2,1)})$.}}
      \end{picture} 
      \end{center}
\caption{Lower bounding models for the (a)~degraded broadcast and 
	 (b)~multiple access channels.}
	 \label{ZZZ-fig:lb}
\end{figure}

Lemma~\ref{ZZZ-lem:lb}, below, shows that $\cRC\in\setR(\cC)$ 
implies $\cC(\cRC)$ is a lower bounding model for $\cC$.  
Applying Lemma~\ref{ZZZ-lem:lb} with existing achievability bounds 
for any network gives immediate lower bounding models 
for that network.  
Figure~\ref{ZZZ-fig:lb} shows two examples.  
Zero capacity bit pipes can carry no bits, so they are not drawn.  

\begin{lem}\label{ZZZ-lem:lb}
If $\cRC\in\setR(\cC)$, then $\cC(\cRC)\subseteq\cC$.  

\Proof
The following argument treats points $\cRC\in\mbox{int}(\setR(\cC))$.  
The result then follows since 
$\setR(\cN(\cRC))\subseteq\setR(\cN)$ 
for all $\cRC\in\mbox{int}(\setR(\cC))$ and 
$\setR(\cN(\cRC))$ is continuous in $\cRC$ by Lemma~\ref{ZZZ-lem:cont} 
together imply that $\setR(\cN(\cRC))\subseteq\setR(\cN)$ 
for all $\cRC\in\setR(\cC)$ by the closure in the definition 
of the network capacity region.  

Consider a pair of networks, 
\begin{eqnarray*}
\cN       & = & (\cXC\times\cXCm,p(\yC|\xC)p(\yCm|\xCm),\cYC\times\cYCm) \\
\cN(\cRC) & = & (\tcXC\times\cXCm,p(\tyC|\txC)p(\yCm|\xCm),\tcYC\times\cYCm).
\end{eqnarray*}
Let $\ucN$ and $\ucN(\cRC)$ be the $N$-fold stacked networks 
for $\cN$ and $\cN(\cRC)$.  
By Theorem~\ref{ZZZ-thm:stacked}, 
it suffices to prove that $\setR(\ucN(\cRC))\subseteq\setR(\ucN)$ 
for $N$ sufficiently large.  
Fix any $\cR\in\mbox{int}(\setR(\ucN(\cRC)))$ and any $\lambda>0$.  
We begin by building a rate-$\cR$ stacked solution $\ucS(\ucN(\cRC))$.  
By Theorem~\ref{ZZZ-thm:stacked}, 
there exists a sequence of stacked solutions $\ucS(\ucN(\cRC))$ 
of some fixed blocklength $n$ (independent of $N$)
but increasing stack size 
such that $\Pr(\ubW\neq\uW)\leq 2^{-N\delta}$ for all $N$ sufficiently large.  
Fix such a sequence of codes.  

Since $\cRC\in\mbox{int}(\setR(\cC))$, $\lambda>0$, and $n$ are fixed, 
there exists a sequence of channel codes 
$\{(\alpha_N,\beta_N)\}_{N=1}^\infty$ for channel $\cC$ 
with encoders $\alpha_N=(\alpha_N^{(i)}:(i)\in V_1)$, 
decoders $\beta_N=(\beta_N^{\ar{\{i\}}{B},j}:(\{i\},B)\in\cM,j\in B)$, 
and average error $P_e^{(N)}<\lambda/(2n)$ 
for all $N$ sufficiently large.\footnote{We here divide by $n$ 
since the channel code will be applied 
across the layers of the stack $n$ times, 
once for each $t\in\{1,\ldots,n\}$ for this blocklength $n$ code.  
Application of the union bound then gives 
an error probability over these $n$ time steps.}
For reasons that are explained below, 
we may wish to use different channel codes at each time $t$.  
We therefore use notation $(\alpha_{N,t},\beta_{N,t})$ 
for the time-$t$ channel code, $t\in\{1,\ldots,n\}$. 

We now build a solution $\cS(\ucN)$ for 
$N$-fold stacked network $\ucN$.  
Solution $\cS(\ucN)$ 
operates $\ucS(\ucN(\cRC))$ across $\ucN$ 
by channel encoding $\uXC_t$ before transmission across $\cC$ 
and channel decoding $\uYC_t$ before use in 
the node encoders and decoders of $\ucS(\ucN(\cRC))$.  
Precisely, at time $t$ 
node $v$ applies the node encoders from $\ucS(\ucN(\cRC))$ as 
\[
\uvtX_t=\uvX(\uvtY_1,\ldots, \uvtY_{t-1},\uWvo,\ldots,\uWvm),
\]
where $\uvtY_t$ is the network output $\uvY_t$ 
channel decoded (if necessary) as 
\[
\uvtY_t=\left\{\begin{array}{ll} 
\left(\left(\beta_{N,t}^{\ar{\{i\}}{B},v}(\uY^{(v,1)}_t):
(\{i\},B)\in\cM, v\in B\right),\uY^{(v,2)}_t\right)
& \mbox{if $v\in V_2$} \\
  \uvY_t & \mbox{otherwise}.  
  \end{array}\right.
\]
Node $v$ then applies channel encoder $\alpha_{N,t}$ (if necessary) as 
\[
\uvX_t = \left\{\begin{array}{ll} 
   \left(\alpha_{N,t}^{(v)}(\utX^{(v,1)}_t),\utX^{(v,2)}\right)
     & \mbox{if $v\in V_1$} \\
   \uvtX_t & \mbox{otherwise},
   \end{array}\right.
\]
and then transmits across the network.  
At time $n$, node $v$ applies the decoder from $\ucS(\ucN(\cRC))$ to give 
\[
\uWuv=\ubWuv(\uvtY_1,\ldots,\uvtY_n,\uWvo,\ldots,\uWvm).
\]

To bound the error probability, 
note that two things can go wrong.  
Either the channel code can fail at one or more times steps 
or all channel codes can succeed but the code can fail anyway.  
If the channel codes $\{(\alpha_{N,t},\beta_{N,t})\}_{t=1}^n$ 
all succeed, then the conditional probability of an error given $\uW=\uw$ 
is precisely what it would have been for the original code.  
Let $E_t$ denote the event that 
the channel code $(\alpha_{N,t},\beta_{N,t})$ employed at time $t$ fails.  
Then we bound the error probability as 
\begin{eqnarray*}
\Pr(\ubW\neq\uW) 
& \stackrel{(a)}{\leq} & \sum_{t=1}^n\Pr(E_t)
        +\sum_{\uw}\Pr(\ubW\neq \uW|\uW=\uw\cap\cap_{t=1}^nE_t^c)
        p(\uw\cap\cap_{t=1}^nE_t^c) \\
& \stackrel{(b)}{\leq} & 
	\left(\sum_{t=1}^n\frac{\lambda}{2n}\right)+2^{-N\delta},
\end{eqnarray*}
which is less than $\lambda$ for all $N$ sufficiently large.  
Inequality $(a)$ follows from the union bound.  
Inequality $(b)$ follows from the channel code's error probability bound 
and the observation that $p(\uw\cap\cap_{t=1}^nE_t^c)\leq p(\uw)$ 
for all $\uw$.  
Bounding the channel code's expected error probability in $(a)$
is slightly subtle since the capacity definition 
guarantees only that the code's average error probability goes to zero.  
An argument suggested by~\cite{ZZZ-Langberg:10}, 
reproduced as Lemma~\ref{ZZZ-lem:avgexp} in Appendix~\ref{ZZZ-app:avgexp}, 
shows that, under careful choice of the channel code's index assignments, 
each channel code $(\alpha_{N,t},\beta_{N,t})$ 
can achieve an expected error probability 
no greater than the code's average error probability $\lambda/(2n)$.  
Since the channel input distribution may vary with time, 
the channel code (or just the channel code's index assignments) 
may likewise need to vary with time.
\IEEEQED
\end{lem}

\begin{rem}
The family of lower bounding models 
described in Lemma~\ref{ZZZ-lem:lb} is tight 
in the sense that there exist networks $\cN$ 
for which the closure of $\cup_{\cRC\in\setR(\cC)}\setR(\cN(\cRC))$ 
is precisely equal to $\setR(\cN)$.  
This observation is immediate 
since network $\cN$ can be the channel $\cC$ in isolation.  
Thus Lemma~\ref{ZZZ-lem:lb} does not necessarily give a tight capacity bound 
for all networks that employ channel $\cC$, 
but we cannot hope to increase the rates in this model 
and still obtain a lower bound 
for any network that contains $\cC$.  
\end{rem}

Just as Lemma~\ref{ZZZ-lem:lb} derives lower bounding models 
by showing that channel coding can be used 
to emulate a collection of noiseless bit pipes across a noisy channel, 
Theorem~\ref{ZZZ-thm:equiv}, below, derives upper bounding models 
by showing that lossy source coding can be used 
to emulate a noisy channel $\cC$ across a bit-pipe model $\cC(\cRC)$.  
Specifically, we prove that $\setR(\cN)\subseteq\setR(\cN(\cRC))$ 
by showing that we can run a solution $\ucS(\ucN)$ 
across network $\ucN(\cRC)$ with similar error probability 
if the source code can emulate the channel to sufficient accuracy.  
We therefore begin by defining source codes 
to run across a generic bit-pipe model $\cC(\cRC)$.  
The source codes introduced here differ from traditional source codes 
in that a good reproduction of $\uXC_t$ is not a value $\uhXC_t$ 
that reproduces it to low distortion 
but a value $\uYC_t$ that is similar statistically 
to the output that would be observed 
if $\uXC$ were transmitted across $N$ independent copies of $\cC$.  
We therefore call the codes channel emulators
and measure performance as emulation accuracy.  

\begin{defi}\label{ZZZ-def:ec}
A random $(2^{N\cRC},N)$ emulator $\hcC=(\alpha_N,\beta_N)$ 
for channel $\cC=(\cXC,p(\yC|\xC),\cYC)$ 
under channel input distribution $p(\xC)$ 
defines a distribution over the family of possible encoders 
$\alpha_N=(\alpha_N^{\ar{A}{B}}:(A,B)\in\cM)$
and decoders $\beta_N=(\beta_N^{(j)}: j\in V_2)$, 
where 
\begin{eqnarray*}
\alpha_N^{\ar{A}{B}}: && \prod_{i\in A}\ucX^{(i,1)}\rightarrow\utcXAB \\
\beta_N^{(j)}: && \prod_{(A,B)\in\cM:j\in B}\utcXAB\rightarrow\ucY^{(i,1)}.  
\end{eqnarray*}
While any instance of code $(\alpha_N,\beta_N)$ 
is deterministic, 
the distribution over codes establishes an emulation distribution 
\[
\hp(\uyC|\uxC)\ \deff\ \Pr(\beta_N(\alpha_N(\uxC))=\uyC).
\]
For any $\nu>0$, we define error probability $P_e^{(N)}(\nu)$ as 
\[
P_e^{(N)}(\nu)=\sum_{\uxC,\uyC}p(\uxC)\hp(\uyC|\uxC)
1\left(\frac1N\log\left(\frac{\hp(\uyC|\uxC)}{p(\uyC|\uxC)}\right)>\nu\right),
\]
where, as usual, $p(\uxC) = \prod_{\ell=1}^Np(\uxC(\ell))$ 
and $p(\uyC|\uxC) = \prod_{\ell=1}^Np(\uyC(\ell)|\uxC(\ell))$.  
\end{defi}

\begin{defi}\label{ZZZ-def:em}
The emulation region $\setE(\cC)$ of channel $\cC$ 
is the closure of all rate vectors $\cRC$ 
such that for any input distribution $p(\xC)$, 
any constant $\nu>0$, 
and all $N$ sufficiently large 
there exists a sequence of $(2^{N\cRC},N)$ emulation codes 
$(\alpha_N,\beta_N)$ 
with $P_e^{(N)}(\nu)<2^{-\eta(\nu) N}$ 
for some positive function $\eta(\nu)$ dependent on $p$ 
such that $\eta(\nu)$ approaches 0 as $\nu$ approaches 0.  
\end{defi}

Theorem~\ref{ZZZ-thm:equiv}, below, demonstrates 
that the standard of accuracy used to define the emulation region 
is sufficient to guarantee that $\cC(\cRC)$ 
is an upper bounding model for $\cC$.  
Whether this condition is also necessary 
remains an open problem.  

\begin{thm}\label{ZZZ-thm:equiv}
If $\cRC\in\mbox{int}(\cE(\cC))$, then $\cC\subseteq\cC(\cRC)$.  

\Proof
Fix rate vector $\cRC\in\mbox{int}(\setE(\cN))$, 
and consider a pair of networks 
\begin{eqnarray*}
\cN       & = & (\cXC\times\cXCm,p(\yC|\xC)p(\yCm|\xCm),\cYC\times\cYCm) \\
\cN(\cRC) & = & (\tcXC\times\cXCm,p(\tyC|\txC)p(\yCm|\xCm),\tcYC\times\cYCm).
\end{eqnarray*}
Next fix $\cR\in\mbox{int}(\setR(\cN))$.  
The argument that follows shows 
that $\cR\in\setR(\ucN(\cRC))$.  
This suffices to prove the desired result by Theorem~\ref{ZZZ-thm:stacked}
and the closure in the definition of $\setR(\cN(\cRC))$.  

\underline{Step 1 - Choose code $\cS(\cN)$ 
and define distribution $p_t(\xC,\yC)$:} \mbox{} \\ 
By Theorem~\ref{ZZZ-thm:stacked}, 
there exists a solution $\cS(\cN)$ 
of some finite blocklength $n$ 
from which we can build a $(2^{-N\delta},\cR)$-$\ucS(\ucN)$ stacked solution 
for $N$-fold stacked network $\ucN$ for all $N$ sufficiently large.  
Each stacked solution applies a random channel code to each message $\uWuv$ 
and then independently applies $\cS(\cN)$ in each layer of $\ucN$.  
For each $t\in\{1,\ldots,n\}$, 
let $p_t(\xC)$ be the input distribution to channel $\cC$ at time $t$ 
under solution $\cS(\cN)$.  
Then $p_t(\uxC,\uyC)\ \deff\ 
\prod_{\ell=1}^Np_t(\uxC(\ell))p(\uyC(\ell)|\uxC(\ell))$ 
is the time-$t$ distribution across 
the $N$ copies of channel $\cC$ in network $\ucN$ 
using solution $\ucS(\ucN)$.  

\underline{Step 2 - Choose channel emulators 
and bound the probability of emulation failure:} \mbox{} \\
For each $t\in\{1,\ldots,n\}$, 
choose $\nu(t)>0$ to satisfy 
\begin{eqnarray*}
\sum_{t'=1}^{t-1}\nu(t')& < & \eta_t(\nu(t))/2 \ \ \forall t\in\{1,\ldots,n\} \\
\sum_{t=1}^n\nu(t') & < & \delta/2,
\end{eqnarray*}
where $\eta_t(\cdot)$ designates the function $\eta$ 
corresponding to channel input distribution $p_t(\xC)$;  
these parameter choices make the error probability vanish 
in Step~5, below.  
We meet these constraints through the following sequence of parameter choices.  
First, set $\nu(n)=\delta/(4n)$.  
Then, in order of decreasing $t$ for each $t<n$, 
set $\nu(t)=\min\{\delta/(4n),\min_{t'>t}\eta_{t'}(\nu(t'))/(4t')\}$.  

Since $\cRC\in\mbox{int}(\setE(\cC))$, 
for each $t\in\{1,\ldots,n\}$ 
there exists a sequence of $(2^{N\cRC},N)$ 
random emulation codes $\hcC_{N,t}=(\alpha_{N,t},\beta_{N,t})$ 
that emulate channel $\cC$ under input distribution $p_t(\xC)$ 
with probability $P_{e,t}^{(N)}(\nu(t))<2^{-N\eta_t(\nu(t))}$ 
for all $N$ sufficiently large.  
Let $\hp_{N,t}(\uyC|\uxC)$ by the emulation distribution 
for $\hcC_{N,t}$, and define 
\begin{eqnarray*}
A_t^{(N)} &  \deff & 
\left\{(\uxC,\uyC):\frac1N\log\left(
\frac{\hp_{N,t}(\uyC|\uxC)}{p(\uyC|\uxC)}\right)\leq\nu(t)\right\}  \\
C_t^{(N)} & \deff & \left\{\uxC:
\hp_{N,t}( (A_t^{(N)})^c|\uxC)>2^{-N\eta_t(\nu(t))/2}\right\},
\end{eqnarray*}
where for any set $\cS\subseteq\cXC\times\cYC$, 
\[
\hp_t(\cS|\uxC)\ \deff\ \sum_{\uyC:(\uxC,\uyC)\in\cS}\hp_t(\uyC|\uxC).
\]
To bound $p_t(C_t^{(N)})=\sum_{\uxC\in C_t^{(N)}}p_t(\uxC)$, 
note that 
\begin{eqnarray*}
2^{-N\eta_t(\nu(t))} 
& \geq & \sum_{\uxC\in C_t^{(N)}}p_t(\uxC)\hp((A_t^{(N)})^c|\uxC)
			+\sum_{\uxC\not\in C_t^{(N)}}p_t(\uxC)\hp((A_t^{(N)})^c|\uxC) \\
& > & 2^{-N\eta_t(\nu(t))/2}\sum_{\uxC\in C_t^{(N)}}p_t(\uxC)
		+0\cdot\sum_{\uxC\not\in C_t^{(N)}}p_t(\uxC), 
\end{eqnarray*}
giving $p_t(C_t^{(N)})<2^{-N\eta_t(\nu(t))/2}$.  

\underline{Step 3 - Define solution $\cS(\ucN(\cRC))$:}\mbox{} \\
Let $\cS(\ucN(\cRC))$ be the code that results 
from operating solution $\ucS(\ucN)$ across network $\ucN(\cRC)$ 
with the aid of emulation codes $\{(\alpha_{N,t},\beta_{N,t})\}_{t=1}^n$.  
Formally, for each $v\in V$, let $\uvtY_t$ 
denote the network output received by node $v$ at time $t$.  
At time $t$, node $v$ 
applies the node encoder from $\ucS(\ucN)$ to obtain 
\[
\uvX_t=\uvX_t(\uvY_1,\ldots,\uvY_{t-1},
\uWvo,\ldots,\uWvm);
\]
here $\uvY_t$ is the channel output $\uvtY_t$ decoded (if necessary) as 
\[
\uvY_t = \left\{\begin{array}{ll} 
  (\beta_{N,t}^{(v)}(\utY^{(v,1)}_t),\utY^{(v,2)}) 
  & \mbox{if $v\in V_2$} \\
  \uvtY_t & \mbox{otherwise.} 
  \end{array}\right.
\]
Node $v$ then encodes $\uvX_t$ (if necessary) to give 
\[
\uvtX_t = \left\{\begin{array}{ll} 
   ((\alpha_{N,t}^{\ar{\{v\}}{B}}(\uX^{(v,1)}_t):B\subseteq V_2),\uX^{(v,2)}_t) 
   	& \mbox{if $v\in V_1$} \\
   (\alpha_{N,t}^{\ar{A}{B}}(\uX^{(v',1)}_t:v'\in A):B\subseteq V_2)
   	& \mbox{if $v=v^A$ for some $A\subseteq V_1$} \\
   \uvX_t & \mbox{otherwise,} 
   \end{array}\right.
\]
which it transmits across the bit-pipe model.  
After time $n$, node $v$ applies the decoders from $\ucS(\ucN)$ as 
\[
\ubWuv=\ubWuv(\uvY_1,\ldots,\uvY_n,\uWvo,\ldots,\uWvm).
\]
Solution $\cS(\ucN(\cRC))$ is not a stacked solution 
since each $(\alpha_{N,t},\beta_{N,t})$ 
operates across the layers of the stack.  

\underline{Step 4 - Characterize the statistical behavior of $\cS(\ucN(\cRC))$:}\mbox{}\\
Under the operation of $\ucS(\ucN)$ on $\ucN$, 
the joint distribution on messages $\uw$, 
network input vectors $\ubx^n=(\ubx_1,\ldots,\ubx_n)$, 
network output vectors $\uby^n=(\uby_1,\ldots,\uby_n)$, 
and message reconstructions $\ubw$ is 
\[
p(\uw,\ubx^n,\uby^n,\ubw)
 =  p(\uw)\left[\prod_{t=1}^np(\ubx_t|\uby^{t-1},\uw)\right]
	    \left[\prod_{t=1}^np(\uyC_t|\uxC_t)p(\uyCm_t|\uxCm_t)\right]p(\ubw|\uby^n,\uw),
\]
where $\ubx_t$ and $\uby_t$ again represent
the full vectors of network inputs and outputs at time $t$;
$p(\uw)$ is the distribution on messages; 
each $p(\ubx_t|\uby^{t-1},\uw)$ 
is a product distribution 
describing the independent operations 
performed by the node encoders at time $t$; 
$p(\uyC_t|\uxC_t)p(\uyCm_t|\uxCm_t)$ 
describes the memoryless network distribution; 
and $p(\ubw|\uby^n,\uw)$ is the product distribution 
describing the independent operation of each node decoder.  
Only the channel distribution changes 
when we run $\cS(\ucN(\cRC))$ on $\ucN(\cRC)$, 
giving 
\begin{eqnarray*}
\hp(\uw,\ubx^n,\uby^n,\ubw)
&  =  & p(\uw)\left[\prod_{t=1}^np(\ubx_t|\uby^{t-1},\uw)\right]
	    \left[\prod_{t=1}^n\hp_t(\uyC_t|\uxC_t)p(\uyCm_t|\uxCm_t)\right]p(\ubw|\uby^n,\uw).  
\end{eqnarray*}

\underline{Step 5 - Bound the expected error probability:} \mbox{} \\
The following error analysis relies on both 
probabilities resulting from running $\ucS(\ucN)$ on $\ucN$ and 
probabilities resulting from running $\cS(\ucN(\cRC))$ on $\ucN(\cRC)$.  
We use $\Pr(\cdot)$ for the former and $\hPr(\cdot)$ for the latter.  

Let 
\begin{equation}
\label{ZZZ-eqn:bad2}
\bad\ \deff\ \left\{(\uxC,\uyC):\Pr\left(\left.\ubW\neq\uW\right|
(\uXC_t,\uYC_t)=(\uxC,\uyC)\right)\geq 2^{-N\delta/2} \right\}.
\end{equation}
denote the set of input-output pairs on channel $\cC$ at time $t$ 
that are most likely to lead to errors in the operation of $\ucS(\ucN)$ on $\ucN$.  
The following error probability bound treats 
$(\uXC_t,\uYC_t)\not\in A_t^{(N)}$ and 
$(\uXC_t,\uYC_t)\in\bad$ for any $t\in\{1,\ldots,n\}$ as error events.  
We therefore define 
\[
G_t\ \deff\ \{(\ubx_t,\uby_t):(\uxC_t,\uyC_t)\in A_t^{(N)}\setminus\bad\}
\]
and bound the expected error probability of code $\cS(\ucN(\cRC))$ as 
\begin{eqnarray*}
\hPr(\ubW\neq\uW)& \leq & 
\sum_{t=1}^n\hPr(\cap_{t'<t}G_{t'}\cap(A_t^{(N)})^c)+
\sum_{t=1}^n\hPr(\cap_{t'<t}G_{t'}\cap A_t^{(N)}\cap\bad)\\
&&+\hPr(\cap_{t'\leq n}G_{t'}\cap\{\ubW\neq\uW\}).
\end{eqnarray*}
To bound the first two terms in the sum, note that for each $\uxC\in\ucXC$, 
\begin{eqnarray*}
\lefteqn{\hPr(\cap_{t'<t}G_{t'}\cap\{\uXC_t=\uxC\})} \\
& = & \sum_{(\uw,\ubx^{t-1},\uby^{t-1},\uxCm_t):
(\uxC_{t'},\uyC_{t'})\in G_{t'}\forall t'<t}
	p(\uw)\left[\prod_{t'=1}^tp(\ubx_{t'}|\uby^{t'-1},\uw)\right]
	    \left[\prod_{t'=1}^{t-1}\hp_{t'}(\uyC_{t'}|\uxC_{t'})p(\uyCm_{t'}|\uxCm_{t'})\right] \\
&\leq& \sum_{(\uw,\ubx^{t-1},\uby^{t-1},\uxCm_t):
	      (\uxC_{t'},\uyC_{t'})\in G_{t'}\forall t'<t}
	p(\uw)\left[\prod_{t'=1}^tp(\ubx_{t'}|\uby^{t'-1},\uw)\right]
	    \left[\prod_{t'=1}^{t-1}2^{N\nu(t')}p_{t'}(\uyC_{t'}|\uxC_{t'})
	    	p(\uyCm_{t'}|\uxCm_{t'})\right] \\
&=& 2^{N\sum_{t'=1}^{t-1}\nu(t')}
	\sum_{(\uw,\ubx^{t-1},\uby^{t-1},\uxCm_t):(\uxC_{t'},\uyC_{t'})\in G_{t'}\ \forall t'<t}
	p(\uw,\ubx^t,\uby^{t-1}) \\
&\leq& 2^{N\sum_{t'=1}^{t-1}\nu(t')}p_t(\uxC) 
\end{eqnarray*}
since $\uxC_{t'}\in G_t$ implies $\uxC_{t'}\in A_t^{(N)}$ 
which implies $p(\uyC_{t'}|\uxC_{t'})<2^{N\nu(t')}p(\uyC_{t'}|\uxC_{t'})$.  
Thus 
\begin{eqnarray*}
\lefteqn{\hPr(\cap_{t'<t}G_{t'}\cap (A_t^{(N)})^c)} \\
&\leq& 2^{N\sum_{t'=1}^{t-1}\nu(t')}\sum_{\uxC\in C_t^{(N)}}1\cdot p_t(\uxC) 
	+2^{N\sum_{t'=1}^{t-1}\nu(t')}\sum_{\uxC\not\in C_t^{(N)}}
	\hp_t((A_t^{(N)})^c|\uxC)p_t(\uxC)	\\
&<& 2^{-N(\eta_t(\nu(t))/2-\sum_{t'=1}^{t-1}\nu(t'))}+
	2^{-N(\eta_t(\nu(t))/2-\sum_{t'=1}^{t-1}\nu(t'))}
\end{eqnarray*}
by the definition of $C_t^{(N)}$ 
and the bound on $p_t(C_t^{(N)})$ from Step 2.
This sum goes to zero as $N$ grows without bound by our earlier parameter choice.  

To bound $\hp_t(\cap_{t'<t} G_{t'}\cap A_t^{(N)}\cap \bad)$, 
recall that $\ucS(\ucN)$ is a $(2^{-N\delta},\cR)$ solution 
and there are fewer than $m^2$ messages to transmit.  
Thus $\Pr(\ubW\neq\uW) < m^2 2^{-N\delta}$ for solution $\ucS(\ucN)$ 
by the union bound, giving 
\begin{eqnarray*}
m^22^{-N\delta} 
&> & \Pr(\ubW\neq\uW) \\
&\geq & \sum_{(\uxC,\uyC)\in\bad}p_t(\uxC,\uyC) 
\Pr(\ubW\neq\uW|\uxC,\uyC)\\
&\geq & 2^{-N\delta/2}p_t(\bad), 
\end{eqnarray*}
giving $p_t(\bad)<m^22^{-N\delta/2}$.  
Thus 
\begin{eqnarray*}
\hPr(\cap_{t'<t}G_{t'}\cap A_t^{(N)}\cap B_t^{(N)}) 
&\leq& 2^{N\sum_{t'=1}^{t-1}\nu(t')}
	\sum_{(\uxC,\uyC)\in A_t^{(N)}\cap\bad}p_t(\uxC)\hp_t(\uyC|\uxC) \\
&\leq& 2^{N\sum_{t'=1}^t\nu(t')}
	\sum_{(\uxC,\uyC)\in A_t^{(N)}\cap\bad}p_t(\uxC)p_t(\uyC|\uxC) \\
&\leq&2^{N\sum_{t'=1}^t\nu(t')}p_t(\bad) \\
&<& m^22^{-N(\delta/2-\sum_{t'=1}^t\nu(t'))},
\end{eqnarray*}
which also goes to zero by our choice of $\nu(1),\ldots,\nu(n)$.  

Finally, 
\begin{eqnarray*}
\lefteqn{\hPr\left(\cap_{t=1}^n G_t\cap\{\ubW\neq\uW\}\right)} \\
& \stackrel{(a)}{<} 
& \sum_{(\uw,\ubx^n,\uby^n,\ubw):\ubw\neq\uw, (\uxC_t,\uyC_t)\in G_t\forall t}
p(\uw)
\left[\prod_{t=1}^n p(\ubx_t|\uby^{t-1},\uw)\right] 
\left[\prod_{t=1}^n 2^{N\sum_{t=1}^n\nu(t)}p_t(\uby_t|\ubx_t)\right] 
      p(\ubw|\uw,\ubx^n,\uby^n) \\
&\stackrel{(b)}{\leq} & 2^{N\sum_{t=1}^n\nu(t)}
      \sum_{(\uw,\ubx^n,\uby^n,\ubw):\uw\neq\ubw,\ (\uxC_1,\uyC_1)
      	 \in A_1^{(N)}\setminus B_1^{(N)}}p(\uw,\ubx^n,\uby^n,\ubw)\\
& = & 2^{N\sum_{t=1}^n\nu(t)}\sum_{(\uxC,\uyC)\in A_1^{(N)}\setminus B_1^{(N)}}
p_1(\uxC,\uyC) \Pr(\ubW\neq\uW|(\uxC,\uyC)) \\
&\stackrel{(c)}{<} & 2^{-N(\delta/2-\sum_{t=1}^n\nu(t))}
p_1(A_1^{(N)}\setminus B_1^{(N)}) \\
&\leq& 2^{-N(\delta/2-\sum_{t=1}^n\nu(t))}.
\end{eqnarray*}
Equation ($a$) follows from our probability characterization 
in Step~4 since $(\uxC_t,\uyC_t)\in A_t^{(N)}$ 
for all $t$ by definition of $G_t$.  
In ($b$), we bound the sum over $A_t^{(N)}\setminus\bad$ 
by the sum over all $\ucXC\times\ucYC$ for all $t>1$.  
Equation ($c$) follows 
from the definition of $\bad$ in (\ref{ZZZ-eqn:bad2}).  
This term also goes to zero as $N$ grows large 
by our choice of $\nu(1),\ldots,\nu(n)$.  
Since the expected error probability 
for our randomly drawn code can be made arbitrarily small 
there exists a single instance that does at least as well.  
Thus $\cR\in\setR(\ucN(\cRC))$.  
\IEEEQED  
\end{thm}

\begin{rem}
Lemma~\ref{ZZZ-lem:cont} can be used 
to show that Theorem~\ref{ZZZ-thm:equiv} also holds 
for all points $\cRC$ on the outer boundary of $\setE(\cN)$ 
with strictly positive coefficients.  
It is not clear whether it holds for all boundary points 
since $\cN_R$ is not known to be continuous at $R=0$ 
for general networks~\cite{ZZZ-GuE:09,ZZZ-Gu:09}. 
\end{rem}

\section{Upper Bounding Models}

While existing achievability results for individual channels 
lead immediately to lower bounding networks (see Lemma~\ref{ZZZ-lem:lb}), 
capacity upper bounds do not generally 
give legitimate upper bounding networks.  
Roughly speaking, there are two causes of this phenomenon.  
First, capacity upper bounds for multi-input channels ($|V_1|>1$) 
assume independent transmissions from their input nodes;  
when the channel is used within a larger network, 
the inputs may be statistically dependent.  
Second, capacity upper bounds assume reliable transmission 
across each component channel;
operating individual channels above their capacities 
sometimes increases the network capacity, 
as shown in Examples~\ref{ZZZ-ex:bcmac} and~\ref{ZZZ-ex:bcmacind}.  
By Theorem~\ref{ZZZ-thm:equiv}, we can build upper bounding models 
by finding points in the emulation region 
described in Definition~\ref{ZZZ-def:ec}.  

We here derive example upper bounding models 
for the broadcast, multiple access, and interference channels.  
All of the results use the bit-pipe models 
defined in Section~\ref{ZZZ-sec:models}, 
removing bit pipes of capacity 0.  
Recall that for each $A\subseteq V_1$, 
internal node $v^{A}$ receives a noiseless description 
of channel inputs $(X^{(v,1)}:v\in A)$.  
These noiseless descriptions are transmitted along internal edges 
of capacity $\log|\vcX|$, as described in the model definitions; 
$\log|\vcX|$ is infinite when $\vcX$ is continuous.  
In Section~\ref{ZZZ-sec:ex}, 
we bound the accuracy of capacity bounds 
derived using these models for a variety of example channel types, 
including channels with continuous alphabets.  

This section derives general form solutions.  
Examples for specific channels appear in Section~\ref{ZZZ-sec:ex}.  
Each result describes a family of upper bounding models 
both because multiple rate vectors satisfy the given bounds 
and because switching the roles of the nodes in asymmetrical solutions 
may yield new bounds.  
Taking the intersection of the rate regions corresponding 
to different bounds may yield a tighter bound.  

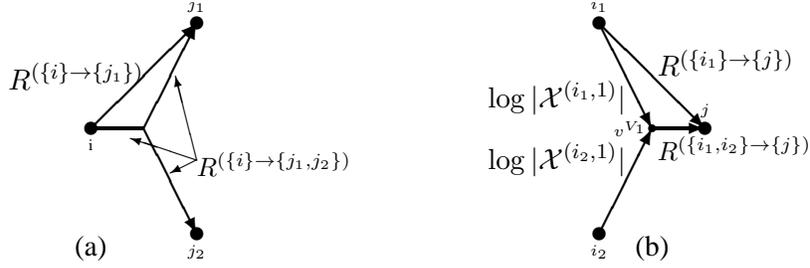
\begin{figure}
  \begin{center}
  \begin{picture}(40,90)(20,-30)
      \thicklines
      \put(20,20){\circle*{5}}
      \put(20,16){\makebox(0,0)[ct]{\tiny i}}
      \put(20,20){\vector(1,1){40}}
      \put(20,20){\line(1,0){20}}
      \put(40,20){\vector(1,2){20}}
      \put(40,20){\vector(1,-2){20}}
      \put(40,35){\makebox(0,0)[rb]{$R^{\ar{\{i\}}{\{j_1\}}}$}}
      \put(60,60){\circle*{5}}
      \put(60,-20){\circle*{5}}
      \put(60,64){\makebox(0,0)[cb]{\tiny $j_1$}}
      \put(60,-24){\makebox(0,0)[ct]{\tiny $j_2$}}
      \put(60,2){\makebox(0,0)[lb]{\small $R^{\ar{\{i\}}{\{j_1,j_2\}}}$}}
      \thinlines
      \put(60,8){\vector(-2,-1){10}}
      \put(60,8){\vector(-1,4){8}}
      \put(60,8){\vector(-3,1){25}}
      \put(20,-30){\makebox(0,0)[cb]{(a)}}
  \end{picture} 
  \hspace{2in}
  \begin{picture}(40,90)(0,-30)
    \thicklines
      \put(0,60){\circle*{5}}
      \put(0,-20){\circle*{5}}
      \put(0,64){\makebox(0,0)[cb]{\tiny $i_1$}}
      \put(0,-24){\makebox(0,0)[ct]{\tiny $i_2$}}
      \put(0,60){\vector(1,-1){40}}
      \put(22,40){\makebox(0,0)[lb]{$R^{\ar{\{i_1\}}{\{j\}}}$}}
      \put(0,60){\vector(1,-2){20}}
      \put(10,38){\makebox(0,0)[rt]{$\log|\cX^{(i_1,1)}|$}}
      \put(10,2){\makebox(0,0)[rb]{$\log|\cX^{(i_2,1)}|$}}
      \put(0,-20){\vector(1,2){20}}
      \put(20,20){\circle*{3}}
      \put(18,20){\makebox(0,0)[rc]{\tiny $v^{V_1}$}}
      \put(20,20){\vector(1,0){20}}
      \put(40,20){\circle*{5}}
      \put(40,24){\makebox(0,0)[cb]{\tiny $j$}}
      \put(22,17){\makebox(0,0)[lt]{\small $R^{\ar{\{i_1,i_2\}}{\{j\}}}$}}
      \put(20,-30){\makebox(0,0)[cb]{(b)}}
      \end{picture} %
  \end{center}
\caption{Upper bounding models for the 
(a)~broadcast channel (Theorem~\ref{ZZZ-thm:bcmain}), 
and (b) multiple access channel (Theorem~\ref{ZZZ-thm:mamain}).}\label{ZZZ-fig:ub}
\end{figure}

Given a broadcast channel with transmitter $V_1=\{i\}$ 
and receivers $V_2=\{j_1,j_2\}$, 
Theorem~\ref{ZZZ-thm:bcmain} derives 
an upper bounding model of the form 
shown in Figure~\ref{ZZZ-fig:ub}(a).  

\begin{thm}\label{ZZZ-thm:bcmain} 
Let 
\begin{eqnarray*}
\cC & = & (\ciXs,p(\jyo,\jyt|\ixs),\cjYo\times\cjYt) \\
\cC(\cRC) & = & (\tciXs,p(\tjyo,\tjyt|\tixs),\tcjYo\times\tcjYt) \
\end{eqnarray*}
be a broadcast channel and its corresponding bit-pipe model 
for some $\cRC$ satisfying 
\begin{eqnarray*} 
R^{\ar{\{i\}}{\{j_1,j_2\}}}  & > & I(\iXs;\jYt) \\
R^{\ar{\{i\}}{\{j_1,j_2\}}}+R^{\ar{\{i\}}{\{j_1\}}} & > & I(\iXs;\jYo,\jYt) 
\end{eqnarray*}
for all distributions $p(\ixs,\jyo,\jyt)= p(\ixs)p(\jyo,\jyt|\ixs)$.  
Then $\cC\subseteq\cC(\cRC)$.  

\Proof  See Appendix~\ref{ZZZ-app:bc}.  
\IEEEQED
\end{thm}

Theorem~\ref{ZZZ-thm:mamain} derives an upper bounding model 
of the form shown in Figure~\ref{ZZZ-fig:ub}(b) 
for a multiple access channel with transmitters $V_1=\{i_1,i_2\}$ 
and receiver $V_2=\{j\}$.  

\begin{thm}\label{ZZZ-thm:mamain} 
Let 
\begin{eqnarray*}
\cC & = & (\ciXo\times\ciXt,p(\jyr|\ixo,\ixt),\cjYr) \\
\cC(\cRC) & = & (\tciXo\times\tciXt,p(\tjyr|\tixo,\tixt),\tcjYr)
\end{eqnarray*}
be a multiple access channel 
and its corresponding bit-pipe model for some $\cRC$.  
If for each distribution $p(\ixo,\ixt)$ 
there exists a distribution $p(u|\ixo)$ on an alphabet 
$\cU$ with $|\cU|\leq|\ciXo|$ such that 
\begin{eqnarray*} 
R^{\ar{\{i_1\}}{\{j\}}} & > & I(\iXo;U) \\
R^{\ar{\{i_1,i_2\}}{\{j\}}} & > & I(\iXo,\iXt;\jYr|U), 
\end{eqnarray*}
then $\cC\subseteq \cC(\cRC)$.  

\Proof  See Appendix~\ref{ZZZ-app:ma}.
\IEEEQED
\end{thm}

\begin{figure}
\begin{center}
  \begin{center}
  \begin{picture}(95,100)(105,-40)
      \thicklines
      \put(105,60){\circle*{5}}
      \put(105,-20){\circle*{5}}
      \put(105,64){\makebox(0,0)[cb]{\tiny $i_1$}}
      \put(105,-24){\makebox(0,0)[ct]{\tiny $i_2$}}
      \put(105,60){\vector(1,-1){40}}
      \put(105,60){\line(1,-4){10}}
      \put(115,20){\line(1,0){10}}
      \put(125,20){\vector(1,1){40}}
      \put(125,20){\vector(1,-1){40}}
      \put(105,60){\vector(1,0){60}}
      \put(105,-20){\vector(1,1){40}}
      \put(145,20){\circle*{3}}
      \put(142,18){\makebox(0,0)[tl]{\tiny $v^{V_1}$}}
      \put(145,20){\vector(1,2){20}}
      \put(145,20){\line(1,0){10}}
      \put(155,20){\vector(1,4){10}}
      \put(155,20){\vector(1,-4){10}}
      \put(165,60){\circle*{5}}
      \put(165,-20){\circle*{5}}
      \put(165,64){\makebox(0,0)[cb]{\tiny $j_1$}}
      \put(165,-24){\makebox(0,0)[ct]{\tiny $j_2$}}
      \thinlines
      \put(55,52){\makebox(0,0)[lc]{\tiny $R^{\ar{\{i_1\}}{V_2}}$}}
      \put(96,52){\vector(1,0){10}}
      \put(55,42){\makebox(0,0)[lc]{\tiny $\log|\cX^{(i_1,1)}|$}}
      \put(96,42){\vector(1,0){26}}
      \put(55,-5){\makebox(0,0)[lc]{\tiny $\log|\cX^{(i_2,1)}|$}}
      \put(96,-5){\vector(1,0){23}}
      \put(135,62){\makebox(0,0)[cb]{\tiny $R^{\ar{\{i_1\}}{\{j_1\}}}$}}
      \put(157,20){\makebox(0,0)[cl]{\tiny $R^{\ar{V_1}{V_2}}$}}
      \put(162,30){\makebox(0,0)[cl]{\tiny $R^{\ar{V_1}{\{j_1\}}}$}}
      \put(161.5,29){\vector(-1,0){11.5}}
      \put(135,-45){\makebox(0,0)[cb]{(a) Model 1}}
  \end{picture} %
  \hspace{1in}
  \begin{picture}(95,100)(105,-40)
      \thicklines
      \put(105,60){\circle*{5}}
      \put(105,-20){\circle*{5}}
      \put(105,64){\makebox(0,0)[cb]{\tiny $i_1$}}
      \put(105,-24){\makebox(0,0)[ct]{\tiny $i_2$}}
      \put(105,60){\vector(3,-4){60}}
      \put(105,60){\line(1,-4){10}}
      \put(115,20){\line(1,0){10}}
      \put(125,20){\vector(1,1){40}}
      \put(125,20){\vector(1,-1){40}}
      \put(105,60){\vector(1,-1){40}}
      \put(105,-20){\vector(1,1){40}}
      \put(145,20){\circle*{3}}
      \put(145,23){\makebox(0,0)[lb]{\tiny $v^{V_1}$}}
      \put(145,20){\vector(1,-2){20}}
      \put(145,20){\line(1,0){10}}
      \put(155,20){\vector(1,4){10}}
      \put(155,20){\vector(1,-4){10}}
      \put(165,60){\circle*{5}}
      \put(165,-20){\circle*{5}}
      \put(165,64){\makebox(0,0)[cb]{\tiny $j_1$}}
      \put(165,-24){\makebox(0,0)[ct]{\tiny $j_2$}}
      \put(135,-45){\makebox(0,0)[cb]{(b) Model 2}}
      \thinlines
      \put(55,52){\makebox(0,0)[lc]{\tiny $R^{\ar{\{i_1\}}{V_2}}$}}
      \put(96,52){\vector(1,0){10}}
      \put(55,42){\makebox(0,0)[lc]{\tiny $R^{\ar{\{i_1\}}{\{j_2\}}}$}}
      \put(102,42){\vector(1,0){16}}
      \put(55,32){\makebox(0,0)[lc]{\tiny $\log|\cX^{(i_1,1)}|$}}
      \put(96,32){\vector(1,0){36}}
      \put(55,-5){\makebox(0,0)[lc]{\tiny $\log|\cX^{(i_2,1)}|$}}
      \put(96,-5){\vector(1,0){23}}
      \put(157,20){\makebox(0,0)[cl]{\tiny $R^{\ar{V_1}{V_2}}$}}
      \put(167,10){\makebox(0,0)[lc]{\tiny $R^{\ar{V_1}{\{j_2\}}}$}}
      \put(166,10){\vector(-1,0){15}}
  \end{picture} %
  \end{center}
\caption{Upper bounding models for the interference channel.}
\label{ZZZ-fig:inequiv}
\end{center}
\end{figure}
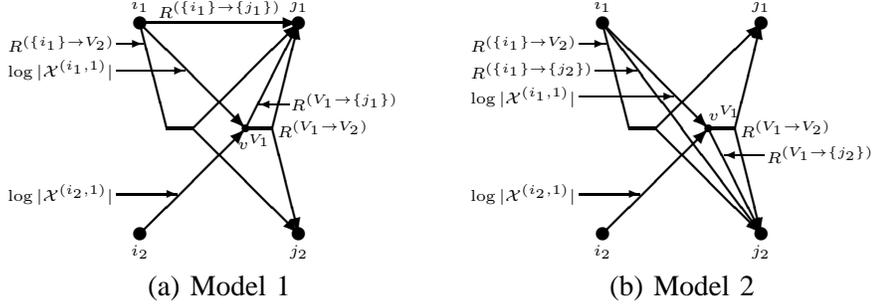

Let 
$\cC = (\ciXo\times\ciXt,p(\jyo,\jyt|\ixo,\ixt),\cjYo\times\cjYt)$ 
be an interference channel 
with transmitters $V_1=\{i_1,i_2\}$ and receivers $V_2=\{j_1,j_2\}$.  
Theorems~\ref{ZZZ-thm:inmaina} and~\ref{ZZZ-thm:inmainb} 
derive upper bounding models for $\cC$ 
of the forms shown in Figures~\ref{ZZZ-fig:inequiv}(a) and~(b), 
respectively.  
In the first case, node $i_1$ transmits two descriptions, 
one to just $j_1$ and the other to both receivers.  
Node $v^{V_1}$ noiselessly receives both channel inputs 
and transmits one description to $j_1$ and the other to both receivers.  

\begin{thm}\label{ZZZ-thm:inmaina} 
Let 
\begin{eqnarray*}
\cC       & = & (\ciXo\times\ciXt,p(\jyo,\jyt|\ixo,\ixt),\cjYo\times\cjYt) \\
\cC(\cRC) & = & (\tciXo\times\tciXt,p(\tjyo,\tjyt|\tixo,\tixt),
\tcjYo\times\tcjYt)
\end{eqnarray*}
be an interference channel and its rate-$\cRC$ bit-pipe model.  
If for each distribution $p(\ixo,\ixt)$ 
there exist conditional distributions  
$p(\ut|\ixo)$ and $p(\uo|\ixo,\ut)$ 
with $|\cU_1\times\cU_2|\leq |\ciXo|$ 
and 
\begin{eqnarray*}
R^{\ar{\{i_1\}}{\{j_1\}}}+R^{\ar{\{i_1\}}{\{j_1,j_2\}}} &>& I(\iXo;\Uo,\Ut) \\
R^{\ar{\{i_1\}}{\{j_1,j_2\}}} &>& I(\iXo;\Ut) \\
R^{\ar{\{i_1,i_2\}}{\{j_1\}}}+R^{\ar{\{i_i,i_2\}}{\{j_1,j_2\}}} &>&
I(\iXo,\iXt;\jYo|\Uo,\Ut,\jYt) \\
&&+I(\iXo,\iXt;\jYt|\Ut) \\
R^{\ar{\{i_1,i_2\}}{\{j_1,j_2\}}} &>& I(\iXo,\iXt;\jYt|\Ut),
\end{eqnarray*}
then $\cC\subseteq\cC(\cRC)$.  

\Proof See Appendix~\ref{ZZZ-app:ina}
\IEEEQED
\end{thm}

In the second bit-pipe model for the interference channel, 
node $i_1$ again transmits two descriptions.
Here the first is delivered to both receivers while the second 
is delivered only to $j_2$.  
Node $v^{V_1}$ noiselessly receivers 
both channel inputs and transmits one description to both receivers 
and the other only to $j_2$.

\begin{thm}\label{ZZZ-thm:inmainb} 
Let 
\begin{eqnarray*}
\cC &= & (\ciXo\times\ciXt,p(\jyo,\jyt|\ixo,\ixt),\cjYo\times\cjYt) \\
\cC(\cRC) & = & 
(\tciXo\times\tciXt,p(\tjyo,\tjyt|\tixo,\tixt),\tcjYo\times\tcjYt) 
\end{eqnarray*}
be an interference channel and its rate-$\cRC$ bit-pipe model.  
If for each distribution $p(\ixo,\ixt)$ 
there exist conditional distributions  
$p(\uo|\ixo)$ and $p(\ut|\uo,\ixo)$ 
with $|\cU_1\times\cU_2|\leq|\ciXo|$
for which 
\begin{eqnarray*}
R^{\ar{\{i_1\}}{\{j_1,j_2\}}} & > & I(\iXo;\Uo) \\
R^{\ar{\{i_1\}}{\{j_1,j_2\}}}+R^{\ar{\{i_1\}}{\{j_2\}}}
& > & I(\iXo;\Uo,\Ut) \\
R^{\ar{\{i_1,i_2\}}{\{j_1,j_2\}}} & > & I(\iXo,\iXt;\jYo|\Uo) \\
R^{\ar{\{i_1,i_2\}}{\{j_1,j_2\}}}+R^{\ar{\{i_1,i_2\}}{\{j_2\}}}
& > & I(\iXo,\iXt;\jYo|\Uo) \\
&&+I(\iXo,\iXt;\jYt|\Uo,\Ut,\jYo) 
\end{eqnarray*}
then $\cC\subseteq\cC(\cRC)$.  

\Proof See Appendix~\ref{ZZZ-app:inb}
\IEEEQED
\end{thm}

\section{Bounding Accuracy}\label{ZZZ-sec:ex}

The equivalence tools 
derived in Section~\ref{ZZZ-sec:equiv} 
yield upper and lower bounding models 
for a single independent channel $\cC$.  
Repeated application of these tools 
on networks containing multiple independent channels 
allows us to bound the capacity of a network of noisy channels 
by bounding the capacity of another network 
in which some or all of the network's stochastic components 
have been replaced by bit-pipe models.  
To make this precise, let $\cN$ be a network 
containing some collection $\cA$ of independent channels.  
Then for any $\cRL=(\cRCL:\cC\in\cA)\in\prod_{\cC\in\cA}\setR(\cC)$ 
and any      $\cRU=(\cRCU:\cC\in\cA)\in\prod_{\cC\in\cA}\setE(\cC)$, 
$\cRCL$ and $\cRCU$ describe lower and upper bounds for $\cC$
(i.e., $\cC(\cRCL)\subseteq\cC\subseteq\cC(\cRCU)$) 
for each $\cC\in\cA$.  
Let $\cN(\cRL)$ denote the network obtained by replacing each $\cC\in\cA$ 
by its lower bounding model $\cC(\cRCL)$ 
and $\cN(\cRU)$ denote the network obtained by replacing each $\cC\in\cA$ 
by it upper bounding model $\cC(\cRCU)$.  
Then Lemma~\ref{ZZZ-lem:lb} and Theorem~\ref{ZZZ-thm:equiv} 
imply 
\[
\setR_L(\cN)\subseteq\setR(\cN)\subseteq\setR_U(\cN),
\]
where 
\begin{eqnarray*}
\setR_L(\cN) & \deff & \bigcup_{\cRL\in\prod_{\cC\in\cA}\setR(\cC)}
                  \setR(\cN(\cRL)) \\
\setR_U(\cN) & \deff & \bigcap_{\cRU\in\mbox{int}(\prod_{\cC\in\cA}\setE(\cC))}
                  \setR(\cN(\cRU)).  
\end{eqnarray*}
The discussion that follows 
finds multiplicative and additive bounds 
on the difference between $\setR_L(\cN)$ and $\setR_U(\cN)$, 
thereby bounding the accuracy of $\setR_L(\cN)$ and $\setR_U(\cN)$ 
as approximations for $\setR(\cN)$. 

Lemma~\ref{ZZZ-lem:ratio}, below, 
shows that there exists a constant $a\in[0,1]$ 
such that $\cR\in\setR_U(\cN)$ implies $a\cR\in\setR_L(\cN)$; 
we henceforth use notation 
\[
\setR_L(\cN)\geq a\setR_U(\cN), 
\]
to specify this relationship.  
Lemma~\ref{ZZZ-lem:ratio}'s strength is that it applies to all demand types 
and does not increase with the network size;  
its weakness that constant $a$ 
is determined by the worst-case channel in $\cA$.  
The following definition is used in that result.  
Recall from Section~\ref{ZZZ-sec:models} that 
the models for vectors 
$\cRCL=(R_{\cC,L}^{\ar{A}{B}}:(A,B)\in\cM)\in\setR(\cC)$ and 
$\cRCU=(R_{\cC,U}^{\ar{A}{B}}:(A,B)\in\cM)\in\setE(\cC)$ 
are identical in their topologies 
(except for possible missing edges corresponding to 
rate-0 entries in $\cRCL$ or $\cRCU$).  
We can therefore define the worst-case ratio 
between individual edges of these  models as 
\[
\rho(\cC)\ \deff\ \sup_{(\cRCL,\cRCU)\in\setR(\cC)\times{\rm int}(\setE(\cC))}
\ \min_{(A,B)\in\cM:\ 
                    R_{\cC,U}^{\ar{A}{B}}\geq R_{\cC,L}^{\ar{A}{B}},\ 
                    R_{\cC,U}^{\ar{A}{B}}>0}
\frac{R_{\cC,L}^{\ar{A}{B}}}{R_{\cC,U}^{\ar{A}{B}}}.  
\]

\begin{lem}\label{ZZZ-lem:ratio}
\[
\setR_L(\cN)\geq \left[\min_{\cC\in\cA}\rho(\cC)\right] \setR_U(\cN) 
\]
\Proof
Let $a=\min_{\cC\in\cA}\rho(\cC)$, 
and for each $\cC\in\cA$ 
fix some sequence $\{(\cRCLk,\cRCUk)\}_{k=1}^\infty$ 
such that $(\cRCLk,\cRCUk)\in\setR(\cC)\times\mbox{int}(\setE(\cC))$ 
for all $k$ and ratio 
\[
a_{\cC,k}\ \deff\ \ \min_{(A,B)\in\cM:
  R_{\cC,U,k}^{\ar{A}{B}}\geq R_{\cC,L,k}^{\ar{A}{B}},\ 
  R_{\cC,U,k}^{\ar{A}{B}}>0}
\frac{R_{\cC,L,k}^{\ar{A}{B}}}{R_{\cC,U,k}^{\ar{A}{B}}}
\]
monitonically approaches $\rho(\cC)$ as $k$ grows without bound.  
Let $\cN_{L,k}$, $\cN_{U,k}$, and $\cN_{a_kU,k}$ 
be the networks that result 
when each channel $\cC\in\cA$ 
is replaced by bit-pipe model 
$\cC(\cR_{\cC,L,k})$, 
$\cC(\cR_{\cC,U,k})$, and 
$\cC(a_k\cR_{\cC,U,k})$, respectively, 
where $a_k=\min_{\cC\in\cA}a_{\cC,k}$.  
Then 
\[
\setR(\cN_{a_kU,k})\subseteq\setR(\cN_{L,k})
\subseteq \setR(\cN)\subset \setR(\cN_{U,k}) 
\]
since $a_k\cRCUk\leq\cRCLk$ for all $\cC$ by definition of $a_k$.  
Network $\cN_{a_kU,k}$ is identical to network $\cN_{U,k}$ 
except that the capacity of each bit-pipe model edge 
has been decreased by factor $a_k$.  
We next employ Theorem~\ref{ZZZ-thm:stacked} 
to bound the difference between 
$\setR(\cN_{a_kU,k})$ and $\setR(\cN_{U,k})$.  
Let $\ucN_{U,k}$ be the $N$-fold stacked network for $\cN_{U,k}$ 
and let $\ucN_{a_kU,k}$ be the $\lceil N/a_k\rceil$-fold 
stacked network for $\cN_{a_kU,k}$.  
We can operate any $(\cR,\lambda)$ solution 
$\cS(\ucN_{U,k})$ for $\ucN_{U,k}$ 
across network $\ucN_{a_kU,k}$ as follows.  
For each $\cC\in\cA$, transmit the $N\cRCUk$ bits 
intended for transmission across $N$ copies of $\cC(\cRCUk)$ 
across the $\lceil N/a_k\rceil$ copies 
of $\cC(a_k\cRCUk)$ in $\ucN_{a_kU,k}$.  
Transmissions across the remainder of the network 
are sent unchanged.  
Applying $\cS(\ucN_{U,k})$ across $\ucN_{a_kU,k}$ in this way 
delivers $N\cR$ bits over $\ceil{N/a_k}$ layers 
with error probability $\lambda$.  
The rate $N\cR/\ceil{N/a_k}$ 
approaches $a_k\cR$ as $N$ grows without bound.  
Letting $k$ grow without bound 
achieves the desired result.  
\IEEEQED
\end{lem}

By~\cite[Corollary 5]{ZZZ-KoetterE:10a}, 
the best upper and lower bound 
for any memoryless point-to-point channel are the same.  
Thus $\rho(\cC)=1$ for memoryless point-to-point channels.  
The following examples bound $\rho(\cC)$ 
for binary broadcast and multiple access channels with additive noise.  

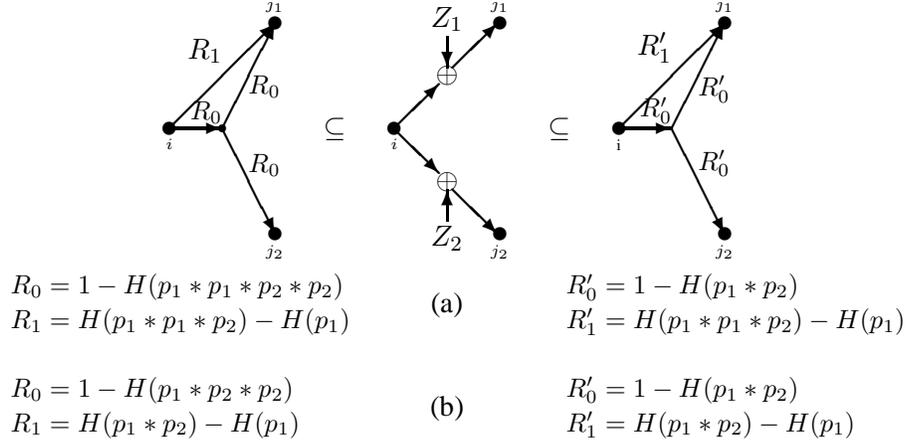
\begin{figure}
  \begin{center}
    \begin{picture}(210,157)(20,-97)
      \thicklines
      \put(20,20){\circle*{5}}
      \put(20,16){\makebox(0,0)[ct]{\tiny $i$}}
      \put(40,20){\circle*{3}}
      \put(20,20){\vector(1,1){40}}
      \put(20,20){\vector(1,0){20}}
      \put(40,20){\vector(1,2){20}}
      \put(40,20){\vector(1,-2){20}}
      \put(40,45){\makebox(0,0)[rb]{$R_1$}}
      \put(60,60){\circle*{5}}
      \put(60,-20){\circle*{5}}
      \put(60,64){\makebox(0,0)[cb]{\tiny $j_1$}}
      \put(60,-24){\makebox(0,0)[ct]{\tiny $j_2$}}
      \put(50,40){\makebox(0,0)[lt]{\small $R_{0}$}}
      \put(34,22){\makebox(0,0)[cb]{\small $R_{0}$}}
      \put(50,2){\makebox(0,0)[lb]{\small $R_{0}$}}
      \put(82.5,20){\makebox(0,0)[cc]{$\subseteq$}}
      \put(105,20){\circle*{5}}
      \put(105,16){\makebox(0,0)[ct]{\tiny $i$}}
      \put(105,20){\vector(1,1){17}}
      \put(105,20){\vector(1,-1){17}}
      \put(125,40){\makebox(0,0)[cc]{$\oplus$}}
      \put(125,0){\makebox(0,0)[cc]{$\oplus$}}
      \put(125,55){\vector(0,-1){12}}
      \put(125,-15){\vector(0,1){12}}
      \put(125,57){\makebox(0,0)[cb]{$Z_1$}}
      \put(125,-17){\makebox(0,0)[ct]{$Z_2$}}
      \put(127,43){\vector(1,1){17}}
      \put(127,-3){\vector(1,-1){17}}
      \put(145,60){\circle*{5}}
      \put(145,-20){\circle*{5}}
      \put(145,64){\makebox(0,0)[cb]{\tiny $j_1$}}
      \put(145,-24){\makebox(0,0)[ct]{\tiny $j_2$}}
      \put(167.5,20){\makebox(0,0)[cc]{$\subseteq$}}
      \put(190,20){\circle*{5}}
      \put(190,16){\makebox(0,0)[ct]{\tiny i}}
      \put(190,20){\circle*{3}}
      \put(190,20){\vector(1,1){40}}
      \put(190,20){\vector(1,0){20}}
      \put(210,20){\vector(1,2){20}}
      \put(210,20){\vector(1,-2){20}}
      \put(210,45){\makebox(0,0)[rb]{$R'_1$}}
      \put(230,60){\circle*{5}}
      \put(230,-20){\circle*{5}}
      \put(230,64){\makebox(0,0)[cb]{\tiny $j_1$}}
      \put(230,-24){\makebox(0,0)[ct]{\tiny $j_2$}}
      \put(220,40){\makebox(0,0)[lt]{\small $R'_{0}$}}
      \put(204,22){\makebox(0,0)[cb]{\small $R'_{0}$}}
      \put(220,2){\makebox(0,0)[lb]{\small $R'_{0}$}}
      \put(-40,-45){\makebox(0,0)[lb]{\small $R_{0}=1-H(p_1*p_1*p_2*p_2)$}}
      \put(-40,-58){\makebox(0,0)[lb]{\small $R_1=H(p_1*p_1*p_2)-H(p_1)$}}
      \put(170,-45){\makebox(0,0)[lb]{\small $R'_{0}=1-H(p_1*p_2)$}}
      \put(170,-58){\makebox(0,0)[lb]{\small $R'_1=H(p_1*p_1*p_2)-H(p_1)$}}
      \put(125,-51.5){\makebox(0,0)[cb]{(a)}}
      \put(-40,-84){\makebox(0,0)[lb]{\small $R_{0}=1-H(p_1*p_2*p_2)$}}
      \put(-40,-97){\makebox(0,0)[lb]{\small $R_1=H(p_1*p_2)-H(p_1)$}}
      \put(170,-84){\makebox(0,0)[lb]{\small $R'_{0}=1-H(p_1*p_2)$}}
      \put(170,-97){\makebox(0,0)[lb]{\small $R'_1=H(p_1*p_2)-H(p_1)$}}
      \put(125,-90.5){\makebox(0,0)[cb]{(b)}}
      \end{picture} 
  \end{center}
  \caption{Example upper and lower bounding models 
           for the binary symmetric broadcast channel 
	   with error probabilities $p_1$ and $p_1*p_2$ 
	   at its two receivers.  
	   The bit-pipe capacities given in (a) and (b) 
	   correspond to the independent noise 
	   and physically degraded cases, respectively.  
           }
  \label{ZZZ-fig:bsbc}
\end{figure}

\begin{exam}\label{ZZZ-ex:bsbc}
Let $\cC=(\{0,1\},p(\jyo,\jyt|\ixs),\{0,1\}^2)$ 
be a binary symmetric broadcast channel.   
Then $\jYo=\iXs\oplus Z_1$ and  $\jYt=\iXs\oplus Z_2$ 
as shown in Figure~\ref{ZZZ-fig:bsbc}. 
Let $p_1=EZ_1$ and $p_1*p_2=p_1(1-p_2)+p_2(1-p_1)=EZ_2$.  
Figure~\ref{ZZZ-fig:bsbc} shows example bounding networks.  
The lower bounding models correspond to points 
$(R_0,R_1)=(1-H(\alpha*p_1*p_2),H(\alpha*p_1)-H(p_1))$ 
on the boundary of the capacity region.  
The upper bounds are obtained 
by evaluating Theorem~\ref{ZZZ-thm:bcmain}.
Thus 
\[
\rho(\cC) \geq \left\{
\begin{array}{ll} 
\frac{1-H(p_1*p_1*p_2*p_2)}{1-H(p_1*p_2)} & 
\mbox{when the noise at the receivers is independent} \\
\frac{1-H(p_1*p_2*p_2)}{1-H(p_1*p_2)} & 
\mbox{when the noise is physically degraded,}  
\end{array}\right.
\]
where the bounds are achieved by setting 
$\alpha=p_1*p_2$ and $\alpha=p_2$, respectively.  
Observing both $Y_1$ and $Y_2$ gives more information 
when $Y_1$ and $Y_2$ are independent, 
so $\rho(\cC)$ is smaller in that case.  
\IEEEQED
\end{exam}

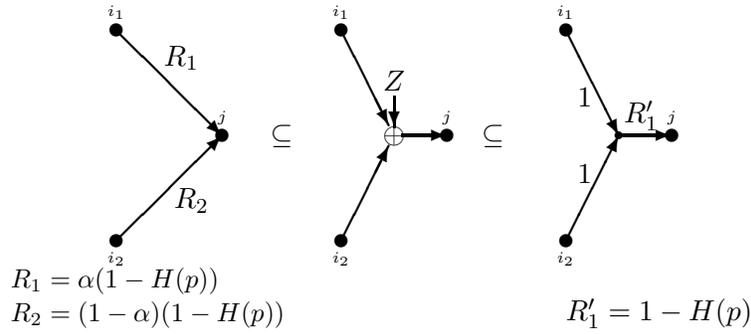
\begin{figure}
  \begin{center}
    \begin{picture}(210,110)(20,-50)
      \thicklines
      \put(60,20){\circle*{5}}
      \put(60,24){\makebox(0,0)[cb]{\tiny $j$}}
      \put(20,60){\vector(1,-1){40}}
      \put(20,-20){\vector(1,1){40}}
      \put(38,45){\makebox(0,0)[lb]{$R_1$}}
      \put(38,0){\makebox(0,0)[lt]{ $R_2$}}
      \put(20,60){\circle*{5}}
      \put(20,-20){\circle*{5}}
      \put(20,64){\makebox(0,0)[cb]{\tiny $i_1$}}
      \put(20,-24){\makebox(0,0)[ct]{\tiny $i_2$}}
      \put(-20,-40){\makebox(0,0)[lb]{\small $R_1=\alpha(1-H(p))$}}
      \put(-20,-53){\makebox(0,0)[lb]{\small $R_2=(1-\alpha)(1-H(p))$}}
      \put(82.5,20){\makebox(0,0)[cc]{$\subseteq$}}
      %
      \put(105,60){\vector(1,-2){18}}
      \put(125,20){\makebox(0,0)[cc]{$\oplus$}}
      \put(105,-20){\vector(1,2){18}}
      \put(125,35){\vector(0,-1){12}}
      \put(125,37){\makebox(0,0)[cb]{$Z$}}
      \put(128,20){\vector(1,0){17}}
      \put(145,20){\circle*{5}}
      \put(145,24){\makebox(0,0)[cb]{\tiny $j$}}
      \put(105,60){\circle*{5}}
      \put(105,-20){\circle*{5}}
      \put(105,64){\makebox(0,0)[cb]{\tiny $i_1$}}
      \put(105,-24){\makebox(0,0)[ct]{\tiny $i_2$}}
      \put(162.5,20){\makebox(0,0)[cc]{$\subseteq$}}
      %
      \put(190,60){\vector(1,-2){20}}
      \put(210,20){\circle*{3}}
      \put(190,-20){\vector(1,2){20}}
      \put(210,20){\vector(1,0){20}}
      \put(230,20){\circle*{5}}
      \put(230,24){\makebox(0,0)[cb]{\tiny $j$}}
      \put(200,38){\makebox(0,0)[rt]{$1$}}
      \put(200,2){\makebox(0,0)[rb]{$1$}}
      \put(212,22){\makebox(0,0)[lb]{$R_1'$}}
      \put(190,60){\circle*{5}}
      \put(190,-20){\circle*{5}}
      \put(190,64){\makebox(0,0)[cb]{\tiny $i_1$}}
      \put(190,-24){\makebox(0,0)[ct]{\tiny $i_2$}}
      \put(190,-53){\makebox(0,0)[lb]{$R_1'=1-H(p)$}}
      \end{picture} 
  \end{center}
  \caption{Example upper and lower bounding models 
           for the binary adder multiple access channel 
	   with error probability $p$.  
           }
  \label{ZZZ-fig:bsma}
\end{figure}

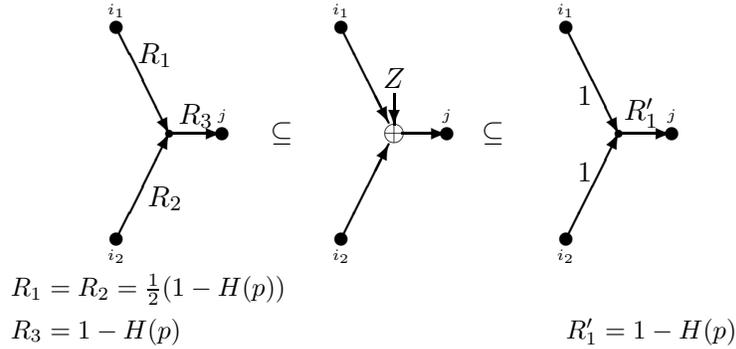
\begin{figure}
  \begin{center}
    \begin{picture}(210,110)(20,-50)
      \thicklines
      \put(60,20){\circle*{5}}
      \put(60,24){\makebox(0,0)[cb]{\tiny $j$}}
      \put(20,60){\vector(1,-2){20}}
      \put(40,20){\circle*{3}}
      \put(20,-20){\vector(1,2){20}}
      \put(40,20){\vector(1,0){20}}
      \put(28,45){\makebox(0,0)[lb]{$R_1$}}
      \put(28,0){\makebox(0,0)[lt]{ $R_2$}}
      	\put(50,22){\makebox(0,0)[cb]{$R_3$}}
      \put(20,60){\circle*{5}}
      \put(20,-20){\circle*{5}}
      \put(20,64){\makebox(0,0)[cb]{\tiny $i_1$}}
      \put(20,-24){\makebox(0,0)[ct]{\tiny $i_2$}}
      \put(-20,-45){\makebox(0,0)[lb]{\small $R_1=R_2=\frac12(1-H(p))$}}
      \put(-20,-60){\makebox(0,0)[lb]{\small $R_3=1-H(p)$}}
      \put(82.5,20){\makebox(0,0)[cc]{$\subseteq$}}
      %
      \put(105,60){\vector(1,-2){18}}
      \put(125,20){\makebox(0,0)[cc]{$\oplus$}}
      \put(105,-20){\vector(1,2){18}}
      \put(125,35){\vector(0,-1){12}}
      \put(125,37){\makebox(0,0)[cb]{$Z$}}
      \put(128,20){\vector(1,0){17}}
      \put(145,20){\circle*{5}}
      \put(145,24){\makebox(0,0)[cb]{\tiny $j$}}
      \put(105,60){\circle*{5}}
      \put(105,-20){\circle*{5}}
      \put(105,64){\makebox(0,0)[cb]{\tiny $i_1$}}
      \put(105,-24){\makebox(0,0)[ct]{\tiny $i_2$}}
      \put(162.5,20){\makebox(0,0)[cc]{$\subseteq$}}
      %
      \put(190,60){\vector(1,-2){20}}
      \put(210,20){\circle*{3}}
      \put(190,-20){\vector(1,2){20}}
      \put(210,20){\vector(1,0){20}}
      \put(230,20){\circle*{5}}
      \put(230,24){\makebox(0,0)[cb]{\tiny $j$}}
      \put(200,38){\makebox(0,0)[rt]{$1$}}
      \put(200,2){\makebox(0,0)[rb]{$1$}}
      \put(212,22){\makebox(0,0)[lb]{$R_1'$}}
      \put(190,60){\circle*{5}}
      \put(190,-20){\circle*{5}}
      \put(190,64){\makebox(0,0)[cb]{\tiny $i_1$}}
      \put(190,-24){\makebox(0,0)[ct]{\tiny $i_2$}}
      \put(190,-60){\makebox(0,0)[lb]{\small $R_1'=1-H(p)$}}
      \end{picture} 
    \end{center}
  \caption{A variation on the lower bounding model 
           from Figure~\ref{ZZZ-fig:bsma}.}
  \label{ZZZ-fig:thm1}
\end{figure}

\begin{exam}\label{ZZZ-ex:bsma}
Let $(\{0,1\}^2,p(\jyr|\ixo,\ixt,\{0,1\})$ be a binary adder multiple access channel 
with $\jYr=\iXo\oplus\iXt\oplus Z$.  Let $EZ=p$.  
Figure~\ref{ZZZ-fig:bsma} shows lower and upper bounding models.  
Each lower bounding model comes from a point on the capacity region.  
The upper bound evaluates Theorem~\ref{ZZZ-thm:mamain} with $U=c$.  
The models for this example are quite intuitive.  
For example, any code designed for network $\cN$ 
can be operated on the given upper bounding model 
by implementing a memoryless binary adder at the central node.  
In this case, the topologies of our upper and lower bounding models 
do not match, but they can be modified to match 
as shown in Figure~\ref{ZZZ-fig:thm1}.  
Thus 
\[
\rho(\cC)\geq \frac{1-H(p)}{2}.
\]
\IEEEQED
\end{exam}

Additive bounds are an alternative to the multiplicative bounds 
described above;  
this approach may be particularly useful when 
$R_{\cC,L}^{\ar{A}{B}}=0$ for some $(A,B)\in\cM$ 
such that $R_{\cC,U}^{\ar{A}{B}}>0$
or when $\cRCU$ incorporates infinite capacity edges 
for some $\cC\in\cA$.  
We here restrict our attention 
to upper and lower bounding networks that are 
entirely deterministic -- 
that is, we assume that the network is comprised of 
independent channels that have all been replaced by 
noiseless bit-pipe models.  
We also focus on demand types 
for which cut-set bounds are tight on networks of noiseless links.  
These include multicast demands, 
multi-source multicast demands, 
non-overlapping demands on single-source networks, 
and two-resolution multicast demands on single-source networks 
(see, for example,~\cite{ZZZ-KoetterM:03}).  

Let $\setR_c(\cN)$ be the set of achievable rate vectors 
for demand types where cut-set bounds are tight on bit-pipe networks, 
and define 
\begin{eqnarray*}
\setR_{c,L}(\cN) & \deff & \bigcup_{\cRL\in\prod_{\cC\in\cN}\setR(\cC)}
                  \setR_c(\cN(\cRL)) \\
\setR_{c,U}(\cN)
& \deff & \bigcap_{\cRU\in\mbox{int}(\prod_{\cC\in\cN}\setE(\cC))}
                  \setR_c(\cN(\cRU)).  
\end{eqnarray*}
For any $b>0$, we use 
\[
\setR_{c,L}(\cN)\geq \setR_{c,U}(\cN)- b
\]
to specify that $\cR\in\setR_{c,U}(\cN)$ 
implies $[\cR-b(1,\ldots,1)]^+\in\setR_{c,L}(\cN)$.  
That is, for any $\cR\in\setR_{c,U}(\cN)$, 
reducing the rate for each demand by $b$ 
yields an achievable rate vector from $\setR_{c,L}(\cN)$.  
For any network $\cN$ of noiseless bit-pipes 
and any $S\subset\{1,\ldots,m\}$, 
define $\mbox{val}(\cN,S)$ to be the sum of the capacities 
of all bit pipes with input in $S$ and output in $S^c$.  
Since bit-pipe models incorporate internal nodes 
not present in the original network 
(and therefore not present in the cut-set definitions), 
we define the value of a cut across a bit-pipe model 
using the assignment of internal nodes that minimizes the cut's value.  
To make this precise, again let $V_o=\{v^A:A\subseteq V_1,|A|>1\}$ 
be the set of internal nodes 
for bit-pipe model $\cC(\cRC)$ for channel $\cC$.  
For any $\cC\in\cN$ and $S\subseteq\{1,\ldots,m\}$, 
define 
\[
\mbox{val}(\cC(\cRC),S)\ \deff\ \left\{\begin{array}{ll}
\min_{S'=S\cup T:T\subseteq V_o}\mbox{val}(\cC(\cRC),S') 
& \mbox{ if $S\cap V_1\neq\emptyset$ and $S^c\cap V_2\neq\emptyset$} \\
0 & \mbox{otherwise.}
\end{array}\right.
\]
Finally, define $\Delta(\cC,S)$ as 
\[
\Delta(\cC,S)=\min_{(\cRCL,\cRCU)\in\setR(\cC)\times\cE(\cC)}
[\mbox{val}(\cC(\cRCU),S)-\mbox{val}(\cC(\cRCL),S)].
\]

\begin{lem}\label{ZZZ-lem:diff}
For any network $\cN$ and any set $S\subseteq\{1,\ldots,m\}$, 
\[
\setR_{c,L}(\cN)\geq\setR_{c,U}(\cN)-\max_{S\subseteq\{1,\ldots,m\}}
\sum_{\cC\in\cN}\Delta(\cC,S)
\]
\Proof
Since cut-set bounds are tight 
for the given demand types by assumption, 
we bound the difference in capacities 
by bounding the difference in each cut-set 
using the best choice of the upper and lower bounding models 
for each cut.  
\IEEEQED
\end{lem}

Given bounds on $\Delta(S,\cC)$ for some family of channels, 
Lemma~\ref{ZZZ-lem:diff} yields immediate bounds on 
the accuracy of the capacity bounds 
resulting from our models.  
These bounds take the same form 
as prior bounds in the literature (e.g.,~\cite{ZZZ-AvestimehrD:08}).  
In particular capacity bounds resulting from our 
upper and lower bounding models 
differ from each other (and therefore from the true capacity) 
by a constant multiple of the number of channels in the network.  
For networks of Gaussian point-to-point, multiple access, 
and broadcast channels 
with independent noise at the receivers, 
this constant is bounded from above by $1/2$, 
as shown by the examples that follow;  
the resulting capacity bounds agree 
precisely with~\cite{ZZZ-AvestimehrD:08} 
for unicast and multicast demands.  
The result here extends to other demand types 
where cut sets are tight, to tighter bounds 
outside the high-SNR region, 
and to corresponding results for networks containing 
broadcast channels with dependent noise at the receivers.  

By~\cite{ZZZ-KoetterE:10a}, 
$\Delta(\cC,S)=0$ for all memoryless point-to-point channels.  
Example~\ref{ZZZ-ex:nbc} bounds $\Delta(\cC,S)$ 
for the Gaussian broadcast channel.

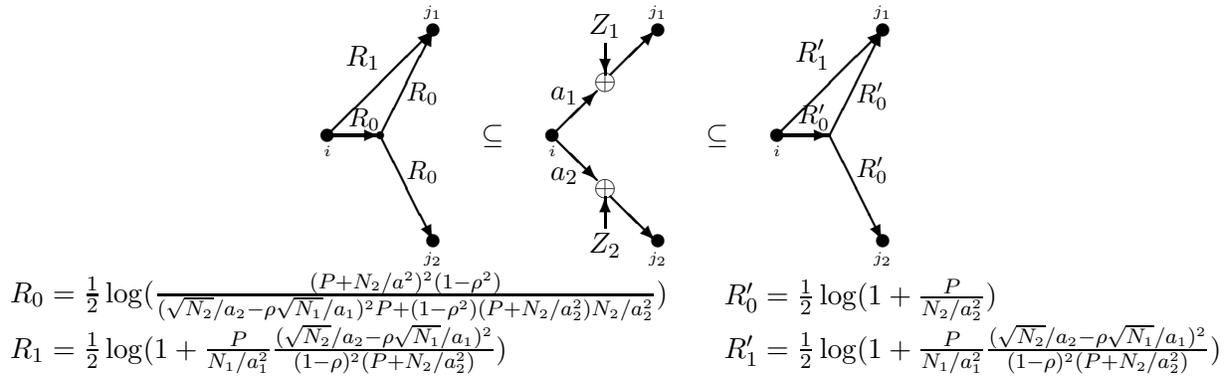
\begin{figure}
  \begin{center}
    \begin{picture}(250,130)(0,-70)
      \thicklines
      \put(20,20){\circle*{5}}
      \put(20,16){\makebox(0,0)[ct]{\tiny $i$}}
      \put(40,20){\circle*{3}}
      \put(20,20){\vector(1,1){40}}
      \put(20,20){\vector(1,0){20}}
      \put(40,20){\vector(1,2){20}}
      \put(40,20){\vector(1,-2){20}}
      \put(40,45){\makebox(0,0)[rb]{$R_1$}}
      \put(60,60){\circle*{5}}
      \put(60,-20){\circle*{5}}
      \put(60,64){\makebox(0,0)[cb]{\tiny $j_1$}}
      \put(60,-24){\makebox(0,0)[ct]{\tiny $j_2$}}
      \put(50,40){\makebox(0,0)[lt]{\small $R_{0}$}}
      \put(34,22){\makebox(0,0)[cb]{\small $R_{0}$}}
      \put(50,2){\makebox(0,0)[lb]{\small $R_{0}$}}
      \put(82.5,20){\makebox(0,0)[cc]{$\subseteq$}}
      \put(105,20){\circle*{5}}
      \put(105,16){\makebox(0,0)[ct]{\tiny $i$}}
      \put(105,20){\vector(1,1){17}}
      \put(105,20){\vector(1,-1){17}}
      \put(115,32){\makebox(0,0)[rb]{$a_1$}}
      \put(115,8){\makebox(0,0)[rt]{$a_2$}}
      \put(125,40){\makebox(0,0)[cc]{$\oplus$}}
      \put(125,0){\makebox(0,0)[cc]{$\oplus$}}
      \put(125,55){\vector(0,-1){12}}
      \put(125,-15){\vector(0,1){12}}
      \put(125,57){\makebox(0,0)[cb]{$Z_1$}}
      \put(125,-17){\makebox(0,0)[ct]{$Z_2$}}
      \put(127,43){\vector(1,1){17}}
      \put(127,-3){\vector(1,-1){17}}
      \put(145,60){\circle*{5}}
      \put(145,-20){\circle*{5}}
      \put(145,64){\makebox(0,0)[cb]{\tiny $j_1$}}
      \put(145,-24){\makebox(0,0)[ct]{\tiny $j_2$}}
      \put(167.5,20){\makebox(0,0)[cc]{$\subseteq$}}
      \put(190,20){\circle*{5}}
      \put(190,16){\makebox(0,0)[ct]{\tiny $i$}}
      \put(190,20){\circle*{3}}
      \put(190,20){\vector(1,1){40}}
      \put(190,20){\vector(1,0){20}}
      \put(210,20){\vector(1,2){20}}
      \put(210,20){\vector(1,-2){20}}
      \put(210,45){\makebox(0,0)[rb]{$R'_1$}}
      \put(230,60){\circle*{5}}
      \put(230,-20){\circle*{5}}
      \put(230,64){\makebox(0,0)[cb]{\tiny $j_1$}}
      \put(230,-24){\makebox(0,0)[ct]{\tiny $j_2$}}
      \put(220,40){\makebox(0,0)[lt]{\small $R'_{0}$}}
      \put(204,22){\makebox(0,0)[cb]{\small $R'_{0}$}}
      \put(220,2){\makebox(0,0)[lb]{\small $R'_{0}$}}
      \put(-100,-50){\makebox(0,0)[lb]{$R_0=
      	\frac12\log(\frac{(P+N_2/a^2)^2(1-\rho^2)}
		   {(\sqrt{N_2}/a_2-\rho\sqrt{N_1}/a_1)^2P
		     +(1-\rho^2)(P+N_2/a_2^2)N_2/a_2^2})$}}
      \put(-100,-70){\makebox(0,0)[lb]{$R_1=\frac12\log(1+
      	\frac{P}{N_1/a_1^2}\frac{(\sqrt{N_2}/a_2-\rho\sqrt{N_1}/a_1)^2}
	     {(1-\rho)^2(P+N_2/a_2^2)})$}}
      \put(170,-50){\makebox(0,0)[lb]
        {$R'_0=\frac12\log(1+\frac{P}{N_2/a_2^2})$}}
      \put(170,-70){\makebox(0,0)[lb]
      	{$R'_1=\frac12\log(1+\frac{P}{N_1/a_1^2}
	  \frac{(\sqrt{N_2}/a_2-\rho\sqrt{N_1}/a_1)^2}
		{(1-\rho)^2(P+N_2/a_2^2)})$}}
      \end{picture} 
    \end{center}
  \caption{Example models for the Gaussian broadcast channel.}
  \label{ZZZ-fig:nbc}
\end{figure}

\begin{exam}\label{ZZZ-ex:nbc}
Let $\cC$ be a two-receiver Gaussian broadcast channel 
$({\rm I\! R},p(\jyo,\jyt|\ixs),{\rm I\! R}^2)$ 
with $\jYo=a_1\iXs+Z_1$ 
and  $\jYt=a_2\iXs+Z_2$ 
for some jointly Gaussian random variables $Z_1$ and $Z_2$ 
with $E[(\iXs)^2]\leq P$, $E[Z_1^2]=N_1$, $E[Z_2^2]=N_2$, 
$E[Z_1Z_2]=\rho\sqrt{N_1N_2}$, and $N_1/a_1^2\leq N_2/a_2^2$.  
Figure~\ref{ZZZ-fig:nbc} shows example upper and lower bounding models.  
The lower bounding model is found by evaluating the broadcast capacity bounds 
\begin{eqnarray*}
R_1 & = & \frac12\log\left(1+\frac{(1-\alpha) P}{N_1/a_1^2}\right) \\
R_2 & = & \frac12\log\left(1+\frac{\alpha P}{(1-\alpha)P+N_2/a_2^2}\right) 
\end{eqnarray*}
at 
\[
1-\alpha
=\frac{(\sqrt{N_2}/a_2-\rho \sqrt{N_1}/a_1)^2}{(1-\rho)^2(a_2^2 P+N_2)}.  
\]
The upper bounding network is obtained 
by evaluating the model from Theorem~\ref{ZZZ-thm:bcmain}.  
This upper and lower bound imply 
\begin{eqnarray*}
\Delta(\cC,S)
&\leq&\frac12\log\left(1+\frac{P}{N_2/a_2^2}\right)-
      	\frac12\log\left(\frac{(P+N_2/a^2)^2(1-\rho^2)}
		   {(\sqrt{N_2}/a_2-\rho\sqrt{N_1}/a_1)^2P
		     +(1-\rho^2)(P+N_2/a_2^2)N_2/a_2^2}\right)\\
&=&\frac12\log\left(1+\frac{P}{P+N_2/a_2^2}
\frac{\left(1-\rho\sqrt{\frac{N_1/a_1^2}{N_2/a_2^2}}\right)^2}{1-\rho^2}
	\right).
\end{eqnarray*}
When $Z_1$ and $Z_2$ are independent, $\rho=0$ and the upper bound is 
\[
\Delta(\cC,S)\leq
\frac12\log\left(1+\frac{P}{P+N_2/a_2^2}\right),
\]
which is at most 1/2 
and signficantly smaller in the low SNR region.  
\IEEEQED
\end{exam}

\begin{figure}
  \begin{center}
    \begin{picture}(210,110)(20,-50)
      \thicklines
      \put(60,20){\circle*{5}}
      \put(60,24){\makebox(0,0)[cb]{\tiny $j$}}
      \put(20,60){\vector(1,-1){40}}
      \put(20,-20){\vector(1,1){40}}
      \put(38,45){\makebox(0,0)[lb]{$R_1$}}
      \put(38,0){\makebox(0,0)[lt]{ $R_2$}}
      \put(20,60){\circle*{5}}
      \put(20,-20){\circle*{5}}
      \put(20,64){\makebox(0,0)[cb]{\tiny $i_1$}}
      \put(20,-24){\makebox(0,0)[ct]{\tiny $i_2$}}
      \put(10,-45){\makebox(0,0)[lb]
	{\small $R_1=\frac12\log(1+\frac{P_1}{N})$}}
      \put(10,-60){\makebox(0,0)[lb]
	{\small $R_2=\frac12\log(1+\frac{P_2}{P_1+N})$}}
      \put(82.5,20){\makebox(0,0)[cc]{$\subseteq$}}
      %
      \put(105,60){\vector(1,-2){18}}
      \put(125,20){\makebox(0,0)[cc]{$\oplus$}}
      \put(105,-20){\vector(1,2){18}}
      \put(125,35){\vector(0,-1){12}}
      \put(122,37){\makebox(0,0)[lb]{$Z\sim{\cN}(0,N)$}}
      \put(128,20){\vector(1,0){17}}
      \put(145,20){\circle*{5}}
      \put(145,24){\makebox(0,0)[cb]{\tiny $j$}}
      \put(105,60){\circle*{5}}
      \put(105,-20){\circle*{5}}
      \put(105,64){\makebox(0,0)[cb]{\tiny $i_1$}}
      \put(105,-24){\makebox(0,0)[ct]{\tiny $i_2$}}
      \put(162.5,20){\makebox(0,0)[cc]{$\subseteq$}}
      \put(190,60){\vector(1,-1){40}}
      \put(190,60){\vector(1,-2){20}}
      \put(210,20){\circle*{3}}
      \put(208,45){\makebox(0,0)[lb]{$R'_1$}}
      \put(190,-20){\vector(1,2){20}}
      \put(210,20){\vector(1,0){20}}
      \put(230,20){\circle*{5}}
      \put(230,24){\makebox(0,0)[cb]{\tiny $j$}}
      \put(200,38){\makebox(0,0)[rt]{$\infty$}}
      \put(200,2){\makebox(0,0)[rb]{$\infty$}}
      \put(212,18){\makebox(0,0)[lt]{$R_2'$}}
      \put(190,60){\circle*{5}}
      \put(190,-20){\circle*{5}}
      \put(190,64){\makebox(0,0)[cb]{\tiny $i_1$}}
      \put(190,-24){\makebox(0,0)[ct]{\tiny $i_2$}}
      \put(115,-45){\makebox(0,0)[lb]{$R_1'=\frac12\log(1+\frac{P_1}{N})$}}
      \put(115,-62){\makebox(0,0)[lb]
	{$R_2'=\frac12\log(1+\frac{2\sqrt{P_1P_2}+P_2}{P_1+N})$}}
      \end{picture} 
    \end{center}
  \caption{Example models for the Gaussian multiple access channel 
	   with power constraints $P_1\geq P_2$ 
	   at transmitters~1 and~2 and variance-$N$ Gaussian noise.  
           }
  \label{ZZZ-fig:nma}
\end{figure}
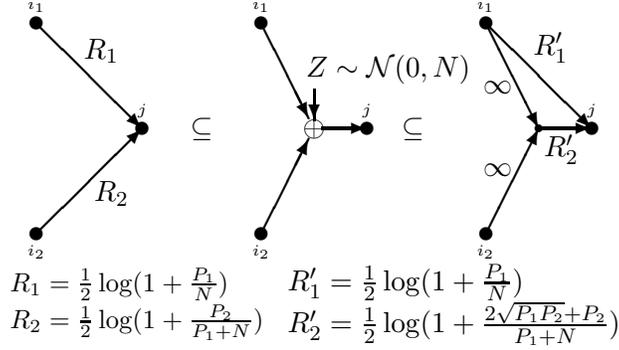

\begin{exam}\label{ZZZ-ex:nma}
Let $\cC=({\rm I\! R}^2,p(\jyr|\ixo,\ixt),{\rm I\! R})$ 
be a Gaussian multiple access channel 
with $\jYr=\iXo+\iXt+Z$, 
$E[(\iXo)^2]\leq P_1$, $E[(\iXt)^2]\leq P_2$, $P_1\geq P_2$, 
and $Z\sim \cN(0,N)$.  
Figure~\ref{ZZZ-fig:nma} shows upper and lower bounding models 
for the given multiple access channel.  
The lower bound is chosen as the corner point 
\begin{eqnarray*}
R_1 & = & \frac12\log\left(1+\frac{P_1}{N}\right) \\
R_2 & = & \frac12\log\left(1+\frac{P_1+P_2}{N}\right)
         -\frac12\log\left(1+\frac{P_1}{N}\right)
         =\frac12\log\left(1+\frac{P_2}{P_1+N}\right)
\end{eqnarray*}
of the multiple access capacity region.  

The upper bounding network is obtained 
by evaluating Theorem~\ref{ZZZ-thm:mamain} 
under the maximizing joint distribution on $(\iXo,\iXt)$ 
using a statistically dependent distortion-$D$ reproduction $U$ 
of $\iXt$ similar to those used in lossy source coding.  
Precisely, 
\begin{eqnarray*}
\iXt & = & \iXo\sqrt{\frac{P_2}{P_1}} \\
U    & = & \frac{1}{(1+\sqrt{P_2/P_1})}\iXo+Z_1 \\
Z    & = & Z_1+Z_2 
\end{eqnarray*}
where $Z_1$ is a Gaussian random variable 
with mean 0 and variance $N/(1+\sqrt{P_2/P_1})^2$, 
$Z_2$ is a Gaussian random variable 
with mean 0 and variance $N(1-1/(1+\sqrt{P_2/P_1})^2)$, 
and $(X_1,X_2)$, $Z_1$, and $Z_2$ 
are mutually independent.  
Using this choice of $U$, the upper bound from Theorem~\ref{ZZZ-thm:mamain} 
is 
\begin{eqnarray*}
R_1 & = & \frac12\log\left(1+\frac{P_1}{N}\right) \\
R_2 & = & \frac12\log\left(\frac{(\sqrt{P_1}+\sqrt{P_2})^2+N}{P_1+N}\right) 
\end{eqnarray*}
Using the given upper and lower bounds yields 
\[
\Delta(\cC,S)\leq
\frac12\log\left(\frac{(\sqrt{P_1}+\sqrt{P_2})^2+N}{P_1+P_2+N}\right),
\]
which is at most 1/2 
(and considerably smaller when the signal-to-noise ratio is small).
\IEEEQED
\end{exam}

Examples~\ref{ZZZ-ex:bsbc},~\ref{ZZZ-ex:bsma},~\ref{ZZZ-ex:nbc} 
and~\ref{ZZZ-ex:nma} 
show that for some network types, the upper and lower bounds 
differ by at most an additive or multiplicative constant 
that depends on the statistics of the network's component channels.  
Given any network $\cN$ built from arbitrary point-to-point channels, 
binary symmetric broadcast channels (Example~\ref{ZZZ-ex:bsbc}), 
and binary adder multiple access channels (Example~\ref{ZZZ-ex:bsma}), 
Lemma~\ref{ZZZ-lem:ratio} shows that the capacities 
of the derived upper and lower bounding networks 
differ from the true capacity and each other by at most 
a multiplicative constant $\rho^*=\max_{C\in\cN}\rho(C)$.  
This constant depends on the channel for which 
the distance between our upper and lower bounds 
is largest but not on the size of the nework.  
Likewise, given any network $\cN$ built from arbitrary point-to-point channels, 
Gaussian broadcast channels (Example~\ref{ZZZ-ex:nbc}), 
and additive Gaussian multiple access channels (Example~\ref{ZZZ-ex:nma}), 
Lemma~\ref{ZZZ-lem:diff} shows that for all demand types for which 
cut-set bounds on the network coding capacity are tight, 
the capacities of the derived upper and lower bounding networks 
differ from the true capacity and each other by at most 
an additive constant equal to a constant multiple 
of the number of channels in the network.  
When the noise at the receivers of each broadcast channel is independent, 
this immediately extends the well-known 1/2-bit per component bounds 
to a variety of other demand types where cut-sets bounds 
on the network coding capacity are tight.  
It also gives tighter bounds outside the high-SNR region 
and derives the corresponding bounds for broadcast channels 
with statistically dependent noise at the receivers.  
Of course, examples~\ref{ZZZ-ex:bcmac} and~\ref{ZZZ-ex:bcmacind} 
demonstrate that the lower and upper bounds for some channels 
are, by necessity, far apart.  
When such large gaps arise, they motivate the investigation 
of larger network components.  
For example, modeling the network from example~\ref{ZZZ-ex:bcmac} 
not as two independent components but instead as a single component 
with one input and one output 
yields matching lower and upper bounding models 
and therefore a precise network equivalence.  

\section{Conclusions}

The equivalence tools introduced in this paper 
are proposed as one step in a new path towards 
the construction of computational tools 
for bounding the capacities of large networks.  
Unlike cut-set strategies, 
which investigate networks in their entirety, 
the approach proposed here 
is to bound capacities of networks 
by carefully characterizing the behaviors 
of the individual components from which they are built.  
As described in Lemma~\ref{ZZZ-lem:lb}, 
the capacity region of an isolated component 
can be used to calculate lower bounds 
on the capacities of all networks in which the component may be employed.  
Since capacity regions of individual components 
cannot be used to derive upper bounds (see Example~\ref{ZZZ-ex:bcmac}), 
Theorem~\ref{ZZZ-thm:equiv} employs an alternative 
component characterization -- 
here offered as a complement to the traditional capacity problem.  
Given an arbitrary channel, 
describe the family of bit-pipe models 
over which accurate channel emulation is possible.  
The question is essentially a source coding problem  -- 
for each vector $\uXC$ at the channel input nodes $V_1$, 
we characterize the family of rate vectors 
$(R^{\ar{A}{B}}:A\subseteq V_1,B\subseteq V_2)$ 
sufficient for constructing a reproduction $\uYC$ 
at the channel output nodes $V_2$ 
such that $\uYC$ appears to result 
from the operation of channel $\cC$ on input $\uXC$.  
The upper bounding models for the 
point-to-point, broadcast, multiple access, and interference channels 
are here offered as examples of this characterization strategy.  
Increasing the library of component models 
offers a route to studying capacities of 
larger and larger families of networks 
using computational tools for 
bounding network coding capacities.  

\appendices

\section{Average vs. Expected Error Probability in Channel Coding}
\label{ZZZ-app:avgexp}

Lemma~\ref{ZZZ-lem:avgexp}, below, 
shows that given a blocklength-$N$ channel code 
with average error probability $P_e^{(N)}$, 
there exists an index assignment such that 
the code's expected error probability is no greater than $P_e^{(N)}$.  
This is obvious for channels with a single transmitter 
but more subtle for channels with multiple transmitters.  
The outline of this proof was suggested by~\cite{ZZZ-Langberg:10}.  
The property is useful since 
messages transmitted across a channel $\cC$ 
in the middle of some large network $\cN$ need not be equally probable, 
which means that the expected error probability 
can equal the code's maximal error probability 
if the codeword indices are poorly assigned.  
We denote the average error probability 
under channel code $(\alpha_N,\beta_N)$ as 
\[
P_e^{(N)}\ \deff\ \sum_{\utxC\in\utcXC}\frac1{|\utcXC|}
      \Pr\left(\left.\beta_N(\uYC)\neq\utxC\right|\uXC=\alpha_N(\utxC)\right) 
\]
and the expected error probability of the same code as 
\[
\sum_{\utxC\in\utcXC}p(\utxC)
      \Pr\left(\left.\beta_N(\uYC)\neq\utxC\right|\uXC=\alpha_N(\utxC)\right). 
\]
This notation hides the independent operation 
of the encoders $\alpha_N=(\alpha^{\ar{\{i\}}{B}}:(\{i\},B)\in\cM)$
and the decoders $\beta_N=(\beta^{\ar{\{i\}}{B},j}:(\{i\},B)\in\cM,j\in B)$.  
We relabel the codeword indices by applying a permutation 
$\phi^{\ar{\{i\}}{B}}$ on each message set.  
Given permutations $\phi=(\phi^{\ar{\{i\}}{B}}:(\{i\},B)\in\cM)$, 
we denote the expected error probability 
after relabeling the codeword indices by 
\[
\sum_{\utxC\in\utcXC}p(\utxC)\Pr\left(\left.\beta_N(\uYC)\neq\phi(\utxC)\right|
      \uXC=\alpha_N(\phi(\utxC))\right),
\]
where 
$\phi(\utxC)=(\phi^{\ar{\{i\}}{B}}(\utx^{\ar{\{i\}}{B}}):(\{i\},B)\in\cM)$.  

\begin{lem}[\cite{ZZZ-Langberg:10}]\label{ZZZ-lem:avgexp}
Let $(\alpha_N,\beta_N)$ be a blocklength-$N$ channel code for channel $\cC$ 
with transmitters $V_1$ and receivers $V_2$.  
For any distribution $p(\cdot)$ on the space 
$\utcXC=\prod_{(\{i\},B)\in\cM}\utcXiB$ of possible transmissions, 
there exist independent permutations 
$\phi=(\phi^{\ar{\{i\}}{B}}:(\{i\},B)\in\cM)$ 
of the transmission indices for which 
\[
\sum_{\utxC\in\utcXC}p(\utxC)
      \Pr\left(\left.\beta_N(\uYC)\neq\phi(\utxC)\right|
      \uXC=\alpha_N(\phi(\utxC))\right) 
\leq P_e^{(N)}.
\]

\Proof 
For each $(\{i\},B)\in\cM$, 
choose permutation $\Phi^{\ar{\{i\}}{B}}$ uniformly at random 
from the space of possible permutations on $\ucW^{\ar{\{i\}}{B}}$.  
Then, using $E_\Phi[\cdot]$ to denote the expectation 
with respect to the random permutation choice, 
the expected error probability of the resulting channel code is 
\begin{eqnarray*}
\lefteqn{E_{\Phi}\left[\sum_{\utxC\in\utcXC}p(\utxC)
      \Pr\left(\left.\beta_N(\uYC)\neq\Phi(\utxC)\right|
      \uXC=\alpha_N(\Phi(\utxC))\right)\right]} \\
& = & \sum_{\utxC\in\utcXC}p(\utxC)E_\Phi\left[
      \Pr\left(\left.\beta_N(\uYC)\neq\Phi(\utxC)\right|
      \uXC=\alpha_N(\Phi(\utxC))\right)\right] \\
& \stackrel{(a)}{=} & \sum_{\utxC\in\utcXC}p(\utxC)
      \left[\sum_{\uhtxC\in\utcXC}\frac1{|\utcXC|}
      \Pr\left(\left.\beta_N(\uYC)\neq\uhtxC\right|
      \uXC=\alpha_N(\uhtxC)\right)\right] \\
& = & P_e^{(N)}
\end{eqnarray*}
where $(a)$ holds 
since all codewords are equally probable 
under the uniform distribution on permutations.  
The result follows since the optimal choice 
of permutations $(\phi^{\ar{\{i\}}{B}}:(\{i\},B)\in\cM)$ 
achieves expected error probability no greater than 
that achieved by the given random permutation choice.  
\IEEEQED
\end{lem}

\section{Typical Set Notation and Tools}
\label{ZZZ-app:typ}

The appendices that follow define typical sets 
for many combinations of random variables and many parameter values.  
The following definitions are useful for streamlining the exposition.  
Given a random variable $Z$ drawn from distribution $p(z)$ 
on alphabet $\cZ$ 
and an $N$-vector $\ub{z}\in\ub{\cZ}$, 
define 
\[
f(\ub{z})\ \deff\ \left|-\frac1N\log p(\ub{z})-H(Z)\right|, 
\]
where $p(\ub{z})\ \deff\ \prod_{\ell=1}^Np(\ub{z}(\ell))$ 
and $H(Z)$ is the (discrete or differential) entropy 
of random variable $Z$.  
The random variable and distribution are implicit, 
with $f(\ux)$ and $f(\uy)$ 
referring to random variables $X$ and $Y$, respectively.  
For example, the usual jointly typical set for $(X,Y)$ 
is here expressed as 
\[
\typ = \{(\ux,\uy):f(\ux)\leq\eps,\ f(\uy)\leq\eps,\ f(\ux,\uy)\leq \eps
\}.
\]
For each collection of random variables 
for which we define a typical set, 
we also define a restricted typical $\rtyp\subseteq\typ$ 
and an indicator function $K(\cdot)$ 
that equals one for values in $\rtyp$ and 0 otherwise.  
The formal definitions for the restricted typical sets 
are given in the appendices that follow.  
When multiple restricted typical sets are in use 
we distinguish between them either by context 
or by adding arguments.  
For example, $(X,Y)\in\rtyp$ and $\rtyp(X,Y)$ 
refer to the same restricted typical set.  
A summary of definitions and results from~\cite{ZZZ-KoetterE:10a} follows.  

Given any distribution $p(u,v)$ 
and any constant $\eps>0$, 
define 
\begin{eqnarray*}
a(\eps) & \deff& (1+\eps)\cdot\inf\left\{\eps'>0:
                 p(f(\uV)>\eps'\vee f(\uU,\uV)>\eps')\leq 2^{-N6\eps}
		 \ \ \forall N\mbox{ suffic. large}\right\} \\
\typ&\deff&\{(\uu,\uv):f(\uu)<\eps,\ f(\uv)<a(\eps),\ f(\uu,\uv)<a(\eps)\}\\
\rtyp & \deff & \{(\uu,\uv)\in\typ:p(f(\uV)>a(\eps)\vee f(\uU,\uV)>a(\eps)
|\uU=\uu)
<2^{-N3\eps}\} \\
K(\ux,\uy) & \deff & \left\{\begin{array}{ll}
1 & \mbox{if $(\ux,\uy)\in\rtyp$} \\
0 & \mbox{otherwise}.\end{array}\right.
\end{eqnarray*}

\begin{lem}\cite[Lemma~6]{ZZZ-KoetterE:10a}\label{ZZZ-lem:typ}
Let $(\uU,\uV)$ be drawn i.i.d.\ $p(u,v)$.  Then 
\[
p((\rtyp(U,V))^c)<2^{-Nc(\eps)}
\]
for some constant $c(\eps)>0$ and all $N$ sufficiently large.  
Constant $c(\eps)$ approaches 0 as $\eps$ approaches 0.  
\end{lem}

Design random source code $(\alpha_N,\beta_N)$ 
by drawing codewords 
$\beta_N(1),\ldots,\beta_N(2^{NR})$ i.i.d.\ $p(\uv)$ 
and choosing $\alpha_N(\uu)$ uniformly at random 
from the indices $w\in\{1,\ldots,2^{NR}\}$ 
for which codeword $\beta_N(w)$ satisfies $(\uu,\beta_N(w))\in\rtyp$;  
$\alpha_N(\uu)$ is set to 1 if no index $w$ satisfies this constraint.  
Define 
\[
\hp(\uv|\uu)\ \deff\ \Pr(\beta_N(\alpha_N(\uu))=\uv), 
\]
and for any $A\subseteq\ub{\cU}\times\ub{\cV}$, 
let $\hp(A|\uu)\ \deff\ \sum_{\uv:(\uu,\uv)\in A}\hp(\uv|\uu)$.  

\begin{lem}\cite[Lemma~9]{ZZZ-KoetterE:10a}\label{ZZZ-lem:conprob}
For any $(\uu,\uv)\in\rtyp$, 
\[
\hp(\uv|\uu)
\leq p(\uv|\uu)2^{N(4a(\eps)+2\eps+1/N)}.
\]
\end{lem}

\begin{lem}\cite[Lemma~10]{ZZZ-KoetterE:10a}\label{ZZZ-lem:a1}
\[
\hp((\rtyp)^c|\uu)\leq 
p((\rtyp)^c|\uu)+e^{-2^{N(R-I(U;V)-2a(\eps)-\eps)}}
\]
\end{lem}

\section{Broadcast Channels}\label{ZZZ-app:bc}

We begin by defining the typical sets 
used in the proof of Theorem~\ref{ZZZ-thm:bcmain}.  
That proof appears later in this section.  
We here employ notation and results developed in Appendix~\ref{ZZZ-app:typ}.  

Given any $p(x,y_1,y_2)$, 
fix $\eps=(\eps_1,\eps_2)$ with $\eps_1,\eps_2>0$, 
and let 
\[
a_1(\eps_1)\ \deff\ (1+\eps_1)\cdot\inf\left\{\eps'>0:
p\left(f(\uY_2)>\eps'\vee f(\uX,\uY_2)>\eps'\right)
    \leq 2^{-N6\eps_1}\ \forall N \mbox{suff.\ large}\right\}.
\]
For $p(x,y_2)$ the typical set is 
\[
\typ\ \deff\ \left\{(\ux,\uy_2)\in\ucX\times\ucY_2:
    f(\ux)\leq \eps_1,\ 
    f(\uy_2)\leq a_1(\eps_1),\ 
    f(\ux,\uy_2)\leq a_1(\eps_1)\right\},
\]
which we restrict to 
\[
\rtyp\ \deff\ \left\{(\ux,\uy_2)\in\typ:
\left.p\left((\typ(X,Y_2))^c\right|\ux\right)
\leq 2^{-3N\eps_1}\right\}.
\]
Let 
\begin{eqnarray*}
a_2(\eps_2) 
& \deff & (1+\eps_2)\inf\left\{\eps'>0:
          p(f(\uY_2)>\eps'\vee f(\uY_1,\uY_2)>\eps'\vee 
	      f(\uX,\uY_2)>\eps'\right.\\
& & \left.\vee f(\uX,\uY_1,\uY_2)>\eps')
	  \leq 2^{-N6\eps_2}\ \forall N\mbox{ suff.\ large}\right\}. 
\end{eqnarray*}
For distribution $p(x,y_1,y_2)$, 
the typical set is 
\[
\typ \deff \left\{(\ux,\uy_1,\uy_2):
f(\uy_2)\leq a_2(\eps_2),\ f(\ux,\uy_2)\leq a_2(\eps_2),\ 
f(\uy_1,\uy_2)\leq a_2(\eps_2),f(\ux,\uy_1,\uy_2)\leq a_2(\eps_2)\right\}\\
\]
which we restrict to 
\begin{eqnarray*}
\rtyp
& \deff & \left\{(\ux,\uy_1,\uy_2)\in\typ:
p\left(\left.(\typ(X,Y_1,Y_2))^c\right|\ux,\uy_2\right)
\leq 2^{-3N\eps_2}\right\}.
\end{eqnarray*}
By Lemma~\ref{ZZZ-lem:bctyp}, the probability 
of observing atypical elements is asymptotically negligible.  

\begin{lem}\label{ZZZ-lem:bctyp}
If $(\uX,\uY_1,\uY_2)$ are drawn i.i.d.\ $p(x,y_1,y_2)$, 
then 
\begin{eqnarray*}
p\left((\rtyp(X,Y_2))^c\right) & \leq & 2^{-Nc_1(\eps_1)} \\
p\left((\rtyp(X,Y_1,Y_2))^c\right)& \leq &  2^{-Nc_2(\eps_2)}
\end{eqnarray*}
for some constants $c_1(\eps_1),c_2(\eps_2)>0$ 
and all $N$ sufficiently large.  
Constants $c_1(\eps_1)$ and $c_2(\eps_2)$ approach zero 
as $\eps_1$ and $\eps_2$, respectively, decay to zero.  

\Proof  
Like Lemma~\ref{ZZZ-lem:typ}, the result follows from Chernoff's bound 
and the definition of $\rtyp$.  
\IEEEQED
\end{lem}

{\bf Proof of Theorem~\ref{ZZZ-thm:bcmain}:}
Since $R^{\ar{\{i\}}{\{j_2\}}}$ is not bounded from below, 
we set it to 0.  
For concision, we further define 
$R_0\ \deff\ R^{\ar{\{i\}}{\{j_1,j_2\}}}$ and 
$R_1\ \deff\ R^{\ar{\{i\}}{\{j_1\}}}$ and use 
$\cC=(\cX,p(y_1,y_2|x),\cY_1\times\cY_2)$ 
in place of $\cC=(\ciXs,p(\jyo,\jyt|\ixs),\cjYo\times\cjYt)$ 
both in this proof and its supporting lemmas.  

Fix $(R_0,R_1)$ to satisfy the theorem constraints.  
Suppose that $R_1>I(X;Y_1|Y_2)$;
for any rate pair satisfying the theorem assumptions 
but not satisfying this bound, 
we can operate the code as if this condition were satisfied 
by using part of the common rate 
to carry private information for receiver $j_1$.  
By Theorem~\ref{ZZZ-thm:equiv} 
it suffices to show that for any channel input distribution $p(x)$ 
there exists a sequence of rate-$(R_0,R_1)$ 
random emulation codes $(\alpha_N,\beta_N)$ 
for which the resulting emulation distribution
\[
\hp(\uy_1,\uy_2|\ux)\ \deff\ \Pr(\beta_N(\alpha_N(\ux))=(\uy_1,\uy_2)) 
\]
satisfies 
\[
P_e^{(N)}(\nu)=\sum_{\ux,\uy_1,\uy_2}p(\ux)\hp(\uy_1,\uy_2|\ux)
1\left(\frac1N\log\left(\frac{\hp(\uy_1,\uy_2|\ux)}{p(\uy_1,\uy_2|\ux)}\right)>\nu\right)
<2^{-N\eta(\nu)}
\]
for some positive function $\eta(\nu)$ dependent on $p(x)$ for which 
$\eta(\nu)$ goes to zero as $\nu$ goes to zero.  

We employ the definitions for the (restricted) typical set $\rtyp$ 
from the beginning of this section.  
We distinguish between these sets either by context   
(e.g., $(\ux,\uy_2)\in\rtyp$ refers to the typical set for $p_t(x,y_2)$) 
or by adding arguments 
(e.g., $\rtyp(X,Y_2)$).  
Typical sets $\rtyp(X,Y_2)$ and $\rtyp(X,Y_1,Y_2)$ 
employ parameters $\eps_1$ and $\eps_2$, respectively.  

Next, we define codes $(\alpha_N,\beta_N)$ 
to emulate the typical behavior of channel $\cC$ 
under input distribution $p(\ux)=\prod_{\ell=1}^Np(\ux(\ell))$.  
Recall that $(\alpha_N,\beta_N)$ has encoders 
\[
\alpha_N=\left(\alpha_N^{\ar{A}{B}}:(A,B)\in\cM\right)
=\left(\alpha_N^{\ar{\{i\}}{\{j_1\}}},\alpha_N^{\ar{\{i\}}{\{j_2\}}},
\alpha_N^{\ar{\{i\}}{\{j_1,j_2\}}}\right)
\] 
at rates $R_1=R^{\ar{\{i\}}{\{j_1\}}}$, $R^{\ar{\{i\}}{\{j_2\}}}=0$, 
and $R_0=R^{\ar{\{i\}}{\{j_1,j_2\}}}$ 
and decoders $\beta_N=(\beta_N^{(j_1)},\beta_N^{(j_2)})$.
Rate 0 requires no encoder.  
We abbreviate the notation for the remaining encoders to 
$\alpha_N^{(1)}=\alpha_N^{\ar{\{i\}}{\{j_1\}}}$ 
and $\alpha_N^{(0)}=\alpha_N^{\ar{\{i\}}{\{j_1,j_2\}}}$ 
and for the decoders to 
$\beta_N^{(j_1)}=\beta_N^{(1)}$  and $\beta_N^{(j_2)}=\beta_N^{(2)}$.  
Thus 
\[
\begin{array}{rlcrl}
\alpha^{(0)}_N: & \ucX\rightarrow\cW_0 &\hspace{.5in}&
\beta^{(1)}_N: & \cW_0\times\cW_1\rightarrow\ucY_1 \\
\alpha^{(1)}_N: & \ucX\rightarrow\cW_1 &&
\beta^{(2)}_N: & \cW_0\rightarrow\ucY_2.  
\end{array}
\]
where $\cW_0=\utcX^{\ar{\{i\}}{\{j_1,j_2\}}} =\{0,1\}^{NR_0}$
and $\cW_1=\utcX^{\ar{\{i\}}{\{j_1\}}} =\{0,1\}^{NR_1}$.  
For the random code design, 
first draw codewords $\{\beta^{(2)}_N(w_0): w_0\in\cW_0\}$ 
i.i.d.\ according to distribution $\prod_{\ell=1}^Np(\uy_2(\ell))$.
Then, for each $w_0\in\cW_0$ draw codewords 
$\{\beta^{(1)}_N(w_0,w_1):w_1\in\cW_1\}$ 
i.i.d.\ according to $\prod_{\ell=1}^N p(\uy_1(\ell)|\beta^{(2)}_N(w_0,\ell))$, 
where $\beta^{(2)}_N(w_0,\ell)$ denotes the $\ell$th 
component of $N$-vector $\beta^{(2)}_N(w_0)$.  
For the random encoder design, 
choose index $\alpha^{(0)}_N(\ux)$ uniformly at random 
from those $w_0\in\cW_0$ for which $(\ux,\beta^{(2)}_N(w_0))\in\rtyp$;  
if there is no such $w_o$, then set $\alpha^{(0)}_N(\ux)$ to 1.  
Let $w_0=\alpha^{(0)}_N(\ux)$;  then 
choose index $\alpha^{(1)}_N(\ux)$ uniformly at random 
from those $w_1\in\cW_1$
for which $(\ux,\beta^{(1)}_N(w_0,w_1),\beta^{(2)}_N(w_0))\in\rtyp$.
If there is no such $w_1$, 
then set $\alpha^{(1)}_N(\ux)$ to 1.  

By Lemma~\ref{ZZZ-lem:bcconprob}, below, 
\[
\begin{array}{rcll}
\hp(\uy_2|\ux) & \leq & 2^{N(4a_1(\eps_1)+2\eps_1+1/N)}p(\uy_2|\ux) 
	& \forall (\ux,\uy_2)\in\rtyp \\
\hp(\uy_1|\ux,\uy_2) & \leq & 2^{N(8a_2(\eps_2)+1/N)}p(\uy_1|\ux,\uy_2)
     & \forall (\ux,\uy_1,\uy_2)\in\rtyp.
\end{array}
\]
Thus 
\[
\hp(\uy_1,\uy_2|\ux) \leq 
2^{N(4a_1(\eps_1)+2\eps_1+8a_2(\eps_2)+2/N)}p(\uy_1,\uy_2|\ux)
\]
for all $(\ux,\uy_1,\uy_2)$ for which $(\ux,\uy_1)\in\rtyp$ 
and $(\ux,\uy_1,\uy_2)\in\rtyp$.  
By Lemma~\ref{ZZZ-lem:bca1}, below, 
\begin{eqnarray*}
\hp((X,Y_2)\not\in\rtyp|\ux) 
& \leq & e^{-2^{N(R_0-I(X;Y_2)-2a_1(\eps_1)-\eps_1)}}
+p((\rtyp(X,Y_2))^c|\ux) \\
\hp((X,Y_1,Y_2)\not\in\rtyp|\ux,\uy_2) 
& \leq &  e^{-2^{N(R_1-I(X;Y_1|Y_2)-4a_2(\eps_2))}}
	 +p((\rtyp(X,Y_1,Y_2))^c|\ux,\uy_2).
\end{eqnarray*}
By Lemma~\ref{ZZZ-lem:bctyp}, above, 
\begin{eqnarray*}
p\left((\rtyp(X,Y_2))^c\right) & \leq & 2^{-Nc_1(\eps_1)} \\
p\left((\rtyp(X,Y_1,Y_2))^c\right)& \leq &  2^{-Nc_2(\eps_2)} 
\end{eqnarray*}
for some constants $c_1(\eps_1)$ and $c_2(\eps_2)$ 
that go to zero as $\eps_1$ and $\eps_2$ go to zero.  

Thus when $\nu=4a_1(\eps_1)+3\eps_1+8a_2(\eps_2)$ 
and $N$ is sufficiently large, 
\begin{eqnarray*}
P_e^{(N)}(\nu)
&\leq& \sum_{(\ux,\uy_1,\uy_2):(\ux,\uy_2)\not\in\rtyp\vee(\ux,\uy_1,\uy_2)\not\in\rtyp}
	p(\ux)\hp(\uy_1,\uy_2|\ux) \\
&\leq& \sum_{\ux}\hp((\rtyp(X,Y_2))^c|\ux)p(\ux) 
	+\sum_{(\ux,\uy_1,\uy_2):(\ux,\uy_2)\in\rtyp\wedge(\ux,\uy_1,\uy_2)\not\in\rtyp}
	\hp(\uy_1|\uy_2,\ux)\hp(\uy_2|\ux)p(\ux) \\
&\leq& \sum_{\ux}\left(e^{-2^{N(R_0-I(X;Y_2)-2a_1(\eps_1)-\eps_1)}}
	+p((\rtyp(X,Y_2))^c|\ux)\right)p(\ux) \\
& & +\sum_{(\ux,\uy_2):(\ux,\uy_2)\in\rtyp}
	\left(e^{-2^{N(R_1-I(X;Y_1|Y_2)-4a_2(\eps_2))}}
		+p((\rtyp(X,Y_1,Y_2))^c|\ux,\uy_2)\right)\hp(\uy_2|\ux)p(\ux) 
\end{eqnarray*}
\begin{eqnarray*}
&\leq& e^{-2^{N(R_0-I(X;Y_2)-2a_1(\eps_1)-\eps_1)}}+2^{-Nc_1(\eps_1)}
		+e^{-2^{N(R_1-I(X;Y_1|Y_2)-4a_2(\eps_2))}}\\
&&	 +2^{N(4a_1(\eps_1)+3\eps_1)}\sum_{(\ux,\uy_2)}
	 	p((\rtyp(X,Y_1,Y_2))^c|\ux,\uy_2)p(\uy_2|\ux)p(\ux) \\
&\leq& e^{-2^{N(R_0-I(X;Y_2)-2a_1(\eps_1)-\eps_1)}}+2^{-Nc_1(\eps_1)}
		+e^{-2^{N(R_1-I(X;Y_1|Y_2)-4a_2(\eps_2))}}
		+2^{N(c_2(\eps_2)-4a_1(\eps_1)-3\eps_1)}.
\end{eqnarray*}
Thus for all $N$ sufficiently large, 
$P_e^{(N)}(\nu)$ can be made to decay exponentially to zero 
by choosing $\eps_1$ such that $2a_1(\eps_1)+\eps_1<R_0-I(X;Y_2)$ and 
$\eps_2$ such that $4a_2(\eps_2)<R_1-I(X;Y_1|Y_2)$ and 
$c(\eps_2)>4a_1(\eps_1)$.  
The resulting exponent decays to zero as $\eps_1$ and $\eps_2$ decay to zero.  
\IEEEQED

Lemmas~\ref{ZZZ-lem:bcconprob} and~\ref{ZZZ-lem:bca1}, below, 
bound the conditional probability of $(\uY_1,\uY_2)$ given $\uX$ 
when we emulate the broadcast channel with the random code 
defined in the proof of Theorem~\ref{ZZZ-thm:bcmain}.  

\begin{lem}\label{ZZZ-lem:bcconprob}
If $(\ux,\uy_2)\in\rtyp$, then 
\[
\hp(\uy_2|\ux) \leq 2^{N(4a_1(\eps_1)+2\eps_1+1/N)}p(\uy_2|\ux);  
\]
if, further, $(\ux,\uy_1,\uy_2)\in\rtyp$, then 
\[
\hp(\uy_1|\ux,\uy_2) \leq 2^{N(8a_2(\eps_2)+1/N)}p(\uy_1|\ux,\uy_2).  
\]

\Proof
The first bound is precisely Lemma~\ref{ZZZ-lem:conprob} 
by the definition of $\rtyp$.  
The proof of the second bound is almost identical 
except in this case codewords are drawn according to $p(\uy_1|\uy_2)$.  
This leads to both the extra variable in the condition 
and the slightly larger exponent in the bound.  
\IEEEQED
\end{lem}

\begin{lem}\label{ZZZ-lem:bca1}
\begin{eqnarray*}
\hp((X,Y_2)\not\in\rtyp|\ux)
& \leq & e^{-2^{N(R_0-I(X;Y_2)-2a_2(\eps_2)-\eps_2)}}
+p((\rtyp(X,Y_2))^c|\ux) \\
\hp((X,Y_1,Y_2)\not\in\rtyp|\ux,\uy_2)
& \leq & e^{-2^{N(R_1-I(X;Y_1|Y_2)-4a_2(\eps_2))}}
+p((\rtyp(X,Y_1,Y_2))^c|\ux,\uy_2)
\end{eqnarray*}

\Proof  
The given code fails to find a jointly typical reproduction $(\uY_1,\uY_2)$ 
for $\uX$ if either stage of its encoder fails.  
The first stage fails with probability 
\[
\hp((\rtyp(X,Y_2))^c|\ux) \leq 
p((\rtyp(X,Y_2))^c|\ux)+e^{-2^{N(R_0-I(X;Y_2)-2a_2(\eps_2)-\eps_2)}}
\]
by Lemma~\ref{ZZZ-lem:a1}.  
Otherwise, let $\uy_2$ be the first-stage codeword 
with $(\ux,\uy_2)\in\rtyp$.  
If $(\ux,\uy_2)$ satisifies 
$p((\typ(X,Y_1,Y_2))^c|\ux,\uy_2)>2^{-N3\eps_2}$,
then $\hp((\rtyp(X,Y_1,Y_2))^c|\ux,\uy_2)
=p((\rtyp(X,Y_1,Y_2))^c|\ux,\uy_2)=1$ by definition of $\rtyp$.    
Otherwise $(\ux,\uy_1,\uy_2)\not\in\rtyp$ implies that 
encoder $\alpha^{(1)}_N$ failed to find 
a jointly typical codeword $\uy_1$ for $(\ux,\uy_2)$.  
Thus  
\begin{eqnarray*}
\hp\left((\rtyp((X,Y_1,Y_2)))^c|\ux,\uy_2\right) 
& \leq & \left(\sum_{\uy_1}p(\uy_1|\uy_2)
(1-K(\ux,\uy_1,\uy_2))\right)^{2^{nR_1}}.  
\end{eqnarray*}
When $K(\ux,\uy_1,\uy_2)=1$, 
the usual bounds on the probabilities of typical strings give 
\[
p(\uy_1|\uy_2) = p(\uy_1|\ux,\uy_2) 
\frac{p(\uy_1,\uy_2)p(\ux,\uy_2)}{p(\uy_2)p(\ux,\uy_1,\uy_2)} 
\geq p(\uy_1|\ux,\uy_2)2^{-N(I(X;Y_1|Y_2)+4a_2(\eps_2))}.  
\]
Therefore, since $(1-ab)^n\leq 1-a+e^{-bn}$, 
\begin{eqnarray*}
\hp((\rtyp(X,Y_1,Y_2))^c|\ux,\uy_2) 
& \leq & \left(1-2^{-N(I(X;Y_1|Y_2)+4a_2(\eps_2))}\sum_{\uy_1}
	  p(\uy_1|\ux,\uy_2)K(\ux,\uy_1,\uy_2)\right)^{2^{nR_1}}\\
& \leq & 1-\sum_{\uy_1}p(\uy_1|\ux,\uy_2)K(\ux,\uy_1,\uy_2)
	  +e^{-2^{N(R_1-I(X;Y_1|Y_2)-4a_2(\eps_2))}} \\
& = & p((\rtyp(X,Y_1,Y_2))^c|\ux,\uy_2)
	  +e^{-2^{N(R_1-I(X;Y_1|Y_2)-4a_2(\eps_2))}}.
\end{eqnarray*}
\IEEEQED
\end{lem}

\section{Multiple Access Channels}\label{ZZZ-app:ma}

The following definitions, 
used in the proof of Theorem~\ref{ZZZ-thm:mamain}, below, 
rely on notation defined in Appendix~\ref{ZZZ-app:typ}.  

Given any $p(u,x_1,x_2,y)=p(u|x_1)p(x_1,x_2)p(y|x_1,x_2)$, 
fix $\eps=(\eps_1,\eps_2)$ with $\eps_1,\eps_2>0$.  
Let 
\begin{eqnarray}
a_1(\eps_1) 
&\deff& (1+\eps_1)\cdot\inf\left\{\eps'>0: 
p( f(\uU,\uX_1)>\eps'\vee f(\uU)>\eps')\leq 2^{-N6\eps_1}
\ \forall N\mbox{ suff large}\right\}.\label{ZZZ-eqn:maatdef1}\\
a_2(\eps_2) & \deff &  (1+\eps_2)\cdot\inf\left\{\eps'>0: 
p\left(f(\uY)>\eps'\vee f(\uU,\uY)>\eps'\vee f(\uU,\uX_1,\uX_2)>\eps'
\right.\right. \nonumber \\
&&\left.\left.\vee f(\uX_1,\uX_2,\uY)>\eps'
              \vee f(\uU,\uX_1,\uX_2,\uY)>\eps'\right)
	      \leq 2^{-N6\eps_2}\ \forall N\mbox{ suff.\ large}\right\}
\end{eqnarray}
The typical sets for $p(u,x_1)$, $p(u,x_1,x_2,y)$, and $p(x_1,x_2,y)$ are 
\begin{eqnarray*}
\typ(U,X_1) & \deff & \left\{(\uu,\ux_1):
f(\ux_1)\leq \eps_1,\ f(\uu)\leq a_1(\eps_1),\ 
f(\uu,\ux_1)\leq a_1(\eps_1)\right\} \\
\typ(U,X_1,X_2,Y)&\deff& \{(\uu,\ux_1,\ux_2,\uy):
f(\uu,\ux_1,\ux_2)\leq a_2(\eps_2), 
f(\uu,\ux_1,\ux_2,\uy)\leq a_2(\eps_2), \\
&&f(\uu,\uy)\leq a_2(\eps_2), 
f(\uy)\leq a_2(\eps_2)\} \\
\typ(X_1,X_2,Y)&\deff& \left\{(\ux_1,\ux_2,\uy):
f(\ux_1,\ux_2)\leq \eps_2,\ 
f(\uy)\leq a_2(\eps_2),\ 
f(\ux_1,\ux_2,\uy)\leq a_2(\eps_2)\right\}, 
\end{eqnarray*}
which we restrict as 
\begin{eqnarray*}
\rtyp(U,X_1) & \deff & \left\{(\uu,\ux_1)\in\typ:
p\left(\left.((\typ(U,X_1)^c\right|\ux_1\right)\leq 2^{-3N\eps_1}\right\} \\
\rtyp(U,X_1,X_2,Y) &\deff & \left\{(\uu,\ux_1,\ux_2,\uy)\in\typ:
p\left((\typ(U,X_1,X_2,Y))^c|(\uu,\ux_1,\ux_2)\right)
\leq 2^{-3N\eps_2}\right\} \\
\rtyp(X_1,X_2,Y) &\deff& \left\{(\ux_1,\ux_2,\uy)\in\typ:
p\left(\left.(\typ(X_1,X_2,Y)^c\right|(\ux_1,\ux_2)\right)
\leq 2^{-3N\eps_2}\right\}.  
\end{eqnarray*}

Lemma~\ref{ZZZ-lem:matyp} bounds the probability 
that i.i.d.\ samples from $p(u,x_1,x_2,y)$ are atypical.  

\begin{lem}\label{ZZZ-lem:matyp}
If $(\uU,\uX_1,\uX_2,\uY)$ 
are drawn i.i.d.\ $p(u,x_1,x_2,y)$, 
then 
\begin{eqnarray*}
p\left((\rtyp(U,X_1))^c\right) &  \leq & 2^{-Nc_1(\eps_1)}\\
p\left((\rtyp(U,X_1,X_2,Y))^c\cup(\rtyp(X_1,X_2,Y))^c\right)
& \leq &  2^{-Nc_2(\eps_2)} 
\end{eqnarray*}
for some $c_1(\eps_1),c_2(\eps_2)>0$ and all $N$ sufficiently large.  
Constants $c_1(\eps_1)$ and $c_2(\eps_2)$ approach 0 
as $\eps_1$ and $\eps_2$, respectively, approach 0.  

\Proof 
Like Lemma~\ref{ZZZ-lem:typ}, 
the result follows Chernoff's bound and the definition of $\rtyp$.  
\end{lem}

{\bf Proof of Theorem~\ref{ZZZ-thm:mamain}:}
Since $R^{\ar{\{i_2\}}{\{j\}}}$ is not bounded from below, 
we set it to 0.  
For concision, we further define 
$R_1\ \deff\ R^{\ar{\{i_1\}}{\{j\}}}$ and 
$R_2\ \deff\ R^{\ar{\{i_1,i_2\}}{\{j\}}}$ and use 
$\cC=(\cX_1\times\cX_2,p(y|x_1,x_2),\cY)$ 
in place of $\cC=(\ciXo\times\ciXt,p(\jyr|\ixo,\ixt),\cjYr)$ 
both in this proof and its supporting lemmas.  

Fix $(R_1,R_2)$ to satisfy the theorem constraints.  
By Theorem~\ref{ZZZ-thm:equiv}, 
it suffices to show that for any channel input distribution $p(x_1,x_2)$ 
there exists a sequence of rate-$(R_1,R_2)$ 
random emulation codes $(\alpha_N,\beta_N)$ 
for which the resulting emulation distribution
\[
\hp(\uy|\ux_1,\ux_2)\ \deff\ \Pr(\beta_N(\alpha_N(\ux_1,\ux_2))=(\uy)) 
\]
satisfies 
\[
P_e^{(N)}(\nu)=\sum_{\ux,\uy_1,\uy_2}p(\ux)\hp(\uy_1,\uy_2|\ux)
1\left(\frac1N\log\left(\frac{\hp(\uy_1,\uy_2|\ux)}{p(\uy_1,\uy_2|\ux)}\right)>\nu\right)
<2^{-N\eta(\nu)}
\]
for some positive function $\eta(\nu)$ dependent on $p(x)$ for which 
$\eta(\nu)$ goes to zero as $\nu$ goes to zero.  

Fix any $p(x_1,x_2)$, and then choose $p(u|x_1)$ 
to satisfy the constraints on $R_1$ and $R_2$.  
Let 
\[
p(u,x_1,x_2,y)\ \deff\ p(u|x_1)p(x_1,x_2)p(y|x_1,x_2).  
\]

Recall that $(\alpha_N,\beta_N)$ has encoders 
\[
\alpha_N=\left(\alpha_N^{\ar{A}{B}}:(A,B)\in\cM\right)
=\left(\alpha_N^{\ar{\{i_1\}}{\{j\}}},\alpha_N^{\ar{\{i_2\}}{\{j\}}},
\alpha_N^{\ar{\{i_1,i_2\}}{\{j\}}}\right)
\] 
at rates $R_1=R^{\ar{\{i_1\}}{\{j\}}}$, $R^{\ar{\{i_2\}}{\{j\}}}=0$, 
and $R_2=R^{\ar{\{i_1,i_2\}}{\{j\}}}$ 
and decoder $\beta_N=\beta_N^{(j)}$.  
Rate 0 requires no encoder.  
We abbreviate the notation for the remaining encoders to 
$\alpha_N^{(1)}=\alpha_N^{\ar{\{i_1\}}{\{j\}}}$ 
and $\alpha_N^{(2)}=\alpha_N^{\ar{\{i_1,i_2\}}{\{j\}}}$.  
The code also relies on a mapping $\gamma_N$.  
Thus the code defines a collection of mappings 
\[
\begin{array}{rlcrl}
\alpha^{(1)}_N: & \ucX_1\rightarrow\cW_1 &\hspace{.5in}&
\beta^{(j)}_N: & \cW_1\times\cW_2\rightarrow\ucY \\
\alpha^{(2)}_N: & \ucX_1\times\ucX_2\rightarrow\cW_2 &&
\gamma_N & \cW_1\rightarrow \ucU,
\end{array}
\]
where $\cW_1=\utcX^{\ar{\{i_1\}}{\{j\}}} =\{0,1\}^{NR_1}$
and $\cW_2=\utcX^{\ar{\{i_1,i_2\}}{\{j\}}} =\{0,1\}^{NR_2}$.  
Encoder $\alpha^{(1)}_N$ operates at node $i_1$.  
Encoder $\alpha^{(2)}_N$ is operates at node $x^{V_1}$ 
using inputs $\uX_1$ and $\uX_2$ 
losslessly received from nodes $i_1$ and $i_2$.  
The decoder is operated at node $j$.  

The random code design draws $\{\gamma_N(w_1):w_1\in\cW_1\}$ 
i.i.d.\ from the distribution $p(\uu)$.  
For each $w_1\in\cW_1$ set $\uu=\gamma_N(w_1)$ and then draw 
$\{\beta_N(w_1,w_2):w_2\in\cW_2\}$ 
i.i.d.\ from $p(\uy|\uu)=\prod_{\ell=1}^N p_t(\uy(\ell)|\uu(\ell))$.  
For the random encoder design, 
choose $\alpha_N^{(1)}(\ux_1)$ uniformly at random 
from the indices $w_1$ for which 
$(\gamma_N(w_1),\ux_1)\in\rtyp$.  
If there is no such $w_1$, 
then set $\alpha_N^{(1)}(\ux_1)$ to 1.  
For each $(\ux_1,\ux_2)$, 
let $w_1=\alpha_N^{(1)}(\ux_1)$ and $\uu=\gamma_N(w_1)$,
and choose $\alpha_N^{(2)}(\ux_1,\ux_2)$ uniformly at random 
from the indices $w_2$ for which 
\begin{eqnarray*}
& (\ux_1,\ux_2,\beta_N(w_1,w_2))\in\rtyp(X_1,X_2,Y) \\
& (\uu,\ux_1,\ux_2,\beta_N(w_1,w_2))\in\rtyp(U,X_1,X_2,Y);
\end{eqnarray*}
if there is no such index, then $\alpha_N^{(2)}(\ux_1,\ux_2)=0$.  

By Lemma~\ref{ZZZ-lem:maconprob}, below, 
\[
\hp(\uy|\ux_1,\ux_2)
\leq 2^{N(4a_1(\eps_1)+2\eps_1+8a_2(\eps_2)+2/N)}p(\uy|\ux_1,\ux_2).  
\]
for all $(\ux_1,\ux_2,\uy)\in\rtyp$.  
By Lemma~\ref{ZZZ-lem:maa1}, below, 
\begin{eqnarray*}
\lefteqn{\hp((\rtyp(X_1,X_2,Y))^c|\ux_1,\ux_2)
\ \leq\ \delta_1+\delta_2+p((\rtyp(U,X_1))^c|\ux_1)} \\ 
&& + 2^{N(2\eps_1+4a_1(\eps_1)+1/N)}
p((\rtyp(U,X_1,X_2,Y))^c\cup(\rtyp(X_1,X_2,Y))^c|\ux_1,\ux_2),
\end{eqnarray*}
where $\delta_1\ \deff\ e^{-2^{N(R_1-I(U;X_1)-\eps_1-2a_1(\eps_1))}}$
and $\delta_2\ \deff\ e^{-2^{N(R_2-I(X_1,X_2;Y|U)-4a_2(\eps_2))}}$.  
By Lemma~\ref{ZZZ-lem:matyp}, above, 
\begin{eqnarray*}
p\left((\rtyp(U,X_1))^c\right) & \leq &  2^{-Nc_1(\eps_1)} \\
p\left((\rtyp(U,X_1,X_2,Y))^c\cup(\rtyp(X_1,X_2,Y))^c\right)
& \leq &  2^{-Nc_2(\eps_2)}
\end{eqnarray*}
for some constants $c_1(\eps_1),c_2(\eps_2)>0$ 
and all $N$ sufficiently large; 
constants $c_1(\eps_1)$ and $c_2(\eps_2)$ 
go to zero as $\eps_1$ and $\eps_2$ go to zero.  

Thus when $\nu=4a_1(\eps_1)+3\eps_1+8a_2(\eps_2)$ and $N$ is sufficiently large, 
\begin{eqnarray*}
P_e^{(N)}(\nu)
&\leq& \sum_{(\ux_1,\ux_2,\uy)\not\in\rtyp}p(\ux_1,\ux_2)
	\hp(\uy|\ux_1,\ux_2) \\
&\leq& \sum_{(\ux_1,\ux_2)}p(\ux_1,\ux_2)
	\left(\delta_1+\delta_2+p((\rtyp(U,X_1))^c|\ux_1)\right. \\ 
&& \left.+ 2^{N(2\eps_1+4a_1(\eps_1)+1/N)}
	p((\rtyp(U,X_1,X_2,Y))^c\cup(\rtyp(X_1,X_2,Y))^c|\ux_1,\ux_2)\right) \\
&\leq& \delta_1+\delta_2+2^{-Nc_1(\eps_1)}
	+2^{N(c_2(\eps_2)-2\eps_1-4a_1(\eps_1)-1/N)}.  
\end{eqnarray*}
Thus for all $N$ sufficiently large, 
$P_e^{(N)}(\nu)$ decays exponentially to zero 
provided that $\eps_1$ is chosen to satisfy 
$2a_1(\eps_1)+\eps_1<R_1-I(U;X_1)$ and 
$\eps_2$ is chosen to satisfy 
$4a_2(\eps_2)<R_2-I(X_1,X_2;Y|U)$ and 
$c(\eps_2)>2\eps_1+4a_1(\eps_1)$.  
The resulting exponent decays to zero 
as $\eps_1$ and $\eps_2$ decay to zero.  

We next derive the bound on $|\cU|$.  
For any fixed conditional distribution $p(x_1|u)$ 
on an alphabet $\cU$ that is arbitrarily large, 
we can express the optimization of $U$ 
as a minimization of the Lagrangian $I(X_1;U)+\nu I(X_1,X_2;Y|U)$ 
over all $p(u)$, $u\in\cU$, 
satisfying the constraints $p(u)\geq0$ for all $u\in\cU$, 
$\sum_{u\in\cU}p(u)=1$, and 
$\sum_{u\in\cU}p(u)p(x_1|u)=p(x_1)$ 
for all but one $x_1\in\cX_1$,\footnote{If $\sum_{u\in\cU}p(u)=1$,
$\sum_{u\in\cU}p(u)p(x_1|u)=p(x_1)$ for all but one $x_1\in\cX_1$, 
then $\sum_{u\in\cU}p(u)p(x_1|u)=p(x_1)$ for the remaining $x_1\in\cX_1$ as well.} 
where $\nu>0$ is the Lagrangian constant.  
The Lagrangian and the constraints are linear in $p(u)$, 
so this is a linear program.  
For every linear program, there exists a solution 
on the boundary of the constrained region.  
Therefore, given $|\cU|$ variables, there exists a minimizing distribution $p(u)$ 
that satisfies $|\cU|$ of the given constraints with equality.  
We have one constraint $\sum_{u\in\cU}p(u)=1$ and 
$|\cX|-1$ constraints of form $\sum_{u\in\cU}p(u)p(x_1|u)=p(x_1)$, 
so at least $|\cU|-|\cX|$ constraints of the form $p(u)\geq 0$ 
are met with equality.  
This implies $p(u)>0$ for at most $|\cX|$ values of $u$, 
which gives the desired bound on $|\cU|$. 
\IEEEQED

\begin{lem}\label{ZZZ-lem:maconprob}
For all $(\uu,\ux_1)\in\rtyp$, 
\[
\hp(\uu|\ux_1)\leq 2^{N(4a_1(\eps_1)+2\eps_1+1/N)}p(\uu|\ux_1);
\]
if, further, $(\uu,\ux_1,\ux_2,\uy)\in\rtyp$ 
and $(\ux_1,\ux_2,\uy)\in\rtyp$, then 
\[
\hp(\uy|\uu,\ux_1,\ux_2)\leq 2^{N(8a_2(\eps_2)+1/N)}p(\uy|\uu,\ux_1,\ux_2).
\]
Thus, for all $(\ux_1,\ux_2,\uy)\in\rtyp$, 
\[
\hat{p}(\uy|\ux_1,\ux_2)
\leq 2^{N(4a_1(\eps_1)+2\eps_1+8a_2(\eps_2)+2/N)}p(\uy|\ux_1,\ux_2).
\]
\Proof  
The first bound follows immediately from Lemma~\ref{ZZZ-lem:conprob}.  
For the second bound, 
recall that the second encoder observes both $\ux_1$ and $\ux_2$ 
and looks for a match among codewords drawn according to $p(\uy|\uu)$.  
The second bound follows an argument similar to the first, 
just accounting for these minor differences.  
Note that $\hp(\uu|\ux_1)=\hp(\uu|\ux_1,\ux_2)$ 
for the given code design.  
Likewise $p(\uu|\ux_1)=p(\uu|\ux_1,\ux_2)$ 
since $U\rightarrow X_1\rightarrow X_2$ forms a Markov chain.  
Note further that each encoder chooses an index 0 
if it fails to find a matching codeword, 
and there is no codeword defined for this index;  
this choice guarantees that source code's $(\ux_1,\ux_2)$ 
and output $\uy_1$ are jointly typical only 
if both encoders succeed in finding jointly typical codewords -- 
that is, if the conditions of the first two inequalities are met.  
Therefore
\begin{eqnarray*}
\hp(\uy|\ux_1,\ux_2) 
& = & \sum_{\uu}\hp(\uy|\uu,\ux_1,\ux_2)\hp(\uu|\ux_1) \\
&\leq & \sum_{\uu}p(\uu,\uy|\ux_1,\ux_2)
        2^{N(4a_1(\eps_1)+2\eps_1+8a_2(\eps_2)+2/N)}.
\end{eqnarray*}
\IEEEQED
\end{lem}

\begin{lem}\label{ZZZ-lem:maa1}
For all $(\ux_1,\ux_2)\in\ucX_1\times\ucX_2$, 
\begin{eqnarray*}
\lefteqn{\hp((\rtyp(X_1,X_2,Y))^c|\ux_1,\ux_2)\ \leq\ 
\delta_1+\delta_2+p((\rtyp(U,X_1))^c|\ux_1)} \\
&&\hspace{1in}+ 2^{N(2\eps_1+4a_1(\eps_1)+1/N)} 
	p((\rtyp(U,X_1,X_2,Y))^c\cup(\rtyp(X_1,X_2,Y))^c|\ux_1,\ux_2),
\end{eqnarray*}
where $\delta_1\ \deff\ e^{-2^{N(R_1-I(U;X_1)-\eps_1-2a_1(\eps_1))}}$
and 
$\delta_2\ \deff\ e^{-2^{N(R_2-I(X_1,X_2;Y|U)-4a_2(\eps_2))}}$.

\Proof  
If $p((\typ(X_1,X_2,Y))^c|(\ux_1,\ux_2))>2^{-3N\eps_2}$,
then $\hp((\rtyp)^c|\ux_1,\ux_2)=p((\rtyp)^c|\ux_1,\ux_2)=1$ 
by the definition of $\rtyp(X_1,X_2,Y)$ and the bound is satisfied.  
Otherwise, $(\ux_1,\ux_2,\uY)\not\in\rtyp$ 
implies that one or both 
of the encoders $\alpha^{(1)}_N$ and $\alpha^{(2)}_N$ 
failed to find a matching codeword for $(\ux_1,\ux_2)$.  
Encoder $\alpha^{(1)}_N$ fails 
if there is no jointly typical codeword for $\ux_1$ 
in codebook $\{\gamma_N(1),\ldots,\gamma_N(2^{NR_1})\}$.  
Otherwise, let $w_1=\alpha^{(1)}_N(\ux_1)$ and $\uu=\gamma_N(w_1)$.  
Then encoder $\alpha^{(2)}_N$ fails if no codeword in 
$\{\beta_N(w_1,1),\ldots,\beta_N(w_1,2^{NR_1})\}$ is 
jointly typical with $(\uu,\ux_1,\ux_2)$.  
Therefore 
\begin{eqnarray*}
\lefteqn{p\left((\uX_1,\uX_2,\uY)\not\in\rtyp|(\uX_1,\uX_2)=(\ux_1,\ux_2)
\right)}\\
&\leq& \left(\sum_{\uu}p(\uu)(1-K(\uu,\ux_1))\right)^{2^{NR_1}}
   +\sum_{\uu:K(\uu,\ux_1)=1}\hp(\uu|\ux_1) 
\left(\sum_{\uy}p(\uy|\uu)(1-K(\uu,\ux_1,\ux_2,\uy))
\right)^{2^{nR_2}}.
\end{eqnarray*}
By the usual probability bounds for elements of the typical set, 
\[
\begin{array}{rcll}
p(\uu) &\geq&p(\uu|\ux_1)2^{-N(I(U;X_1)+\eps_1+2a_1(\eps_1))} 
	& \mbox{when $K(\uu,\ux_1)=1$} \\
p(\uy|\uu)&\geq& p(\uy|\uu,\ux_1,\ux_2) 
		 2^{-N(I(X_1,X_2;Y|U)+4a_2(\eps_2))} 
		 & \mbox{when $K(\uu,\ux_1,\ux_2,\uy)=1$}.   
\end{array}
\]
Applying these bounds, the bound $(1-ab)^n\leq 1-a+e^{-bn}$, 
and Lemma~\ref{ZZZ-lem:maconprob} gives 
\begin{eqnarray*}
\lefteqn{p\left((\uX_1,\uX_2,\uY)\not\in\rtyp|
(\uX_1,\uX_2)=(\ux_1,\ux_2)\right)}\\
&\leq& 1-\sum_{\uu}p(\uu|\ux_1)K(\uu,\ux_1) 
+e^{-2^{N(R_1-I(U;X_1)-\eps_1-2a_1(\eps_1))}} 
+\sum_{\uu}K(\uu,\ux_1)\hp(\uu|\ux_1) \\
&&\cdot\left(1-\sum_{\uy}K(\uu,\ux_1,\ux_2,\uy)K(\ux_1,\ux_2,uy)
p(\uy|\uu,\ux_1,\ux_2)
+e^{-2^{N(R_2-I(X_1,X_2;Y|U)-4a_2(\eps_2))}}\right) \\
&\leq& p((\rtyp(U,X_1))^c|\ux_1) 
+e^{-2^{N(R_1-I(U;X_1)-\eps_1-2a_1(\eps_1))}} 
+e^{-2^{N(R_2-I(X_1,X_2;Y|U)-4a_2(\eps_2))}} \\
& & +2^{N(2\eps_1+4a_1(\eps_1)+1/N)}p((\rtyp(U,X_1,X_2,Y))^c|\ux_1,\ux_2).  
\end{eqnarray*}
\IEEEQED
\end{lem}

\section{Interference Channels:  Model 1}\label{ZZZ-app:ina}

The following definitions, 
used in the proof of Theorem~\ref{ZZZ-thm:inmaina}, below, 
rely on notation defined in Appendix~\ref{ZZZ-app:typ}.  

Given any distribution
$p(u_1,u_2,x_1,x_2,y_1,y_2)=p(u_2|x_1)p(u_1|u_2,x_1)p(x_1,x_2)
p(y_1,y_2|x_1,x_2)$, 
fix $\eps=(\eps_1,\eps_2,\eps_3,\eps_4)$ with
$\eps_1,\eps_2,\eps_3,\eps_4>0$.  
Define
\begin{eqnarray*}
a_1(\eps_1) & \deff & (1+\eps_1)\cdot\inf\left\{\eps'>0:
	p\left(f(\uU_2)>\eps'\vee f(\uU_2,\uX_1)>\eps'\right)
	\leq 2^{-N6\eps_1}\ \forall N\mbox{ suff.\ large}\right\} \\
a_2(\eps_2) & \deff & (1+\eps_2)\cdot\inf\left\{\eps'>0:
p\left(f(\uU_1,\uU_2)>\eps'\vee f(\uU_1,\uU_2,\uX_1)>\eps'\right) 
\leq 2^{-N6\eps(t)} \ \forall N\mbox{ suff.\ large}\right\} \\
a_3(\eps_3)
&\deff& (1+\eps_3(t))\cdot\inf\left\{\eps'>0: \Pr\left(
f(\uU_2)>\eps'\vee
f(\uU_2,\uY_2)>\eps'\vee
f(\uU_2,\uX_1,\uX_2)>\eps'\vee\right.\right.\\
&&\left.\left.
f(\uU_2,\uX_1,\uX_2,\uY_2)>\eps'\right)\leq 2^{-N6\eps_3(t)}
\ \forall N\mbox{ suff.\ large} \right\} \\
a_4(\eps_4)& \deff& (1+\eps_4(t))\cdot\inf\{\eps'>0:
\Pr(f(\uU_1,\uU_2,\uY_2)>\eps'\vee
    f(\uU_1,\uU_2,\uY_1,\uY_2)>\eps'\vee\\
&&  f(\uU_1,\uU_2,\uX_1,\uX_2,\uY_2)>\eps'\vee
    f(\uU_1,\uU_2,\uX_1,\uX_2,\uY_1,\uY_2)>\eps'\vee
    f(\uY_1,\uY_2)>\eps'\vee \\
&&  f(\uX_1,\uX_2,\uY_1,\uY_2)>\eps')
	 \leq 2^{-N6\eps_4(t)}\ \forall N\mbox{\ suff.\ large}\}. 
\end{eqnarray*}
The corresponding typical sets are  
\begin{eqnarray*}
\typ(U_2,X_1) & \deff & \left\{(\uu_2,\ux_1):
	f(\ux_1)\leq \eps_1,\ 
	f(\uu_2), f(\uu_2,\ux_1)\leq a_1(\eps_1) \right\} \\
\typ(U_1,U_2,X_1)& \deff& \left\{(\uu_1,\uu_2,\ux_1):
f(\uu_2), f(\uu_1,\uu_2), f(\ux_1,\uu_2), f(\uu_1,\uu_2,\ux_1)
\leq a_2(\eps_2) \right\} \\
\typ(U_2,X_1,X_2,Y_2)& \deff& \left\{(\uu_2,\ux_1,\ux_2,\uy_2):
f(\uu_2), f(\uu_2,\uy_2), f(\uu_2,\ux_1,\ux_2), \right. \\
&&\left.f(\uu_2,\ux_1,\ux_2,\uy_2)\leq a_3(\eps_3)\right\} \\
\typ(U_1,U_2,X_1,X_2,Y_1,Y_2)
& \deff & \left\{(\uu_1,\uu_2,\ux_1,\ux_2,\uy_1,\uy_2):
f(\uu_1,\uu_2,\uy_2),f(\uu_1,\uu_2,\uy_1,\uy_2),\right.\\
&&\left.f(\uu_1,\uu_2,\ux_1,\ux_2,\uy_2),
f(\uu_1,\uu_2,\ux_1,\ux_2,\uy_1,\uy_2)\leq a_4(\eps_4) \right\} \\
\typ(X_1,X_2,Y_1,Y_2) 
& \deff & \left\{(\ux_1,\ux_2,\uy_1,\uy_2):
f(\ux_1,\ux_2)\leq \eps_4(t),\ \right.\\
&&\left.f(\uy_1,\uy_2),
f(\ux_1,\ux_2,\uy_1,\uy_2)\leq a_4(\eps_4)\right\},
\end{eqnarray*}
which we restrict as 
\begin{eqnarray*}
\rtyp(U_2,X_1) 
& \deff & \left\{(\uut,\ux_1)\in\typ:
	\Pr\left((\typ(U_2,X_1))^c|\ux_1\right)\leq 2^{-3N\eps_1}\right\} \\
\rtyp(U_1,U_2,X_1)
& \deff & \left\{(\uuo,\uut,\ux_1)\in\typ:
	\Pr\left((\typ(U_1,U_2,X_1))^c|\uu_2,\ux_1\right)
	\leq 2^{-3N\eps_2}\right\} \\
\rtyp(U_2,X_1,X_2,Y_2) 
& \deff & \left\{(\uut,\ux_1,\ux_2,\uy_2)\in\typ:
	\Pr\left((\typ)^c|\uu_2,\ux_1,\ux_2\right)
	\leq 2^{-3N\eps_3(t)}\right\} \\
\rtyp(U_1,U_2,X_1,X_2,Y_1,Y_2)
& \deff& \left\{(\uuo,\uut,\ux_1,\ux_2,\uy_1,\uy_2)\in\typ:
	\Pr\left((\typ)^c|\uu_1,\uu_2,\ux_1,\ux_2,\uy_2\right)\right.\\
&&	\left.\leq 2^{-3N\eps_4(t)}\right\} \\
\rtyp(X_1,X_2,Y_1,Y_2)
& \deff& \left\{(\ux_1,\ux_2,\uy_1,\uy_2)\in\typ:
\Pr\left((\typ)^c|\ux_1,\ux_2\right)\leq 2^{-3N\eps_4(t)}\right\}.  
\end{eqnarray*}

\begin{lem}\label{ZZZ-lem:intypa}
If $(\uU_1,\uU_2,\uX_1,\uX_2,\uY_1,\uY_2)$ 
are drawn i.i.d.\ $p(u_1,u_2,x_1,x_2,y_1,y_2)$, 
then there exist positive constants 
$c_1(\eps_1)$, $c_2(\eps_2)$, $c_3(\eps_3)$, and $c_4(\eps_4)$  
for which 
\begin{eqnarray*}
\Pr\left((\rtyp(U_2,X_1))^c\right)& \leq & 2^{-Nc_1(\eps_1)}  \\
\Pr\left((\rtyp(U_1,U_2,X_1))^c\right)&\leq&2^{-Nc_2(\eps_2)}  \\
\Pr\left((\rtyp(U_2,X_1,X_2,Y_2))^c\right)&\leq&2^{-Nc_3(\eps_3)}  \\
\Pr\left((\rtyp(U_1,U_2,X_1,X_2,Y_1,Y_2))^c\cup
	(\rtyp(X_1,X_2,Y_1,Y_2))^c	\right) & \leq & 2^{-Nc_4(\eps_4)}
\end{eqnarray*}
for all $N$ sufficiently large.  
Constant $c_k(\eps,t)$ approaches 0 as $\eps_k(t)$ decays to 0.  

\Proof 
Like Lemma~\ref{ZZZ-lem:typ}, 
the result follows from Chernoff's bound 
and the definiton of $\rtyp$.  
\IEEEQED
\end{lem}

{\bf Proof of Theorem~\ref{ZZZ-thm:inmaina}:}
We set the rates 
$R^{\ar{\{i_1\}}{\{j_2\}}}$, 
$R^{\ar{\{i_2\}}{\{j_1\}}}$, 
$R^{\ar{\{i_2\}}{\{j_2\}}}$,
$R^{\ar{\{i_2\}}{\{j_1,j_2\}}}$, and 
$R^{\ar{\{i_1,i_2\}}{\{j_2\}}}$ 
for which no bounds are given to zero, 
simplify remaining notation as 
$R_{11}\ \deff\ R^{\ar{\{i_1\}}{\{j_1\}}}$,
$R_{12}\ \deff\ R^{\ar{\{i_1\}}{\{j_1,j_2\}}}$,
$R_{21}\ \deff\ R^{\ar{\{i_1,i_2\}}{\{j_1\}}}$, and
$R_{22}\ \deff\ R^{\ar{\{i_1,i_2\}}{\{j_1,j_2\}}}$ 
and use $\cC=(\cX_1\times\cX_2,p(y_1,y_2|x_1,x_2),\cY_1\times\cY_2)$
instead of 
$(\ciXo\times\ciXt,p(\jyo,\jyt|\ixo,\ixt),\cjYo\times\cjYt)$
in this proof and its supporting lemmas.  

Fix $(R_{11},R_{12},R_{21},R_{22})$ to satisfy the theorem constraints.  
Let $p(x_1,x_2)$ be arbitrary, and choose $p(\ut|x_1)$ and $p(\uo|x_1,\ut)$ 
to satisfy the given bounds.  
Let 
\[
p(u_1,u_2,x_1,x_2,y_1,y_2)\ 
\deff\ p(u_2|x_1)p(u_1|x_1,u_2)p(x_1,x_2)p(y_1,y_2|x_1,x_2).
\]
We define corresponding (restricted) typical sets in Appendix~\ref{ZZZ-app:ina}.  

Excluding the rate-0 codes, four encoders and two decoders are required.  
We simplify their notation as 
\[
\begin{array}{rclcrclcrcl}
\alpha^{(11)}_N&=&\alpha_N^{\ar{\{i_1\}}{\{j_1\}}} &\hspace{.5in}&
\alpha^{(21)}_N&=&\alpha_N^{\ar{\{i_1,i_2\}}{\{j_1\}}} &\hspace{.5in}&
\beta^{(1)}_N&=&\beta_N^{(j_1)} \\
\alpha^{(12)}_N&=&\alpha_N^{\ar{\{i_1\}}{\{j_1,j_2\}}} &&
\alpha^{(22)}_N&=&\alpha_N^{\ar{\{i_1,i_2\}}{\{j_1,j_2\}}} &&
\beta^{(2)}_N&=&\beta_N^{(j_2)},
\end{array}
\]
where
\[
\begin{array}{lclcl}
\alpha^{(11)}_N:\ucX_1\rightarrow\cW_{11} & \hspace{.25in}&
\alpha^{(21)}_N :\ucX_1\times\ucX_2\rightarrow\cW_{21}&\hspace{.25in}&
\beta^{(1)}_N:\cW_{11}\times\cW_{12}\times\cW_{21}\times\cW_{22}
\rightarrow\ucY_1\\
\alpha^{(12)}_N:\ucX_1\rightarrow\cW_{12} &&
\alpha^{(22)}_N :\ucX_1\times\ucX_2\rightarrow\cW_{22}&&
\beta^{(2)}_N:\cW_{12}\times\cW_{22}\rightarrow\ucY_2
\end{array}
\]
and 
\[
\begin{array}{rclcrcl}
\cW_{11}&=&\utcX^{\ar{\{i_1\}}{\{j_1\}}}=\{0,1\}^{NR_{11}} &&
\cW_{12}&=&\utcX^{\ar{\{i_1\}}{\{j_1,j_2\}}}=\{0,1\}^{NR_{12}} \\
\cW_{21}&=&\utcX^{\ar{\{i_1,i_2\}}{\{j_1\}}}=\{0,1\}^{NR_{21}} &&
\cW_{22}&=&\utcX^{\ar{\{i_1,i_2\}}{\{j_1,j_2\}}}=\{0,1\}^{NR_{22}}
\end{array}
\]
Encoder $(\alpha^{(11)}_N,\alpha^{(12)}_N)$ operates at node $i_1$, 
transmitting its rate $R_{11}$ and $R_{12}$ descriptions 
to node $j_1$ and both nodes, respectively.  
Encoder $(\alpha^{(21)}_N,\alpha^{(22)}_N)$ operates at node $v^{V_1}$, 
receiving noiseless descriptions 
of $\ux_1$ and $\ux_2$ from nodes $i_1$ and $i_2$ 
and transmitting its rate $R_{21}$ output to node $j_1$ 
and its $R_{22}$ to both nodes.  
The code also employs mappings 
$\gamma_N^{(1)}:\cW_{11}\times\cW_{12}\rightarrow\cU_1$
and $\gamma_N^{(2)}:\cW_{12}\rightarrow\cU_2$

The random code design draws codewords 
$\{\gamma^{(2)}_N(w_{12}):w_{12}\in\cW_{12}\}$
i.i.d.\ from distribution $\prod_{\ell=1}^N p(\uut(\ell))$.  
For each $w_{12}\in\uhcY_{12}$, 
let $\uUt=\gamma^{(2)}_N(w_{12})$ and draw codewords 
$\{\gamma^{(1)}_N(w_{11},w_{12}):w_{11}\in\cW_{11}\}$
i.i.d.\ from $\prod_{\ell=1}^N p(\uuo(\ell)|\uUt(\ell))$
and codewords 
$\{\beta^{(2)}_N(w_{12},w_{22}):w_{22}\in\cW_{22}\}$ 
i.i.d.\ from $\prod_{\ell=1}^Np(\uy_2(\ell)|\uUt(\ell))$. 
Finally, for each 
$(w_{11},w_{12},w_{22})\in\uhcY_{11}\times\uhcY_{12}\times\uhcY_{22}$, 
let 
\[
(\uUo,\uUt,\uY_2)=(\gamma^{(1)}_N(w_{11},w_{12}),
       \gamma^{(2)}_N(w_{12}),
       \beta^{(2)}_N(w_{12},w_{22})),
       \]
and draw $\{\beta^{(1)}_N(w_{11},w_{12},w_{21},w_{22}):w_{21}\in\cW_{21}\}$
i.i.d.\ from 
$\prod_{\ell=1}^N p(\uy_1(\ell)|\uUo(\ell),\uUt(\ell),\uY_2(\ell))$.
For the encoder design, 
choose $\alpha^{(12)}_N(\ux_1)$ uniformly at random from 
those $w_{12}\in\uhcX_{12}$ for which $(\gamma^{(2)}(w_{12}),\ux_1)\in\rtyp$; 
if there is no such $w_{12}$, 
then set $\alpha^{(12)}_N(\ux_1)$ to 0.  
Let $w_{12}$ be the chosen index and $\uUt=\gamma^{(2)}_N(w_{12})$. 
Choose $\alpha^{(11)}_N(\ux_1)$ uniformly at random 
from the set of $w_{11}\in\uhcX_{11}$ for which 
$(\gamma^{(1)}(w_{11},w_{12}),\uUt,\ux_1)\in\rtyp$;  
if there is no such $w_{11}$, 
then set $\alpha^{(11)}_N(\ux_1)$ to 0.  
Let $w_{11}$ be the chosen index and $\uUo=\gamma^{(1)}_N(w_{11},w_{12})$.
Then choose $\alpha^{(22)}_N(\ux_1,\ux_2)$ uniformly at random 
from the set of $w_{22}\in\uhcX_{22}$ for which 
$(\uUt,\ux_1,\ux_2,\beta^{(2)}_N(w_{12},w_{22}))\in\rtyp$;
if this set is empty, set $\alpha^{(22)}_N(\ux_1,\ux_2)=0$.  
Then let $w_{22}$ be the chosen index and 
$\uY_2= \beta^{(2)}_N(w_{12},w_{22})$, 
and choose $\alpha^{(21)}_N(\ux_1,\ux_2)$  uniformly at random 
from the set of $w_{21}\in\uhcX_{21}$ for which 
\[
(\uUo,\uUt,\ux_1,\ux_2,\beta^{(1)}(w_{11},w_{12},w_{21},w_{22}),\uY_2)\in\rtyp;
\]
if this set is empty, set $\alpha^{(21)}_N(\ux_1,\ux_2)$  to 0.  

By Lemma~\ref{ZZZ-lem:inconproba}, below, 
\[
\hp(\uy_1,\uy_2|\ux_1,\ux_2)
\leq 2^{N(2\sum_{k=1}^4b_k(\eps_k)+4/N)} \hspace{.5in}\forall
(\ux_1,\ux_2,\uy_1,\uy_2)\in\rtyp,
\]
where $b_1(\eps_1)=\eps_1+2a_1(\eps_1)$, 
and $b_k(\eps_k)=4a_k(\eps_k)$ for $k\in\{2,3,4\}$.  
By Lemma~\ref{ZZZ-lem:ina1a}, below, 
\begin{eqnarray*}
\lefteqn{\hp((\rtyp(X_1,X_2,Y_1,Y_2))^c|\ux_1,\ux_2)} \\
& \leq & \delta_{11}+\delta_{12}+\delta_{21}+\delta_{22}
         +p((\rtyp(U_2,X_1))^c|\ux_1)
	 +2^{N(2b_1(\eps_1)+1/N)}p((\rtyp(U_1,U_2,X_1))^c|\ux_1) \\
&&	 +2^{N(2\sum_{k=1}^2b_k(\eps_k)+2/N)}p((\rtyp(U_2,X_1,X_2,Y_2))^c) \\
&& +2^{N(2\sum_{k=1}^3b_k(\eps,t)+3/N)} 
	p((\rtyp(U_1,U_2,X_1,X_2,Y_1,Y_2))^c\cup\rtyp(X_1,X_2,Y_1,Y_2))^c),
\end{eqnarray*}
where
\begin{eqnarray*}
\delta_{11} = e^{-2^{N(R_{11}-I(X_1;U_1|U_2)-b_2(\eps_2))}}  &\hspace{.5in}& 
\delta_{21} = e^{-2^{N(R_{21}-I(X_1,X_2;Y_1|U_1,U_2,Y_2)-b_4(\eps_4))}} \\
\delta_{12} = e^{-2^{N(R_{12}-I(X_1;U_2)-b_1(\eps_1))}} &&
\delta_{22} = e^{-2^{N(R_{22}-I(X_1,X_2;Y_2|U_2)-b_3(\eps_3))}}.  
\end{eqnarray*}
By Lemma~\ref{ZZZ-lem:intypa}, above, 
\begin{eqnarray*}
p\left((\rtyp(U_2,X_1))^c\right) & \leq & 2^{-Nc_1(\eps_1)} \\
p\left((\rtyp(U_1,U_2,X_1))^c\right) & \leq & 2^{-Nc_2(\eps_2)} \\
p\left((\rtyp(U_2,X_1,X_2,Y_2))^c\right) & \leq & 2^{-Nc_3(\eps_3)} \\
p\left((\rtyp(U_1,U_2,X_1,X_2,Y_1,Y_2))^c\cup(\rtyp(X_1,X_2,Y_1,Y_2))^c\right)
& \leq & 2^{-Nc_4(\eps_4)} 
\end{eqnarray*}
for all $N$ sufficiently large, 
where each $c_k(\eps_k)$ approaches 0 as $\eps_k$ approaches 0.  
So, if $\nu=3\sum_{k=1}^4b_k(\eps_k)$, 
\begin{eqnarray*}
P_e^{(N)}(\nu) 
& \leq & \delta_{11}+\delta_{12}+\delta_{21}+\delta_{22}
         +2^{-Nc_1(\eps_1)}
	 +2^{-N(c_2(\eps_2)-2b_1(\eps_1)-1/N)} \\
&&	 +2^{-N(c_3(\eps_3)-2\sum_{k=1}^2b_k(\eps_k)-2/N)} 
	 +2^{-N(c_4(\eps_4)-2\sum_{k=1}^3b_k(\eps_t)-3/N)} 
\end{eqnarray*}
for $N$ sufficiently large.  
Thus sequentially choosing $\eps_4$, $\eps_3$, $\eps_2$, and $\eps_1$ 
to satisfy 
\begin{eqnarray*}
b_4(\eps_4) & < & R_{21}-I(X_1,X_2;Y_1|U_1,U_2,Y_2) \\
b_3(\eps_3) & < & \min\{R_{22}-I(X_1,X_2;Y_2|U_2),c_4(\eps_4)/6\} \\
b_2(\eps_2) & < & \min\{R_{11}-I(X_1;U_1|U_2),c_4(\eps_4)/6,c_3(\eps_3)/4\} \\
b_1(\eps_1) & < & \min\{R_{12}-I(X_1;U_2),c_4(\eps_4)/6,
                  c_3(\eps_3)/4,c_2(\eps_2)/2\}
\end{eqnarray*}
yields an error probability $P_e^{(N)}(\nu)$ 
that decays exponentially to zero.  
The exponent approaches 0 as $\eps_1$, $\eps_2$, $\eps_3$, and $\eps_4$ 
approach 0, 
which gives the desired result by Theorem~\ref{ZZZ-thm:equiv}.  
\IEEEQED

Lemmas~\ref{ZZZ-lem:inconproba} and~\ref{ZZZ-lem:ina1a} bound 
the emulation distribution 
and the conditional probability 
of observing atypical strings using the code 
defined in the proof of Theorem~\ref{ZZZ-thm:inmaina}.  

\begin{lem}\label{ZZZ-lem:inconproba}
For all $(\uu_2,\ux_1)\in\rtyp$, 
\[
\hp(\uu_2|\ux_1)\leq 2^{N(4a_1(\eps_1)+2\eps_1+1/N)}p(\uu_1|\ux_1);
\]
if, in addition, $(\uu_1,\uu_2,\ux_1)\in\rtyp$, then 
\[
\hp(\uu_1|\uu_2,\ux_1)\leq 2^{N(8a_2(\eps_2)+1/N)}p(\uu_1|\uu_2,\ux_1).
\]
if, further, $(\uu_2,\ux_1,\ux_2,\uy_2)\in\rtyp$,  then 
\[
\hp(\uy_2|\uu_2,\ux_1,\ux_2)
\leq 2^{N(8a_3(\eps_3)+1/N)}p(\uy_2|\uu_2,\ux_1,\ux_2).
\]
if, also, $(\uu_1,\uu_2,\ux_1,\ux_2,\uy_1,\uy_2)\in\rtyp$ and 
and $(\ux_1,\ux_2,\uy_1,\uy_2)\in\rtyp$, then 
\begin{eqnarray*}
\hp(\uy_1|\uu_1,\uu_2,\ux_1,\ux_2,\uy_2) 
& \leq & 2^{N(8a_4(\eps_4)+1/N)}p(\uy_1|\uu_1,\uu_2,\ux_1,\ux_2,\uy_2).
\end{eqnarray*}
For all $(\ux_1,\ux_2,\uy_1,\uy_2)\in\rtyp$,
\[
\hp(\uy_1,\uy_2|\ux_1,\ux_2)
\leq 2^{N(4a_1(\eps_1)+2\eps_1+\sum_{k=2}^4 8a_k(\eps,t)+4/N)}
p(\uy_1,\uy_2|\ux_1,\ux_2).
\]

\Proof 
Applying Lemma~\ref{ZZZ-lem:conprob} 
as in Lemmas~\ref{ZZZ-lem:bcconprob} and~\ref{ZZZ-lem:maconprob} gives 
the first four bounds.  
We then apply the Markov structure imposed on $\hp(\cdot)$ by the code design 
and the Markovity of the underlying distribution 
\[
p(u_1,u_2,x_1,x_2,y_1,y_2) 
= p(x_1,x_2)p(u_1,u_2|x_1)p(y_1,y_2|x_1,x_2) 
\]
to obtain 
\[
\hp(\uy_1,\uy_2|\ux_1,\ux_2) 
\leq \sum_{\uu_1,\uu_2}
p(\uu_1,\uu_2,\uy_2,\uy_1|\ux_1,\ux_2)
2^{N(4a_1(\eps_1)+2\eps_1+8\sum_{k=2}^4a_k(\eps,t)+4/N)}.
\]
\IEEEQED
\end{lem}

\begin{lem}\label{ZZZ-lem:ina1a}
Let $b_1(\eps_1)=\eps_1+2a_1(\eps_1)$ 
and $b_k(\eps_k)=4a_k(\eps_k)$ for $k=2,3$.  Then 
\begin{eqnarray*}
\lefteqn{\hp((\rtyp(X_1,X_2,Y_1,Y_2))^c|\ux_1,\ux_2)} \\
& \leq & \delta_{11}+\delta_{12}+\delta_{21}+\delta_{22}
        +p((\rtyp(U_2,X_1))^c|\ux_1)
	+2^{N(2b_1(\eps_1)+1/N)}p((\rtyp(U_1,U_2,X_1))^c|\ux_1) \\
&&	+2^{N(2\sum_{k=1}^2 b_k(\eps_k)+2/N)}
	p((\rtyp(U_2,X_1,X_2,Y_2))^c|\ux_1,\ux_2) \\
&&+2^{N(2\sum_{k=1}^3 b_k(\eps_k)+3/N)}
	p((\rtyp(U_1,U_2,X_1,X_2,Y_1,Y_2))^c\cup(\rtyp(X_1,X_2,Y_1,Y_2))^c|\ux_1,\ux_2) 
\end{eqnarray*}
where
\[
\begin{array}{rclcrcl}
\delta_{11} &=& e^{-2^{N(R_{11}-I(X_1;U_1|U_2)-b_2(\eps_2))}}  &&
\delta_{12} &= &e^{-2^{N(R_{12}-I(X_1;U_2)-b_1(\eps_1))}} \\
\delta_{21} &= &e^{-2^{N(R_{21}-I(X_1,X_2;Y_1|U_1,U_2,Y_2)-b_4(\eps_4))}}&&
\delta_{22} &= &e^{-2^{N(R_{22}-I(X_1,X_2;Y_2|U_2)-b_3(\eps_3))}}  
\end{array}
\]

\Proof 
For notational brevity, let 
\[
\begin{array}{rclcrcl}
K_1& \deff& K(\uu_2,\ux_1) && 
K_3& \deff& K(\uu_2,\ux_1,\ux_2,\uy_2) \\
K_2& \deff& K(\uu_1,\uu_2,\ux_1) &&
K_4& \deff& K(\uu_1,\uu_2,\ux_1,\ux_2,\uy_1,\uy_2)\cdot K(\ux_1,\ux_2,\uy_1,\uy_2);
\end{array}
\]
we rely on context to specify the values of arguments.  
$(\ux_1,\ux_2,\uy_1,\uy_2)$ not jointly typical 
implies that one of the four encoders failed 
to find a jointly typical codeword.  
We bound the probability of such a failure 
for each encoder in turn 
and then apply Lemma~\ref{ZZZ-lem:inconproba} 
to bound $\hp(\cdot)$, giving
\begin{eqnarray*}
\lefteqn{\hp((\rtyp(X_1,X_2,Y_1,Y_2))^c|\ux_1,\ux_2)} \\
&\leq & \left(\sum_{\uu_2}p(\uu_2)(1-K_1)\right)^{2^{NR_{12}}} 
        + \sum_{\uu_2}K_1\hp(\uu_2)\left(\sum_{\uu_1}
	p(\uu_1|\uu_2)(1-K_2)\right)^{2^{NR_{11}}} \\
& & +\sum_{\uu_2}K_1K_2\hp(\uu_2)\left(\sum_{\uy_2}
	p(\uy_2|\uu_2)(1-K_3)\right)^{2^{NR_{22}}} \\
& & +\sum_{\uu_1,\uu_2,\uy_2}K_1K_2K_3\hp(\uu_1,\uu_2,\uy_2) 
\left(\sum_{\uy_1}p(\uy_1|\uu_1,\uu_2,\uy_2)(1-K_4)\right)^{2^{NR_{21}}} \\
&\leq& p((\rtyp(U_2,X_1))^c|\ux_1) 
       +e^{-2^{N(R_{12}-I(X_1;U_2)-b_1(\eps_1))}}   \\
&&     +2^{N(2b_1(\eps_1)+1/N)}p((\rtyp(U_1,U_2,X_1))^c|\ux_1) 
       +e^{-2^{N(R_{11}-I(X_1;U_1|U_2)-b_2(\eps_2))}} \\
&&     +2^{N(2\sum_{k=1}^2 b_k(\eps_k)+2/N)}
       p((\rtyp(U_2,X_1,X_2,Y_2))^c|\ux_1,\ux_2) 
       +e^{-2^{N(R_{22}-I(X_1,X_2;Y_2|U_2)-b_3(\eps_3))}}  \\
&&     +2^{N(2\sum_{k=1}^3b_k(\eps,t)+3/N)}
       p((\rtyp(U_1,U_2,X_1,X_2,Y_1,Y_2))^c
       \cup(\rtyp(X_1,X_2,Y_1,Y_2))^c|\ux_1,\ux_2) \\
&&     +e^{-2^{N(R_{21}-I(X_1,X_2;Y_1|U_1,U_2,Y_2)-b_4(\eps_4))}}.  
\end{eqnarray*}
\IEEEQED  
\end{lem}

\section{Interference Channels:  Model 2}\label{ZZZ-app:inb}

The following definitions, 
used in the proof of Theorem~\ref{ZZZ-thm:inmainb}, below, 
rely on notation defined in Appendix~\ref{ZZZ-app:typ}.  

Given any distribution 
$p(u_1,u_2,x_1,x_2,y_1,y_2)
=p(u_1|x_1)p(u_2|u_1,x_1)p(x_1,x_2)p(y_1,y_2|x_1,x_2)$, 
fix $\eps=(\eps_1,\eps_2,\eps_3,\eps_4)$ with
$\eps_1,\eps_2,\eps_3,\eps_4>0$.  
Fix $\eps=(\eps_1,\eps_2,\eps_3,\eps_4)$ 
with $\eps_k>0$ for all $k$.  
Let 
\begin{eqnarray*}
a_1(\eps_1) & \deff & (1+\eps_1)\cdot\inf\{\eps'>0: 
	\Pr(f(\uU_1)>\eps'\vee f(\uU_1,\uX_1)>\eps')
	\leq 2^{-N6\eps_1}\ \forall N\mbox{suff.\ large} \} \\
a_2(\eps_2) & \deff & (1+\eps_2)\cdot\inf\{\eps'>0:
	\Pr(f(\uU_1)>\eps'\vee f(\uU_1,\uX_1)>\eps'
	\vee f(\uU_1,\uU_2)>\eps'\\
&&	\vee f(\uU_1,\uU_2,\uX_1)>\eps')\leq 2^{-N6\eps(t)}
	\ \forall N\mbox{ suff.\ large} \} \\
a_3(\eps_3) & \deff & (1+\eps_3(t))\cdot\inf\{\eps'>0: 
	\Pr(f(\uU_1)>\eps'\vee
	f(\uU_1,\uY_1)>\eps'\vee 
	f(\uU_1,\uX_1,\uX_2)>\eps'\vee \\
&&f(\uU_1,\uX_1,\uX_2,\uY_1)>\eps')\leq 2^{-N6\eps_3(t)}
	\ \forall N\mbox{ suff.\ large}\} \\
a_4(\eps_4)& \deff& (1+\eps_4(t))\cdot\inf\{\eps'>0:
	\Pr(f(\uU_1,\uU_2,\uY_1)>\eps'\vee
         f(\uU_1,\uU_2,\uY_1,\uY_2)>\eps'\vee\\
&&    f(\uU_1,\uU_2,\uX_1,\uX_2,\uY_1)>\eps'\vee
	    f(\uU_1,\uU_2,\uX_1,\uX_2,\uY_1,\uY_2)>\eps'\vee
	    f(\uY_1,\uY_2)>\eps'\vee \\
&&\left.\left.
	    f(\uX_1,\uX_2,\uY_1,\uY_2)>\eps'\right)\leq 2^{-N6\eps_4(t)}
	    \ \forall N\mbox{ suff.\ large} \right\}.
\end{eqnarray*}
The typical sets are defined as 
\begin{eqnarray*}
\typ(U,X_1) & \deff& \left\{(\uu_1,\ux_1):
f(\ux_1)\leq \eps_1,\ 
f(\uu_1),f(\uu_1,\ux_1)\leq a_1(\eps_1) \right\} \\
\typ(U_1,U_2,X)& \deff& \{(\uu_1,\uu_2,\ux_1):
f(\uu_1),
f(\uu_1,\uu_2),
f(\uu_1,\ux_1),
f(\uu_1,\uu_2,\ux_1)\leq a_2(\eps_2)\} \\
\typ(U_1,X_1,X_2,Y_1)& \deff& \{(\uu_1,\ux_1,\ux_2,\uy_1):
f(\uu_1),
f(\uu_1,\uy_1),
f(\uu_1,\ux_1,\ux_2),\\
&&f(\uu_1,\ux_1,\ux_2,\uy_1)\leq a_3(\eps_3)\} \\
\typ(U_1,U_2,X_1,X_2,Y_1,Y_2)
& \deff& \{(\uu_1,\uu_2,\ux_1,\ux_2,\uy_1,\uy_2):
f(\uu_1,\uu_2,\uy_1),
f(\uu_1,\uu_2,\uy_1,\uy_2), \\
&&f(\uu_1,\uu_2,\ux_1,\ux_2,\uy_1),
f(\uu_1,\uu_2,\ux_1,\ux_2,\uy_1,\uy_2)\leq a_4(\eps_4)\} \\
\typ(X_1,X_2,Y_1,Y_2)& \deff& \{(\ux_1,\ux_2,\uy_1,\uy_2):
f(\ux_1,\ux_2)\leq \eps_4(t),  
f(\uy_1,\uy_2), \\
&&f(\ux_1,\ux_2,\uy_1,\uy_2)\leq a_4(\eps_4)\},
\end{eqnarray*}
which we restrict as 
\begin{eqnarray*}
\rtyp(U_1,X_1)& \deff& \left\{(\uuo,\ux_1)\in\typ:
\Pr\left((\typ(U_1,X_1))^c|\ux_1\right)\leq 2^{-3N\eps_1}\right\} \\
\rtyp(U_1,U_2,X_1)& \deff& \left\{(\uuo,\uut,\ux_1)\in\typ:
\Pr\left((\typ(U_1,U_2,X_1))^c|\uu_1,\ux_1\right)
\leq 2^{-3N\eps_2}\right\}.  \\
\rtyp(U_1,X_1,X_2,Y_1) & \deff & \left\{(\uuo,\ux_1,\ux_2,\uy_1)\in\typ:
\Pr\left((\typ(U_1,X_1,X_2,Y_1)^c|\uu_1,\ux_1,\ux_2\right)\right. \\
&&\left.\leq 2^{-3N\eps_3(t)}\right\} \\
\rtyp(U_1,U_2,X_1,X_2,Y_1,Y_2)& \deff& \left\{(\uuo,\uut,\ux_1,\ux_2,\uy_1,\uy_2)
\in\typ:\Pr\left((\typ)^c|\uu_1,\uu_2,\ux_1,\ux_2,\uy_1\right)\right. \\
&&\left.\leq 2^{-3N\eps_4(t)}\right\}  \\
\rtyp(X_1,X_2,Y_1,Y_2)& \deff& \left\{(\ux_1,\ux_2,\uy_1,\uy_2)\in\typ:
\Pr\left((\typ)^c|\ux_1,\ux_2\right)\leq 2^{-3N\eps_4(t)}\right\}.  
\end{eqnarray*}

Lemma~\ref{ZZZ-lem:intypb} bounds the probability of observing elements 
outside of those typical sets.  
We omit the proof, 
which follows the same outline as the corresponding 
examples in prior sections.  

\begin{lem}\label{ZZZ-lem:intypb}
If $(\uU_1,\uU_2,\uX_1,\uX_2,\uY_1,\uY_2)$ 
are drawn i.i.d.\ $p(u_1,u_2,x_1,x_2,y_1,y_2)$, 
then there exist positive constants $c_1(\eps_1)$ 
and $c_2(\eps_2)$ for which 
\begin{eqnarray*}
\Pr\left((\rtyp(U_1,X_1))^c\right) & \leq & 2^{-Nc_1(\eps_1)} \\
\Pr\left((\rtyp(U_1,U_2,X_1))^c\right) & \leq & 2^{-Nc_2(\eps_2)}  \\
\Pr\left((\rtyp(U_1,X_1,X_2,Y_1))^c\right)& \leq & 2^{-Nc_3(\eps_3)}  \\
\Pr\left((\rtyp(U_1,U_2,X_1,X_2,Y_1,Y_2))^c
\cup(\rtyp(X_1,X_2,Y_1,Y_2))^c \right)& \leq & 2^{-Nc_4(\eps_4)}
\end{eqnarray*}
for all $N$ sufficiently large.  
Constant $c_k(\eps,t)$ approaches 0 as $\eps_k(t)$ approaches 0.  
\IEEEQED
\end{lem}

{\bf Proof of Theorem~\ref{ZZZ-thm:inmainb}:}
All rates not bounded in the theorem statement are set to zero.  
We simplify the remaining notation as 
$R_{11}\ \deff\ R^{\ar{\{i_1\}}{\{j_1,j_2\}}}$,
$R_{12}\ \deff\ R^{\ar{\{i_1\}}{\{j_2\}}}$,
$R_{21}\ \deff\ R^{\ar{\{i_1,i_2\}}{\{j_1,j_2\}}}$, and
$R_{22}\ \deff\ R^{\ar{\{i_1,i_2\}}{\{j_2\}}}$.  
We use $\cC=(\cX_1\times\cX_2,p(y_1,y_2|x_1,x_2),\cY_1\times\cY_2)$
in place of the formal channel definition 
$(\ciXo\times\ciXt,p(\jyo,\jyt|\ixo,\ixt),\cjYo\times\cjYt)$
in this proof and its supporting lemmas.  

Fix $(R_{11},R_{12},R_{21},R_{22})$ to satisfy the theorem constraints.  
Let $p(x_1,x_2)$ be arbitrary, and choose $p(\uo|x_1)$ and $p(\ut|\uo,x_1)$ 
to satisfy the given bounds.  
Let 
\[
p(u_1,u_2,x_1,x_2,y_1,y_2)\ 
\deff\ p(u_1|x_1)p(u_2|x_1,u_1)p(x_1,x_2)p(y_1,y_2|x_1,x_2).
\]
We apply the typical set definitions given above.  

Excluding the rate-0 codes, four encoders and two decoders are required.  
We simplify their notation as 
\[
\begin{array}{rclcrclcrcl}
\alpha^{(11)}_N&=&\alpha_N^{\ar{\{i_1\}}{\{j_1,j_2\}}} &\hspace{.5in}&
\alpha^{(21)}_N&=&\alpha_N^{\ar{\{i_1,i_2\}}{\{j_1,j_2\}}} &\hspace{.5in}&
\beta^{(1)}_N&=&\beta_N^{(j_1)} \\
\alpha^{(12)}_N&=&\alpha_N^{\ar{\{i_1\}}{\{j_2\}}} &&
\alpha^{(22)}_N&=&\alpha_N^{\ar{\{i_1,i_2\}}{\{j_2\}}} &&
\beta^{(2)}_N&=&\beta_N^{(j_2)},
\end{array}
\]
where
\[
\begin{array}{lclcl}
\alpha^{(11)}_N:\ucX_1\rightarrow\cW_{11} & \hspace{.25in}&
\alpha^{(21)}_N :\ucX_1\times\ucX_2\rightarrow\cW_{21}&\hspace{.25in}&
\beta^{(1)}_N:\cW_{11}\times\cW_{21}\rightarrow\ucY_1\\
\alpha^{(12)}_N:\ucX_1\rightarrow\cW_{12} &&
\alpha^{(22)}_N :\ucX_1\times\ucX_2\rightarrow\cW_{22}&&
\beta^{(2)}_N:\cW_{11}\times\cW_{12}\times\cW_{21}\times\cW_{22}
\rightarrow\ucY_2
\end{array}
\]
and 
\[
\begin{array}{rclcrcl}
\cW_{11}&=&\utcX^{\ar{\{i_1\}}{\{j_1\}}}=\{0,1\}^{NR_{11}} &&
\cW_{12}&=&\utcX^{\ar{\{i_1\}}{\{j_1,j_2\}}}=\{0,1\}^{NR_{12}} \\
\cW_{21}&=&\utcX^{\ar{\{i_1,i_2\}}{\{j_1\}}}=\{0,1\}^{NR_{21}} &&
\cW_{22}&=&\utcX^{\ar{\{i_1,i_2\}}{\{j_1,j_2\}}}=\{0,1\}^{NR_{22}}
\end{array}
\]
Encoder $(\alpha^{(11)}_N,\alpha^{(12)}_N)$ operates at node $i_1$, 
transmitting its rate $R_{11}$ and $R_{12}$ descriptions 
to both nodes and only $j_2$, respectively.  
Encoder $(\alpha^{(21)}_N,\alpha^{(22)}_N)$ operates at node $v^{V_1}$, 
receiving noiseless descriptions 
of $\ux_1$ and $\ux_2$ from nodes $i_1$ and $i_2$ 
and transmitting its rate $R_{21}$ output to both nodes 
and its $R_{22}$ output to only $j_2$.  
The code also employs mappings 
$\gamma_N^{(1)}:\cW_{11}\rightarrow\cU_1$ 
and $\gamma_N^{(2)}:\cW_{11}\times\cW_{12}\rightarrow\cU_2$.  

The random code design draws 
$\{\gamma^{(1)}_N(w_{11}):w\in\cW_{11}\}$
i.i.d.\ from the distribution $\prod_{\ell=1}^N p(\uuo(\ell))$.  
For each $w_{11}\in\uhcY_{11}$, 
let $\uUo=\gamma^{(1)}_N(w_{11})$ and draw codewords 
$\{\gamma^{(2)}_N(w_{11},w_{12}):w_{12}\in\cW_{12}\}$ 
i.i.d.\ from $\prod_{\ell=1}^N p(\uut(\ell)|\uUo(\ell))$ 
and codewords $\{\beta^{(1)}_N(w_{11},w_{21}):w_{21}\in\cW_{21}\}$
i.i.d.\ from $\prod_{\ell=1}^Np(\uy_1(\ell)|\uUo(\ell))$.  
Finally, for each 
$(w_{11},w_{12},w_{21})\in\uhcY_{11}\times\uhcY_{12}\times\uhcY_{21}$, 
let 
\[
(\uUo,\uUt,\uY_1) = 
    (\gamma^{(1)}_N(w_{11}),
     \gamma^{(2)}_N(w_{11},w_{12}),
     \beta^{(1)}_N(w_{11},w_{21})),
\]
and draw
$\{\{\beta^{(2)}_N(w_{11},w_{12},w_{21},w_{22}):w_{22}\in\cW_{22}\}$ 
i.i.d.\ from 
$\prod_{\ell=1}^N p(y_2(\ell)|\uUo(\ell),\uUt(\ell),\uY_1(\ell))$.
Choose $\alpha^{(11)}_N(\ux_1)$ uniformly at random 
from the indices $w_{11}\in\uhcX_{11}$ for which 
$(\gamma^{(1)}(w_{11}),\ux_1)\in\rtyp$; 
if there is no such index, then set $\alpha^{(11)}_N(\ux_1)$ to 1.  
Let $w_{11}$ be the chosen index and $\uUo=\gamma^{(1)}(w_{11})$.  
Choose $\alpha^{(12)}_N(\ux_1)$ uniformly at random 
from the indices $w_{12}\in\uhcX_{12}$ for which 
$(\uUo,\gamma^{(2)}(w_{11},w_{12}),\ux_1)\in\rtyp$;  
if there is no such index $w_{12}$, 
then set $\alpha^{(12)}_N(\ux_1)$ to 1.  
Let $w_{12}$ be the chosen index, 
and let $\uUt=\gamma^{(2)}_N(w_{11},w_{12})$.
Choose $\alpha^{(21)}_N(\ux_1,\ux_2)$  uniformly at random 
from the set of $w_{21}\in\uhcX_{21}$ for which 
\[
(\uUo,\ixo,\ixt,\beta^{(1)}_N(w_{11},w_{21n}))\in\rtyp;
\]
if this set is empty, then $\alpha^{(21)}_N(\ux_1,\ux_2)$  is set to 0.  
Let $w_{21}$ be the chosen index and set 
$\uY_1 = \beta^{(1)}_N(w_{11},w_{21})$;  
choose $\alpha^{(22)}_N(\ux_1,\ux_2)$  uniformly at random 
from the set of $w_{22}\in\uhcX_{22}$ for which 
\[
(\uUo,\uUt,\ux_1,\ux_2,\uY_1,\beta^{(2)}(w_{11},w_{12},w_{21},w_{22}))
\]
is typical;  
if this set is empty, then $\alpha^{(22)}_N(\ux_1,\ux_2)$  is set to 0.  

For all $(\ux_1,\ux_2,\uy_1,\uy_2)\in\rtyp$, 
Lemma~\ref{ZZZ-lem:inconprobb}, below,  
\[
\hp(\uy_1,\uy_2|\ux_1,\ux_2)
\leq 2^{N(2\sum_{k=1}^4b_k(\eps_k)+4/N)},
\]
where $b_1=2a_1(\eps_1)+\eps_1$ and $b_k=4a_k(\eps_k)$, $k\in\{2,3,4\}$. 
By Lemma~\ref{ZZZ-lem:ina1b}, below, 
\begin{eqnarray*}
\lefteqn{\hp(\rtyp((X_1,X_2,Y_1,Y_2))^c|\ux_1,\ux_2)} \\
&\leq& \delta_{11}+\delta_{12}+\delta_{21}+\delta_{22}+
    p((\rtyp(U_1,X_1))^c|\ux_1)
   +2^{N(2b_1(\eps_1)+1/N)}p((\rtyp(U_1,U_2,X_1))^c|\ux_1) \\
&& +2^{N(2\sum_{k=1}^2b_k(\eps_k)+2/N)} 
   p((\rtyp(U_1,X_1,X_2,Y_1))^c|\ux_1,\ux_2) \\
&& +2^{N(2\sum_{k=1}^3b_k(\eps_k)+3/N)} 
   p((\rtyp(U_1,U_2,X_1,X_2,Y_1,Y_2))^c|\ux_1,\ux_2),
\end{eqnarray*}
where 
\begin{eqnarray*}
\delta_{11} = e^{-2^{N(R_{11}-I(X_1;U_1)-b_1(\eps_1))}}   & & 
\delta_{12} = e^{-2^{N(R_{12}-I(X_1;U_2|U_1)-b_2(\eps_2))}}  \\
\delta_{21} = e^{-2^{N(R_{21}-I(X_1,X_2;Y_1|U_1)-b_3(\eps_3))}}  & & 
\delta_{22} = e^{-2^{N(R_{22}-I(X_1,X_2;Y_2|U_1,U_2,Y_1)-b_4(\eps_4))}}.  
\end{eqnarray*}
Lemma~\ref{ZZZ-lem:intypb}, above, gives 
\begin{eqnarray*}
p\left((\rtyp(U_1,X_1))^c\right) & \leq & 2^{-Nc_1(\eps_1)}  \\
p\left((\rtyp(U_1,U_2,X_1))^c\right)& \leq & 2^{-Nc_2(\eps_2)}  \\
p\left((\rtyp(U_1,X_1,X_2,Y_1))^c\right) & \leq & 2^{-Nc_3(\eps_3)}  \\
p\left((\rtyp(U_1,U_2,X_1,X_2,Y_1,Y_2))^c\cup
         (\rtyp(X_1,X_2,Y_1,Y_2))^c\right)& \leq & 2^{-Nc_4(\eps_4)}
\end{eqnarray*}
for all $N$ sufficiently large, 
where each $c_k(\eps,t)$ approaches 0 as $\eps_k(t)$ approaches 0.  
Thus setting $\nu=3\sum_{k=1}^4b_k(\eps_k)$ gives 
\begin{eqnarray*}
P_e^{(N)}(\nu) 
& \leq & \delta_{11}+\delta_{12}+\delta_{21}+\delta_{22}
    +2^{-Nc_1(\eps_1)}
    +2^{-N(c_2(\eps_2)-2b_1(\eps_1)-1/N)}\\
&&  +2^{-N(c_3(\eps_3)-2\sum_{k=1}^2b_k(\eps_k)-2/N)} 
    +2^{-N(c_4(\eps_4)-2\sum_{k=1}^3b_k(\eps_k)+3/N)} 
\end{eqnarray*}
for $N$ sufficiently large.  
Thus sequentially choosing $\eps_4$, $\eps_3$, $\eps_2$, and $\eps_1$ 
to satisfy 
\begin{eqnarray*}
b_4(\eps_4) & < & R_{22}-I(X_1,X_2;Y_2|U_1,U_2,Y_1) \\
b_3(\eps_3) & < & \min\{R_{21}-I(X_1,X_2;Y_1|U_1),c_4(\eps_4)/6\} \\
b_2(\eps_2) & < & \min\{R_{12}-I(X_1;U_2|U_1),c_4(\eps_4)/6,c_3(\eps_3)/4\} \\
b_1(\eps_1) & < & \min\{R_{11}-I(X_1;U_1),c_4(\eps_4)/6,
                  c_3(\eps_3)/4,c_2(\eps_2)/2\}
\end{eqnarray*}
yields an error probability $P_e^{(N)}(\nu)$ 
that decays exponentially to zero.  
The exponent approaches 0 as $\eps_1$, $\eps_2$, $\eps_3$, and $\eps_4$ 
approach 0, 
which gives the desired result by Theorem~\ref{ZZZ-thm:equiv}.  
\IEEEQED

\begin{lem}\label{ZZZ-lem:inconprobb}
For all $(\uu_1,\ux_1)\in\rtyp$, 
\[
\hp(\uu_1|\ux_1)\leq 2^{N(4a_1(\eps_1)+2\eps_1+1/N)}p(\uu_1|\ux_1);
\]
if, further, $(\uu_1,\uu_2,\ux_1)\in\rtyp$ then 
\[
\hp(\uu_2|\uu_1,\ux_1)\leq 2^{N(8a_2(\eps_2)+1/N)}p(\uu_2|\uu_1,\ux_1);
\]
if, in addition, $(\uu_1,\ux_1,\ux_2,\uy_1)\in\rtyp$ 
\[
\hp(\uy_1|\uu_1,\ux_1,\ux_2)
\leq 2^{N(8a_3(\eps_3)+1/N)}p(\uy_1|\uu_1,\ux_1,\ux_2).
\]
if also $(\uu_1,\uu_2,\ux_1,\ux_2,\uy_1,\uy_2)\in\rtyp$, then 
\begin{eqnarray*}
\hp(\uy_2|\uu_1,\uu_2,\ux_1,\ux_2,\uy_1) 
& \leq & 2^{N(8a_4(\eps_4)+1/N)}
p(\uy_2|\uu_1,\uu_2,\ux_1,\ux_2,\uy_1).
\end{eqnarray*}
Thus, if $(\ux_1,\ux_2,\uy_1,\uy_2)\in\rtyp$,
\[
\hp(\uy_1,\uy_2|\ux_1,\ux_2)
\leq 2^{N(4a_1(\eps_1)+\eps_1+\sum_{k=2}^48a_k(\eps,t)+4/N)}.  
\]
\Proof  The proof follows the same outline as the preceding examples. 
\IEEEQED
\end{lem}

Lemma~\ref{ZZZ-lem:ina1b} bounds the probability 
of observing atypical strings using the code designed 
in Theorem~\ref{ZZZ-thm:inmainb}.  

\begin{lem}\label{ZZZ-lem:ina1b}
Let $b_1(\eps_1)=4a_1(\eps_1)+2\eps_1+1/N$
and $b_k(\eps_k)=8a_k(\eps_k)+1/N$, $k\in\{2,3\}$.  Then 
\begin{eqnarray*}
\lefteqn{\hp(\rtyp((X_1,X_2,Y_1,Y_2))^c|\ux_1,\ux_2)} \\
&\leq& \delta_{11}+\delta_{12}+\delta_{21}+\delta_{22}+
    p((\rtyp(U_1,X_1))^c|\ux_1)
    	+2^{Nb_1(\eps_1)}p((\rtyp(U_1,U_2,X_1))^c|\uu_1,\ux_1) \\
&& +2^{N\sum_{k=1}^2b_k(\eps_k)}p((\rtyp(U_1,X_1,X_2,Y_1))^c|\ux_1,\ux_2)  \\
&& +2^{N\sum_{k=1}^3b_k(\eps_k)}p((\rtyp(U_1,U_2,X_1,X_2,Y_1,Y_2))^c|\ux_1,\ux_2).
\end{eqnarray*}
where 
\[
\begin{array}{rclcrcl}
\delta_{11}&=&e^{-2^{N(R_{11}-I(X_1;U_1)-b_1(\eps_1))}} &&
\delta_{12}&=&e^{-2^{N(R_{12}-I(X_1;U_2|U_1)-b_2(\eps_2))}} \\
\delta_{21}&=&e^{-2^{N(R_{21}-I(X_1,X_2;Y_1|U_1)-b_3(\eps_3))}}  &&
\delta_{22}&=&e^{-2^{N(R_{22}-I(X_1,X_2;Y_2|U_1,U_2,Y_1)-b_4(\eps_4))}}.
\end{array}
\]
\IEEEQED
\end{lem}

\bibliographystyle{ieeetr}

\end{document}